\documentclass[prd,amsmath,aps,floats,amssymb, floatfix, superscriptaddress,
twocolumn, nofootinbib]{revtex4-1}

\usepackage{tabularx}
\usepackage{natbib,latexsym,times}
\usepackage{multirow}
\usepackage{physics}
\usepackage{subfig, graphicx}
\usepackage{hyperref}   
\usepackage[dvipsnames]{xcolor}
\usepackage{dcolumn}    
\usepackage{bm}         
\usepackage{booktabs}   
\usepackage{url}
\usepackage{newtxtext,newtxmath}
\DeclareGraphicsExtensions{.pdf,.jpeg,.jpg, .png, .eps, .tiff}
\usepackage[T1]{fontenc}
\usepackage{xlmacros}
\usepackage{natbib}
\usepackage{tcolorbox}
\usepackage{orcidlink}

\begin{document}

\title{Hyper Suprime-Cam Year 3 Results: Cosmology from Cosmic Shear Two-point
Correlation Functions}
\author{Xiangchong~Li\orcidlink{0000-0003-2880-5102}}
\email{xiangchl@andrew.cmu.edu}
\affiliation{
    McWilliams Center for Cosmology, Department of Physics, Carnegie Mellon
    University, 5000 Forbes Avenue, Pittsburgh, PA 15213, USA
}
\affiliation{
    Kavli Institute for the Physics and Mathematics of the Universe (WPI), The
    University of Tokyo Institutes for Advanced Study (UTIAS), The University
    of Tokyo, Chiba 277-8583, Japan
}
\author{Tianqing~Zhang\orcidlink{0000-0002-5596-198X}}
\affiliation{
    McWilliams Center for Cosmology, Department of Physics, Carnegie Mellon
    University, 5000 Forbes Avenue, Pittsburgh, PA 15213, USA
}

\author{Sunao~Sugiyama\orcidlink{0000-0003-1153-6735}}
\affiliation{
    Kavli Institute for the Physics and Mathematics of the Universe (WPI), The
    University of Tokyo Institutes for Advanced Study (UTIAS), The University
    of Tokyo, Chiba 277-8583, Japan
}
\affiliation{
    Department of Physics, The University of Tokyo, Bunkyo, Tokyo 113-0031,
    Japan
}

\author{Roohi~Dalal\orcidlink{0000-0002-7998-9899}}
\affiliation{
    Department of Astrophysical Sciences, Princeton University, Princeton, NJ
    08544, USA
}

\author{Ryo~Terasawa\orcidlink{0000-0002-1193-623X}}
\affiliation{
    Kavli Institute for the Physics and Mathematics of the Universe (WPI), The
    University of Tokyo Institutes for Advanced Study (UTIAS), The University
    of Tokyo, Chiba 277-8583, Japan
}
\affiliation{
    Department of Physics, The University of Tokyo, Bunkyo, Tokyo 113-0031,
    Japan
}

\author{Markus~M.~Rau\orcidlink{0000-0003-3709-1324}}
\affiliation{
    High Energy Physics Division, Argonne National Laboratory, Lemont, IL
    60439, USA
}
\affiliation{
    McWilliams Center for Cosmology, Department of Physics, Carnegie Mellon
    University, 5000 Forbes Avenue, Pittsburgh, PA 15213, USA
}

\author{Rachel~Mandelbaum\orcidlink{0000-0003-2271-1527}}
\affiliation{
    McWilliams Center for Cosmology, Department of Physics, Carnegie Mellon
    University, 5000 Forbes Avenue, Pittsburgh, PA 15213, USA
}

\author{Masahiro~Takada\orcidlink{0000-0002-5578-6472}}
\affiliation{
    Kavli Institute for the Physics and Mathematics of the Universe (WPI), The
    University of Tokyo Institutes for Advanced Study (UTIAS), The University
    of Tokyo, Chiba 277-8583, Japan
}

\author{Surhud~More\orcidlink{0000-0002-2986-2371}}
\affiliation{
    The Inter-University Centre for Astronomy and Astrophysics, Post bag 4,
    Ganeshkhind, Pune 411007, India
}
\affiliation{
    Kavli Institute for the Physics and Mathematics of the Universe (WPI), The
    University of Tokyo Institutes for Advanced Study (UTIAS), The University
    of Tokyo, Chiba 277-8583, Japan
}

\author{Michael~A.~Strauss\orcidlink{0000-0002-0106-7755}}
\affiliation{
    Department of Astrophysical Sciences, Princeton University, Princeton, NJ
    08544, USA
}

\author{Hironao~Miyatake\orcidlink{0000-0001-7964-9766}}
\affiliation{
    Kobayashi-Maskawa Institute for the Origin of Particles and the Universe
    (KMI), Nagoya University, Nagoya, 464-8602, Japan
}
\affiliation{
    Institute for Advanced Research, Nagoya University, Nagoya 464-8601, Japan
}
\affiliation{
    Kavli Institute for the Physics and Mathematics of the Universe (WPI), The
    University of Tokyo Institutes for Advanced Study (UTIAS), The University
    of Tokyo, Chiba 277-8583, Japan
}

\author{Masato~Shirasaki\orcidlink{0000-0002-1706-5797}}
\affiliation{
    National Astronomical Observatory of Japan, National Institutes of Natural
    Sciences, Mitaka, Tokyo 181-8588, Japan
}
\affiliation{
    The Institute of Statistical Mathematics, Tachikawa, Tokyo 190-8562, Japan
}

\author{Takashi~Hamana}
\affiliation{
    National Astronomical Observatory of Japan, National Institutes of Natural
    Sciences, Mitaka, Tokyo 181-8588, Japan
}

\author{Masamune~Oguri\orcidlink{0000-0003-3484-399X}}
\affiliation{
    Center for Frontier Science, Chiba University, 1-33 Yayoi-cho, Inage-ku,
    Chiba 263-8522, Japan
}
\affiliation{
    Research Center for the Early Universe, The University of Tokyo, Bunkyo,
    Tokyo 113-0031, Japan
}
\affiliation{
    Department of Physics, The University of Tokyo, Bunkyo, Tokyo 113-0031,
    Japan
}
\affiliation{
    Kavli Institute for the Physics and Mathematics of the Universe (WPI), The
    University of Tokyo Institutes for Advanced Study (UTIAS), The University
    of Tokyo, Chiba 277-8583, Japan
}

\author{Wentao~Luo\orcidlink{0000-0001-6579-2190}}
\affiliation{
    School of Physical Sciences, University of Science and Technology of China,
    Hefei, Anhui 230026, China
}
\affiliation{
    CAS Key Laboratory for Researches in Galaxies and Cosmology/Department of
    Astronomy, School of Astronomy and Space Science, University of Science and
    Technology of China, Hefei, Anhui 230026, China
}

\author{Atsushi~J.~Nishizawa\orcidlink{0000-0002-6109-2397}}
\affiliation{
    Gifu Shotoku Gakuen University, Gifu 501-6194, Japan
}
\affiliation{
    Kobayashi-Maskawa Institute for the Origin of Particles and the Universe
    (KMI), Nagoya University, Nagoya, 464-8602, Japan
}
\affiliation{
    Institute for Advanced Research, Nagoya University, Nagoya 464-8601, Japan
}

\author{Ryuichi~Takahashi\orcidlink{0000-0001-6021-0147}}
\affiliation{
    Faculty of Science and Technology, Hirosaki University, 3 Bunkyo-cho,
    Hirosaki, Aomori 036-8561, Japan
}

\author{Andrina~Nicola\orcidlink{0000-0003-2792-6252}}
\affiliation{
    Argelander Institut f\"ur Astronomie, Universit\"at Bonn, Auf dem H\"ugel
    71, 53121 Bonn, Germany
}
\affiliation{
    Department of Astrophysical Sciences, Princeton University, Princeton, NJ
    08544, USA
}

\author{Ken~Osato\orcidlink{0000-0002-7934-2569}}
\affiliation{
    Center for Frontier Science, Chiba University, 1-33 Yayoi-cho, Inage-ku,
    Chiba 263-8522, Japan
}
\affiliation{
    Department of Physics, Graduate School of Science, Chiba University,1-33
    Yayoi-cho, Inage-ku, Chiba 263-8522, Japan
}

\author{Arun~Kannawadi\orcidlink{0000-0001-8783-6529}}
\affiliation{
    Department of Astrophysical Sciences, Princeton University, Princeton, NJ
    08544, USA
}

\author{Tomomi~Sunayama\orcidlink{0009-0004-6387-5784}}
\affiliation{
    Department of Astronomy and Steward Observatory, University of
    Arizona, 933 North Cherry Avenue, Tucson, AZ 85719, USA
}
\affiliation{
    Kobayashi-Maskawa Institute for the Origin of Particles and the Universe
    (KMI), Nagoya University, Nagoya, 464-8602, Japan
}

\author{Robert~Armstrong}
\affiliation{Lawrence Livermore National Laboratory, Livermore, CA 94551, USA}

\author{James~Bosch\orcidlink{0000-0003-2759-5764}}
\affiliation{
    Department of Astrophysical Sciences, Princeton University, Princeton, NJ
    08544, USA
}

\author{Yutaka~Komiyama\orcidlink{0000-0002-3852-6329}}
\affiliation{
    Department of Advanced Sciences, Faculty of Science and Engineering, Hosei
    University, 3-7-2 Kajino-cho, Koganei-shi, Tokyo 184-8584, Japan
}

\author{Robert~H.~Lupton\orcidlink{0000-0003-1666-0962}}
\affiliation{
    Department of Astrophysical Sciences, Princeton University, Princeton, NJ
    08544, USA
}

\author{Nate~B.~Lust\orcidlink{0000-0002-4122-9384}}
\affiliation{
    Department of Astrophysical Sciences, Princeton University, Princeton, NJ
    08544, USA
}

\author{Lauren~A.~MacArthur}
\affiliation{
    Department of Astrophysical Sciences, Princeton University, Princeton, NJ
    08544, USA
}

\author{Satoshi Miyazaki\orcidlink{0000-0002-1962-904X}}
\affiliation{
    Subaru Telescope,  National Astronomical Observatory of Japan, 650 North
    Aohoku Place Hilo, HI 96720, USA
}

\author{Hitoshi~Murayama\orcidlink{0000-0001-5769-9471}}
\affiliation{
    Berkeley Center for Theoretical Physics, University of California,
    Berkeley, CA 94720, USA
}
\affiliation{
    Theory Group, Lawrence Berkeley National Laboratory, Berkeley, CA 94720,
    USA
}
\affiliation{
    Kavli Institute for the Physics and Mathematics of the Universe (WPI), The
    University of Tokyo Institutes for Advanced Study (UTIAS), The University
    of Tokyo, Chiba 277-8583, Japan
}

\author{Takahiro~Nishimichi\orcidlink{0000-0002-9664-0760}}
\affiliation{
    Center for Gravitational Physics and Quantum Information, Yukawa Institute
    for Theoretical Physics, Kyoto University, Kyoto 606-8502, Japan
}
\affiliation{
    Kavli Institute for the Physics and Mathematics of the Universe (WPI), The
    University of Tokyo Institutes for Advanced Study (UTIAS), The University
    of Tokyo, Chiba 277-8583, Japan
}
\affiliation{
    Department of Astrophysics and Atmospheric Sciences, Faculty of Science,
    Kyoto Sangyo University, Motoyama, Kamigamo, Kita-ku, Kyoto 603-8555, Japan
}

\author{Yuki~Okura\orcidlink{0000-0001-6623-4190}}
\affiliation{
    National Astronomical Observatory of Japan, National Institutes of Natural
    Sciences, Mitaka, Tokyo 181-8588, Japan
}

\author{Paul~A.~Price\orcidlink{0000-0003-0511-0228}}
\affiliation{
    Department of Astrophysical Sciences, Princeton University, Princeton, NJ
    08544, USA
}

\author{Philip~J.~Tait}
\affiliation{
    Subaru Telescope,  National Astronomical Observatory of Japan, 650 N Aohoku
    Place Hilo, HI 96720, USA
}

\author{Masayuki~Tanaka}
\affiliation{
    National Astronomical Observatory of Japan, National Institutes of Natural
    Sciences, Mitaka, Tokyo 181-8588, Japan
}

\author{Shiang-Yu~Wang}
\affiliation{
    Institute of Astronomy and Astrophysics, Academia Sinica, Taipei 10617,
    Taiwan
}

\date{\today}

\begin{abstract}
We perform a blinded cosmology analysis with cosmic shear two-point correlation
functions (2PCFs) measured from more than 25 million galaxies in the Hyper
Suprime-Cam three-year shear catalog in four tomographic redshift bins ranging
from $0.3$ to $1.5$. After conservative masking and galaxy selection, the
survey covers $416~\mathrm{deg}^2$ of the northern sky with an effective galaxy
number density of $15~\mathrm{arcmin}^{-2}$ over the four redshift bins. The
2PCFs adopted for cosmology analysis are measured in the angular range: $7.1 <
\theta/\mathrm{arcmin} < 56.6$ for $\xi_+$ and $31.2 <\theta/\mathrm{arcmin} <
248$ for $\xi_-$, with a total signal-to-noise ratio of $26.6$\,. We apply a
conservative, wide, flat prior on the photometric redshift errors on the last
two tomographic bins, and the relative magnitudes of the cosmic shear amplitude
across four redshift bins allow us to calibrate the photometric redshift
errors. With this flat prior on redshift errors, we find
$\Omega_\mathrm{m}=0.256_{-0.044}^{+0.056}\,$ and $S_8\equiv \sigma_8
\sqrt{\Omega_\mathrm{m}/0.3}=0.769_{-0.034}^{+0.031}$ (both 68\% CI) for a flat
$\Lambda$ cold dark matter cosmology. We find, after unblinding, that our
constraint on $S_8$ is consistent with the Fourier space cosmic shear and the
3$\times$2pt analyses on the same HSC dataset. We carefully study the potential
systematics from astrophysical and systematic model uncertainties in our
fiducial analysis using synthetic data, and report no biases (including
projection bias in the posterior space) greater than $0.5\sigma$ in the
estimation of $S_8$. Our analysis hints that the mean redshifts of the two
highest tomographic bins are higher than initially estimated. In addition, a
number of consistency tests are conducted to assess the robustness of our
analysis. Comparing our result with \textit{Planck}-2018 cosmic microwave
background observations, we find a $\sim$$2\sigma$ tension for the $\Lambda$CDM
model.
\end{abstract}

\maketitle

\section{INTRODUCTION}

The flat $\Lambda$ Cold Dark Matter ($\Lambda$CDM) model, which is now
considered as concordance cosmology model, explains a diverse set of
observations with a non-zero cosmological constant $\Lambda$ (which drives the
accelerating expansion of the late-time Universe) and cold dark matter (which
drives large-scale structure formation). The observations include the Hubble
diagram of type Ia supernovae \citep[e.g.,][]{IaSNA_SDSS_Betoule2014}, Big Bang
nucleosynthesis \citep[e.g.,][]{BBNPostplanck_Fields2020}, fluctuations in the
cosmic microwave background radiation (CMB; e.g., \citep{cmb_WMAP9_Hinshaw2013,
cmb_Planck2018_Cosmology}), cosmic shear
\citep[e.g.,][]{cosmicShear_HSC1_Hamana2019, KiDS1000_CS_Asgari2020,
DESY3_CS_Secco2022} and galaxy clustering
\citep[e.g.,][]{HSC1_2x2pt_Sugiyama2022, HSC1_2x2pt_Miyatake2022,
KiDS1000_3x2pt_Heymans2021, DESY3_3x2pt2022,2022PhRvD.105h3517K}. As the
precision of these observations has grown, we are now in the era of precision
cosmology, focusing on possible small discrepancies between different
observations when interpreted by the flat $\Lambda$CDM cosmology model. One
such tension is the so-called $\sigma_8$ or $S_8$ tension, which refers to the
fact that the $\Lambda$CDM models inferred from large-scale structure probes
consistently exhibit a lower value of $\sigma_8$ or $S_8$ \citep[see][for a
recent review]{2022JHEAp..34...49A}, which characterizes the clustering
amplitude in the present-day universe, than do cosmological models inferred
from the \textit{Planck}-2018 CMB measurements
\citep{cmb_Planck2018_Cosmology}. A statistically significant discrepancy after
marginalizing over the known systematic uncertainties could be an indication of
physics beyond the flat $\Lambda$CDM cosmology. However, the discrepancy could
also be a sign of unknown systematics in some of the observations or the
analyses.

Weak gravitational lensing is one of the most important observations of
large-scale structure at low redshifts. It refers to the small but coherent
distortion of images of background galaxies due to the deflection of light when
it travels through an inhomogeneous foreground matter density field
\citep{wlRevBartelmann}. Since weak lensing is caused by gravity, it is
sensitive to the projected total matter (both dark matter and baryons)
distribution along the line of sight \citep{rev_cosmicShear_Kilbinger15}.
Cosmic shear, namely the two-point statistics of lensing-shear distortion
measured from background galaxy images, are related to the two-point statistics
(i.e., the power spectrum) of the projected foreground matter density field.
Cosmic shear measurements are particularly sensitive to the combination of
cosmology parameters $S_8 \equiv \sigma_8\sqrt{\Omega_\mathrm{m}/0.3}$, where
$\Omega_\mathrm{m}$ is the total matter density parameter.

The ongoing Stage-III large-scale multi-band photometric surveys which have
weak lensing among their primary science targets include the Kilo-Degree Survey
\citep[KiDS;][]{KIDS13}, the Dark Energy Survey
\citep[DES;][]{DES_overview_2016}, and the Hyper Suprime-Cam survey
\citep[HSC;][]{HSC_SSP2018} which is the subject of this paper. The HSC survey
is an optical imaging survey covering about $1,100~\mathrm{deg}^2$ using a 1.77
deg$^2$ field-of-view imager mounted on the 8.2-meter Subaru telescope
\citep{HSC_hardware_Miyazaki2018, HSC_hardware_Komiyama2018,
HSC_hardware_Kawanomoto2018, HSC_hardware_Furusawa2018}. The HSC survey is able
to measure cosmic shear signals up to $z$$\sim$$2$ from its $i$-band coadded
images thanks to the combination of its depth ($5\sigma$ point-source magnitude
of $i$$\sim$$26$) and good seeing (mean seeing size of
$\sim$$0.6~\mathrm{arcsec}$) for the HSC wide layer.  In this paper, we focus
on the Year 3 results of HSC (HSC-Y3), based on roughly 430 deg$^2$ of sky.

\citet{HSC3_catalog_Li2021} presented the HSC-Y3 shear catalog for weak-lensing
science. We conducted a number of null tests on the shear catalog against many
possible systematics such as modeling errors in the point-spread function (PSF)
and shear estimation biases thus demonstrating that the HSC-Y3 shear catalog
meets the requirements for weak-lensing science. \citet{HSC3_photoz_Rau2022}
performed a joint redshift distribution inference on the sample, combining
photometric redshift information with clustering redshifts from the CAMIRA
luminous red galaxy sample \citep[CAMIRA-LRG;][]{CAMIRA_Oguri2014,
CAMIRA_HSC_Oguri2018, HSC1_2D3Dmassmap}. \citet{HSC3_PSF} developed a technique
to correct for systematic bias in cosmic shear analysis from fourth-order PSF
modeling error and shape leakage to shear estimation.

In this paper, we present results from a tomographic cosmic shear analysis
using the HSC-Y3 shear catalog. We measure the two-point correlation functions
(2PCFs) from the HSC-Y3 shear catalog. Then we model the 2PCFs with
twenty-three cosmological, astrophysical, and nuisance parameters. With a
nested Bayesian sampling analysis, we constrain the cosmological parameters,
especially focusing on $S_8$, in the context of the flat $\Lambda$CDM
cosmology.

In our likelihood model, we carefully marginalize over various nuisance
parameters quantifying systematic errors in the cosmic shear analysis
\citep{rev_wlsys_Mandelbaum2017}. The systematic errors we consider include
systematic errors due to imperfect PSF modeling and PSF shape leakage
\citep{HSC3_PSF}; shear calibration uncertainties \citep{HSC3_catalog_Li2021};
and the photometric redshift (photo-$z$) uncertainties
\citep{HSC3_photoz_Rau2022}. In addition to systematic errors, we study the
modeling uncertainties in the matter power spectrum at small scales, e.g., the
model uncertainties in the nonlinear power spectrum \citep{halofit_mead21,
cosmicEmu2022} and baryonic physics from star formation, supernovae, and AGN
feedback \citep{baryon_osato15, baryon_chen22, baryon_troster22}. Specifically,
we use our fiducial pipeline to analyze various mock 2PCFs simulated using
different nonlinear \citep{cosmicEmu2022} and baryonic models to quantify the
systematic uncertainties on the $\Omega_\mathrm{m}$ and $S_8$ constraints.  In
addition, we adopt a conservative model to marginalize intrinsic shape
correlations due to tidal alignment \citep{nla_hirata07, nla_bridle07} and
tidal torquing \citep{tatt_blazek17}.

In order to obtain robust cosmological constraints, we perform a blinded
analysis to avoid confirmation biases affecting our results. In particular, we
conduct various blinded internal consistency tests by analyzing data in the
context of the flat $\Lambda$CDM cosmology in different subfields, with
different angular scale cuts, and removing each of the redshift bins to check
the robustness of our results. Furthermore, we look for sensitivity of the
central value and uncertainty in our $S_8$ constraint for analyses with flat
priors on different cosmological parameters and prior ranges; analyses with
different models for the linear and nonlinear matter power spectrum as well as
different models for baryonic physics; analyses with different intrinsic
alignment models; and analyses with different systematics models. After we
confirm that there is no internal inconsistency in our cosmic shear 2PCFs
analysis, we unblind our analysis and check the consistency of our constraints
with the \textit{Planck}-2018 CMB analysis \citep{cmb_Planck2018_Cosmology} and
other lensing surveys such as DES \citep{DESY3_CS_Secco2022,
cosmicShear_DESY3_Amon2021} and KiDS \citep{KiDS1000_CS_Asgari2020}.

Our paper is organized as follows. In section~\ref{sec:data} we describe the
basic characteristics of the HSC-Y3 dataset (including galaxy shear, photo-$z$,
star shape and mock catalogs) that we use for the real space cosmic shear
analysis. In section~\ref{sec:2pt} we measure the two-point correlation
functions from the HSC shear catalog and the covariance from HSC mock catalogs.
In section~\ref{sec:model} we provide a brief overview of the theoretical model
used in our likelihood. In section~\ref{sec:res_inter}, we conduct internal
consistency checks. In section~\ref{sec:res_exter}, we present our main results
and compare them with constraints from external datasets. Throughout this
paper, we report the mode of the 1D projected posterior distribution, along
with 68\% credible interval (CI) for parameter values and uncertainties. In
addition, we also quote the maximum a posteriori (MAP) estimate from the Monte
Carlo (MC) chain.

We note that this paper is one of a series of HSC-Y3 cosmological analysis
papers, alongside:
\begin{itemize}
    \item A cosmic shear analysis using pseudo-$C_\ell$ measurement
        \citep{HSC3_cosmicShearFourier};
    \item A 3$\times$2pt analysis combining galaxy clustering, cosmic shear, and
        galaxy-galaxy lensing \citep{HSC3_3x2pt_meas, HSC3_3x2pt_ls,
        HSC3_3x2pt_ss}.
\end{itemize}
Those three cosmology analyses are conducted without any comparison between the
cosmology constraints before unblinding. However, when performing model
validation tests on synthetic data vectors, we make sure that the two cosmic
shear analyses, from 2PCFs and pseudo-$C_\ell$, are subject to the same
criteria when making decisions on analysis choices.

\begin{figure*}
\includegraphics[width=0.95\textwidth]{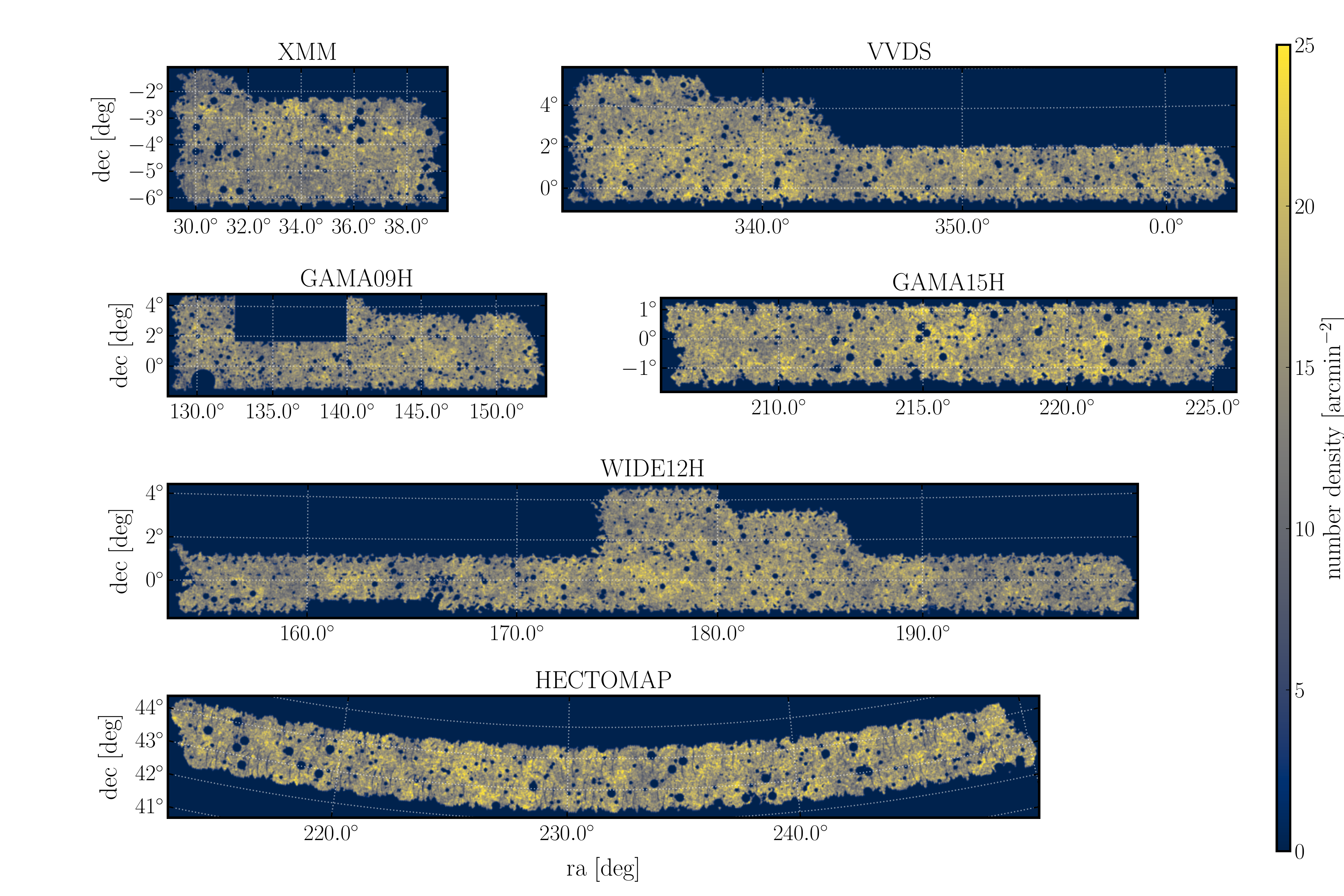}
\caption{
    The map of effective number density $n_\text{eff}$ of galaxies across four
    redshift bins. A rectangular region in GAMA09H ($132.5 < \mathrm{ra}< 140$
    [deg], $1.6<\mathrm{dec}<5$ [deg]) with very good seeing, a smaller number
    of input single exposures, and significant fourth-order PSF shape residual
    is removed from the original catalog.
}
\label{fig:data_neffMap}
\end{figure*}

\section{HSC-Y3 DATA}
\label{sec:data}

In this section we briefly introduce the HSC-Y3 data for the cosmic shear
analysis. The data is based on the S19A internal data release, which was
released in September 2019 and was acquired between March 2014 and April 2019.
First we introduce the galaxy shear catalog \citep{HSC3_catalog_Li2021} that is
used to measure the cosmic shear two-point correlation functions (2PCFs) is
introduced in Section~\ref{subsec:data_shear}; the shear catalog blinding is
discussed in Section~\ref{subsec:data_blinding}; the photometric redshift
(photo-$z$) catalog \citep{HSC3_photoz_Nishizawa2020} that is used to separate
source galaxies into four tomographic bins and infer the galaxy redshift
distribution is introduced in Section~\ref{subsec:data_photoz}. The star
catalog that is used to quantify PSF systematics is introduced in
Section~\ref{subsec:data_stars}. Finally, we introduce the mock catalogs that
are used to estimate the statistical uncertainties on our 2PCFs measurement in
Section~\ref{subsec:data_mock}.

\subsection{Weak-lensing Shear Catalog}
\label{subsec:data_shear}

\begin{table*}
\caption{
    The area and effective number density $n_\text{eff}^{(i)}$ ($i=1,\dots, 4$)
    \citep{WLsurvey_neffective_Chang2013}  in each tomographic bin and in six
    different subfields (i.e., XMM, VVDS, GAMA09H, WIDE12H, GAMA15H, HECTOMAP);
    and those for the whole HSC-Y3 footprint across the four redshift bins.
}
\label{tab:catsummary}
\setlength{\tabcolsep}{12pt}
\begin{center}
\begin{tabular}{lllllll}
\hline\hline
Fields      &Area ($\scriptstyle{\mathrm{deg}^2}$)
            & $n_{\scriptstyle{\text{eff}}}^{(1)}~\scriptstyle{(\mathrm{arcmin}^{-2})}$
            & $n_{\scriptstyle{\text{eff}}}^{(2)}~\scriptstyle{(\mathrm{arcmin}^{-2})}$
            & $n_{\scriptstyle{\text{eff}}}^{(3)}~\scriptstyle{(\mathrm{arcmin}^{-2})}$
            & $n_{\scriptstyle{\text{eff}}}^{(4)}~\scriptstyle{(\mathrm{arcmin}^{-2})}$
            & All   ($\scriptstyle{\mathrm{arcmin}^{-2}}$)  \\ \hline
XMM         &33.17  & 3.44  & 4.46  & 3.66  & 1.94  & 13.51 \\
VVDS        &96.18  & 3.82  & 5.13  & 4.21  & 2.13  & 15.30 \\
G09H        &82.36  & 3.99  & 4.74  & 3.81  & 2.11  & 14.65 \\
W12H        &121.32 & 3.61  & 5.20  & 3.96  & 2.06  & 14.82 \\
G15H        &40.87  & 3.92  & 5.38  & 4.27  & 2.24  & 15.81 \\
HECT        &43.06  & 3.74  & 5.34  & 4.05  & 2.32  & 15.44 \\
All         &416.97 & 3.77  & 5.07  & 4.00  & 2.12  & 14.96 \\
\hline\hline
\end{tabular}
\end{center}
\end{table*}

\subsubsection{Basic characterization}
\label{subsubsec:data_shear_basic}

The original HSC-Y3 shape catalog \citep{HSC3_catalog_Li2021} contains more
than 35 million source galaxies covering $433~\mathrm{deg}^2$ of the northern
sky. The galaxy sample is conservatively selected for the weak-lensing science
with a magnitude cut on extinction-corrected \cmodel{} magnitude at $i <
24.5$\,, a \cmodel{} signal-to-noise ratio (SNR) cut at $\mathrm{SNR}>10$ and a
\reGauss{} resolution cut at $R_2 > 0.3$ \citep{HSC3_catalog_Li2021}.

After the production of the shear catalog, a few additional cuts are applied to
improve the data quality. In particular, we follow
\citep{KiDS450_cs_Hildebrandt2017} to remove objects with extremely large
$i$-band ellipticity which are potentially unresolved binary stars. To be more
specific, we remove objects with large ellipticity, $\abs{e} > 0.8$ and
$i$-band determinant radius $r_\mathrm{det} < 10^{- 0.1 r + 1.8}$ arcsec (where
$r$ in the exponent is the r-band magnitde), amounting to $0.46\%$ of the
galaxy sample \citep{HSC3_catalog_Li2021}.

In addition, we remove a region in GAMA09H with $132.5 < \mathrm{ra}<
140$~[deg], $1.6<\mathrm{dec}<5$~[deg],  containing an area of
$\sim$$20~\mathrm{deg}^2$. This region has very good seeing size
$\sim$$0.4~\mathrm{arcsec}$, but it has a smaller number of single-frame
exposures contributing to the coadded images. In addition, we find significant
PSF fourth moment modeling errors in this region \citep{HSC3_PSF}. We find that
including galaxy shapes in this region causes significant $B$-modes in 2PCFs at
high redshifts and large scales.

Additionally, a number of galaxies are found to have secondary solutions at
very high redshifts in their estimated photo-$z$ posterior distributions, due
to redshift template degeneracies. These secondary solutions are outside the
redshift coverage of our CAMIRA-LRG sample \citep{CAMIRA_HSC_Oguri2018} making
it difficult to calibrate with the cross-correlation technique
\citep{HSC3_photoz_Rau2022}. The details will be discussed in
Section~\ref{subsec:data_photoz}.

After these cuts, the final shear catalog contains 25 million galaxies covering
416 deg$^2$ of the northern sky. The catalog is split into six subfields: XMM,
GAMA09H, WIDE12H, GAMA15H, VVDS and HECTOMAP. The area and effective galaxy
number densities, $n_\text{eff}$ \citep[as defined in
Ref.][]{WLsurvey_neffective_Chang2013}, in different redshift bins of the
subfields are summarized in Table~\ref{tab:catsummary}. The number density maps
for six subfields are shown in Fig.~\ref{fig:data_neffMap}. The effective
standard deviation of the error on the per-component shear per galaxy is
$\sigma_\gamma = 0.236$\,.

\subsubsection{Galaxy shear}
\label{ssubec:shearcat_galshapes}

The HSC-Y3 shear catalog contains galaxy shapes, estimated with the
re-Gaussianization (\reGauss{}) PSF correction method \citep{Regaussianization}
from the HSC $i$-band wide-field coadded images \citep{HSC1_pipeline}. The
\reGauss{} estimator measures the two components of galaxy ellipticity:
\begin{equation}
    (e_1,e_2)=\frac{1-(r_b/r_a)^2}{1+(r_b/r_a)^2} (\cos 2\phi,\sin 2\phi),
\end{equation}
where $r_b/r_a$ is the axis ratio, and $\phi$ is the position angle of the
major axis with respect to the equatorial coordinate system. The lensing shear
distortion, denoted as $\gamma$, coherently changes the galaxy ellipticities.

To control the shear estimation bias below $1\%$ of the shear distortion, the
galaxy shapes are calibrated with realistic image simulations downgrading
galaxy images from Hubble Space Telescope \citep{HST_shapeCatalog_Alexie2007}
to the HSC observing conditions \citep{HSC1-GREAT3Sim}. In the shear
calibration, we modeled the biases, including multiplicative ($m$) and
additive ($c$) biases from shear estimation, galaxy selection and galaxy
detection as functions of galaxy properties (i.e. galaxy resolution, galaxy
SNR, and galaxy redshift). For a galaxy sample distorted by a constant shear,
the multiplicative bias and additive bias are given by
\begin{equation}
\begin{split}
    \hat{m}&=\frac{\sum_i w_i m_i}{\sum_i w_i},\\
    \hat{c}_\alpha&=\frac{\sum_i w_i a_i e^\text{psf}_{\alpha;i}}{\sum_i w_i},
\end{split}
\end{equation}
respectively. Here, $i$ refers to the galaxy index, and $w_i$, $m_i$, $a_i$,
$e^\text{psf}_\alpha$ are the galaxy shape weight, multiplicative bias,
fractional additive bias, and PSF ellipticity for the galaxy with index $i$\,.
$\alpha = 1, 2$ are the two components of spin-$2$ properties (e.g.,
ellipticity, shear and additive bias). The galaxy shape weight for each galaxy
is defined as
\begin{equation}
    w_{i} =\frac{1}{\sigma_{e;i}^2+e_{\text{rms};i}^2},
\end{equation}
where $e_{\text{rms};i}$ is the root-mean-square ($\texttt{RMS}$) of the
intrinsic ellipticity per component for the $i$th galaxy. $e_{\rm{rms}}$ and
$\sigma_e$ are modeled and estimated for each galaxy using the image
simulations. The estimated shear for the galaxy ensemble after calibration is
\begin{equation}\label{eq:shear_ensemble_pre}
    \check{\gamma}_{\alpha}=\frac{\sum_i w_i e_{\alpha;i}}
        {2\, \mathcal{R} (1+\hat{m})\sum_i w_i}
    -\frac{\hat{c}_\alpha}{1+\hat{m}}\,,
\end{equation}
where $\alpha = 1,2$\,, and ${\cal R}$ is the shear responsivity for the galaxy
population, defined as
\begin{equation}\label{eq:response}
    \res=1-\frac{\sum_i w_i e^2_{\mathrm{rms};i}}{\sum_i w_i}\,.
\end{equation}

\subsubsection{Selection bias}
\label{ssubec:shearcat_selbias}

Selection bias refers to a bias induced by selection cuts correlated with the
true lensing shear and/or anisotropic systematics (e.g. PSF anisotropy). As a
result, the selected galaxies that are sufficiently close to the edge of the
cuts coherently align in a direction that correlates with the lensing shear
and/or the systematics. The correlation with lensing shear leads to
multiplicative shear estimation bias, whereas the correlation with anisotropic
systematics leads to additive shear estimation bias.

We quantify selection bias in terms of multiplicative bias ($m^\text{sel}$) and
fractional additive bias ($a^\text{sel}$) and estimate these biases from image
simulations \citep{HSC3_catalog_Li2021}. The estimated shear is corrected
as
\begin{equation}\label{eq:shear_ensemble}
    \hat{\gamma}_\alpha =
    \frac{\check{\gamma}_\alpha - \hat{c}^\text{sel}_\alpha}{1+m^\text{sel}},
\end{equation}
where
\begin{equation}
\hat{c}^\text{sel}_\alpha =
    \frac{ a^\text{sel} \sum_{i}
    w_i e^\text{psf}_{\alpha;i}}{\sum_i w_i}
\end{equation}
is the estimated additive selection bias \citep{HSC3_catalog_Li2021}.

Finally, the per-object shear ($\gamma_{\alpha;i}$) for a single galaxy is
defined as
\begin{equation}
    \label{eq:shear_single}
    \gamma_{\alpha;i} =\frac{1}{1+m^\text{sel}}\left(
        \frac{e_{\alpha;i}/(2\mathcal{R})
        - a_i e^\text{psf}_{\alpha;i}}{1+\hat{m}}
        - a^\text{sel} e^\text{psf}_{\alpha;i}
    \right).
\end{equation}
The shear estimation from an galaxy ensemble defined in
equation~\eqref{eq:shear_ensemble} is the weighted average of the per-galaxy
shear.

In addition to the distortion of galaxy images from lensing shear,
the lensing convergence, denoted as $\kappa$, isotropically distorts galaxy
images and changes galaxy sizes and fluxes. Since the intrinsic galaxy sizes
are unknown, we can only observe the reduced shear, denoted as $g_\alpha \equiv
\gamma_\alpha/(1-\kappa)$ from distorted galaxy images. In our work, we do not
distinguish between the lensing shear and the reduced shear since it is a
higher-order systematic bias, as shown in \citep{DESY3_highOrder_Krause2021}.
The bias caused by the reduced shear is less than $0.15 \sigma_{2\mathrm{D}}$,
where $\sigma_{2\mathrm{D}}$ is the $1\sigma$ contour in the 2D
($\Omega_\mathrm{m}$, $S_8$) plane for the DES fiducial cosmic shear analysis.

\subsection{Shear-Catalog Blinding}
\label{subsec:data_blinding}

In order to avoid confirmation bias in our cosmic shear analyses, we conduct
our analysis with catalog-level blinding and analysis-level blinding. That is,
our results are masked while conducting the analysis before unblinding.

For the catalog-level blinding, we measure 2PCFs and constrain cosmology using
three blinded catalogs. Each catalog is blinded by adding a random additional
multiplicative bias with a two-level catalog blinding scheme
\citep{HSC3_catalog_Li2021}. The first is a user-level blinding to prevent an
accidental comparison of blinded catalogs between different cosmological
analyses (i.e.,\ the cosmic shear 2PCFs analysis in this paper, the cosmic
shear Fourier space analysis \citep{HSC3_cosmicShearFourier} and the
3$\times$2pt analysis \citep{HSC3_3x2pt_ls, HSC3_3x2pt_ss}), whereas the second
is a collaboration-level blinding to prevent analysers knowing which catalog of
the three is the true catalog.

For the user-level blinding, a random additional multiplicative bias
$\mathrm{d}m_1$ is generated for each catalog. The values of $\mathrm{d}m_1$
are different for each analysis team, and they are encrypted with the
public keys from the principal investigators of the corresponding analysis
teams. This single value of $\mathrm{d}m_1$ is decrypted and subtracted from
the multiplicative bias values for each catalog entry to remove the user-level
blinding before the cosmic shear analysis.

For the collaboration-level blinding, three blinded catalogs are generated with
indexes $j=0, 1, 2$. The additional multiplicative biases $\mathrm{d}m_2^j$ for
these three blinded catalogs are randomly selected from the following three
different choices of ($\mathrm{d}m_2^1$, $\mathrm{d}m_2^2$, $\mathrm{d}m_2^3$):
$(-0.1,-0.05,0)$, $(-0.05,0,0.05)$, $(0,0.05,0.1)$\,. Note, we set the
difference in multiplicative bias between three catalogs to be $0.05$,
corresponding to a shift in $S_8$ by $\sim$$0.05$, in order to cover the $S_8$
tension between weak-lensing and CMB observations.  The additional
multiplicative biases are listed in an ascending order, in each case, while the
true catalog (with $\mathrm{d}m_2 = 0$) has a different index for the three
options. The values of $\mathrm{d}m_2^{1, 2, 3}$ are encrypted by a public key
from one designated person who is not involved in any cosmology analysis.

The final blinded multiplicative bias values for the galaxies in each of these
three catalogs are
\begin{equation}
    m_{{\rm blind;i}}^j=
    m_{{\rm true;i}}+\mathrm{d}m_{1}^j+\mathrm{d}m_{2}^j,
\end{equation}
where ${\rm i}$ is the galaxy index in each blinded catalog indexed by $j$. We
carry out the same analysis for all three catalogs for internal consistency
checks (see Section~\ref{sec:res_inter}) after decrypting and subtracting the
$\mathrm{d}m_1$ from the multiplicative bias for each catalog.

We adopt an analysis-level blinding for the internal consistency tests in
Section~\ref{sec:res_inter}. Specifically, we shift the posteriors along each
cosmological parameter, e.g.,\ $S_8$ and $\OmegaM$, by the corresponding
projected mode estimate from the fiducial chain. As a result, we only show the
difference between the internal tests and the fiducial chain. In addition, we
do not compare the measured 2PCFs with predictions of any known cosmology.
Moreover, the analysis team did not compare the posterior of cosmology
parameters with any external results (e.g., \textit{Planck} CMB, DES and KiDS's
constraints) before unblinding.

The analysis team agreed that, once the results were unblinded, they would be
published regardless of the outcome. In addition, the analysis method could not
be changed or modified after unblinding.

\subsection{Photometric Redshift Catalog}
\label{subsec:data_photoz}

\begin{figure*}
\includegraphics[width=0.98\textwidth]{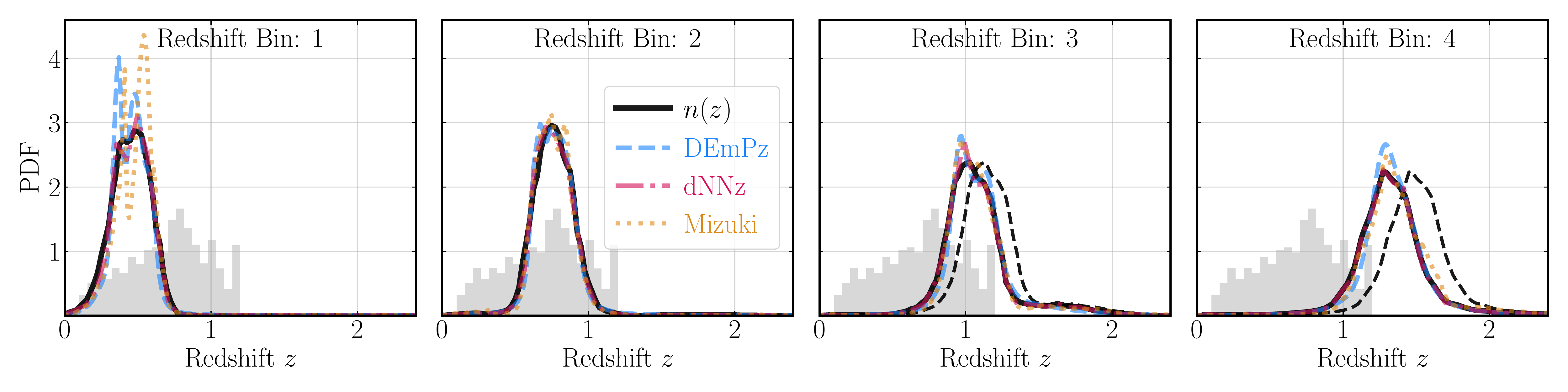}
\caption{
    The comparison between $n(z)$ distributions (solid line) estimated by the
    joint calibration with CAMIRA-LRG sample \citep{HSC3_photoz_Rau2022} and
    those estimated by stacking the \dempz{} (dashed lines), \dnnz{}
    (dot-dashed lines) and \mizuki{} (dotted lines) photo-$z$ posteriors from
    individual galaxies. The shaded grey histogram is the number density as a
    function of redshift of CAMIRA-LRGs used to calibrate the $n(z)$
    distributions of the solid lines. The median redshifts for the four
    redshift bins (solid black lines) are 0.44, 0.75, 1.03 and 1.31, and the
    median redshift for the overall sample is 0.80\,. The dashed black lines in
    the last two redshift bins are the $n(z)$ distributions after after the
    self-calibration in parameter inference (see text for details).
    }
\label{fig:data_pzs}
\end{figure*}

In the following, we briefly summarize the three methods for photometric
redshift (photo-$z$) estimation at the individual galaxy level. We refer the
readers to \citet{HSC3_photoz_Nishizawa2020} for more details.

\dnnz{} is a photo-$z$ conditional density estimation algorithm based on a
neural network. Its architecture consists of multi-layer perceptrons with five
hidden layers. The code uses \cmodel{} fluxes, convolved fluxes, PSF fluxes,
galaxy sizes and galaxy shapes for the training. The photo-$z$ conditional
density is constructed with 100 nodes in the output layer, and each node
represents a redshift histogram bin spanning from $z=0$ to redshift $z=7$
(Nishizawa et al. in prep.).

The Direct Empirical Photometric redshift code (\dempz{}) is an empirical
algorithm for photo-$z$ conditional density estimation \citep{dempz_Hsieh2014}.
It uses quadratic polynomial interpolation of 40 nearest neighboring galaxies
in a training set, with a distance estimated in a 10 dimensional feature space
(5 magnitudes, 4 colors, and 1 size information). \dempz{} estimates the error
for the constructed photo-$z$ conditional densities with resampling procedures.

\mizuki{} \citep{mizuki_Tanaka2015} is a photo-$z$ algorithm adopting a
Spectral Energy Distribution (SED) fitting technique. The method uses an SED
template set constructed with Bruzual-Charlot models \citep{SPS_Bruzual2003}, a
stellar population synthesis code using an Initial Mass Function following
\citet{IMF_Chabrier2003}, emission-line modeling assuming solar metallicity
\citep{metaGal_Inoue2011}, and a dust attenuation model from
\citet{StarFormGal_Calzetti2000}. It applies a set of redshift-dependent
Bayesian priors on the photo-$z$ estimation, and, to improve the accuracy, the
photo-$z$ posteriors of galaxies are calibrated with the specXphot dataset
\citep{specXphot_Bordoloi2010}.

We divide the galaxies in the shear catalog introduced in
Section~\ref{subsec:data_shear} into four tomographic redshift bins by
selecting galaxies using the best estimate, minimizing the estimation risk (see
\citet{HSC3_photoz_Nishizawa2020} for more details), of the \dnnz{} photo-$z$
algorithm within four redshift intervals --- (0.3, 0.6], (0.6, 0.9], (0.9, 1.2]
and (1.2, 1.5]. We find $\sim$$31\%$ and $\sim$$8\%$ galaxies in the first and
second redshift bins, respectively, have double peaks in the \mizuki{} and
\dnnz{} photo-$z$ probability density function (PDF), and the secondary peak
corresponds to a significant fraction of outliers at $z\gtrsim3.0$\,. We remove
these galaxies from our sample for 2PCFs measurement since the secondary peaks
are outside the redshift coverage of the CAMIRA-LRGs
\citep{HSC3_photoz_Rau2022} (see Section~\ref{subsec:model_photoz}) that is
used to calibrate the galaxy redshift distribution, and therefore can
potentially produce large systematic uncertainties.

To be more specific, galaxies with secondary peaks are identified with the
following selection criteria based on the distance between the $0.025$ and
$0.975$ quantiles of the \mizuki{} and \dnnz{} photo-$z$ PDF estimates
\begin{equation}
    \left(z_{\text{0.975}; i}^\text{mizuki}
    - z_{\text{0.025}; i}^\text{mizuki}\right) < 2.7
    \quad \text{and} \quad
    \left(z_{\text{0.975}; i}^\text{dnnz}
    - z_{\text{0.025}; i}^\text{dnnz}\right) < 2.7 \,,
    \label{eq:selection}
\end{equation}
where $z_{\text{0.975}; i}^\text{mizuki (dnnz)}$ and $z_{\text{0.025};
i}^\text{mizuki (dnnz)}$ denote the $97.5$ and $2.5$ percentiles for galaxy $i$
derived with the \mizuki{} (\dnnz{}) photo-$z$ PDF estimates, respectively. We
do not find a significant number of double solutions for \dempz, thus we do not
include it in the criteria above. In Fig.~\ref{fig:data_pzs}, we show the
stacked photo-$z$ posteriors from individual galaxies in each redshift bin for
these three photo-$z$ estimators, after rejecting galaxies with double
solutions. The $n(z)$ obtained by combining multiple photo-$z$'s and calibrated
with CAMIRA LRGs \citep{HSC3_photoz_Rau2022} is used for our fiducial analysis.
The calibrated $n(z)$ is shown in Fig.~\ref{fig:data_pzs}\,.

\subsection{Star Catalog}
\label{subsec:data_stars}

The HSC-Y3 star catalog used to quantify the PSF systematics in the estimation
of the 2PCFs is selected from the star samples described in Section~5.1 of
\citet{HSC3_catalog_Li2021}, which covers the same footprint as the galaxy
shear catalog described in Section~\ref{subsec:data_shear}. We briefly
summarize the star sample we used in this paper, and refer the readers to
\citep{HSC3_catalog_Li2021} for more details.

The PSF models in the HSC-Y3 coadded images are constructed by stacking the PSF
models estimated in each CCD exposure contributing to the coadded pixels, and
the PSF models in a CCD exposure are constructed by interpolating star images
on the same CCD. The selection of stars used for PSF modeling is based on the
$k$-means clustering of high-SNR (i.e., SNR$>50$) objects in size, typically
resulting in $\sim 80$ star candidates per CCD chip (an area of $\sim
60$~arcmin$^2$; see \citealt{HSC1_pipeline} for more details). In the single
exposure CCD processing, $\sim20\%$ of the stars in a given single exposure are
randomly selected and reserved for cross-validation, and are not used for PSF
modeling. Since the star sample used in PSF modeling is derived on individual
exposures, different exposures will not necessarily select the same set of
reserved stars. At the coadded image level, stars that were used by $\geq 20\%$
of the input exposures are labelled as having been used in the modeling, namely
``\texttt{i$\_$calib$\_$psf$\_$used}$==$True''.

The star sample that is used to quantify PSF systematics on 2PCFs is selected
by ``\texttt{i$\_$extendedness$\_$value}$==0$'', a cut indicating whether an
object is an extended galaxy or a point-like star. After that, we apply an
$i$-band magnitude cut at $22.5$ to select a star sample with high SNR.
\citet{HSC3_catalog_Li2021} further divide this magnitude limited star sample
into two subsamples: those flagged by ``\texttt{i$\_$calib$\_$psf$\_$used} $==$
True'' are PSF stars; and the others are defined as non-PSF stars.

In this paper, we use the PSF star sample to estimate the additive bias on
2PCFs from PSF systematics, since as shown in \citet{HSC3_PSF}, the additive
bias on 2PCFs estimated from PSF stars is consistent with that estimated with
non-PSF stars. In addition, the estimation of the PSF systematic error from PSF
stars has higher SNR since there are more stars in the PSF star sample. We give
the details of how we use PSF stars to estimate the additive PSF systematic
error and marginalize over it in our cosmological analysis in
Section~\ref{subsec:model_psfsys}\,.

\subsection{Mock Catalogs}
\label{subsec:data_mock}

In this subsection, we introduce the HSC-Y3 galaxy mock shear catalogs, which
are used to accurately quantify the uncertainties of our measured 2PCFs (both
galaxy-galaxy and galaxy-star shape correlations) due to cosmic variance,
galaxy shape noise, measurement errors due to photon noise, and photometric
redshift uncertainties. The mock catalogs are generated following
\citet{HSC1_mock_Shirasaki2019} with updates to incorporate the survey
footprint, galaxy shape noise, shape measurement error, and photometric
redshift error of the HSC-Y3 shear catalog.

The mock shear catalog uses simulations of the full-sky shear map at $38$
redshifts generated by the ray-tracing simulation
\citep{raytracingTakahashi2017} with $108$ $N$-body simulations of the WMAP9
cosmology ($H_0 = 70$ km/s/Mpc, $\Omega_\mathrm{m} = 0.279$, $\Omega_\mathrm{b}
= 0.046$, $\sigma_8=0.82$) \citep{cmb_WMAP9_Hinshaw2013}. The ray-tracing
simulation calculates the light-ray deflection on the celestial sphere using
the projected matter density field at the spherical shells
\citep{clusterCount_Hamana2015, clusterCount_Shirasaki2015}. Each shell has a
radial ``thickness'' of $150~h^{-1}\mathrm{Mpc}$\,. The angular resolution of
the shear map is $0.43~\mathrm{arcmin}$\,.

In order to increase the number of realizations of the mock catalogs, we
extract $13$ separate regions with the same HSC three-year survey geometry from
each full-sky shear map, obtaining $108 \times 13 = 1404$ mock catalogs in
total. These 1404 lensing-shear maps at 38 redshift planes are combined with
the observed angular positions, photo-$z$s, and shapes of real galaxies
\citep{HSC3_catalog_Li2021} to generate mock shear catalogs. To be more
specific, source galaxies are populated on the lensing-shear maps using the
original angular positions and the \dnnz{} ``best-fit'' redshift estimates of
the galaxies in the HSC shear catalog. Each galaxy is assigned a source
redshift estimate in the mock following the \dnnz{} photo-$z$ posterior
distribution. The shape noise on each galaxy is generated with a random
rotation of the galaxy's intrinsic shape following the intrinsic shape
dispersion estimated in the HSC shear catalog, and the measurement error is
generated as a zero-mean Gaussian random number with the standard deviation
measured in the HSC shear catalog. We distort each galaxy's intrinsic shape
with the shear value on the shear map and add measurement error to the
distorted shape to generate the final galaxy shape \citep[see Section~4.2
in Ref.][]{HSC1_mock_Shirasaki2019}.

We note that our simulations use source galaxy positions from the the real HSC
data but unlike the real universe, the positions are not correlated with the
density field in the simulations. The correlation between the source galaxy
clustering and the shear signal \citep{bmode_zcluster_schneider2002} are
neglected in the mock.

\section{TWO-POINT STATISTICS}
\label{sec:2pt}

The 2PCFs of galaxy shear \citep{wlRevBartelmann}, denoted as
$\xi_{\pm}(\theta)$, are two-point statistics that are widely used to constrain
cosmological parameters. In Section~\ref{subsec:meas_2pcf}, we measure the
2PCFs from the galaxy shear catalog introduced in
Section~\ref{subsec:data_shear}; in Section~\ref{subsec:meas_cov}, we derive
the covariance matrix of the 2PCFs using mock shear catalogs introduced in
Section~\ref{subsec:data_mock}; in Section~\ref{subsec:meas_bmode}, we measure
the $B$-modes on 2PCFs to test the systematics in our measurement.

\subsection{Two-point Correlation Functions}
\label{subsec:meas_2pcf}

\begin{figure*}
\includegraphics[width=0.98\textwidth]{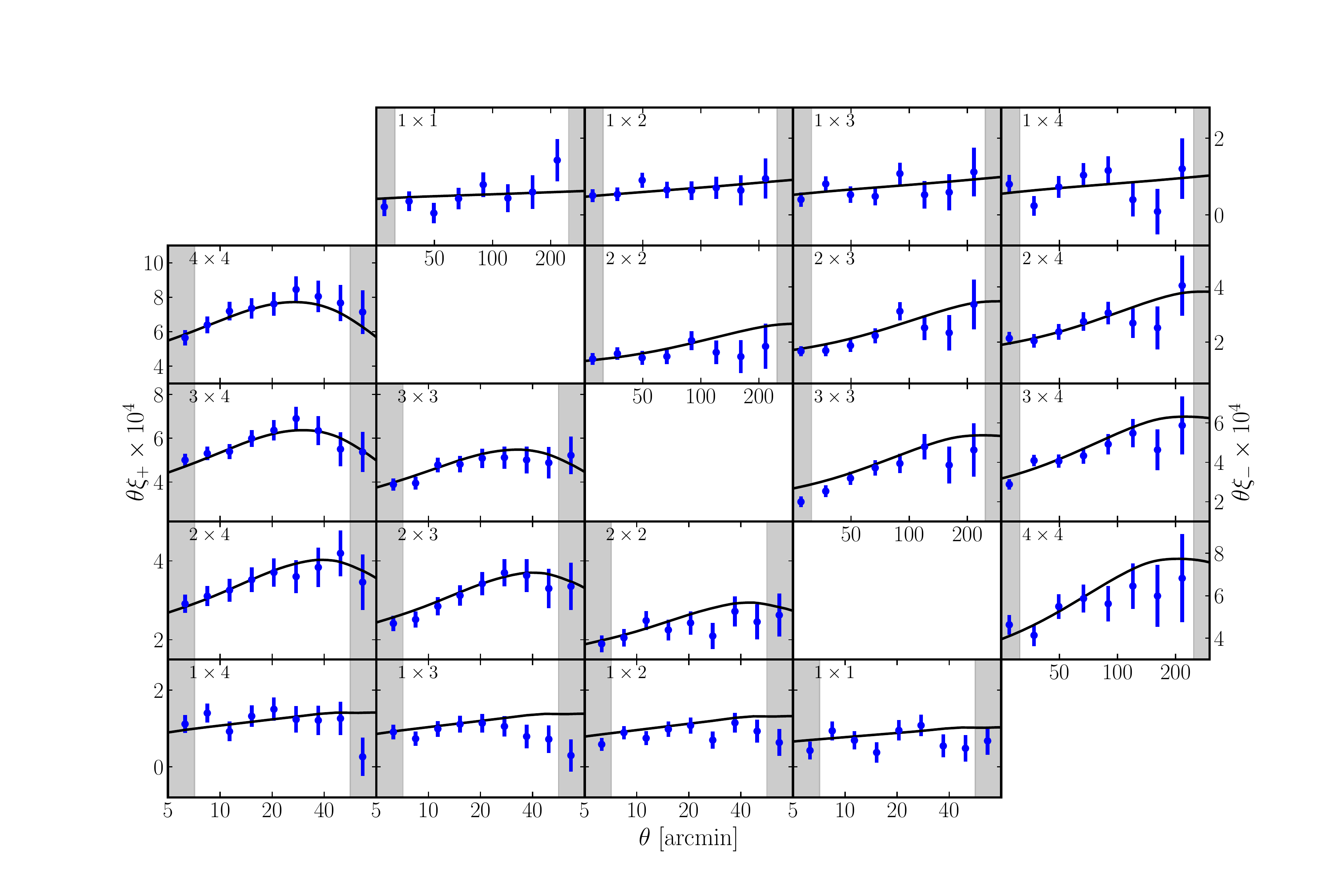}
\caption{
    The ten 2PCFs including four autocorrelations and six cross-correlations
    between the four tomographic redshift bins (labeled with 1--4). This plot
    shows the 2PCFs on scales $5.3 < \theta < 76$ [arcmin] for $\xi_+$, and
    $23.2 < \theta < 248$ [arcmin] for $\xi_-$. The unshaded region refers to
    the fiducial scale cut: $7.1 < \theta < 56.6$ [arcmin] for $\xi_+$, and
    $31.2 < \theta < 248$ [arcmin] for $\xi_-$\,. The errorbars are estimated
    with mock catalogs. The total SNR of the measured 2PCFs is $26.6$\,. The
    solid lines are the best-fit model of our fiducial analysis, as discussed
    in Section~\ref{subsec:inter_fid}.
    }
\label{fig:data_2pcf}
\end{figure*}

The 2PCFs can be measured from the shear catalog using the per-object shear
defined in equation~\eqref{eq:shear_single}:
\begin{equation}
\label{eq:meas_xipm}
\begin{split}
\widehat{\xi}_\pm(\theta)
&=
\frac{
    \sum_{i,j} w(\vec{r}_i) \gamma_{+}(\vec{r}_i)\,
    w(\vec{r}_j)\gamma_{+}(\vec{r}_j)}
    {\sum_{i,j} w(\vec{r}_i) w(\vec{r}_j)
    } \\
&\pm
\frac{
    \sum_{i,j} w(\vec{r}_i) \gamma_{\times}(\vec{r}_i)\,
    w(\vec{r}_j) \gamma_{\times}(\vec{r}_j)}
    {\sum_{i,j} w(\vec{r}_i) w(\vec{r}_j)
    }
\end{split}
\end{equation}
where the summation is over every galaxy pair $(i, j)$ with angular separation
$\theta$\,. For each galaxy pair, we decompose the per-object shear estimates
$\gamma_{\alpha}(\vec{r}_i)$ into tangential components,
$\gamma_{+}(\vec{r}_i)$, and cross components, $\gamma_{\times}(\vec{r}_i)$,
with respect to the direction connecting the two galaxies in a pair.

We use the public software \treecorr{}\footnote{
    \url{https://github.com/rmjarvis/TreeCorr}} to measure both the auto- and
cross-correlations from the four tomographic redshift bins in equal
log-intervals of $\Delta \log(\theta) = 0.29$ in the range $7.1 <
\theta/\mathrm{arcmin} < 56.6$ for $\xi_+$, and $31.2 < \theta/\mathrm{arcmin}
< 248$ for $\xi_-$. The small-scale cut is determined by the requirement to
control the modeling error on the matter power spectrum at small scales due to
baryonic physics (Section~\ref{subsec:model_ps}); and the large-scale cut is
determined by the $B$-mode systematics (Section~\ref{subsec:meas_bmode}). For
different redshift bins, we use consistent scale cuts in $\theta$ for each of
the measured auto- and cross-correlations $\xi_+$ and $\xi_-$\,. It is worth
mentioning that the DES cosmic shear analysis \citep{DESY3_CS_Amon2021,
DESY3_CS_Secco2022} adopts a redshift-dependent scale cut. Given that we have
not observed compelling evidence suggesting a specific scale at any particular
redshift introduces significant bias, we choose to fix the scale cut across
different bins, which simplifies our decision-making process regarding scale
cuts. Fig.~\ref{fig:data_2pcf} shows the 2PCFs (i.e., $\widehat{\xi}^{ij}_+$
and $\widehat{\xi}^{ij}_-$) measured from the galaxy shear catalog in four
tomographic bins. The $i$ and $j$ specify the galaxy samples in two tomographic
bins (note, in the case of $i=j$, the same tomographic bin) from which the
correlation function is calculated. The unshaded region denotes the scales used
for our fiducial analysis. We have 7 angular bins for both $\xi_+$ and
$\xi_-$\,. In total, we have $(7+7) \times 10 = 140$ data points for the $10$
auto- and cross-redshift bins, and the SNR of the 2PCFs is $26.6$ including the
Hartlap correction when estimating the inverse of the covariance matrix
\citep{covariance_Hartlap2007}.

We note that although we focus on 2PCFs in this paper, several alternative
cosmic shear two-point statistics have been used in the literature. These
two-point statistics include the angular power spectrum in Fourier space
\citep[e.g.\ ][]{pseudoCl_Camacho2021, pseudoCl_Nicola2021, pseudoCl_Singh2021}
and the Complete Orthogonal Sets of $E$/$B$-Integrals \citep[e.g.,\
][]{Schneider2010, HSC1_CS_cosebis2022}. In particular,
\citet{HSC3_cosmicShearFourier} carry out analysis in parallel to this paper
using the angular power spectrum in Fourier space with the same catalog.

\subsection{Covariance}
\label{subsec:meas_cov}

\begin{figure}
\includegraphics[width=0.45\textwidth]{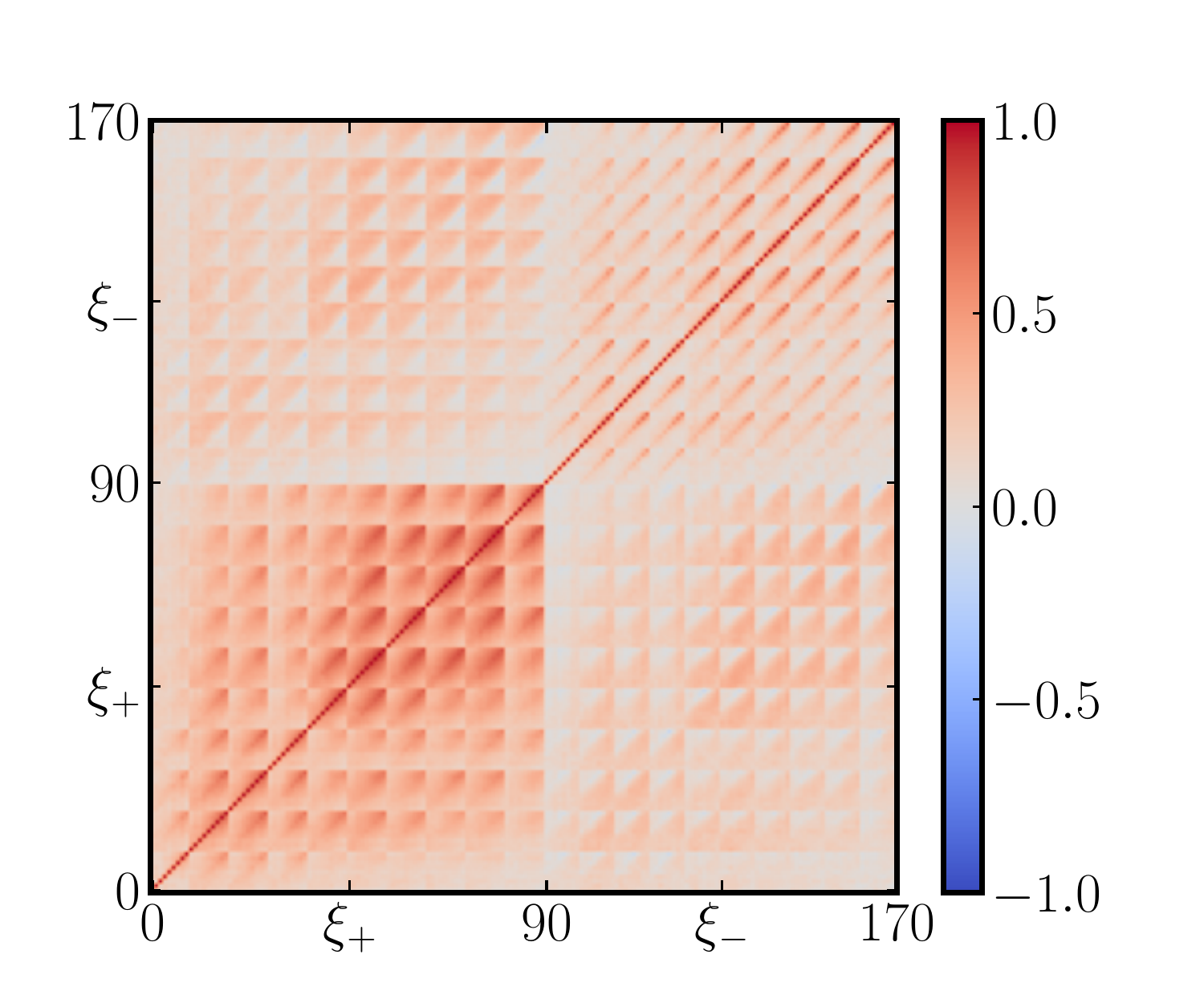}
\caption{
    The normalized covariance matrix (correlation coefficients) estimated with
    mock catalogs. Note that this plot shows the coefficients on scales $5.3 <
    \theta < 76$ [arcmin] for $\xi_+$, and $23.2 < \theta < 248$ [arcmin] for
    $\xi_-$. The fiducial scale cut --- $7.1 < \theta < 56.6$ [arcmin] for
    $\xi_+$, and $31.2 < \theta < 248$ [arcmin] for $\xi_-$ --- is a subset of
    this scale range.
    }
\label{fig:data_cov}
\end{figure}

We derive a covariance matrix of the estimated 2PCFs using the $1404$ HSC mock
shear catalogs summarized in Section~\ref{subsec:data_mock} with different
realizations of galaxy intrinsic shape, measurement error from image noise, and
cosmic shear signal \citep{HSC1_mock_Shirasaki2019}. We measure the 2PCFs from
all $1404$ realizations of mock catalogs in the same manner as the measurement
from the real HSC shear catalog and calculate the covariance matrix from these
$1404$ measurements. The covariance matrix is denoted as $\bm{C}$, and the
correlation coefficients, defined as $\rho_{ij} \equiv C_{ij}/ \sqrt{C_{ii}
C_{jj}}$, are shown in Fig.~\ref{fig:data_cov}. We inspect the diagonal
covariance elements with bootstrap resampling and confirm that each element of
the covariance matrix has SNR greater than 21 ($\lesssim$5\% statistical
uncertainty), which indicates that the covariance is minimally affected by the
finite number of realizations.

Since the cosmic shear signal in the mock catalogs are obtained from a large
number of full-sky ray-tracing simulations of the WMAP9 cosmology which takes
into account nonlinear structure formation \citep{raytracingTakahashi2017}, the
derived cosmic variance includes both Gaussian and non-Gaussian information.
Also, the galaxy positions and survey geometry in the mock catalogs mimic those
of the real data; therefore the derived covariance includes super-survey
covariance \citep{2013PhRvD..87l3504T,HSC1_mock_Shirasaki2019}. Moreover, we
generate random shape noise and measurement error using the galaxy intrinsic
shapes and measurement error from the real shear catalog
\citep{HSC3_catalog_Li2021}. We find that the shape noise covariance is
prominent at the smallest angular bins, while the cosmic variance dominates the
covariance at the largest angular bins.

The accuracy of the covariance matrix from the mocks was studied in detail by
\citet{HSC1_mock_Shirasaki2019}. They found that multiplicative bias of 10\%
can lead to a $\sim$$20\%$ difference in the covariance from shape noise and
measurement error. We already adopted the real value of multiplicative bias in
the shear catalog, and thus have corrected for its effect. In addition, we
correct for the bias from the effects of shell thickness, finite angular
resolution and finite redshfit resolution in the ray-tracing simulations
\citep[for more details, see Ref][]{HSC1_mock_Shirasaki2019}. Since we find the
average 2PCFs measured from our simulations are lower than the theory
prediction, and the ratio is approximately constant ($0.81$ on average) within
our scale cuts for each redshift bin, we divide the 2PCF from each realization
of mocks by the ratio in each bin.

One caveat in our covariance estimation is that we do not include dependence of
the covariance on the cosmological parameters, since our mock catalogs are
generated from a set of ray-tracing simulations adopting only one WMAP9
cosmology \citep{raytracingTakahashi2017}. \citet{cosmoCov_Kodwani2019} used
a Fisher analysis to study the dependence of the covariance matrix on the
cosmological parameters and the resulting bias in the cosmology constraints
when assuming a cosmology-independent covariance. They reported that the
cosmology dependence of the covariance matrix does not significantly impact the
cosmology constraints (to be more specific, parameters are only biased by $\leq
1\%$ of statistical uncertainties) for any current and future weak-lensing
surveys. Following \citet{cosmoCov_Kodwani2019}, we neglect the parameter
dependence of the covariance matrix in our analysis.

\subsection{$B$-modes}
\label{subsec:meas_bmode}

\begin{figure*}
\includegraphics[width=0.98\textwidth]{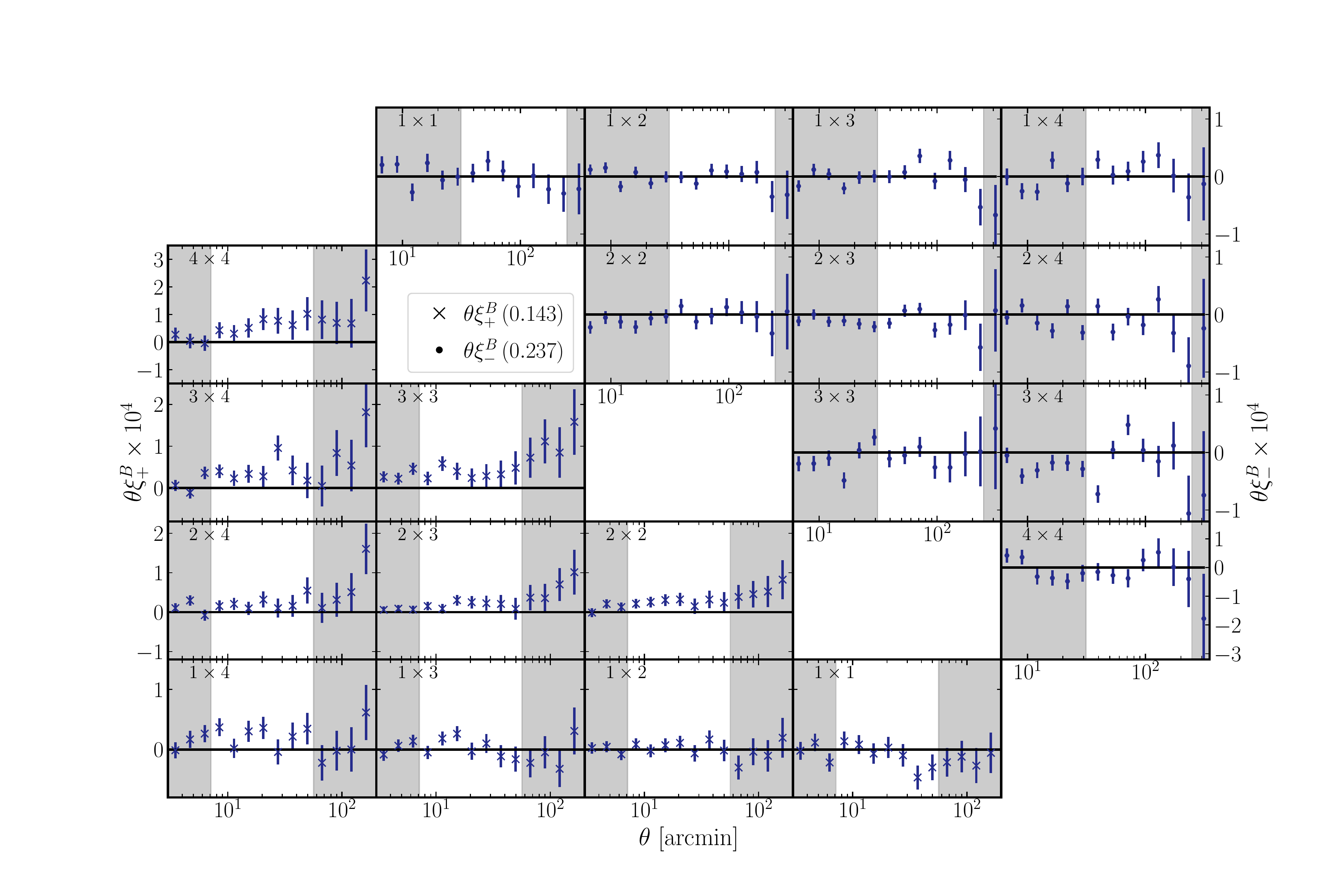}
\caption{
    $B$-modes on 2PCFs measured from the HSC-Y3 catalog in four tomographic
    bins. The $p$-value of the measured $B$-modes relative to a model of
    exactly zero is 0.1143 for $\xi_+$ and 0.1237 for $\xi_-$, as shown in the
    legend. The errorbars are estimated with mock catalogs. The unshaded region
    refers to the fiducial scale cut.
    }
    \label{fig:data_bmodes}
\end{figure*}

The measured 2PCFs $\xi_{\pm}$ include contributions from both curl-free
gradient component ($E$-mode) and curl component ($B$-mode). However, the
physical $B$-mode from a gravitational lensing potential, which can be caused
by second-order lensing deflection \citep{highorderDeflect_Krause2010},
intrinsic alignments \citep{tatt_blazek17} and redshift clustering of source
galaxies \citep{bmode_zcluster_schneider2002}, is expected to be orders of
magnitude smaller than the $E$-modes. Therefore, an estimate of the $B$-mode
component can be used as a test for systematic errors. Following
\citet{bmode_zcluster_schneider2002}, we separate the $E$-mode and $B$-mode
components as
\begin{equation}
\label{eq:ebmodes}
\begin{split}
    \xi_{+}^E(\theta) = \frac{1}{2} \left[ \xi_+(\theta)+ \xi_-(\theta)
    + \int_\theta^\infty  \frac{\mathrm{d}\phi}{\phi}
    \xi_-(\phi)\left(4-12\frac{\theta^2}{\phi^2}\right)\right],\\
    \xi_{-}^E(\theta) = \frac{1}{2} \left[ \xi_+(\theta)+ \xi_-(\theta)
    + \int_0^\theta \frac{\mathrm{d}\phi \phi}{\theta^2}
    \xi_+(\phi)\left(4-12\frac{\phi^2}{\theta^2}\right)\right],\\
    \xi_{+}^B (\theta)= \frac{1}{2} \left[ \xi_+(\theta) - \xi_-(\theta)
    - \int_\theta^\infty  \frac{\mathrm{d}\phi}{\phi}
    \xi_-(\phi)\left(4-12\frac{\theta^2}{\phi^2}\right)\right],\\
    \xi_{-}^B(\theta) = \frac{1}{2} \left[\xi_+(\theta) - \xi_-(\theta)
    + \int_0^\theta \frac{\mathrm{d}\phi \phi}{\theta^2}
    \xi_+(\phi)\left(4-12\frac{\phi^2}{\theta^2}\right)\right],\\
\end{split}
\end{equation}
where $\xi_+(\theta) = \xi^E_+(\theta) + \xi^B_+(\theta)$ and $\xi_-(\theta) =
\xi^E_-(\theta) - \xi^B_-(\theta)$\,. In order to compute the integrals in
equation~\eqref{eq:ebmodes}, we use a Riemann sum and measure $\xi_\pm$ with
much finer log-intervals of $\Delta \log(\theta)=0.02$ ranging from
$0.2~\mathrm{arcmin}$ to $415~\mathrm{arcmin}$. To compute the integral in
$\xi_+^B$ ($\xi_-^B$) beyond (below) $\theta=415~\mathrm{arcmin}$
($\theta=0.2~\mathrm{arcmin}$), we extrapolate the measured $\xi_-$ ( $\xi_+$)
beyond the interval with a WMAP9 cosmology. We confirm the result is not
sensitive to the cosmology model (WMAP or \textit{Planck} cosmology) for the
extrapolation. We use the 1404 HSC-Y3 mock catalogs (introduced in
Section~\ref{subsec:data_mock}) to calculate the errors on the estimated
$B$-modes. Specifically, we conduct the same measurement on each mock
realization, and derive the covariance matrix from the 1404 B-modes
measurement.

As seen in Fig.~\ref{fig:data_bmodes}, the $B$-modes on the $\xi_+$ measurement
are significant at large angular scales, especially in the high redshift bins.
To reduce the influence of the $B$-modes on our cosmology constraints, we apply
a scale cut on $\xi_+$ at scales with $\theta \geq 56.6$~arcmin. Although we do
not find significant $B$-modes on $\xi_-$, we also apply a scale cut on $\xi_-$
to remove scales with $\theta \geq 248$~arcmin since the data at such large
scales is dominated by cosmic variance and contributes little to the SNR of the
2PCFs. Note that the cuts at small scales are imposed to reduce the modeling
uncertainties of baryonic physics as will be shown in
Section~\ref{subsec:model_valid}. After the cuts at large scales, we find that
the probability that the $B$-modes in the fiducial scale range is consistent
with zero is $p=0.143$ for $\xi_+$ and $0.237$ for $\xi_-$, respectively.

\section{MODEL}
\label{sec:model}

\begin{table}
\caption{
Model parameters and priors used in our fiducial cosmological parameter
inference. The label ${\mathcal U}(a,b)$ denotes a noninformative flat prior
between $a$ and $b$, and ${\mathcal N}(\mu, \sigma)$ denotes a normal
distribution with mean $\mu$ and width $\sigma$\,.
}
\label{tab:parameters}
\setlength{\tabcolsep}{20pt}
\begin{center}
\begin{tabular}{ll}  \hline\hline
Parameter & Prior \\ \hline
\multicolumn{2}{l}{\hspace{-1em}\bf Cosmological parameters
(Section~\ref{subsec:model_ps})}\\
$\Omega_\mathrm{m}$                 & ${\cal U}(0.1, 0.7)$\\
$A_\mathrm{s} \,(\times 10^{-9})$   & ${\cal U}(0.5, 10)$\\
$n_\mathrm{s}$                      & ${\cal U}(0.87, 1.07)$\\
$h_0$                               & ${\cal U}(0.62, 0.80)$\\
$\omega_\mathrm{b}$                 & ${\cal U}(0.02, 0.025)$\\
\hline
\multicolumn{2}{l}{\hspace{-1em}\bf Baryonic feedback parameters
(Section~\ref{subsec:model_ps})}\\
$A_\mathrm{b}$                      & ${\cal U}(2, 3.13)$ \\
\multicolumn{2}{l}{\hspace{-1em}\bf Intrinsic alignment parameters
(Section~\ref{subsec:model_ia})}\\
$A_1$                               & ${\cal U}(-6, 6)$ \\
$\eta_1$                            & ${\cal U}(-6, 6)$ \\
$A_2$                               & ${\cal U}(-6, 6)$ \\
$\eta_2$                            & ${\cal U}(-6, 6)$ \\
$b_\mathrm{ta}$                     & ${\cal U}(0, 2)$ \\
\hline
\multicolumn{2}{l}{\hspace{-1em}\bf Photo-$z$ systematics
(Section~\ref{subsec:model_photoz})}\\
$\Delta z_{1}$                    & ${\cal N}(0, 0.024)$ \\
$\Delta z_{2}$                    & ${\cal N}(0, 0.022)$ \\
$\Delta z_{3}$                    & ${\cal U}(-1, 1)$ \\
$\Delta z_{4}$                    & ${\cal U}(-1, 1)$ \\
\multicolumn{2}{l}{\hspace{-1em}\bf Shear calibration biases
(Section~\ref{subsec:model_dm})}\\
$\Delta m_{1}$                    & ${\cal N}(0.0,0.01)$ \\
$\Delta m_{2}$                    & ${\cal N}(0.0,0.01)$ \\
$\Delta m_{3}$                    & ${\cal N}(0.0,0.01)$ \\
$\Delta m_{4}$                    & ${\cal N}(0.0,0.01)$ \\
\multicolumn{2}{l}{\hspace{-1em}\bf PSF systematics
(Section~\ref{subsec:model_psfsys})}\\
$\alpha'^{(2)}$                   & ${\cal N}(0, 1)$\\
$\beta'^{(2)}$                    & ${\cal N}(0, 1)$\\
$\alpha'^{(4)}$                   & ${\cal N}(0, 1)$\\
$\beta'^{(4)}$                    & ${\cal N}(0, 1)$\\ \hline
\hline
\end{tabular}
\end{center}
\end{table}

In this section, we introduce the model, containing twenty-three free
parameters as shown in Table~\ref{tab:parameters}, to predict the tomographic
cosmic shear 2PCFs $\xi^{ij}_{\pm}(\theta)$\,. Note that we coordinate with the
Fourier space cosmic shear analysis \citep{HSC3_cosmicShearFourier} when making
the decision on model choices, and our fiducial model is the same as the Fourier
space analysis. The parameters can be divided into two categories: eleven
physical parameters and twelve systematic parameters. The physical parameters
include five cosmological parameters, one baryonic feedback parameter and five
intrinsic alignment parameters; the systematic parameters include four
photo-$z$ error parameters, four shear calibration bias parameters and four PSF
systematic parameters. Our model is implemented in the public software:
\cosmosis{} \citep{cosmosis_Zuntz2015}. We note that the model choices were set
entirely before unblinding.

We coordinate our model choices with the parallel cosmic shear analysis using
the pseudo-$C_\ell$ \cite{HSC3_cosmicShearFourier}. The analysis tests and
choices described below have also been also adopted and described by
\cite{HSC3_cosmicShearFourier}.

\begin{figure}
\includegraphics[width=0.48\textwidth]{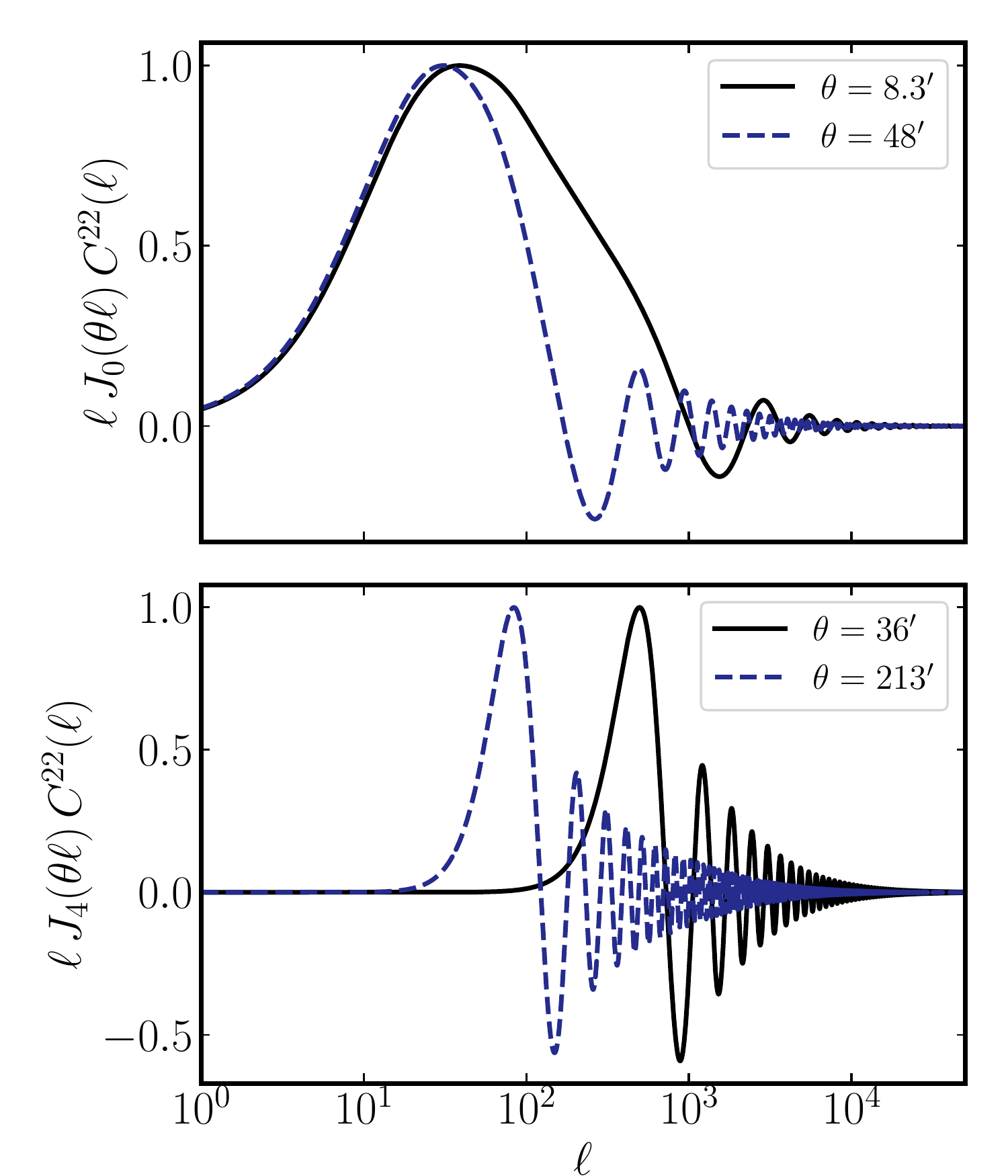}
\caption{
    The integrands in equation~\eqref{eq:model_hankel} which transform angular
    power spectra to correlation functions. We show the integrands for the
    smallest and largest angular bins for $\xi_+$ and $\xi_-$, respectively,
    for the correlation functions of the second redshift bin with itself.
}
\label{fig:model_hankel}
\end{figure}

With the flat-sky approximation, the 2PCFs can be expressed as the $E$ and $B$
modes of angular power spectra $C^{E/B}(\ell)$ via the Hankel transform:
\begin{equation}
\label{eq:model_hankel}
\xi^{ij}_{+/-}(\theta) = \frac{1}{2\pi}\int d\ell \,
    \ell J_{0/4}(\theta \ell) \,
    \left( C^{E;ij}(\ell) \pm C^{B;ij}(\ell) \right),
\end{equation}
where $J_{0/4}$ are the $0$th / $4$th-order Bessel functions of the first kind.
In our analysis, the Hankel transform is computed with \texttt{FFTLog}
\citep{fftlog2020} implemented in \cosmosis{}. In Fig.~\ref{fig:model_hankel},
we demonstrate the integrands in equation~\eqref{eq:model_hankel} for different
scales of $\xi_\pm$\,. As shown, $\xi_\pm$ in one angular bin corresponds to an
integral over a wide range of $\ell$s; therefore, our cosmic shear analysis
based on 2PCFs is sensitive to information on different scales from the Fourier
space analysis \citep{HSC3_cosmicShearFourier}. In a companion paper
\citep{HSC3_cosmicShearFourier}, we measure pseudo-$C_\ell$s and reconstruct
angular power spectra $C_\ell^{E/B;ij}$ from the HSC-Y3 shear catalog.

The observed galaxy shapes are determined by both foreground lensing shear and
the intrinsic shapes induced by the torques from the local environment. The spatial
correlation between intrinsic shapes is known as intrinsic alignment (IA)
\citep{IA_rev_Troxel2015}. Therefore, the $E$-mode angular power spectra in
equation~\eqref{eq:model_hankel} can be decomposed into lensing-lensing
auto-spectra ($C^{E;ij}_\mathrm{GG}$), intrinsic-intrinsic auto-spectra
($C^{E;ij}_\mathrm{II}$) and lensing-intrinsic cross power spectra between
lensing and IA ($C^{E;ij}_\mathrm{GI}$) \citep{IA_rev_Troxel2015}. Although the
$B$-mode induced by lensing shear is negligible, a significant $B$-mode angular
power spectrum can be produced by high-order IA models.
\begin{equation}
\begin{split}
\label{eq:model_Cl_decomp}
   C^{E;ij} &= C^{E;ij}_\mathrm{GG} +  C^{E;ij}_\mathrm{II}
   + C^{E;ij}_\mathrm{GI} + C^{E;ji}_\mathrm{GI}\,, \\
   C^{B;ij} &= C^{B;ij}_\mathrm{II}
\end{split}
\end{equation}

As shown in Section~\ref{subsec:meas_bmode}, the measured $B$-mode signal
within our fiducial scale cuts is not significant, so we set this component to
zero in our analysis. The $E$-mode lensing angular power spectra
($C^{E}_{GG;ij}$) is related to the matter power spectrum. Our implemented
model for the matter power spectra is introduced in
Section~\ref{subsec:model_ps}. Our implemented IA model is introduced in
Section~\ref{subsec:model_ia}. Systematics are described by twelve parameters
in our model, which include uncertainties in photo-$z$ estimation (see
Section~\ref{subsec:model_photoz}), uncertainties in shear calibration (see
Section~\ref{subsec:model_dm}) and PSF related systematic uncertainties (see
Section~\ref{subsec:model_psfsys}). Finally, the Monte Carlo Bayesian analysis
used to constrain the free parameters is introduced in
Section~\ref{subsec:model_baysian}.

\subsection{Matter Power Spectra}
\label{subsec:model_ps}

We first connect the lensing angular power spectrum $C^{ij}_\mathrm{GG}(\ell)$
in equation~\eqref{eq:model_Cl_decomp} to the power spectrum $P_\text{m}(k, z)$
of the matter distribution in the universe at different redshifts. In a
spatially flat universe, the lensing angular power spectrum encodes information
of the matter power spectrum, $P_\text{m}(k, \chi)$ according to the Limber
approximation \citep{Limber1953, Loverde2008}:
\begin{equation}
\label{eq:Cl}
C^{ij}_\mathrm{GG}(\ell) =
\int_{0}^{\chi_{H}} d\chi \frac{q_{i}(\chi)q_{j}(\chi)}{\chi^2}
P_\text{m}\!\left(k=\frac{\ell + 1/2}{\chi}; \chi \right),
\end{equation}
where $\chi$ is the radial comoving distance,
$\chi_{H}$ is the distance to the
horizon (the maximum distance one could possibly observe \footnote{We note
that, in this paper, we only model the structure up to $z=4$.}), and
$q_i(\chi)$ is the lensing efficiency in the $i$th redshift bin defined as
\begin{equation}
q_{i}(\chi) = \frac{3}{2} \Omega_{\rm m} \left( \frac{H_{0}}{c}\right)^{2}
\frac{\chi}{a(\chi)} \int_{\chi}^{\chi_{H}}d\chi ' n_{i}(\chi') \frac{\chi' -
\chi}{\chi'},
\label{eq:lensing_efficiency}
\end{equation}
where $\Omega_{\rm m}$ and $H_0$ are the matter density and Hubble parameter
($H_0 = 100\,h_0~[\mathrm{km/s/Mpc}]$) at redshift zero, $a$ is the cosmology
scale factor, and $n_{i}(\chi)$ is the normalized redshift distribution
of the galaxy in the $i$th redshift bin (see section~\ref{subsec:model_photoz}
for the modeling of $n_i(z)$). Note that $\chi$ is given as a function
of redshift as $\chi=\chi(z)$, so we compute the nonlinear matter power
spectrum $P_{\rm m}(k;z)$ for an input set of $k$ and $z$ given
$\chi=\chi(z)$\,.

\subsubsection{Linear and nonlinear power spectra}

At large scales in the early Universe, the structure grows according to linear
perturbation theory. For a flat $\Lambda$CDM cosmology, the linear matter power
spectrum is determined by the five cosmological parameters in
Table~\ref{tab:parameters}, including the matter density parameter ($\OmegaM$),
the amplitude ($A_s$) and the tilt ($n_s$) parameters of the power spectrum of
the primordial curvature perturbations, the dimensionless Hubble parameter
($h$) and $\omega_\mathrm{b} \equiv \Omega_\mathrm{b} h^2$, where
$\Omega_\mathrm{b}$ is the baryon density parameter. In our analyses, we set
the sum of neutrino mass $\sum m_\nu = 0.06~\mathrm{eV}$\,. The linear power
spectrum can be accurately computed by solving the Einstein-Boltzmann equations
which describe the co-evolution of the different components in the universe
(e.g., dark energy, dark matter, baryonic matter, radiation). The linear power
spectrum of matter density field can be computed with public codes such as
\camb{} \citep{CAMB2000}, and \class{} \citep{CLASSI2011,CLASSII2011}. These
public codes solve the coupled set of differential equations at first order
according to linear perturbation theory, and compute the linear matter power
spectrum.

At small scales, structure growth is nonlinear and cannot be described by a
linear perturbation theory. Therefore, one has to resort to cosmological
$N$-body simulations to model the matter power spectrum at nonlinear scales.
Many empirical models calibrated against high-resolution $N$-body simulations have
been proposed to calculate the nonlinear matter power spectrum, including
\halofit{} \citep{halofitT2012} and \hmcode{} \citep{halofit_mead15,
halofit_mead16, halofit_mead21}. In modern cosmology analysis pipelines,
emulators are broadly adopted to improve the computational speed of the matter
power spectrum. They are constructed by running a large number of cosmological
simulations with different input cosmological and astrophysical parameters, and
interpolating the power spectrum between these parameters. These emulators can
efficiently compute both linear \citep{baccoEmuLin_Arico2021,
cosmicNet2022, darkemu_Nishimichi2019} and nonlinear
\citep{darkemu_Nishimichi2019, baccoEmuNL_Angulo2021, euclidEmu2,
cosmicEmu2022} power spectra with percent-level accuracy.

In our fiducial analysis, we adopt the public \bacco{} emulator
\citep{baccoEmuLin_Arico2021} (version 1.0.0)
\footnote{\url{https://bitbucket.org/rangulo/baccoemu/src/master/}} to compute
the linear matter power spectrum. \bacco{} is a neural network emulator trained
with more than $200,000$ linear matter power spectra computed with \class{} in
the wave-number range between $10^{-4}$ and $50~[h \mathrm{Mpc}^{-1}]$\,. The
supported range of the cosmological parameters of the \bacco{} emulator is shown
in Table~\ref{tab:bacco_support}. The \bacco{} emulator is not limited by
boundaries in $A_s$ and $n_s$ since it emulates the transfer function of the
linear power spectrum \citep{baccoEmuLin_Arico2021}. In order to model the
nonlinear matter power spectrum, we use \hmcode{} \citep{halofit_mead16}~2016,
which is a variant of the halo model with physically motivated parameters
calibrated with N-body and hydrodynamical simulations \citep{halofit_mead15,
halofit_mead16, halofit_mead21}.

\begin{table}
    \caption{
    The supported range for five cosmological parameters in \bacco{} emulator,
    where $\Omega_\mathrm{b} = \omega_\mathrm{b}/h^2$.
    }
\label{tab:bacco_support}
\setlength{\tabcolsep}{20pt}
\begin{center}
\begin{tabular}{lc}  \hline\hline
Parameter & Supported range \\ \hline
$\Omega_\mathrm{m}$                 & ${\cal U}(0.06, 0.7)$\\
$A_\mathrm{s} \,(\times 10^{-9})$   & $\cdot \cdot \cdot$ \\
$h_0$                               & ${\cal U}(0.5, 0.9)$\\
$\Omega_\mathrm{b}$                 & ${\cal U}(0.03, 0.07)$\\
$n_s$                               & $\cdot \cdot \cdot$\\
\hline
\end{tabular}
\end{center}
\end{table}

In our analysis pipeline, we adopt wide flat priors on the five cosmological
parameters: $\OmegaM$, $A_s$, $n_s$, $h$ and $\omega_\text{b}$. However, we
note that a flat prior on $A_s$ leads to an informative prior on $\OmegaM$ and
$S_8$ due to the degeneracies between these parameters. In order to obtain a
chain with uniform prior on the $(\OmegaM, S_8)$ plane, we follow
\citet{HSC1_2x2pt_Sugiyama2022} to apply a weight $w = \sigma_8 / A_s$, which
is the determinant of the Jacobian for the coordinate transform from
($\OmegaM$, $A_s$, \dots) to ($\OmegaM$, $S_8$, \dots), to the MC chain sampled
with the flat prior on $A_s$ to obtain a chain with uniform prior on $\OmegaM$
and $S_8$\,. We refer the readers to \citet{HSC1_2x2pt_Sugiyama2022} for more
details. Our priors are coordinated to be the same with
\citet{HSC3_cosmicShearFourier}, and the reasons for adopting flat prior on
$A_s$ are discussed in details in \citet{HSC3_cosmicShearFourier}.

\subsubsection{Baryonic feedback}

The matter power spectrum at small scales is significantly supressed by
baryonic effects such as feedbacks from supernova and active galactic nuclei
(AGN) (at $k\sim 10~h/\mathrm{Mpc}$) as well as cooling and star formation.

We follow \citet{KiDS1000_CS_Asgari2020} to adopt \hmcode{}~2016
\citep{halofit_mead16} to empirically model baryonic effects on the matter
power spectrum. \hmcode{}~2016 parameterizes the effect of baryonic feedback
with a halo bloating parameter $\eta_\mathrm{b}$ and the amplitude of the halo
mass-concentration relation $A_\mathrm{b}$ \citep{halofit_mead15,
halofit_mead16}. This baryonic model is calibrated with hydrodynamical
simulations. We follow \citet{Joachimi2020} to set the bloating parameter as a
function of the amplitude parameter:
\begin{equation}
    \eta_\mathrm{b} = 0.98 - 0.12A_\mathrm{b}\,.
\end{equation}
To be more specific, we use \hmcode{}~2016 to model the supression from
baryonic feedbacks on small scales of power spectrum with a flat prior:
$A_\mathrm{b}\in [2, 3.13]$ as shown in Table~\ref{tab:parameters}. We
marginalize over the amplitude parameter $A_\mathrm{b}$ when constraining our
cosmological parameters. In \hmcode{}~2016, $A_{\rm b}=3.13$ corresponds to the
matter power spectrum without baryonic feedback (i.e. the spectrum obtained
from dark matter only simulations). The latest version of \hmcode{} is
\hmcode{}~2020 \citep{halofit_mead21}, which improves the modeling of the
nonlinear matter power spectrum with very large neutrino mass, i.e.,\ $m_\nu >
0.5~\mathrm{eV}$\,. However, our analysis focuses on the $\Lambda$CDM cosmology
with $\sum m_\nu = 0.06~\mathrm{eV}$, and the computational speed of
\hmcode{}~2016 is about 1.5 times faster than \hmcode{}~2020. Therefore, we
adopt \hmcode{}~2016 as our fiducial model.

It is worth noting that that there are other approaches to model the baryonic
effects on the matter power spectrum, including baryonic correction models
\citep{baronCorrect_Schneider2019} and approaches based on Principal Component
Analysis (PCA) \citep{baryonPCA_Huang2021}.

\subsection{Intrinsic Alignment}
\label{subsec:model_ia}

In a spatially flat universe, the IA angular power spectrum between two
redshift bins $i$ and $j$ is related to the integrated 3D IA power spectrum via
the Limber approximation:
\begin{equation}
\label{eq:CIIGI}
\begin{split}
    C^{ij}_{\rm II} (\ell) &=
    \int_0^{\chi_{H}} \mathrm{d}\chi \frac{n_i(\chi) n_j(\chi)}{\chi^2}
    P_{\rm II}\!\left ( k=\frac{\ell+1/2}{\chi}; \chi \right ),\\
    C^{ij}_{\rm GI} (\ell) &=
    \int_0^{\chi_{H}} \mathrm{d}\chi \frac{q_i(\chi) n_j(\chi)}{\chi^2}
    P_{\rm GI}\!\left ( k=\frac{\ell+1/2}{\chi}; \chi \right ),
\end{split}
\end{equation}
where $q_i$ is the lensing efficiency defined in
equation~\eqref{eq:lensing_efficiency}. The II (GI) refers to the correlation
between intrinsic shape and intrinsic shape (lensing shear and intrinsic
shape). There are many ways to model the II and GI power spectra, and, in this
paper, we consider two model choices:
\begin{enumerate}
    \item the tidal alignment and tidal torque model (TATT;
        \citep{tatt_blazek17});
    \item the nonlinear alignment model (NLA; \citep{nla_hirata07,
        nla_bridle07}).
\end{enumerate}

TATT is built on nonlinear perturbation theory assuming the intrinsic galaxy
shapes are determined by the tidal field and the density field of matter.
Following \citet{DESY3_CS_Secco2022}, we only keep the quadratic perturbation
terms, and the IA power spectra are given by
\begin{equation}
\label{eq:tatt_gi}
\begin{split}
P^E_{\rm GI}(k)
    &= c_1 P_\delta(k) + b_\mathrm{ta} c_1 P_{0|0E}(k) + c_2 P_{0|E2}(k)\,, \\
P^E_{\rm II}(k)
    &= c^2_1 P_\delta(k) + 2 b_\mathrm{ta} c_1^2 P_{0|0E}(k) \\
    &+ b^2_\mathrm{ta} c_1^2 P_{0E|0E}(k) + c_2^2 P_{E2|E2}(k) \\
    &+ 2c_1c_2 P_{0|E2}(k) + 2b_\mathrm{ta}c_1 c_2 P_{0E|E2}(k)\,,\\
P^B_{\rm II}(k)
    &= b^2_\mathrm{ta}(k) c_1^2P_{0B|0B}(k) + c_2^2 P_{B2|B2}(k) \\
    &+ 2b_\mathrm{ta} c_1 c_2 P_{0B|B2}(k)\,.
\end{split}
\end{equation}
The subscripts of the tidal field power spectra on the right-hand side indicate
correlations between different order terms in the expansion of the matter
field, and these power spectra are calculated to one-loop order using the
public software: \texttt{FAST-PT}~v2.1 \citep{fastpt2016, fastpt2017}
\footnote{\url{https://github.com/JoeMcEwen/FAST-PT}}. We refer the readers to
\citet{tatt_blazek17} for more details. The redshift-dependent amplitudes $c_1$
and $c_2$ are defined as
\begin{equation}
\label{eq:tatt_c1c2}
\begin{split}
    c_1(z) &= -A_1  \frac{\bar{C} \rho_{\rm c}
    \Omega_\mathrm{m}}{D(z)} \left ( \frac{1+z}{1+z_0} \right )^{\eta_1},\\
    c_2(z) &= 5 A_2  \frac{\bar{C} \rho_{\rm c}
    \Omega_\mathrm{m}}{D^2(z)} \left ( \frac{1+z}{1+z_0} \right )^{\eta_2}\,,
\end{split}
\end{equation}
where $D(z)$ is the growth function, $\rho_\mathrm{crit}$ is the critical
density, $z_0=0.62$ is the pivot redshift, and $\bar{C} = 5\times10^{-14} M_\odot
h^{-2} \mathrm{Mpc}^2$ is obtained from \texttt{SuperCOSMOS}
\citep{supercosmos2002}.  The TATT IA model has five free parameters: $A_1,
A_2, \eta_1, \eta_2, b_\mathrm{ta}$. The power-law terms in
equation~\eqref{eq:tatt_c1c2} with two free parameters $(\eta_1, \eta_2)$ are
used to model the possible redshift evolution beyond what is already encoded in
the model; $A_1$ and $A_2$ capture the IA power spectra that scale linearly and
quadratically with the tidal field. The bias parameter $b_\mathrm{ta}$
models the fact that galaxies are over-sampled in the highly clustered regions.
In this paper, we adopt wide flat priors on the TATT model parameters: $A_1,
A_2, \eta_1, \eta_2 \in [-6, 6]$, $b_\mathrm{ta} \in [0,2]$. This is because
the IA signal is very sensitive to the properties (e.g., color, magnitude) of
the galaxy sample \citep{iaSDSS_Singh2015, iaBulgeDisk_Jagvaral2022}, thus it
is very difficult to derive reliable Gaussian priors on the TATT model
parameters for the galaxy sample in the shear catalog.

NLA is a more commonly used IA model, which is a subspace of TATT with $A_2=0$
and $b_\mathrm{ta}=0$\,. The NLA model is built upon the assumption that
intrinsic galaxy shapes are aligned linearly with the tidal field. Under this
assumption, the GI and II power spectra are
\begin{equation}
    \label{eq:nla_PE}
    P^E_\mathrm{GI} = c_1(z) P_\delta, \qquad
    P^E_\mathrm{II} = c^2_1(z) P_\delta,
\end{equation}
where the redshift-dependent amplitude $c_1(z)$ is defined in
equation~\eqref{eq:tatt_c1c2}. Our implementation of the NLA model has two free
parameters, $A_1$ and $\eta_1$, and we adopt wide flat priors on them:
$A_1,\eta_1 \in [-6, 6]$. The NLA model here is different from the original
linear alignment model \citep{nla_bridle07,nla_hirata07} as $P_\delta$ in
equation~\eqref{eq:nla_PE} is not the linear matter power spectrum but the full
matter power spectrum including nonlinear structure growth and baryonic
feedback (in our fiducial analysis, the matter power spectrum is predicted by
\hmcode{}). Another difference to the original model is that our implementation
of NLA also includes a redshift evolution described by a power law in $c_1(z)$
to capture additional redshift evolution as shown in
equation~\eqref{eq:tatt_c1c2}.

\citet{IAmodel_Campos2022} proposed to select the proper IA model with an
empirical approach based on the difference in $\chi^2$ between models applied
to the real data. We analyze all the blinded data vectors with both TATT and
NLA. For each setup, we look at the difference in the $S_8$ estimates and the
resulting $\chi^2$ of the analyses. We do not see a significant difference in
the projected posterior of $S_8$ ($\Delta S_8 \sim 0.4 \sigma$) nor significant
difference in $\chi^2$ between these two setups. Furthermore, TATT and NLA also
give comparable errors on the projected posterior of $S_8$. We decide to use
TATT as our fiducial model, since it is a more complete model of IA, and it
does not degrade our constraints.

\subsection{Photometric Redshift}
\label{subsec:model_photoz}

As shown in equation~\eqref{eq:lensing_efficiency}, the redshift distributions
in four redshift bins, $n_i(z)~(i=1,2,3,4)$, of the source galaxies are
essential ingredients for modeling the shear-shear angular power spectra
$C_\ell^{ij}$, where $i$ and $j$ are the indices of the tomographic redshift
bins.
The HSC-Y3 redshift distributions and their uncertainties are inferred jointly
by the photometric redshift estimation, described in
Section~\ref{subsec:data_photoz}, and by spatial cross-correlations between the
HSC-Y3 shape catalog and the CAMIRA-LRG catalog \cite{CAMIRA_Oguri2014,
CAMIRA_HSC_Oguri2018, CAMIRA3}. Here we provide a brief overview of the
inference process, and we refer the readers to \citet{HSC3_photoz_Rau2022} for
the details.

The redshift distribution of each tomographic bin is modeled as a discrete
probability density function on redshift grids ranging from $z=0$ to $z=4$,
with a $0.025$ step size. In Fig.~\ref{fig:data_pzs} (black lines), we show the
$n_i(z)$ distributions of the joint redshift estimation. The redshift
distributions shown are modeled as a logistic Gaussian process, of which the
parameters are inferred by
\begin{enumerate}
    \item the \dnnz{} photo-$z$ estimation and a model for the cosmic variance
        for all redshifts;
    \item cross-correlation between the photometric samples and the CAMIRA-LRG
        samples between $z= 0.0$ and 1.2\,.
\end{enumerate}
We note that the CAMIRA-LRG sample (the grey histogram in
Fig.~\ref{fig:data_pzs}), which is used for cross-correlation calibration,
covers only part of the redshift range of bin~3, and does not cover any of
bin~4.

In order to quantify and marginalize over the redshift distribution
uncertainty, we allow the mean redshift distribution of each bin to shift by
$\Delta z_i$, namely,
\begin{equation}
\label{eq:mdoel_nz_shift}
    n_i(z) \longrightarrow n_i(z + \Delta z_i).
\end{equation}
\citet{HSCY3_NZ_ERR} demonstrated that this shift model is sufficient for
capturing the uncertainty in redshift distribution for the HSC-Y3 cosmic shear
analysis, and is computationally inexpensive, thus we use it here. As a
result, four extra redshift parameters corresponding to four tomographic bins
are included in the fiducial analysis of this work.

\citet{HSC3_photoz_Rau2022} derived the priors on $\Delta z_i$ using the model
difference, i.e., differences of the inferred $n_i(z)$ between three
photometric redshift models (see Fig.~\ref{fig:data_pzs}) and the reference
CAMIRA-LRG sample. We refer the readers to Section~5.7 of
\citet{HSC3_photoz_Rau2022} for more details on how the priors are determined.

Taking into account the fact that the redshift distributions of galaxies in
bin~3 and bin~4 are only partially calibrated by spatial cross-correlation with
CAMIRA-LRGs, we compare the cosmological constraints with two types of priors
on the redshift shifting errors $\Delta z_i$ in Section~\ref{subsec:inter_sys}.
These two types of priors are summarized as follows:
\begin{enumerate}
    \item Informative Gaussian priors estimated by \citet{HSC3_photoz_Rau2022}
        for four redshift bins. The priors for the first two redshift bins are
        shown in Table~\ref{tab:parameters}, and the last two redshift bins are
        ${\cal N}(0, 0.031)$ and ${\cal N}(0, 0.034)$, respectively;
    \item Informative Gaussian priors \citep{HSC3_photoz_Rau2022} for bin~1 and
        bin~2; and uninformative flat priors between $-1$ and $1$ for bin~3 and
        bin~4.
\end{enumerate}

As will be shown in Section~\ref{subsec:inter_sys}, we find that when adopting
the uninformative flat prior on bin~3 and bin~4, the posteriors on $\Delta z_3$
and $\Delta z_4$ are not consistent with zero, indicating that the true
redshift distributions of the last two redshift bins are higher by $\Delta z
\sim 0.1$ than that estimated by \cite{HSC3_photoz_Rau2022}.  This result is
seen in each of the three blinded catalogs. In Section~\ref{subsec:inter_sys},
we simulate noisy mock data vectors for the three blinded catalogs and find
that such large positive values for $\Delta z_3$ and $\Delta z_4$ are not
likely be due to statistical uncertainties, and thus are likely to be real.
Furthermore, the estimate of $S_8$ is $\sim1~\sigma$ higher for the analysis
using the informative Gaussian priors on bin~3 and bin~4. Therefore, we suspect
that the photo-$z$ inferred redshift distributions of the last two redshift
bins are systematically biased. We leave the calibration of high redshift bins
and their impact on cosmic shear to future studies. The fiducial priors on the
$\Delta z_i$ are listed in Table~\ref{tab:parameters}.

\subsection{Shear Calibration Bias}
\label{subsec:model_dm}

As presented in \citep{HSC3_catalog_Li2021}, our shear catalog is calibrated
with image simulations downgrading the HST images in F814W band to the HSC
observational conditions. Specifically, we model and calibrate the shear
estimation bias including galaxy model bias \citep{modelBias_Bernstein10},
noise bias \citep{noiseBiasRefregier2012}, selection bias \citep{kaiserFlow},
and detection bias \citep{metaDet_Sheldon2020} using realistic image
simulations. In addition, we confirm that the shear estimation bias due to the
blending of galaxies located at different redshifts is small for the HSC-Y3
weak-lensing science.

We also model and marginalize over the uncertainties from the multiplicative
bias residual after the aforementioned calibration due to the assumptions and
the limited galaxy number in the simulations. To be more specific, we follow
\citep{cosmicShear_DESY3_Amon2021} to introduce a nuisance parameter $\Delta
m^{(i)}$ to the $i$th redshift bin (where $i=1,\dots4$) to represent the
redshift-dependent multiplicative bias residual. The theoretical prediction for
the cosmic shear 2PCFs is modified as
\begin{equation}\label{eq:dmtransform}
    \xi^{ij}(\theta) \longrightarrow (1 + \Delta m_i)
    (1+\Delta m_j) \xi^{ij}(\theta)
\end{equation}
The prior range of $\Delta m_i$ is taken to be Gaussian with zero mean and a
standard deviation of $0.01$, which is motivated by the calibration of the
HSC-Y3 shear catalog based on image simulations \citep{HSC3_catalog_Li2021}
since it is confirmed that the multiplicative bias residual is controlled below
the $1\%$ level.

\subsection{PSF Systematics Model}
\label{subsec:model_psfsys}

\begin{figure*}
\includegraphics[width=0.90\textwidth]{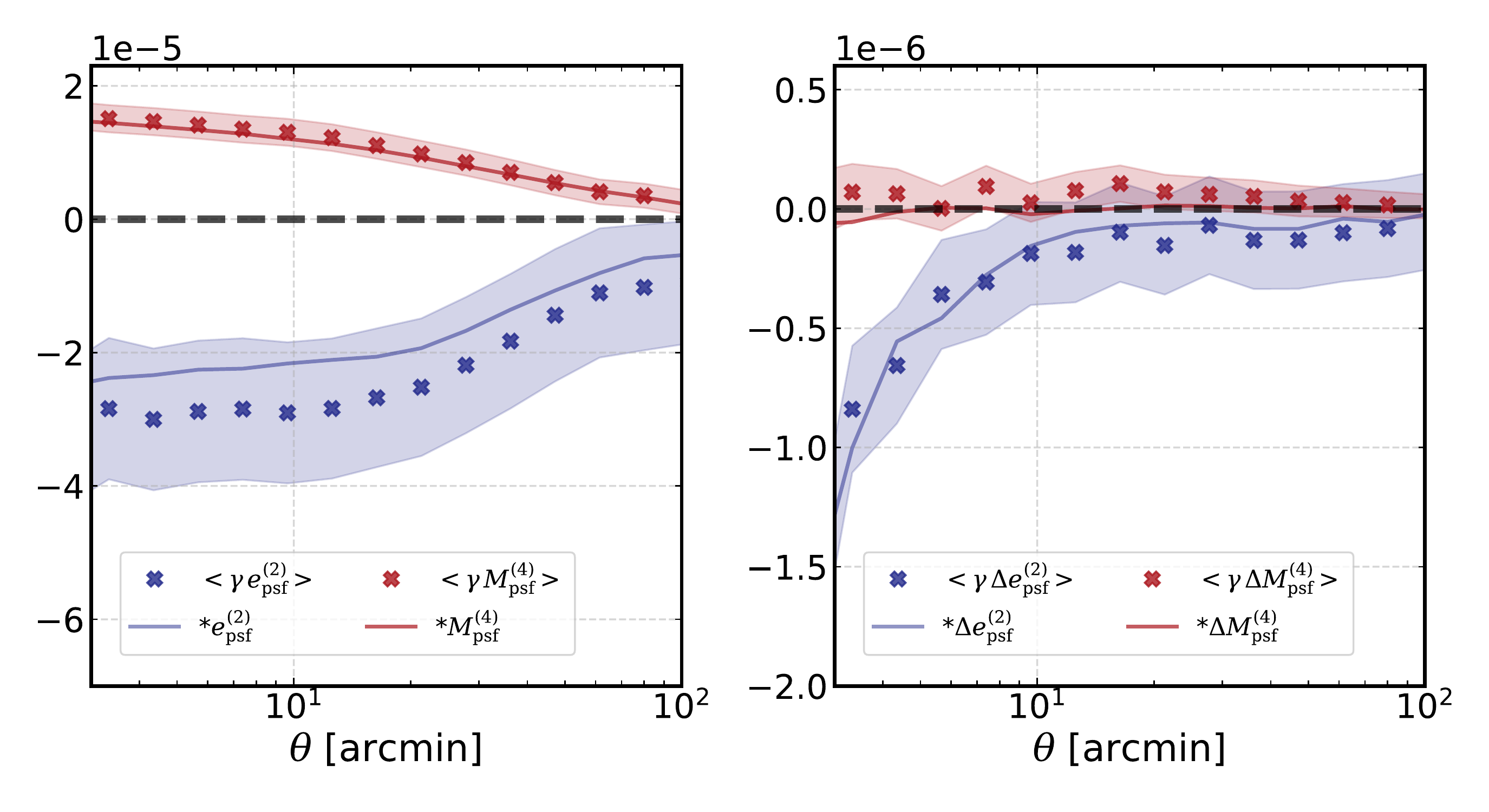}
\caption{
    Galaxy-star correlations, including cross-correlations between galaxy shape
    and star shape (left panel), and also between galaxy shape and star shape
    errors (right panel). We show the correlations to both second-
    ($e^{(2)}_\text{psf}$) and fourth-order ($M^{(4)}_\text{psf}$) star shapes
    and star shape errors. The points are the measurements from HSC data, and
    the solid lines are the best-fit model using star-star correlations (see
    equations~\eqref{eq:null1}--\eqref{eq:null4}).
}
\label{fig:model_psf_gepsf}
\end{figure*}

In this section, we describe the model for PSF-related additive systematics in
this work. The additive bias changes the shear signal as $\gamma \rightarrow
\gamma + \gamma_\text{sys}$\,. As \citet{HSC3_PSF} found, in addition to the
second-order radial moments of the PSF, the spin-2 component of the
fourth-order PSF moments can also cause significant leakage and modeling errors
in shear for the \reGauss{} shear estimator. Therefore, we include both
second-order and fourth-order PSF shapes in the model for the PSF additive bias
as
\begin{equation}
\label{eq:PSF_additive}
\gamma_\mathrm{sys} = \alpha^{(2)} e^{(2)}_\text{psf} + \beta^{(2)} \Delta e^{(2)}_\text{psf}
    + \alpha^{(4)} M^{(4)}_\text{psf} + \beta^{(4)} \Delta M^{(4)}_\text{psf},
\end{equation}
where the first and third terms are the PSF leakage bias by the PSF second- and
fourth-order moments, and the second and fourth terms are the PSF modeling
error in the second- and fourth-order moments.

We find the prior on the PSF systematics parameters by cross-correlating the
shapes and the shape modeling errors of PSF stars with the galaxy shapes. The
cross-correlations are modeled by
\begin{widetext}
\begin{align}
\label{eq:null1}
\langle \hat{\gamma}_\mathrm{gal} \, e^{(2)}_\text{psf}\rangle
&= \alpha^{(2)} \langle e^{(2)}_\text{psf} e^{(2)}_\text{psf} \rangle
+ \beta^{(2)}\langle \Delta e^{(2)}_\text{psf}  e^{(2)}_\text{psf} \rangle
+ \alpha^{(4)} \langle M^{(4)}_\text{psf}  e^{(2)}_\text{psf} \rangle
+ \beta^{(4)} \langle \Delta M^{(4)}_\text{psf}  e^{(2)}_\text{psf}\rangle \\
\label{eq:null2}
\langle \hat{\gamma}_\mathrm{gal} \, \Delta e^{(2)}_\text{psf} \rangle
&= \alpha^{(2)} \langle  e^{(2)}_\text{psf} \Delta e^{(2)}_\text{psf}\rangle
+ \beta^{(2)} \langle \Delta e^{(2)}_\text{psf} \Delta e^{(2)}_\text{psf}\rangle
+ \alpha^{(4)} \langle M^{(4)}_\text{psf} \Delta e^{(2)}_\text{psf} \rangle
+ \beta^{(4)} \langle \Delta M^{(4)}_\text{psf} \Delta e^{(2)}_\text{psf} \rangle \\
\label{eq:null3}
\langle \hat{\gamma}_\mathrm{gal} \, M^{(4)}_\text{psf}\rangle
&= \alpha^{(2)} \langle  e^{(2)}_\text{psf}  M^{(4)}_\text{psf} \rangle
+ \beta^{(2)} \langle \Delta e^{(2)}_\text{psf} M^{(4)}_\text{psf} \rangle
+ \alpha^{(4)} \langle  M^{(4)}_\text{psf}  M^{(4)}_\text{psf} \rangle
+ \beta^{(4)} \langle \Delta M^{(4)}_\text{psf}  M^{(4)}_\text{psf} \rangle \\
\label{eq:null4}
\langle \hat{\gamma}_\mathrm{gal} \, \Delta M^{(4)}_\text{psf}\rangle
&= \alpha^{(2)} \langle  e^{(2)}_\text{psf} \Delta M^{(4)}_\text{psf} \rangle
+ \beta^{(2)} \langle \Delta e^{(2)}_\text{psf} \Delta M^{(4)}_\text{psf} \rangle
+ \alpha^{(4)} \langle  M^{(4)}_\text{psf} \Delta M^{(4)}_\text{psf} \rangle
+ \beta^{(4)} \langle \Delta M^{(4)}_\text{psf} \Delta M^{(4)}_\text{psf} \rangle \,.
\end{align}
\end{widetext}
The left hand side of equation~\eqref{eq:null1}--\eqref{eq:null4} are
correlation functions between the galaxy shape and PSF moments, which we call
the ``g-p correlation''. The correlation functions on the  right hand side are
PSF-PSF correlation functions, which we call the ``p-p correlation''. We show
the measurements and the best-fit models of all four g-p correlations in
Fig.~\ref{fig:model_psf_gepsf} using the catalog with blinding ID 0, which
incidentally happened to be the true shear catalog after unblinding. To find
the prior for the PSF systematics parameters, we calculate covariance matrices
for the g-p correlations of all three blinded catalogs using the mock catalogs
described in section~\ref{subsec:data_mock}. The prior of the true catalog
(blinded catalog 0) are listed in Table~\ref{tab:parameters}. Since
\citet{HSC3_PSF} found that $\alpha^{(2)}$ and $\alpha^{(4)}$ are correlated,
we include the correlation between all PSF parameters in the prior. The
correlation between the PSF parameters does not impact the results
significantly, which is consistent with the finding in \citet{HSC3_PSF}.

\citet{HSC3_PSF} find significant bias on $\xi_+$ due to PSF systematics,
whereas the bias on $\xi_-$ to be negligible. We model the impact of PSF
additive bias on $\xi_+$ by
\begin{equation}
\label{eq:expand-esys_esys}
\xi_+(\theta) \longrightarrow \xi_+(\theta) +
\sum_{k=1}^4 \sum_{q=1}^4 p_k p_q \langle S_k S_q \rangle,
\end{equation}
where $\vec{p} = (\alpha^{(2)}, \beta^{(2)}, \alpha^{(4)}, \beta^{(4)})$ is the
parameter vector, and $\vec{S} = (e^{(2)}_\text{psf}, \Delta
e^{(2)}_\text{psf}, M^{(4)}_\text{psf}, \Delta M^{(4)}_\text{psf})$ is the PSF
moments vector. We note that the PSF additive systematics are added to the 2PCFs
after the rescaling from multiplicative bias.

In order to account for the correlation in the prior of PSF systematic
parameters, we sample four uncorrelated parameters, $\vec{p}' =
\left(\alpha'^{(2)}, \beta'^{(2)}, \alpha'^{(4)}, \beta'^{(4)} \right)$, with
uncorrelated Gaussian priors. We then transform these parameters into our
original parameters by the following invertible transform:
\begin{equation}
    \vec{p} = \mathbf{T} \cdot \vec{p}' + \vec{\bar{p}},
\end{equation}
where $\vec{\bar{p}}$ is the average of the original PSF systematic parameters,
$\mathbf{T} = \mathbf{V}^{\frac{1}{2}} \mathbf{U}$, $\mathbf{V}$ is a diagonal
matrix with eigenvalues of $\vec{p} - \vec{\bar{p}}$'s covariance matrix as the
diagonal elements, and each column of $\mathbf{U}$ is a eigenvector of the
covariance matrix. In section~\ref{subsec:inter_sys}, we show that the bias
from not including the correlation between $\alpha^{(2)}$ and $\alpha^{(4)}$ is
negligible for HSC-Y3 cosmic shear analysis.

\citet{HSC3_PSF} conducted other extensive mock tests, where they investigated
PSF systematics models taking into account redshift dependency, second-order
terms, and PSF versus non-PSF stars. They found that the above modeling of PSF
additive systematics is sufficient for the HSC-Y3 cosmic shear analysis. We
refer the readers to \citet{HSC3_PSF} for further details about PSF additive
systematics in the HSC-Y3 shear catalog.

\subsection{Bayesian Inference}
\label{subsec:model_baysian}

We use a Monte Carlo Bayesian analysis to sample the posterior in the
23-dimensional space of the cosmological, astrophysical and systematic
parameters. We denote the vector of parameters as $\Theta = (\Omega_\mathrm{m},
A_s, h_0, \dots)$, and the model prediction of 2PCFs, with 140 dimensions (14
angular bins for each of the 10 correlation functions across 4 redshift bins),
as $\xi_\pm(\Theta)$. We adopt a Gaussian likelihood $\mathcal{L}$:
\begin{equation}
    \mathrm{ln}\, \mathcal{L} (\widehat{\xi}_\pm \,|\, \Theta ) =
    -\frac{1}{2} \left(\widehat{\xi}_\pm - \xi_\pm (\Theta)\right)^T
    \mathbf{C}^{-1}
    \left( \widehat{\xi}_\pm - \xi_\pm(\Theta)\right)\,,
\end{equation}
where $\widehat{\xi}_\pm$ is the measured 2PCFs as shown in
Fig.~\ref{fig:data_2pcf} and $\mathbf{C}$ is the covariance matrix estimated
from 1404 mock catalogs with the WMAP9 cosmology, which is shown in
Fig.~\ref{fig:data_cov}. Note, as discussed in Section~\ref{subsec:meas_cov},
we neglect the dependency of the covariance matrix on cosmological parameters.
$\mathbf{C}^{-1}$ is the precision matrix, namely the inverse of the covariance
matrix. When estimating the inverse matrix, we correct for noise bias by
multiplying the numerical inverse of the noisy estimate of covariance by the
Hartlap factor \citep{covariance_Hartlap2007}: $(1404 - 140 -2) / (1404 - 1)
\sim 0.9$\,. With Bayesian inference,  we construct a posterior probability
distribution, denoted as $\mathcal{P}(\Theta\,|\, \widehat{\xi}_\pm)$ for the
parameters $\Theta$, given the data vector $ \widehat{\xi}_\pm$:
\begin{equation}
    \mathcal{P}(\Theta \,|\, \widehat{\xi}_\pm)
    \propto \mathcal{L}(\widehat{\xi}_\pm \,|\, \Theta)
    \Pi(\Theta),
\end{equation}
where $\Pi(\Theta)$ is the prior distribution of $\Theta$\,.

Markov Chain Monte Carlo (MCMC) and Nested Sampling are widely used in the
cosmology community to sample posteriors in high-dimensional parameter space.
MCMC methods directly generate samples from the posterior in high-dimensional
parameter space, whereas nested samplers map the high-dimensional posterior
onto a one-dimensional space and divide the posterior into many nested
``slices''. After generating samples from the ``slices'', they recombine the
samples with appropriate weights to reconstruct the posterior. In this paper,
we compare the constraints from three different samplers, \emcee{}
\citep{emcee_Foreman2013}, \multinest{} \citep{multinest_Feroz2009} and
\polychord{} \citep{polychord_Handley2015} implemented in \cosmosis{}:
\begin{enumerate}
    \item \emcee{} is an affine-invariant ensemble sampler for MCMC;
    \item \multinest{} is a Nested Sampler using a k-means clustering
        algorithm with ellipsoid bounds;
    \item \polychord{} is a Nested Sampler using slice sampling to sample
        within the nested isolikelihoods contours.
\end{enumerate}

We use \polychord{} for our fiducial analysis since, as pointed out by
\citet{DESY3_sampler_Lemos}, the marginalized posterior widths for
$\Omega_\mathrm{m}$ and $\sigma_8$ estimated by \multinest{} are 10\% smaller
and are probably under-estimated. Moreover, \multinest{} gives a biased
estimation of evidence. However, \multinest{} is about five times faster than
\polychord{}, so we utilize \multinest{} for our internal consistency tests.
Since the estimated posterior widths from \multinest{} are systematically
smaller, it is conservative to use the posteriors from \multinest{} for
internal consistency tests. In addition, we compare the posteriors estimated
from \polychord{} and \multinest{} to the estimation from \emcee{}.

To assess the convergence of our chains, we check that, at the end of the
chains, the normalized nested weight (the weight, at each estimation, divided
by the maximum weight in the chains) has stopped increasing and is close to
zero. In addition, we also use \texttt{nestcheck} \citep{nestcheck2019}, a
public software \footnote{\url{https://github.com/ejhigson/nestcheck}}, to
confirm that the posterior mass has peaked out, which indicates that most of
the posterior mass contribution is well sampled. Also, we confirm that the
uncertainty of the $S_8$ posterior distribution is reasonably small. The setups
for the three MC samplers are shown in Table~\ref{tab:sampler_setups}.

\begin{table}
\setlength{\tabcolsep}{1.0em}
\begin{center}
\caption{
    The setups of the three MC samplers. \polychord{} has three
    hyper-parameters: the number of live points, number of repetitions
    $n_\text{repeat}$, and tolerance \texttt{tol}. \texttt{efr} is the sampling
    efficiency for \multinest{}. \emcee{} uses $10,000$ samples with $80$
    walkers and \texttt{nstep}$=5$\,.
}
\label{tab:sampler_setups}
\begin{tabular}{|l|l|l|}
\hline
\polychord{}           & \multinest{}           & \emcee{} \\ \hline
$n_\text{live} = 500$  & $n_\text{live} = 500$  & $n_\text{sample} = 10^{4}$\\
$n_\text{repeat}=20$   & $n_\text{efr}=0.3$     & $n_\text{walker} = 80$\\
\texttt{tol} $=0.01$   & \texttt{tol} $=0.05$   & $n_\text{step}  =5$\\
\hline
\end{tabular}
\end{center}
\end{table}

We report the 1D marginalized mode and its asymmetric $\pm 34\%$ confidence
intervals, together with the MAP estimated as the maximum of the posterior in
the chain returned by the nested sampler:
\begin{equation}
    \label{eq:report_format}
    \text{1\text{D} mode}^{+34\%\,\text{CI}}_{-34\%\,\text{CI}}
    ~(\text{MAP from nested chain})\,.
\end{equation}
We note that the DES cosmic shear analysis \citep{DESY3_CS_Secco2022,
DESY3_CS_Amon2021} reported a projected mean estimated from the nested chain
sampled with \polychord{}; whereas KiDS reported the MAP estimated with
\textsc{MaxLike} implemented in \cosmosis{} as their reported point estimation,
and they report a hybrid confidence interval that is estimated on the joint,
multi-dimensional highest posterior density region, but projected onto the
marginal posterior of the parameter under consideration
\citep{KiDS1000_3x2pt_model}. We choose to report the projected mode since it
is less sensitive than the projected mean to the tails of the projected 1D
posterior. In addition, the projected mode is more stable than the MAP, and its
confidence interval is mathematically well-defined and simple to estimate.

\subsection{Model Validation}
\label{subsec:model_valid}

\begin{figure}
\includegraphics[width=0.48\textwidth]{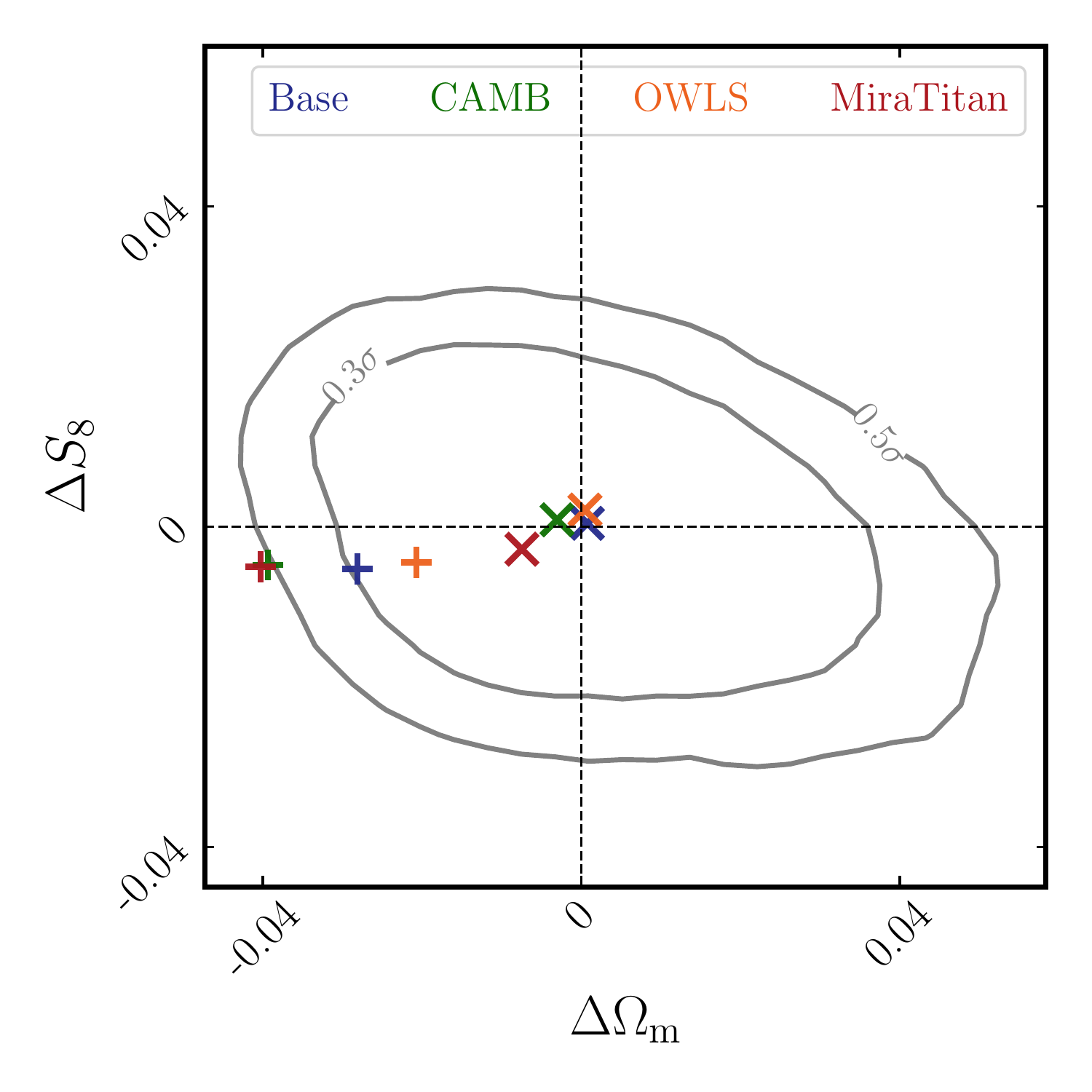}
\caption{
    The modeling errors (estimated parameters - true parameters) in the MAP
    (`$\cross$') and projected 1D mode (`$+$') when applying our model to
    different synthetic mocks of 2PCFs including (from left to right in the
    legend) the baseline simulation, simulation with the \camb{} linear power
    spectrum, \owlsAgn{} suppression, and the\texttt{MiraTitan~II} nonlinear
    power spectrum. The grey lines are $0.3\,\sigma$ and $0.5\,\sigma$ contours
    of the 2D projected posterior in the parameter space of ($\OmegaM$, $S_8$).
}
\label{fig:model_tests}
\end{figure}
In our fiducial analysis, we use the \bacco{} emulator to model the linear
matter power spectrum (Section~\ref{subsec:model_ps}); $\hmcode$ 2016 for
nonlinear matter power spectrum and baryonic feedback
(Section~\ref{subsec:model_ps}); and TATT for intrinsic alignment
(Section~\ref{subsec:model_ia}). The redshift distributions are calibrated by
cross-correlating with CAMIRA-LRGs, and the redshift estimation error is
modelled with shifting errors $\Delta z$, adopting a flat prior in the last two
redshift bins (Section~\ref{subsec:model_photoz}). Additionally, we use
redshift-dependent multiplicative bias residuals
(Section~\ref{subsec:model_dm}) and a PSF systematic model with fourth-order
shape leakage and shape error (Section~\ref{subsec:model_psfsys}). We perform
Bayesian analysis using the nested sampler \polychord{}
(Section~\ref{subsec:model_baysian}).

We validate our model with noiseless synthetic 2PCFs simulated with different
models for the matter power spectrum. We first make a baseline
(systematics-free) simulation using the fiducial model and the $\OmegaM$ and
$A_s$ from the WMAP9 cosmology, and other parameters are from the MAP of the
cosmology constraint using the blinded catalog 0 with our fiducial setup. Then
we change the models in the simulation pipeline to simulate ``contaminated''
data vectors, and check the biases of analyses on $\OmegaM$ and $S_8$ for these
``contaminated'' data. Specifically, we test the following ``contaminated''
models:
\begin{enumerate}
    \item Simulation with \camb{} linear power spectrum instead of the \bacco{}
        emulator;
    \item Simulation with baryonic rescaling from \texttt{OWLS-AGN}
        \citep{OWLS_Schaye2010, OWLS_vanDaalen2011} instead of \hmcode{} 2016;
    \item Simulation with nonlinear power spectrum from \texttt{MiraTitan~II}
        emulator \citep{cosmicEmu2022} and baryonic rescaling from \owlsAgn{}
        instead of \hmcode{} 2016;
\end{enumerate}

We add the baryonic feedback into our synthetic data vectors using a rescaling
scheme proposed by \citet{DESY3_CS_Amon2021}. Specifically, the power spectrum
with baryonic physics is simulated by multiplying a scale dependent suppression
factor to the dark-matter-only power spectrum --- namely the \hmcode{} 2016
nonlinear power spectrum without baryonic feedback ($A_\text{b} = 3.13$) for
our case:
\begin{equation}
    P_{\text{m,b}}(k, z) = \frac{P_\text{hydro}(k, z)}{P_\text{DM}(k,z)}
    P_m(k,z \,|\, A_\text{b} = 3.13),
\end{equation}
where $P_\text{hydro}(k, z)$ is the power spectrum measured from hydrodynamic
simulations (e.g., \owlsAgn{} \citep{OWLS_Schaye2010}, \cowls{}
\citep{cowls_LeBrun2014}, \illustris{} \citep{Illustris_Vogelsberger2014},
\massiveblackII{} \citep{massiveBlackII_Khandai2015}, \eagle{}
\citep{eagle_Crain2015}, \horizonAgn{} \citep{horizonAgn_Kaviraj2017}, and
\illustrisTNG{} \citep{IllustrisTNG_Nelson2019}), and $P_\text{DM}(k,z)$ is the
power spectrum measured from dark matter only simulations of the same suite.

In order to assure that the modeling errors from both baryonic physics and
nonlinear structures for our fiducial scale cut defined in
Section~\ref{subsec:meas_2pcf} are not significant in our analysis, we follow
the DES-Y3 cosmic shear analysis to check whether the amplitude of the 2D
MAP estimation bias (denoted as $\vec{b}_\mathrm{2D}$) on the plane of
($\OmegaM$, $S_8$) is less than $0.3~\sigma$, when applying our fiducial model
to the \owlsAgn{}, CAMB and \texttt{MiraTitan~II} simulations:
\begin{equation}
    \sqrt{\vec{b}_\mathrm{2D}^{T} \,\mathbf{\Sigma}^{-1}\,
    \vec{b}_\mathrm{2D}} < 0.3\,,
\end{equation}
where $\mathbf{\Sigma}$ is the covariance matrix of the 2D posterior on the
plane of ($\OmegaM$, $S_8$) estimated from the first blinded catalog. In
addition, to make sure that our model is not significantly influenced by the
modeling, we test our fiducial model and scale cut with the synthetic
simulation using both \owlsAgn{} baryonic suppression and the \texttt{MiraTitan
II} nonlinear power spectrum. In this paper, we adopt the \cosmosis{}
implementation of \textsc{MaxLike}
\footnote{https://github.com/joezuntz/cosmosis/tree/main/cosmosis/samplers/maxlike},
which is a wrapper of the \texttt{scipy} minimizer, to estimate MAPs. To be
more specific, we use the Nelder-Mead minimizer \citep{NeldMead65} with
tolerance set to $10^{-6}$. In order to reduce the numerical error in the MAP
estimation, we follow \citet{KiDS1000_3x2pt_model} to run the minimizer with 50
different starting points and take the final MAP to be the result with the
largest posterior. The starting points are varied in the parameters of interest
in relation to the model choice, including three cosmology parameters
$\OmegaM$, $A_s$, $A_\text{b}$ and four intrinsic alignment parameters, $A_1$,
$A_2$, $\eta_1$ and $\eta_2$\,. Note, all 23 parameters vary during each MAP
estimation, although the starting points are randomly sampled in only these
seven dimensions. We find that the MAP estimate does not change when adding
additional points for $\sim$$30$ starting points. Our results are shown as
`$+$' points in Fig.~\ref{fig:model_tests}, and the maximum 2D bias we found is
$\sim$$0.2\,\sigma$ for the \owlsAgn{} $+$ \texttt{MiraTitan~II} simulation,
which is less than the requirement threshold of $0.3\,\sigma$\,.

Furthermore, we check the influence of modeling errors on the 1D projected
mode since it is the point estimation we will report as discussed in
Section~\ref{subsec:model_baysian}. The results are shown as ``$\times$'' points
in Fig.~\ref{fig:model_tests}. We find that the 1D biases on $S_8$ are
about $-0.14\,\sigma$ for all the simulations with our fiducial scale cut;
however, the 1D biases on $\OmegaM$ range from $-0.4\,\sigma$ to $-0.8\,\sigma$
for different simulations. The differences between the biases on MAPs and the
biases on projected modes are mainly caused by projection effects in the
projected point estimation. As shown in \citet{KiDS1000_3x2pt_model}, the
projection effects can cause about a $1\,\sigma$ bias on the 1D projected point
estimation. Based on the tests shown here we conclude that the bias (due to
both modeling error and projection effects) on the 1D projected mode of $S_8$
is less than $0.15\,\sigma$, which is not significant; however, the systematic
uncertainty (from modeling error and projection effects) on the 1D projected mode
of $\OmegaM$ is significant. We note that the projected 1D mode of
$\OmegaM$ can be biased low by up to $-0.8\,\sigma$. Therefore, when reporting
cosmology constraint, we do not focus on $\OmegaM$\,.

\begin{figure}
\includegraphics[width=0.48\textwidth]{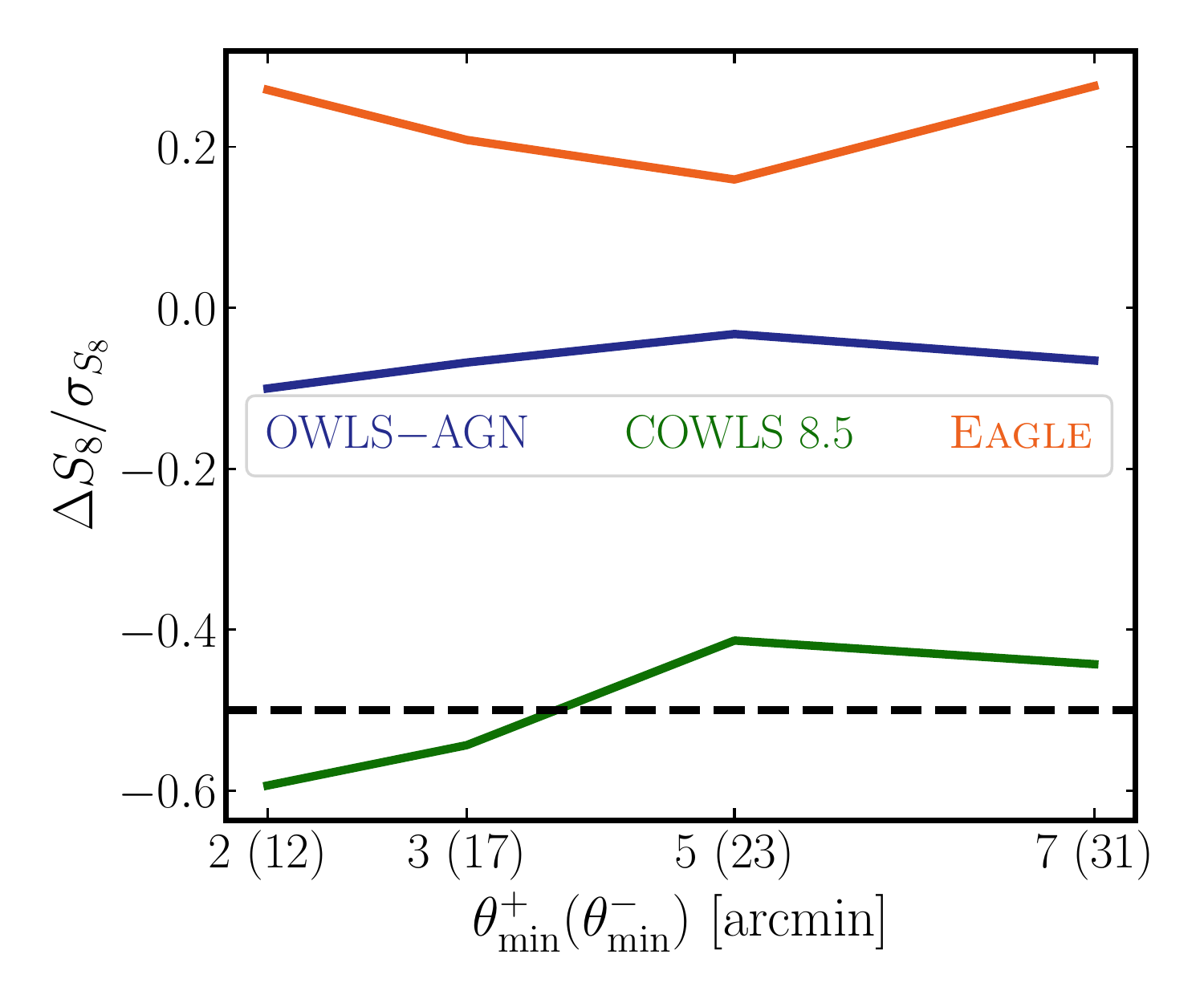}
\caption{
    Tests applying our fiducial model to synthetic mocks of 2PCFs from
    different hydrodynamic simulations. The $y$-axis is the modeling errors in
    the projected 1D mode (estimation - truth) relative to the $1\sigma$
    uncertainty in the projected mode estimated by analyzing real data. The
    $x$-axis is the small-scale cut that we applied. We simultaneously vary the
    cut on $\xi_+$ (denoted as $\theta^{+}_\mathrm{min}$) and the cut on
    $\xi_-$ (denoted as $\theta^{-}_\mathrm{min}$)\,.
    }
    \label{fig:model_tests_hydro}
\end{figure}

In addition to \owlsAgn{}, we also show the bias in projected 1D mode of $S_8$
with a few other extreme simulations (e.g., \cowls{}~8.5 and \eagle{}) as a
function of small-scale cuts on $\xi_{\pm}$ in
Fig.~\ref{fig:model_tests_hydro}. As shown, the \cowls{}~8.5 simulation has a
heating temperature for AGN feedback of $\mathrm{log}(\Theta_\mathrm{AGN}) =
8.5$, and the bias in $S_8$ is about $0.5\sigma$ but larger than \owlsAgn{} for
our fiducial scale cut. The bias may be caused by the extreme baryonic physics
model of the \cowls{}~8.5 simulation.

\begin{figure*}
\subfloat{
    \includegraphics[width=0.48\textwidth]{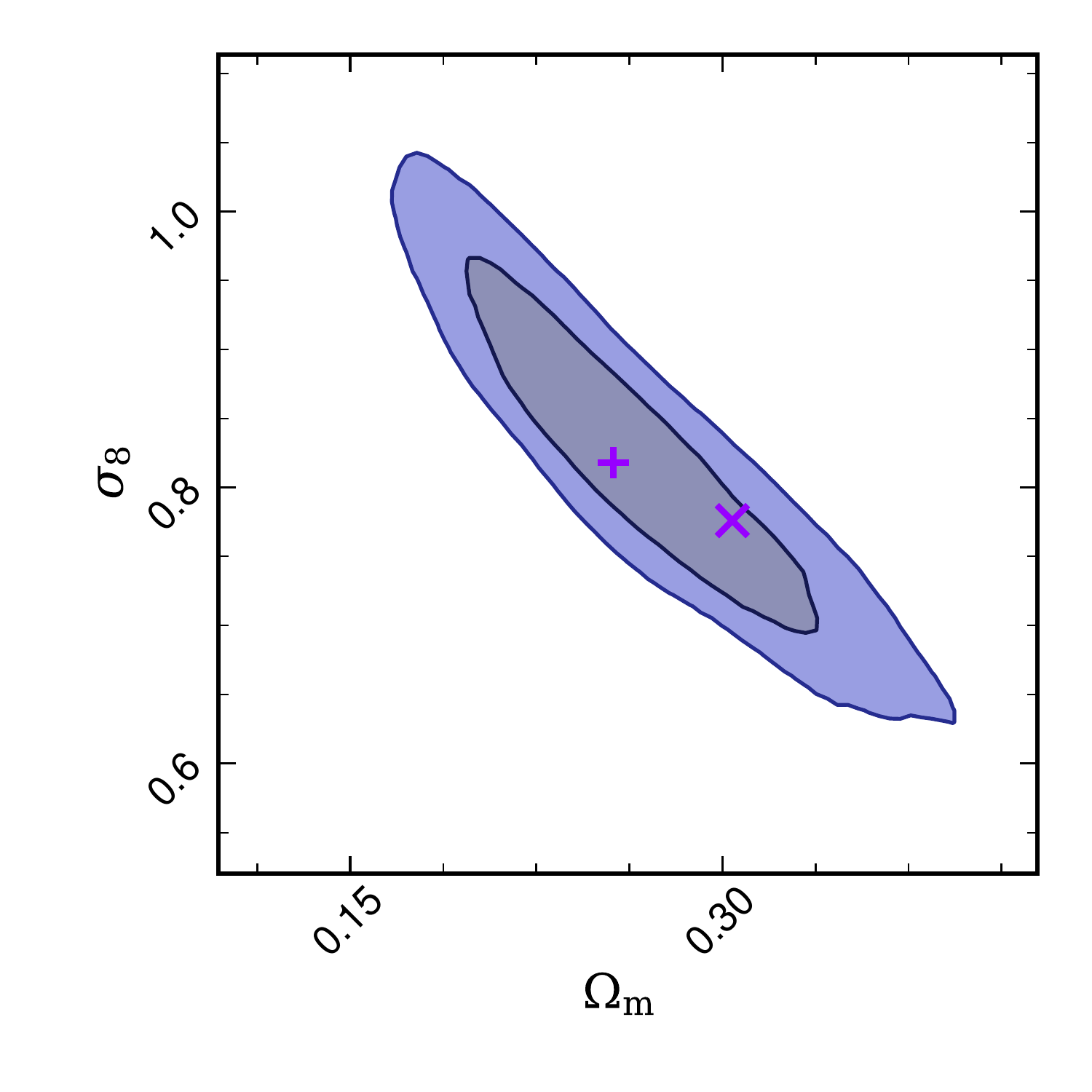}
}
\subfloat{
    \includegraphics[width=0.48\textwidth]{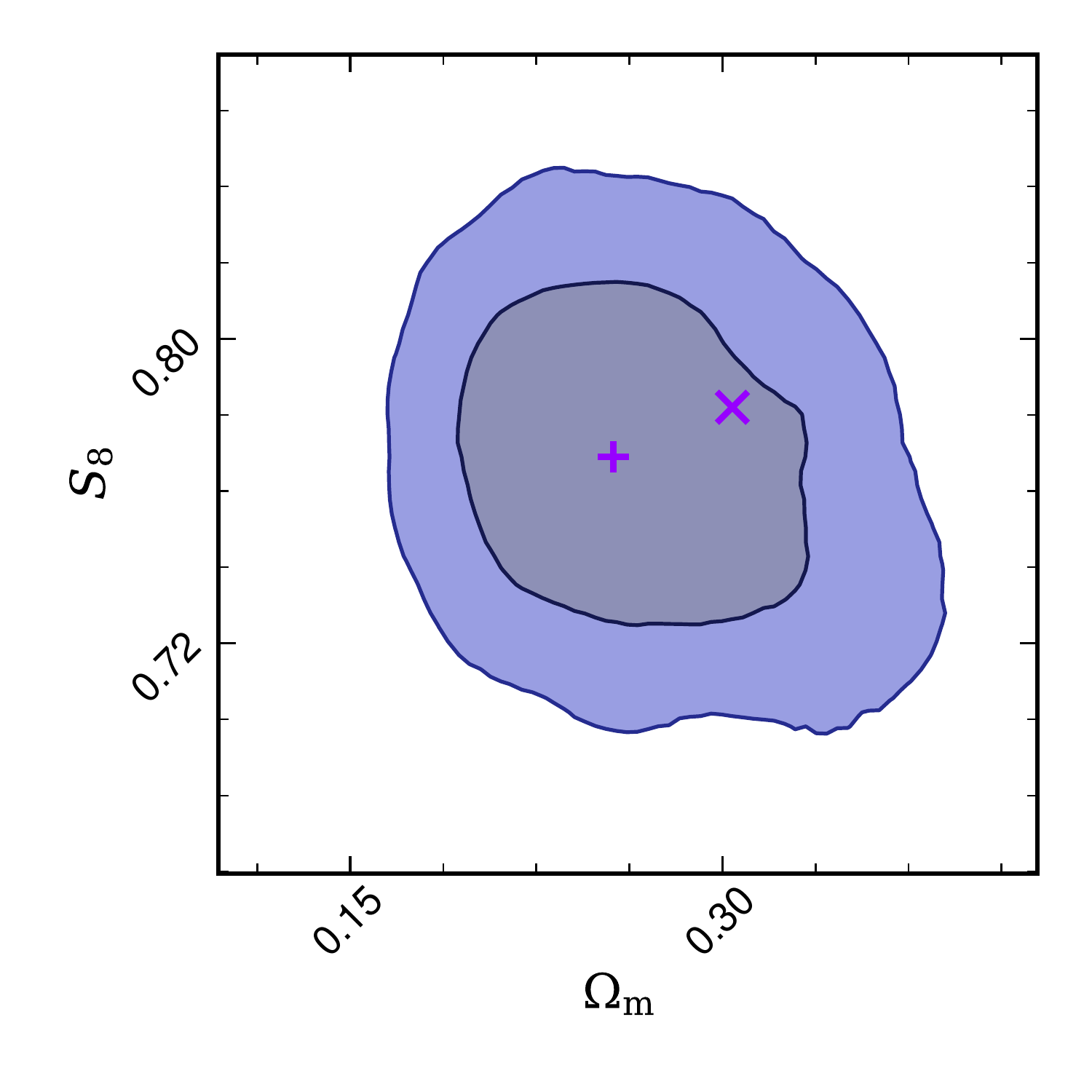}
}
\caption{
    Posterior contours ($68\%$ and $95\%$ CI [For all the 2D posteriors shown
    in this paper, we plot the $68\%$ and $95\%$ CI.]) of the 2D projected
    posterior in the ($\OmegaM$, $\sigma_8$) plane (left panel) and ($\OmegaM$,
    $S_8$) plane (right panel) for our fiducial analysis, where $S_8 = \sigma_8
    \sqrt{\OmegaM/0.3}$\,. In addition, we show the projected 1D mode (`$+$')
    and the MAP estimated from \polychord{} chains (`$\cross$').
    }
    \label{fig:inter_fid_contour}
\end{figure*}

\section{FIDUCIAL CONSTRAINT AND INTERNAL CONSISTENCY}
\label{sec:res_inter}

In this section, we present our cosmology constraints with the 2PCFs measured
in Section~\ref{sec:2pt} and the models introduced in Section~\ref{sec:model}.
First we report our cosmology constraint with our fiducial setup in
Section~\ref{subsec:inter_fid}.

In order to make sure that our fiducial analysis is robust, we conduct various
internal consistency tests by analyzing different subsets of our catalog and
with different analysis setups (e.g., different astrophysical and systematic
models) in the context of the flat $\Lambda$CDM cosmology. First we look into
the differences in our cosmology constraints, especially focusing on $S_8$, for
analyses with different samplers (Section~\ref{subsec:inter_sampler}); flat
priors on different cosmological parameters (Section~\ref{subsec:inter_prior});
different astrophysical models (Section~\ref{subsec:inter_phys}); and different
systematic models (Section~\ref{subsec:inter_sys}). Then we analyze data in
different subfields, with different angular scale cuts, and removing each of
the redshift bins to check the robustness of our results in
Section~\ref{subsec:inter_data}. In addition, we test the influence of $B$-mode
errors in our cosmology constraint.

Note that, we specifically focus on the 1D projected modes of $S_8$ but not
$\OmegaM$ in these tests since, as we saw in Section~\ref{subsec:model_valid},
the constraint on $\OmegaM$ is sensitive to projection effects and modeling
errors in Section~\ref{subsec:inter_bmode}.

Even though we report our constraints obtained from the \polychord{} sampler in
Section~\ref{subsec:inter_fid}, we adopt \multinest{} for most of the
consistency tests in the rest of this section since \multinest{} is much faster
than \polychord{}.

\subsection{Fiducial Constraint}
\label{subsec:inter_fid}

\begin{figure}
\includegraphics[width=0.48\textwidth]{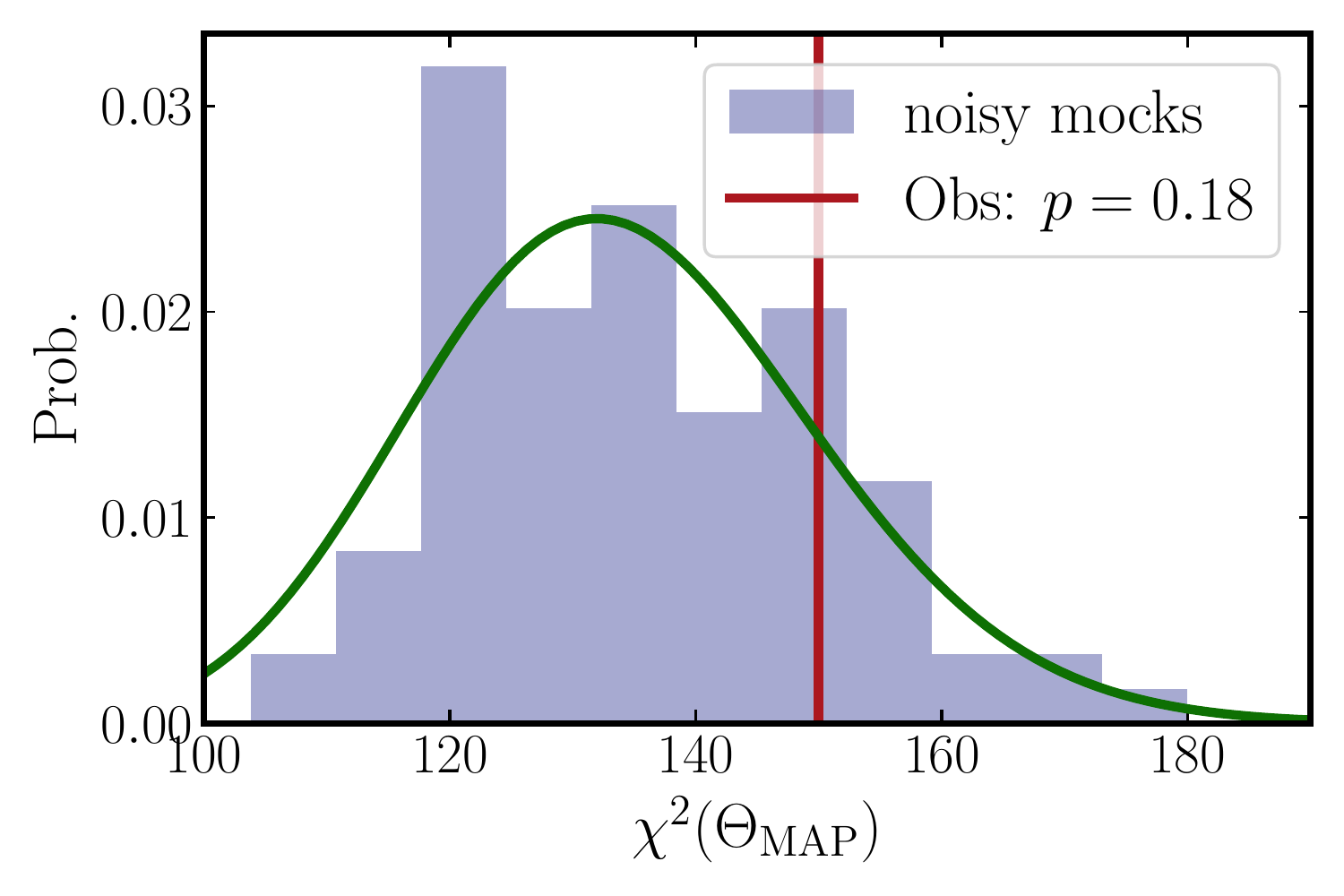}
\caption{
    The evaluation of goodness-of-fit with the $\chi^2$ value at the maximum a
    posteriori (MAP) obtained from the chain of the fiducial analysis (red
    vertical line). The reference distribution (blue histogram) is obtained by
    analyzing the $100$ noisy mock data vectors. The $p$-value is $0.18$\,. The
    green cure is the best-fit $\chi^2$ distribution with effective degrees of
    freedom of $134$.
    }
\label{fig:inter_fid_pvalue}
\end{figure}

Our fiducial cosmology constraint is conducted with the setup outlined at the
start of Section~\ref{subsec:model_valid}. First we report the constraints for
the cosmological parameters,  from our fiducial analysis, following the format
defined in equation~\eqref{eq:report_format}:
\begin{equation}
\begin{split}
    \OmegaM: & \qquad 0.256_{-0.044}^{+0.056} \quad (0.304)\,,\\
    \sigma_8: & \qquad 0.818_{-0.091}^{+0.089} \quad (0.776)\,,\\
    S_8: & \qquad 0.769_{-0.034}^{+0.031} \quad (0.782) \,.
\end{split}
\end{equation}
The marginalized 2D posterior and the point estimates (including the projected
mode and the MAP) of these cosmological parameters are shown in
Fig.~\ref{fig:inter_fid_contour}. We find the projected mode of $\OmegaM$
($S_8$) is less than the MAP by $\sim$$0.9\,\sigma$ ($\sim$$0.4\,\sigma$). It
is consistent with what we found in Fig.~\ref{fig:model_tests} using noiseless
mock 2PCFs that the projected modes are lower than the MAP. In addition to the
cosmological parameters, the redshift shifting errors for the last two redshift
bins estimated with a wide flat prior are:
\begin{equation}
\begin{split}
    \Delta z_3: & \qquad -0.115_{-0.058}^{+0.052} \quad (-0.120)\,,\\
    \Delta z_4: & \qquad -0.192_{-0.088}^{+0.088} \quad (-0.190)\,.
\end{split}
\end{equation}

In both Fourier \citep{HSC3_cosmicShearFourier} and real space cosmic shear
analyses, we employ the \texttt{ChainConsumer} package \citep{ChainConsumer} to
analyze the MC chains and visualize the marginalized posteriors. After
unblinding, we found that the outcomes from \texttt{ChainConsumer} differ
mildly with those from \texttt{GetDist} \citep{GetDist2019}. This discrepancy
arises because \texttt{ChainConsumer} lacks corrections for boundary effects
and biases stemming from chain smoothing. Specifically, for
\texttt{ChainConsumer}, boundary effects lead to inaccuracies in the 1D
marginalized posterior near parameter boundaries predominantly influenced by
top-hat priors. The high-level summary of the significance for our main
reported results on $S_8$ is that the mode value does not change but the
estimated uncertainty on $S_8$ is approximately $10\%$ larger in
\texttt{ChainConsumer} due to the kernel density estimation smoothing.
Nonetheless, we retain in this paper the original parameters and plots, as
unblinded, for transparency. We direct readers to
Appendix~\ref{app:chainconsumer} for an in-depth discussion of these effects.

We evaluate the goodness-of-fit with the value of $\chi^2$ at MAP obtained from
the fiducial MC chain returned by \polychord{}, denoted as
$\chi(\Theta_\text{MAP})$. Since many of the parameters are prior-dominated
(see figures in Appendix~\ref{app:inter}), the calculation of the number of
degrees of freedom is not straightforward. Therefore, we use noisy mocks of
2PCFs simulated according to the covariance matrix for the goodness-of-fit
estimation. As the cosmological parameters were blinded when we did this test,
we use the matter amplitude and matter density parameters from the WMAP9
cosmology but other parameters are from the MAP estimation of the first blinded
catalog. We find, after unblinding, that our best-fit cosmology is very close
to the WMAP9 cosmology. Noises with different realizations are added to the
data vector according to the covariance (corrected by the Hartlap factor) of
the blinded catalog estimated from mocks. We analyze these $50$ mocks using our
fiducial model and, to save computational time, we sample them with
\multinest{}. We obtain the reference $\chi^2$ distribution from the histogram
of the MAPs estimated from the $100$ \multinest{} chains as shown in
Fig.~\ref{fig:inter_fid_pvalue}. By comparing the $\chi^2$ value of 150
obtained from the real data, to the reference $\chi^2$ distribution, we find
the $p$-value $p = 0.18$\,. In conclusion, our measured 2PCFs can be well
described by the best-fit model. In addition, we fit a $\chi^2$ distribution to
the histogram and find that the best-fit effective degrees of freedom amount to
134. Given that the number of data points is $140$, the effective number of
free parameters is calculated to be $140-134=6$\,.

Due to an oversight in the code design, the MAPs of $A_s$ and $\OmegaM$ were
accidentally not blinded when using them as inputs to generate the noisy mocks.
However, we note that the estimated $S_8$, on which our analysis has the
strongest constraining power, was blinded in the analysis process.

\subsection{Priors}
\label{subsec:inter_prior}

\begin{figure*}[!t]
\includegraphics[width=0.95\textwidth]{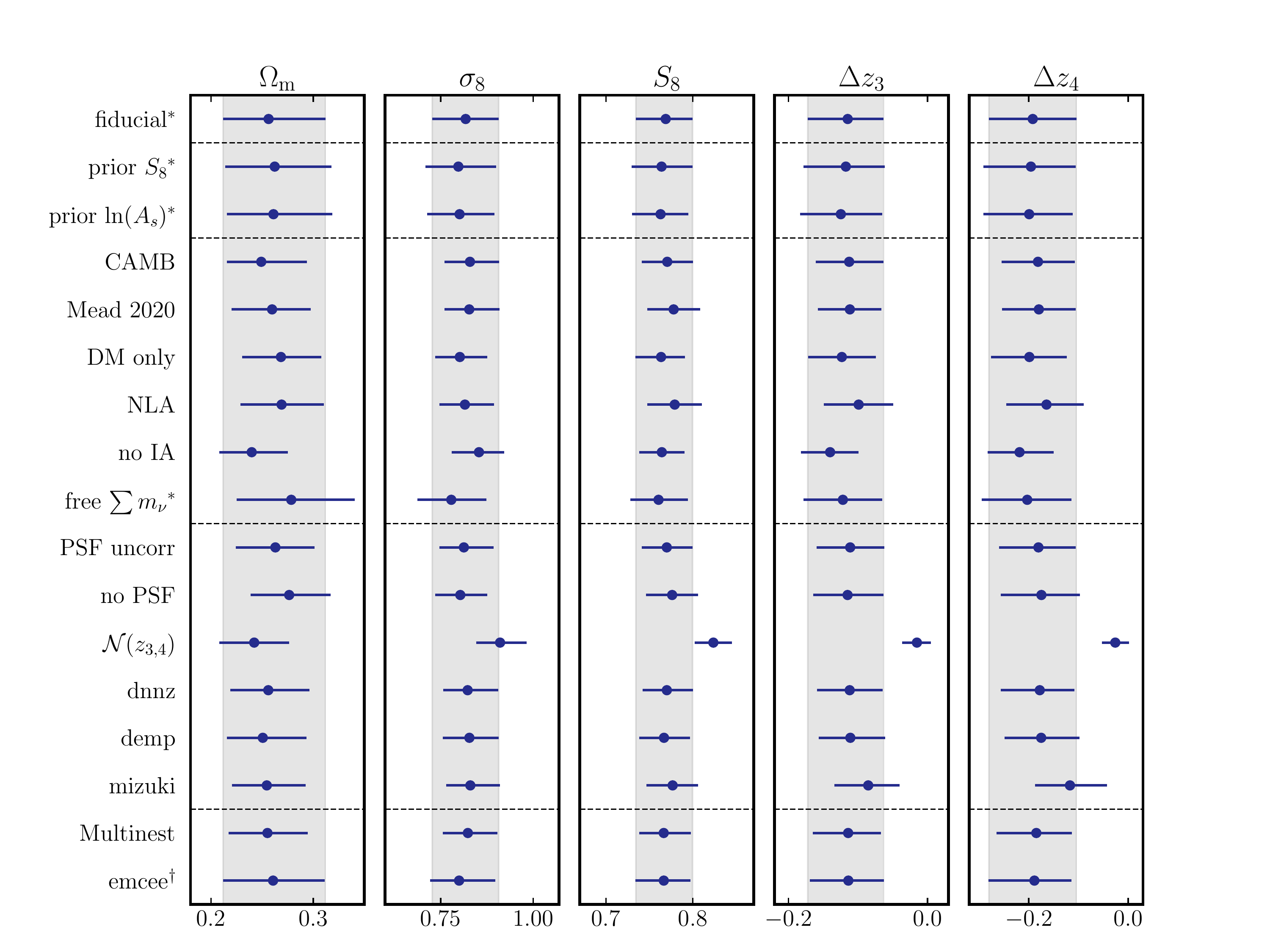}
\caption{
    The $68\%$ CI of the 1D projected posterior on each of the parameters
    $\OmegaM$, $\sigma_8$, $S_8$, $\Delta z_3$ and $\Delta z_4$  for different
    analysis setups. In total we have four groups divided by horizontal dashed
    lines: The first group is the fiducial setup; the second group is for
    different physical models; the third group is for different systematic
    models; the fourth group is for different samplers. The results with
    ``$*$'' (``$\dagger$'') are sampled with \polychord{} (\emcee{}) and the
    others are sampled with \multinest{}.
    }
    \label{fig:inter_model_summary}
\end{figure*}

In our fiducial analysis, we apply a wide top-hat prior on the normalization
parameter of the linear power spectrum: $A_s \in [0.5, 10] \times 10^{-9}$\,.
Different cosmic shear analyses apply top-hat priors on different normalization
parameters, including $A_s$ \citep[see][]{DESY3_CS_Secco2022},
$\mathrm{ln}(A_s)$, $\mathrm{log}(A_s)$
\citep[see][]{cosmicShear_HSC1_Chiaki2019, cosmicShear_HSC1_Hamana2019,
KiDS450_cs_Hildebrandt2017} and $S_8$ \citep[see][]{KiDS1000_CS_Asgari2020}.
Here, we compare the fiducial analysis with the following two priors:
\begin{enumerate}
    \item $\mathrm{ln}(A_s \times 10^{10}) \in [1.7, 5.0]$;
    \item $S_8 \in [0.1,  2.0]$\,.
\end{enumerate}

After obtaining an MC chain from the nested sampling for the analyses with flat
priors on $A_s$ and $\mathrm{ln}(A_s)$, we reweight the chain in order to
obtain a flat prior on the 2D plane of $(\OmegaM, S_8)$ as discussed in
Section~\ref{subsec:model_ps}. Following \citet{HSC1_2x2pt_Sugiyama2022}, the
corrections for flat prior on $A_s$ and $\mathrm{ln}(A_s)$ involve multiplying
the weight of each sample by $\sigma_8/A_s$ and $\sigma_8$, respectively. We
refer the readers to \citet{HSC1_2x2pt_Sugiyama2022} for the derivation of the
correction factors. We find, for a flat prior on $A_s$, the reweighting shifts
the posterior to larger $\OmegaM$\,.

In Appendix~\ref{app:inter_prior}, we show the marginalized 2D posteriors for
different priors using both \multinest{} and \polychord{}. When using the
\multinest{} sampler, the uncertainties in $\OmegaM$ for the analyses with the
$A_s$ ($\mathrm{ln}(A_s)$) prior are smaller by $15\%$ ($10\%$) compared to the
analysis with the $S_8$ prior; while the constraints on $S_8$ show little
difference. This is consistent with the finding of
\citet{wlreana_catalogs_Longley2023}. In addition, for the analyses with
\polychord{}, the marginalized 2D posteriors for these three different priors
are more consistent with each other than with to the analyses with
\multinest{}. It is likely that \multinest{} neglects the tails of  the 2D
posteriors for $A_s$ and $\mathrm{log}(A_s)$ priors, leading to under-estimated
uncertainties for these two analyses. We show the 1D summary statistics of the
\polychord{} posteriors between analyses with different priors in the second
group of Fig.~\ref{fig:inter_model_summary}. As shown, our constraints on $S_8$
and $\OmegaM$ are insensitive to the prior and the choices of sampling
cosmological parameters.

\subsection{Physical Models}
\label{subsec:inter_phys}

We now compare the cosmology constraints as we vary the model, especially
focusing on $S_8$ between different physical models: the linear and nonlinear
matter power spectrum, baryonic feedback (see Section~\ref{subsec:model_ps})
and IA (see Section~\ref{subsec:model_ia}). The 1D summary statistics for the
comparisons are shown in Fig.~\ref{fig:inter_model_summary}, and the 2D contour
plots are shown in Section~\ref{app:inter_phys}.

\subsubsection{Power spectrum and baryonic feedback}

\begin{figure}
    \includegraphics[width=0.48\textwidth]{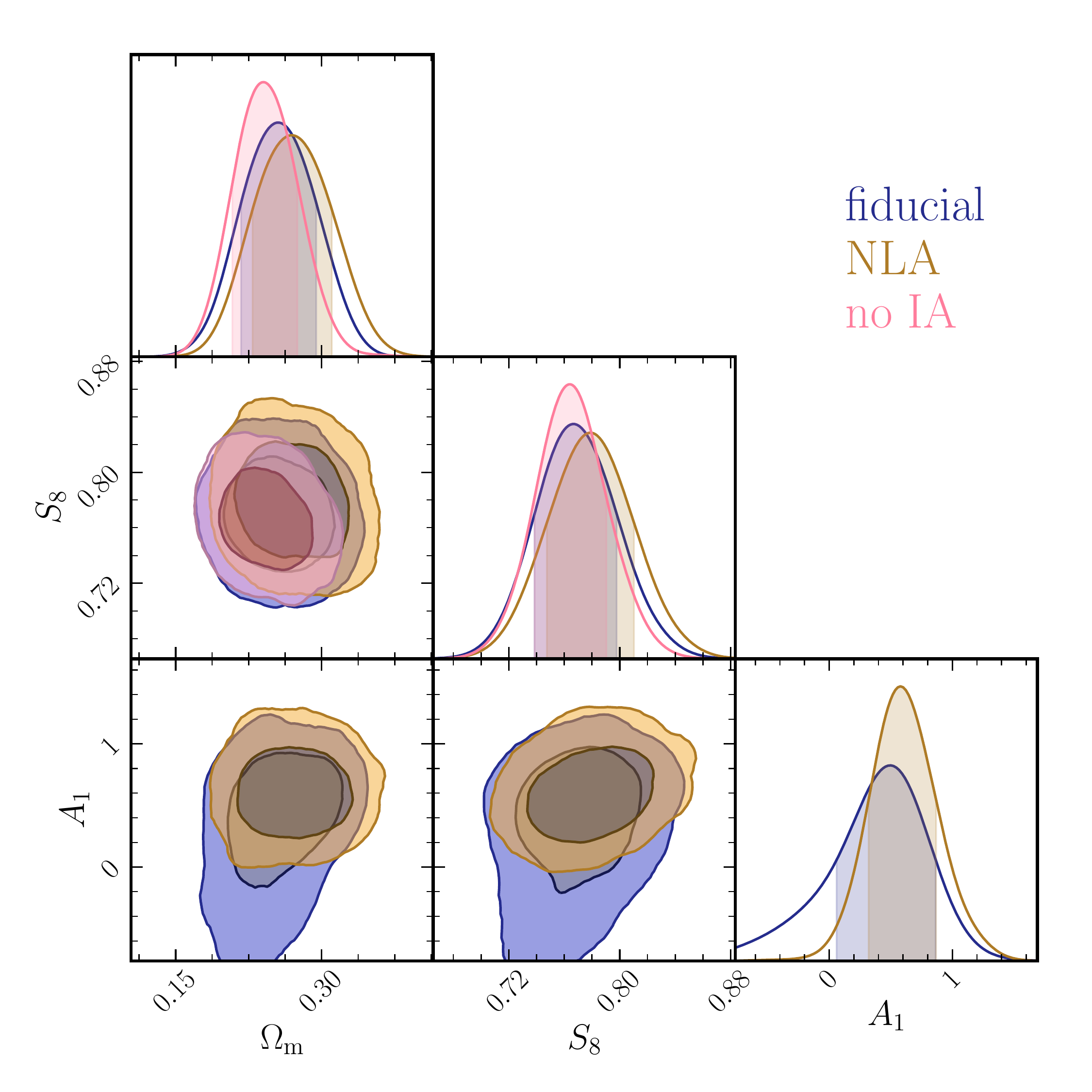}
    \caption{
        The marginalized 2D posteriors of analyses with the TATT model
        (fiducial), the NLA model (NLA) and without modeling of IA (no IA).
    }
    \label{fig:inter_ia_main}
\end{figure}

In our fiducial analysis, we use the BACCO emulator
\citep{baccoEmuLin_Arico2021} to model the linear matter power spectrum, and
\hmcode{}~2016 \citep{halofit_mead16}, implemented in \texttt{pyhmcode}
\citep{pyhmcode_Troster2022}, to model the nonlinear power spectrum and
baryonic feedback. Here, we test the modeling uncertainties by comparing the
constraints on cosmology parameters with other models for the matter power
spectrum, including (i) changing the linear power spectrum modeling to CAMB
\citep{CAMB2000} and (ii) changing the nonlinear modeling to \hmcode{}~2020
\citep{halofit_mead21} with a flat prior on the baryonic feedback parameter:
$\Theta_\text{AGN} \in [7.3, 8.3]$\,. In addition, we test the impact of not
modeling baryonic feedback by (iii) continuing to use the \bacco{} emulator and
\hmcode{}~2016, but fixing the $A_\mathrm{b}$ parameter to $3.13$\,.

The 1D summary statistics of the constraints are shown in the third group of
Fig.~\ref{fig:inter_model_summary}, where the tests (i)---(iii) are labelled as
``CAMB'', ``Mead 2020'' and ``DM only'', respectively. In addition, the
marginalized 2D posteriors are shown in Appendix~\ref{app:inter_phys}.
We find that the shifts in cosmology parameters, i.e., $\OmegaM$, $\sigma_8$
and $S_8$, are less than $0.5\sigma$, and we conclude that the errors due to
uncertainties in matter power spectrum modeling are not significant. This is
consistent with our finding in Section~\ref{subsec:model_valid} that our
analysis is not sensitive to modeling errors in the matter power spectrum.

\subsubsection{Intrinsic alignments}

\begin{figure*}
\includegraphics[width=0.98\textwidth]{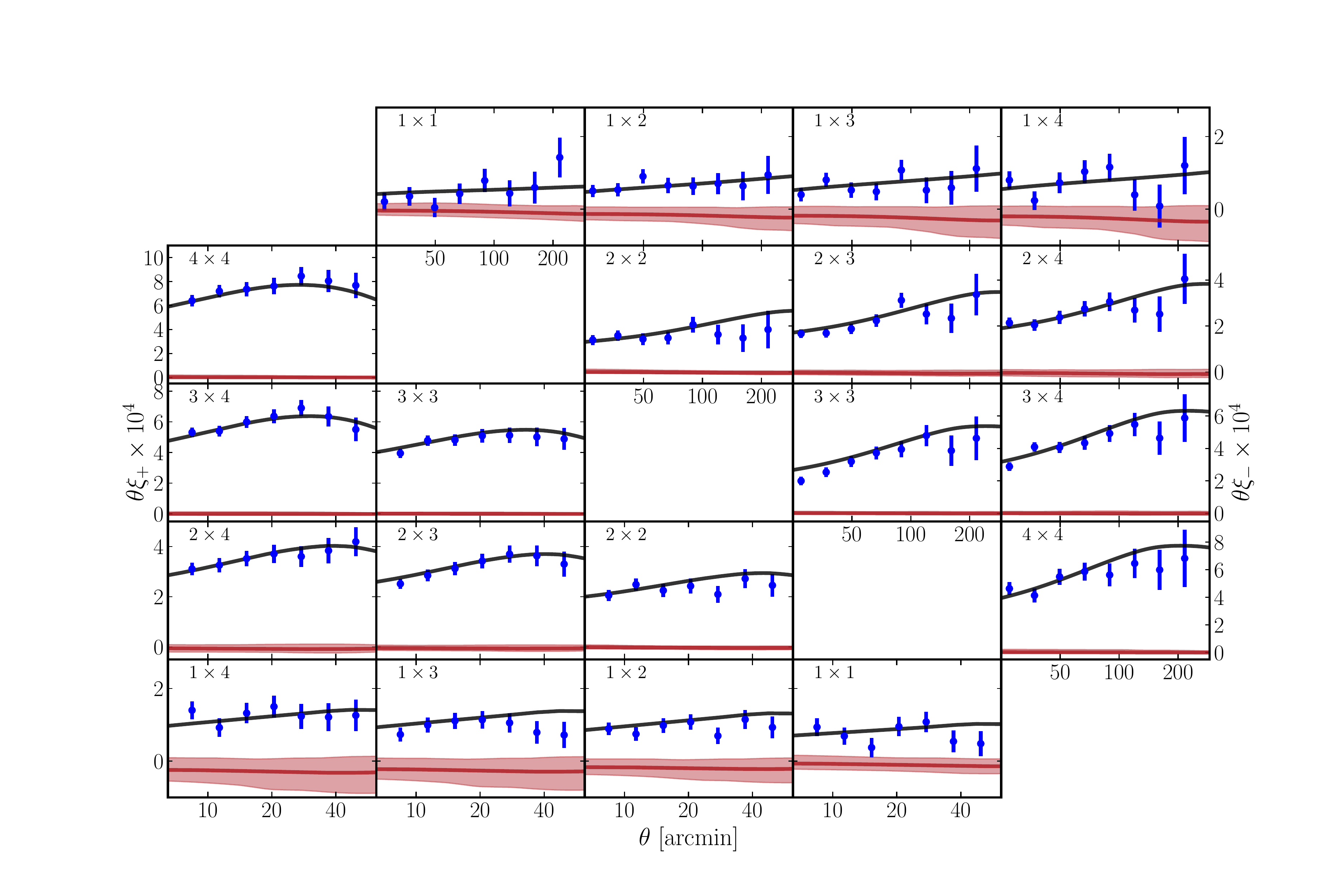}
\caption{
    The contribution from IA (including both the GG and GI terms), based on
    resampling from our fiducial posterior, to our 2PCFs. The IA signal is
    shown in red, where the solid lines are the mean, and the shaded reg
    regions are the 95\% confidence intervals. The blue points are the measured
    2PCFs and the black lines are the model prediction with the MAP, which are
    the same as in Fig.~\ref{fig:data_2pcf}.
}
\label{fig:data_2pcf_ia}
\end{figure*}

In our fiducial analysis, we use the TATT \citep{tatt_blazek17} to model the
intrinsic alignment effect (see Section~\ref{subsec:model_ia}). In order to
test the robustness of our cosmological constraints to the IA modeling errors,
we compare our fiducial analysis with the cosmology constraints obtained with
two simpler models: (i) the NLA model \citep{nla_bridle07}, which is a subset
of TATT (see Section~\ref{subsec:model_ia} for a detailed description); (ii) No
intrinsic alignment model is used (``no IA'') i.e.,\ intrinsic alignments are
assumed to be negligible. The other parts of our analysis pipeline are the same
as the fiducial analysis. For the NLA model, we use the same priors for the IA
parameters ($A_1 \in [-6, 6]$ and $\eta_1 \in [-6, 6]$) as summarized in
Table~\ref{tab:parameters}. The ``no IA'' configuration is a non-physical case,
which is used to test the difference in cosmology constraint under the extreme
condition that IA effect is fully neglected.

The marginalized 2D posteriors are shown in Fig.~\ref{fig:inter_ia_main}, and
the 1D summary statistics are shown in the third group of
Fig.~\ref{fig:inter_model_summary}, where the tests (i)---(ii) are labelled as
``NLA'' and ``no IA'', respectively. We find no significant difference in our
$S_8$ constraint when changing the IA model, although the constraints are
stronger when using these simpler models. We find a smaller $\OmegaM$ and
larger $\sigma_8$ for the constraint without modeling the IA effect; however,
the shifts in these two parameters are less than $0.5\sigma$; therefore, we
conclude that $\OmegaM$ and $\sigma_8$ are not significantly influenced by the
IA modeling error. In Fig.~\ref{fig:inter_ia_main}, we only show the leading
order amplitudes ($A_1$) of the IA model for TATT. As shown, the $A_1$
parameter is detected with only $1.1\sigma$ and $2.1\sigma$ significance for
TATT and NLA, respectively. We show the 95\% confidence intervals of the IA
signal in the 2PCFs using our fiducial model in Fig.~\ref{fig:data_2pcf_ia}.
The contribution from IA to the 2PCFs is not significant. Therefore, we
conclude that we do not find a significant detection of the IA signal. We note
that the conclusion can be different for a higher-order IA model (e.g.,
\citealt{ia_EFT_Bakx2023}). In addition to model-dependent analysis, numerous
model-independent methods exist for detecting the IA signal in cosmic shear
analysis \citep{ia_multiEst_Leonard2018, ia_multiEst_MacMahon2023,
ia_selfcal_Zhang2010, ia_selfcal_Yao2019}. We defer the validation of our
conclusions using these methods on HSC data to future studies.

\subsubsection{Massive neutrinos}

\begin{figure}
    \includegraphics[width=0.48\textwidth]{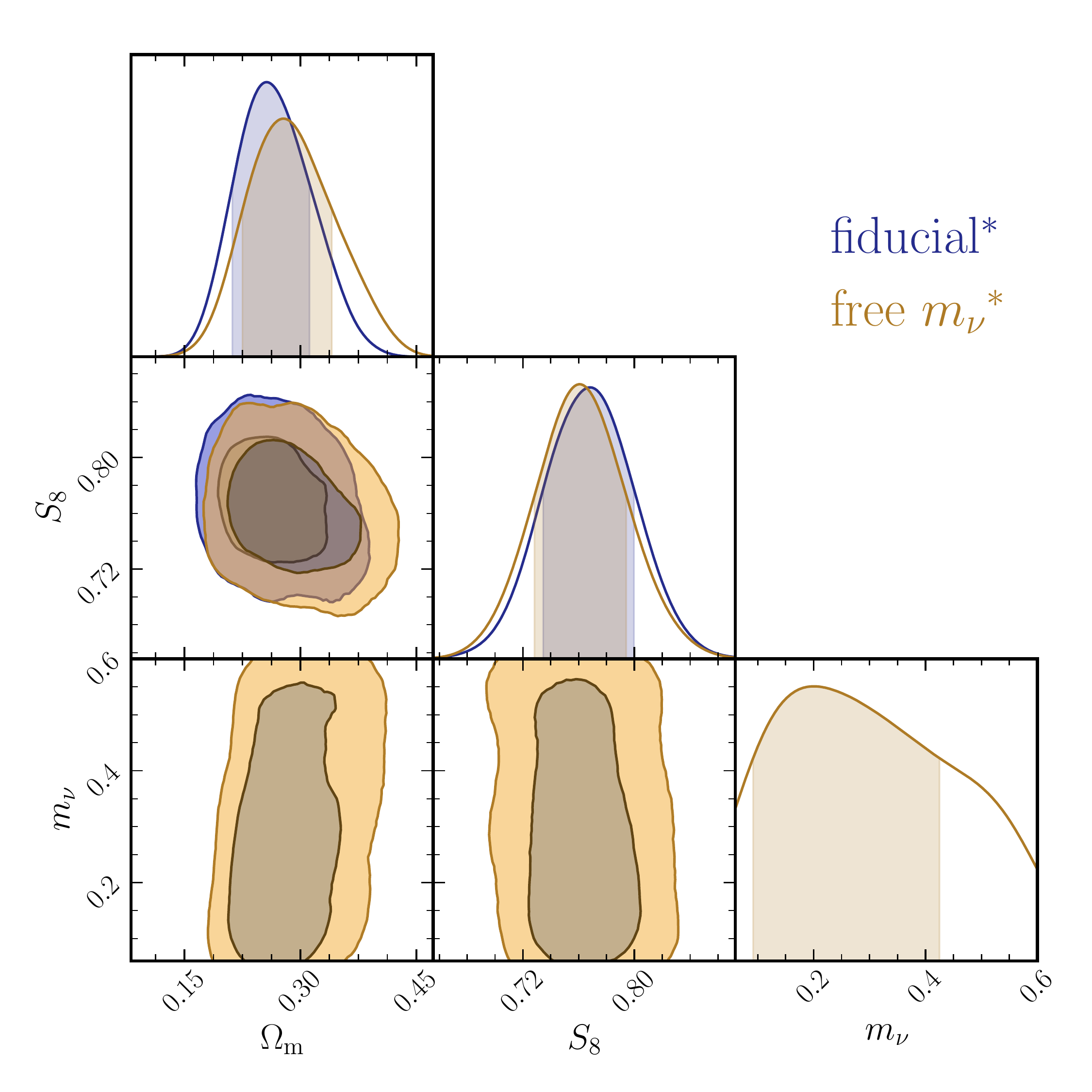}
    \caption{
        The marginalized 2D posteriors of the flat $\Lambda$CDM cosmology with
        fixed neutrino mass (fiducial; $m_\nu=0.06$) and free neutrino mass:
        $m_\nu \in [0.06, 0.6]\,$.
    }
    \label{fig:inter_mnu}
\end{figure}

Unlike the internal tests above, the test shown here on cosmology model with
free neutrino mass is performed after unblinding. Massive neutrinos suppress
the structure growth by smoothing the matter density field and changing the
matter power spectrum at small scales; therefore, neutrino mass influences the
constraint on cosmology parameters e.g., $S_8$\,. In our fiducial analysis, we
fix the total neutrino mass --- $\sum m_\nu=0.06~\mathrm{eV}$, which is the
lower limit obtained by neutrino oscillation experiments
\citep{neutrino_Capozzi2014, neutrino_Esteban2019}. We note that in the DES-Y3
analysis \citep{DESY3_CS_Secco2022}, they constrain neutrino mass with a flat
prior. To make sure that our analysis is not sensitive to the difference in the
prior on the sum of neutrino mass, we run an analysis with flat prior on the
sum of neutrino mass, namely $\sum m_\nu \in [0.06~\mathrm{eV},
0.6~\mathrm{eV}]$. The results are shown in Fig.~\ref{fig:inter_mnu}.

The marginalized 2D posteriors are shown in Fig.~\ref{fig:inter_mnu}, and the
1D summary statistics are shown as ``free $\sum m_\nu$'' in the third group of
Fig.~\ref{fig:inter_model_summary}. We find very little change in our $S_8$
constraint when changing the prior on the neutrino mass. Our constraint on
neutrino mass is weak and degenerate with $\OmegaM$. This is because
weak-lensing 2PCFs are only sensitive to the projected mass along the line of
sight, and, due to the limited number of redshift bins, we lose the information
on the redshift evolution of the large-scale structure. We note that the
constraint on neutrino mass can be significantly improved by combining
weak lensing, CMB and BAO observations \citep{neutrinoWLCMB_Ichiki2009}, but
that is beyond the scope of this work.

\subsection{Models of Systematic Effects}
\label{subsec:inter_sys}

\begin{figure}
\includegraphics[width=0.48\textwidth]{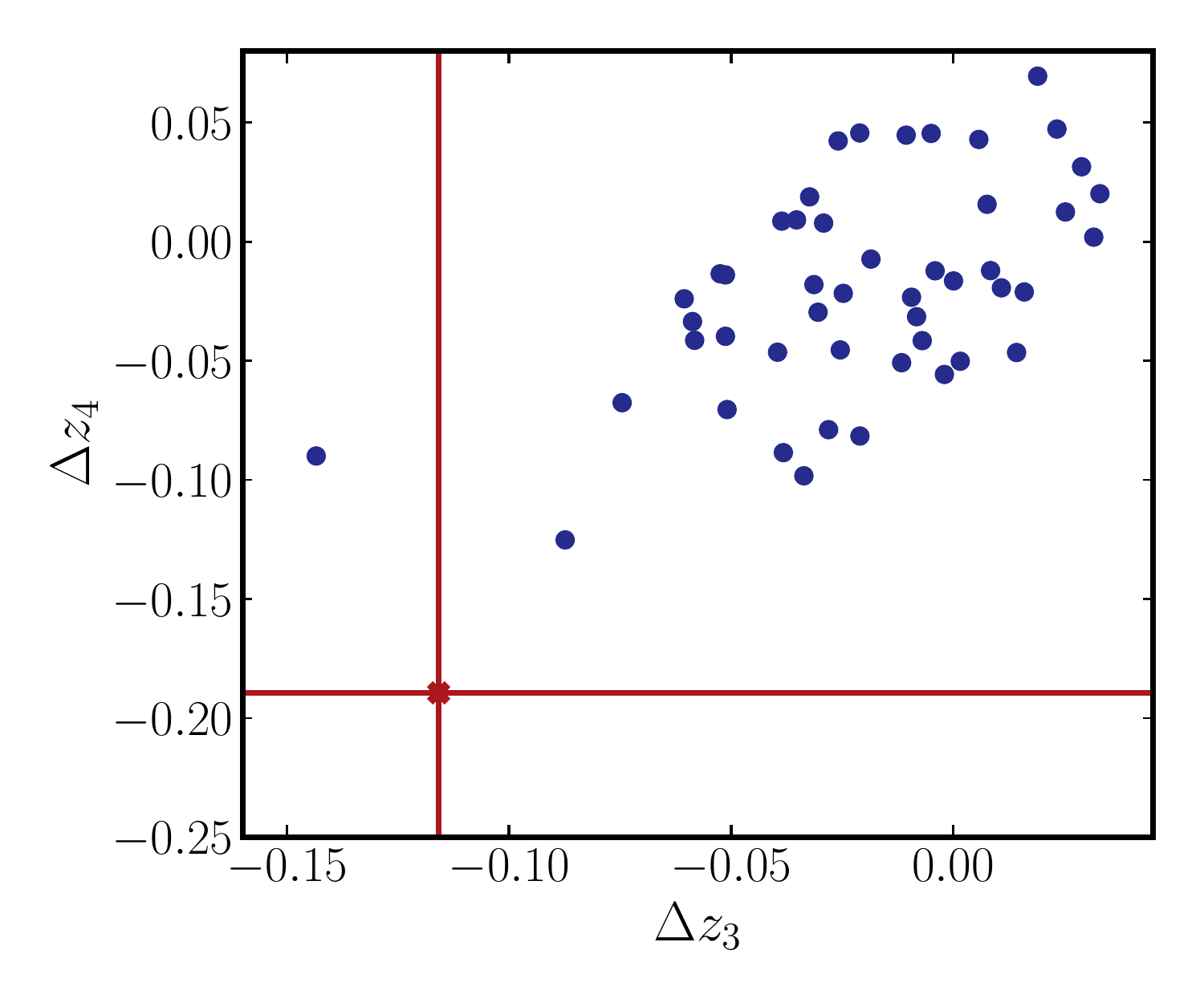}
\caption{
    The evaluation of statistical significance of the shifts in $\Delta
    z_{3,4}$. Blue points show the $\Delta z_{3,4}$ estimations by conducting
    our fiducial analysis on noisy mock 2PCFs. The red point is the estimation
    from our fiducial analysis on HSC-Y3 real data.
}
\label{fig:inter_dz3dz4}
\end{figure}

We compare the cosmology constraints, especially focusing on $S_8$, between
different models for PSF systematics and systematics in modeling of redshift
distributions.

\subsubsection{PSF systematics}
\label{subsub:inter_psf}

In our fiducial analysis, we adopt a PSF systematic model including PSF
modeling error and PSF leakage from fourth-order PSF shapes. Additionally, we
fully take into account the correlation between PSF systematic parameters.
Here, we check the dependence of our cosmology constraint on the model choice
of PSF systematics with the following two tests: (i) We determine the
constraint without taking into account the correlation between the original PSF
systematic parameters by sampling the correlated parameters with the
uncorrelated prior; (ii) we determine the constraint without modeling PSF
systematics at all.

The marginalized 2D posteriors are shown in Appendix~\ref{app:inter_sys}, and
the 1D summary statistics are shown in the fourth group of
Fig.~\ref{fig:inter_model_summary}, where the tests (i)---(ii) are labelled as
``psf uncorr'' and ``no PSF'', respectively. We find that the ``no PSF''
analysis shows $\sim$$0.15\sigma$ and $\sim$$0.3\sigma$ increases from the
baseline analysis in $S_8$ and $\OmegaM$, respectively. We emphasize that the
``no PSF'' test is not realistic; we know that the PSF systematics are
important for the analysis.  The increase in $\OmegaM$ is consistent with our
finding in \citet{HSC3_PSF} using noiseless mock 2PCFs, and the increase in
$S_8$ is too small to be statistically significant. In addition, we do not find
significant difference in cosmology constraints from the ``psf uncorr'' test.
Therefore, we conclude that, within the choice of models we have considered,
our cosmology constraint is not sensitive to the choice of PSF systematics
model.

\subsubsection{Photo-$z$ systematics}
\label{subsub:inter_photoz}

Our fiducial analysis uses the redshift distribution of galaxies in four
redshift bins obtained from a joint estimation using photo-$z$ and
cross-correlations between weak-lensing galaxies and CAMIRA-LRGs. The
uncertainties on the redshift number densities were estimated based on
comparison of different photo-$z$ methods, and do not encompass the full range
of systematic uncertainty. As described in Section~\ref{subsec:model_photoz},
there are potential biases on the redshift estimations in our third and fourth
tomographic redshift bins, since the third bin is only partially calibrated,
and the fourth bin is not calibrated by the CAMIRA-LRGs since the LRG sample
extends only to $z=1.2$\,. To be conservative, in our fiducial analysis, we
adopt an uninformative, wide flat prior ranging from $-1$ to $1$ for the mean
redshift shifts in these two tomographic redshift bins.

Here, we compare the fiducial cosmology constraint with the one using the
Gaussian prior recommended by \citet{HSC3_photoz_Rau2022} for $\Delta z_3$ and
$\Delta z_4$.  The result is labeled as ``$\mathcal{N}(\Delta z_{3,4})$'' in
Fig.~\ref{fig:inter_model_summary}. As shown, we find a $\sim$$2\sigma$
difference in $S_8$ between these two setups. Furthermore, the constraints on
$\Delta z_{3,4}$ show shifts in the mean redshift estimation of the last two
redshift bins. To assess the possibility that the shifts in $\Delta z_{3,4}$
are caused by statistical errors, we use 50 noisy mock 2PCFs generated using
the WMAP9 cosmology and the best-fit fiducial model for nuisance parameters
with $\Delta z_3$ and $\Delta z_4$ set to $0$\,. We run the fiducial analysis
on these mocks (see Fig.~\ref{fig:inter_dz3dz4}) and find that it is not likely
to obtain redshift shifts as large as our fiducial analysis --- $\Delta z_3=
-0.115$ and $\Delta z_4 = -0.192$\,. Note that according to the definition in
equation~\eqref{eq:mdoel_nz_shift} \textit{negative} values of $\Delta z_{3,4}$
indicate that the true mean redshift is \textit{higher} than the mean redshift
estimated by the joint calibration. Our results indicate that these shifts in
the third and fourth redshift bin are statistically significant. We leave the
study and calibration of these biases to our future work. Comparing
``$\mathcal{N}(z_{3,4})$'' with the fiducial constraints in
Fig.~\ref{fig:inter_model_summary}, the conservative flat priors on $\Delta
z_{3,4}$ are a large part of the reason why our constraints on $S_8$ are not
improved over the HSC-Y1 analyses \citep{cosmicShear_HSC1_Hamana2019}, although
the HSC-Y3 2PCF measurement has significantly higher SNR.

Additionally, we test the analysis by comparing the fiducial cosmology
constraint with analyses using straight-up stackings of photometric posteriors
(without deconvolution of photo-$z$ error and calibration from LRGs as our
redshift number density. In this test, we still use flat priors on $\Delta
z_{3,4}$\,. As shown by ``\dnnz{}'', ``\dempz'' and ``\mizuki'' in the fourth
group of Fig.~\ref{fig:inter_model_summary}, we find that the difference in
$S_8$ constraint is smaller than $0.5\sigma$\,. Although directly using the
stacked photo-$z$ posterior is not mathematically correct
\citep{stackpz_Malz2020}, this test attempts to assess the possible bias from
the error in the shape of $n(z)$s. However, we note that none of these $n(z)$s
are well-calibrated at high redshift; therefore, this test is not able to
capture potential bias for galaxies at $z>1.2$\,.

\subsection{Samplers}
\label{subsec:inter_sampler}

We adopt \polychord{} as our fiducial sampler, and several internal consistent
tests are conducted with \multinest{} to save computational resources. Here, we
check the consistency of the analysis between different samplers. Specifically,
we compare the cosmology constraints between the three samplers summarized in
Section~\ref{subsec:model_baysian}, namely \polychord{}, \multinest{} and
\emcee{}. The 1D marginalized posteriors are shown in the fifth group of
Fig.~\ref{fig:inter_model_summary}, and the 2D marginalized posteriors are
shown in Appendix~\ref{app:inter_samplers}. As shown, the confidence intervals on
$\OmegaM$ and $\sigma_8$ from \multinest{} are $\sim$$15\%$ smaller than
\polychord{} and \emcee{}, while the confidence intervals from \polychord{} and
\emcee{} are consistent; \citet{DESY3_sampler_Lemos} found similar results. In
addition, we find the 1D projected modes from these three samplers are
consistent. Therefore, we conclude that it is conservative to use \multinest{}
for our internal consistency tests as any inconsistency is more significant for
results from \multinest{} due to the smaller confidence intervals.

\subsection{Subfields, Tomographic Bins, Scales}
\label{subsec:inter_data}

\begin{figure*}[!t]
\includegraphics[width=0.95\textwidth]{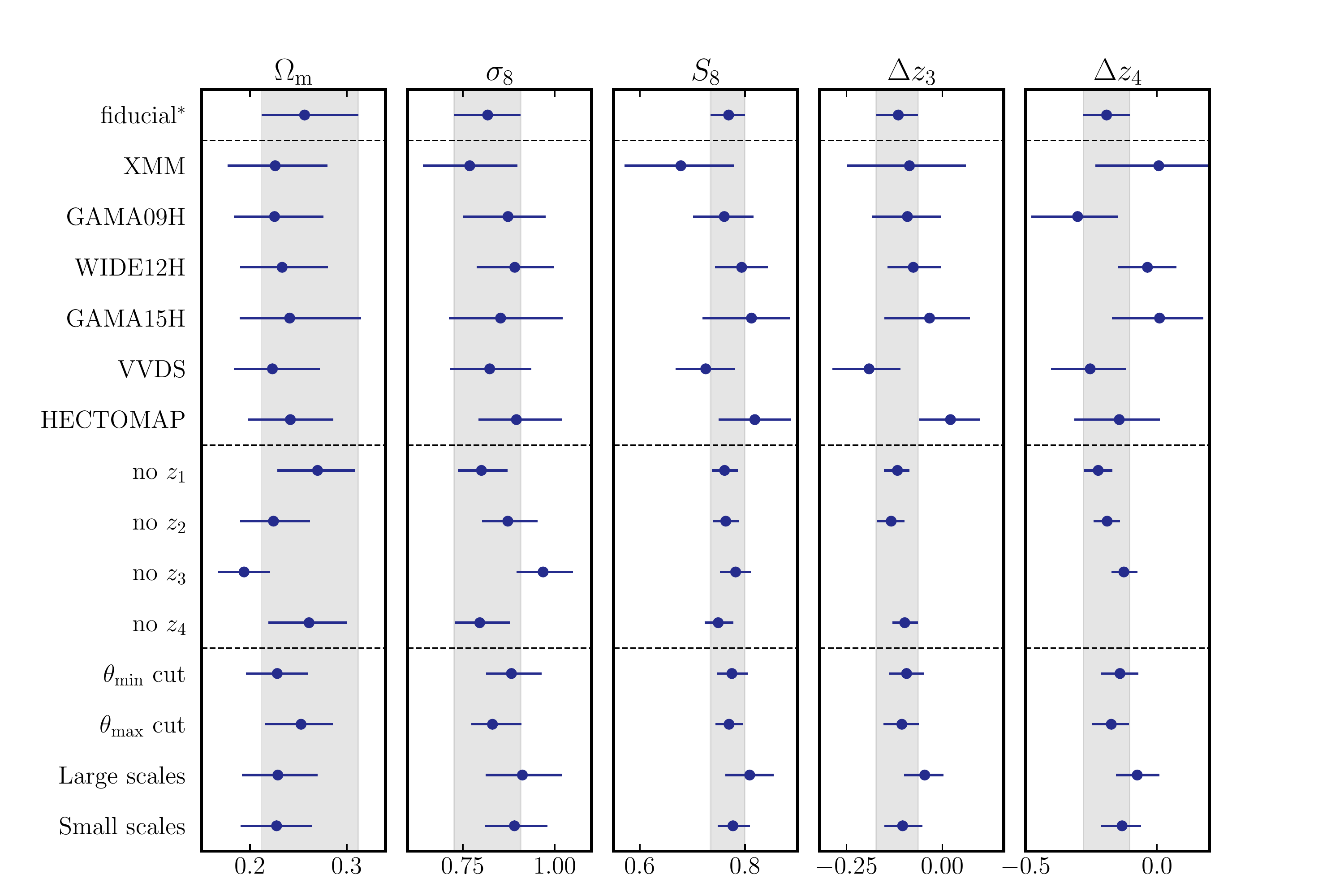}
\caption{
    The $68\%$ CIs of 1D projected modes on the parameters $\OmegaM$,
    $\sigma_8$, $S_8$, $\Delta z_3$ and $\Delta z_4$ for different splits of
    the HSC-Y3 data. We have four groups divided by horizontal dashed lines.
    The first group is the fiducial setup; the second group examines different
    subfields, the third group is for removal of different redshift bins, and
    the fourth group is for different scale cuts. The fiducial results (marked
    with ``$*$'') are sampled with \polychord{} and the others are sampled with
    \multinest{}. The $\Delta z_3$ and $\Delta z_4$ for ``no $z_3$'' and ``no
    $z_4$'' are missing, respectively, since the redshift bins are removed from
    the analysis.
    }
    \label{fig:inter_data_summary}
\end{figure*}

\subsubsection{Subfields}

The HSC-Y3 survey footprint has six different subfields as summarized in
Table~\ref{tab:catsummary}. Here, we assess the consistency in the cosmology
constraints, especially focusing on $S_8$, by performing analyses on each
subfield separately with our fiducial model.

The 1D summary statistics for the six subfields are shown in
Fig.~\ref{fig:inter_data_summary}, which are labelled as ``XMM'', ``GAMA09H'',
``GAMA09H'', ``WIDE12H'', ``VVDS'' and ``HECTOMAP'', respectively. In addition,
the marginalized 2D posteriors are shown in Appendix~\ref{app:inter_fields}.

We note that these fields have very different areas (see
Table~\ref{tab:catsummary} for details); therefore, the $1\sigma$ errors are
different among individual fields, and they are different from the fiducial
analysis using all of the fields.  Additionally, the constraints from these
individual fields are approximately independent of each other since they are
from different regions of sky; therefore, the errors are not significantly
correlated. As shown in Fig.~\ref{fig:inter_data_summary}, XMM, GAMA15H, VVDS
and HECTOMAP show $\sim$$1\,\sigma$ differences in $S_8$ from the average.
However, we note that each shift in $S_8$ is offset by less than $1\sigma$
given the uncertainty in the corresponding individual field. Taking into
account that the errors are not significantly correlated, it is not likely for
the scatters to be a flag of systematic errors.

\subsubsection{Tomographic bins}

\begin{figure}[!t]
\includegraphics[width=0.48\textwidth]{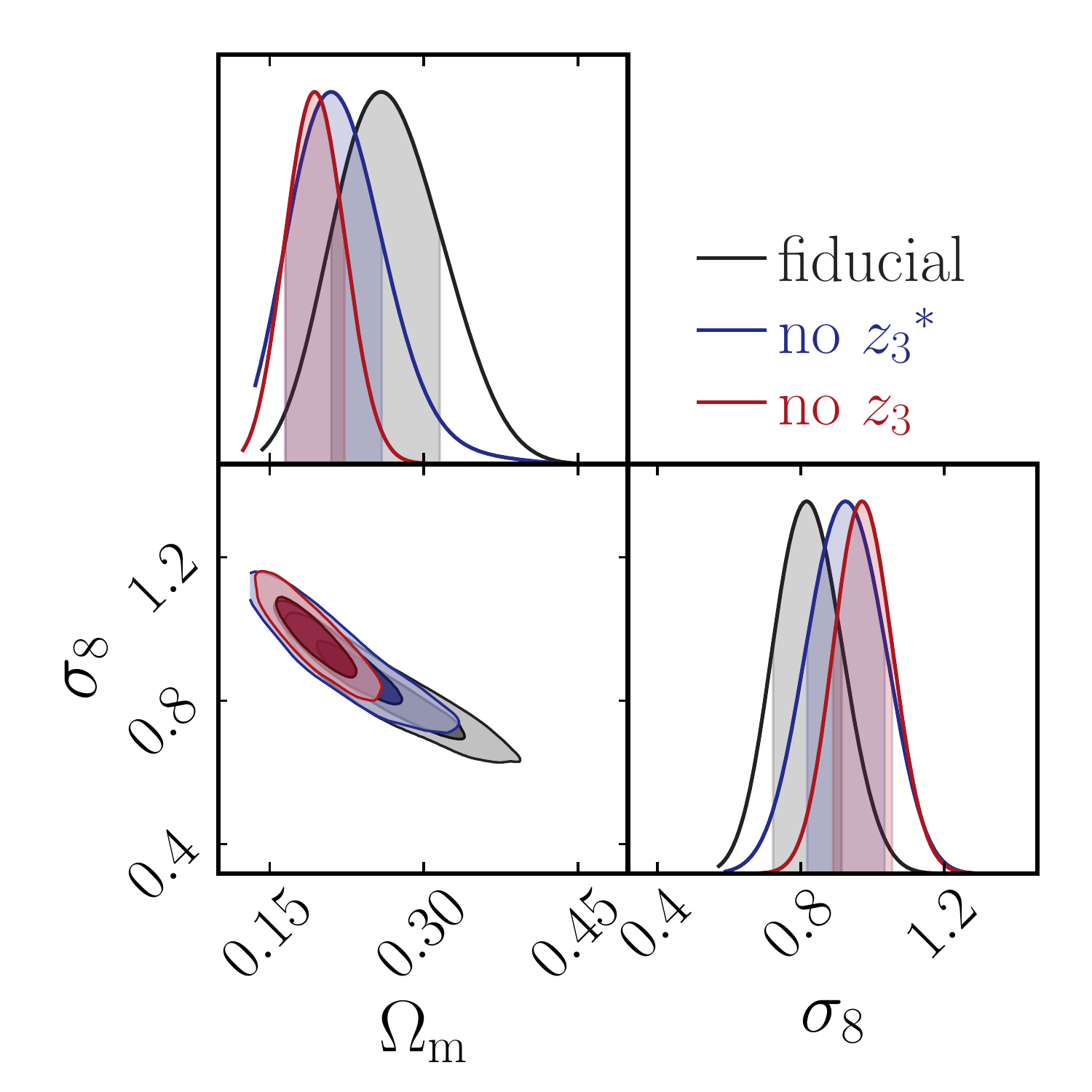}
\caption{
    The $2$D posterior in the ($\OmegaM, \sigma_8$) plane for the fiducial
    analysis sampled with \polychord{}, ``no $z_3$'' analysis sampled with both
    \polychord{} (with $^*$) and \multinest{} (without $^*$). The \multinest{}
    sampler significantly underestimates the error on the parameters,
    particularly $\OmegaM$.
    }
    \label{fig:inter_rmz3_2D}
\end{figure}

Here, we exclude one tomographic redshift bin at a time and check whether the
constraints are consistent with the fiducial constraint with all the four
redshift bins. Our fiducial analysis adopts wide, flat priors on $\Delta z_3$
and $\Delta z_4$\,.

We find that when removing one of the first two redshift bin, the constraining
power of the test is significantly degraded since we apply conservative flat
priors on $\Delta z_{3,4}$ and use the measurements in the first two redshift
bins to calibrate the redshift density estimations in the last two bins.
Therefore, we adopt Gaussian priors on $\Delta z_3$ and $\Delta z_4$, taken
from the posteriors of these parameters from the fiducial analysis, $\Delta z_3
= -0.115 \pm 0.055$ and $\Delta z_4 = -0.192 \pm 0.088$. This test is used to
assess the robustness to our cosmology constraint across different redshift
bins with the re-calibrated redshift densities. When we remove each redshift
bin in turn, we have no constraints on the corresponding $\Delta z$. In
Fig.~\ref{fig:inter_data_summary}, we do not plot any $\Delta z$ posterior when
that bin is removed.

The 1D summary statistics are shown in Fig.~\ref{fig:inter_data_summary}, which
are labelled as ``no $z_1$'', ``no $z_2$'', ``no $z_3$'', ``no $z_4$'',
respectively. The marginalized 2D posteriors are shown in
Appendix~\ref{app:inter_zbins}. We find that the maximum shifts in $S_8$
constraints from removing each redshift bin are $\sim$$0.5\sigma$ of the
fiducial $S_8$ constraint. We note that these constraints with tomographic-bin
removal are not independent. However, the differences in the $S_8$ constraints
are less than $1.7\%$, which is small compared to the statistical error on our
fiducial $S_8$ constraint.

While the ``no $z_3$'' case shows a significant shift in $\OmegaM$\,, our
analysis primarily emphasizes $1$D constraints on $S_8$, rather than on
$\OmegaM$, similar to \citet{HSC3_cosmicShearFourier}. This preference is due
to the strong degeneracy between $\OmegaM$ and $\sigma_8$\,. As depicted in
Fig.~\ref{fig:inter_rmz3_2D}, for both the ``no $z_3$'' and fiducial scenarios,
there is a pronounced uncertainty ellipse in the ($\OmegaM$, $\sigma_8$) plane.
Moreover, the 1D posteriors of $\OmegaM$ do not follow a Gaussian distribution.
In addition, we sample the ``no $z_3$'' case with both \multinest{} and
\polychord{}. As shown in Fig.~\ref{fig:inter_rmz3_2D}, the confidence interval
on $\OmegaM$ from \multinest{} is significantly underestimated compared to
\polychord{} due to the assumptions in \multinest{}
\citep{DESY3_sampler_Lemos}. The ``no $z_3$'' results shown in
Fig.~\ref{fig:inter_data_summary} are based on \multinest{}.

When we compare the data vector for the ``no $z_3$'' case against model
predictions based on our fiducial MAP constraint, the chi-squared value is
72.1\,. In contrast, the chi-squared value is 69.8 for the ``no $z_3$'' MAP.
Given that the effective degrees of freedom are around 79, both sets of
parameters fit the data well. Therefore, even though the ``no $z_3$'' scenario
indicates a noticeable deviation in $\OmegaM$ compared to the fiducial
constraint, the data {\it does not strongly} favor either scenario over the
other.

\subsubsection{Scale cuts}

\begin{figure}[!t]
\includegraphics[width=0.48\textwidth]{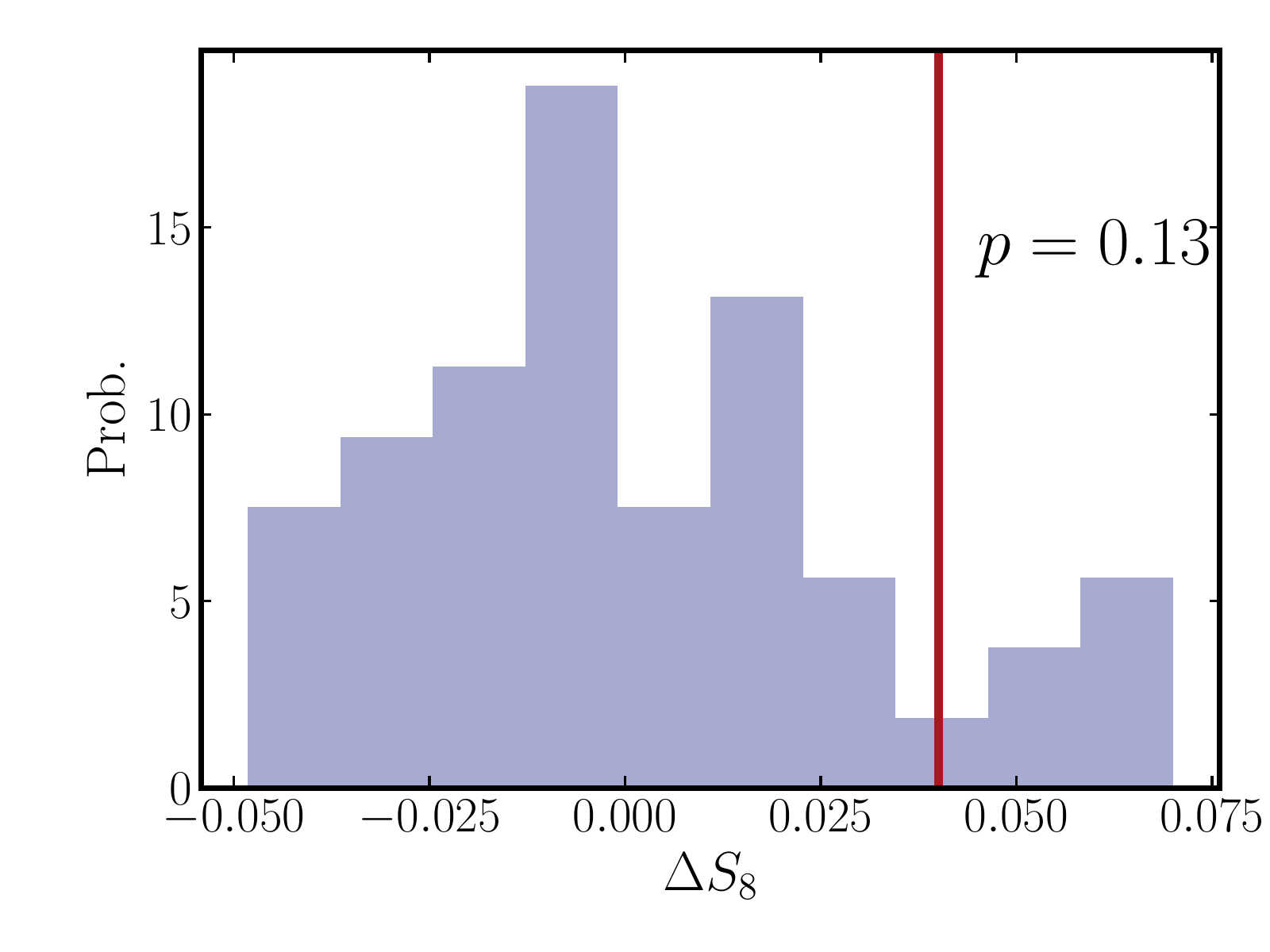}
\caption{
    The evaluation of the statistical significance of the shift in $S_8$ for the
    test with large scale cut using $50$ mock 2PCFs. The reference distribution
    (blue histogram) is obtained by analyzing the noisy mocks with the fiducial
    scale cut, and keeping large scales only. The red vertical line is the
    estimation of $\Delta S_8$ from real data. The probability of finding
    $\Delta S_8$ larger than the real analysis is $13\%$\,.
    }
    \label{fig:inter_dS8_LS}
\end{figure}

Our fiducial scale cuts are $7.1 < \theta/{\rm arcmin} < 56.6$ for $\xi_+$ and
$31.2 < \theta/{\rm arcmin} < 248$ for $\xi_-$\,. Here, we change the angular
scale cuts and check the consistency of the cosmology constraints, especially
focusing on constraint on $S_8$\,. The scale cuts we test include:
\begin{enumerate}
    \item $\bm{\theta_\text{max}}$ \textbf{cut:}
        $\theta_+ \in [7.1, 75.9]$ and $\theta_- \in [31.3, 247.8]$
    \item $\bm{\theta_\text{min}}$ \textbf{cut:}
        $\theta_+ \in [5.3, 56.5]$ and $\theta_- \in [23.3, 247.8]$
    \item \textbf{Large scales only:}
        $\theta_+ \in [17.3, 56.5]$ and $\theta_- \in [75.9, 247.8]$
    \item \textbf{Small scales only:}
        $\theta_+ \in [7.1, 23.3]$ and $\theta_- \in [31.3, 102.1]$
\end{enumerate}
where all numbers are in units of arcminutes. The 1D summary statistics of the
tests are shown in Fig.~\ref{fig:inter_data_summary}, which are labelled as
``$\theta_\text{max}$ cut'', ``$\theta_\text{min}$ cut'', ``Large scales'',
``Small scales'', respectively. The marginalized 2D posteriors are shown in
Appendix~\ref{app:inter_scales}. We find a $\sim$$1\sigma$ difference in $S_8$
for the analysis using large scale data only. In order to quantify the
statistical significance of this difference, we perform our fiducial analysis
on $50$ noisy mock 2PCFs using the fiducial scale cut and the large scales
only. To be more specific, the noisy mocks are generated with the  WMAP9
cosmology but the best-fit nuisance parameters from the fiducial analysis. We
record the difference in $S_8$ estimation for each noisy mock realization, and
the probability distribution of the difference in $S_8$, denoted as $\Delta
S_8$ is shown in Fig.~\ref{fig:inter_dS8_LS}. As shown, there is a $13\%$
probability of $\Delta S_8$ being larger than the real analysis; therefore,
this difference is not statistically sufficient to be a bias. We find that the
differences in $S_8$ constraints are negligible for the other tests on scale
cuts. The shifting error is less significant at large scales, and this is also
the case for the pseudo-$C_\ell$ analysis in \citet{HSC3_cosmicShearFourier}.

\subsection{$B$-mode errors}
\label{subsec:inter_bmode}

\begin{figure*}[!t]
\includegraphics[width=0.95\textwidth]{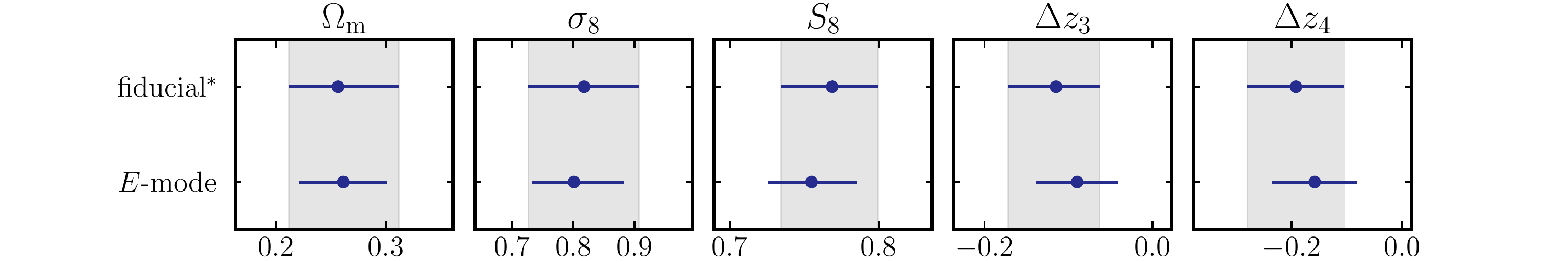}
\caption{
    Similar to Fig.~\ref{fig:inter_model_summary} and
    \ref{fig:inter_data_summary}\,. The $68\%$ CI of the 1D projected posterior
    on each of the parameters $\OmegaM$, $\sigma_8$, $S_8$, $\Delta z_3$ and
    $\Delta z_4$ for the fiducial constraint (first row) and the constraint
    with $E$-mode-only 2PCFs (second row) after removing the $B$-mode signal in
    Fig.~\ref{fig:data_bmodes}\,.
    }
    \label{fig:inter_emode}
\end{figure*}

To assess the robustness of our cosmology constraint, we test the influence of
$B$-mode residuals shown in Fig.~\ref{fig:data_bmodes} to our constraint.
Specifically, we subtract the estimated $B$-mode residuals in
Fig.~\ref{fig:data_bmodes} from our 2PCFs and analysis the data vector with our
fiducial setup. This test is performed after the unblinding. To save the
computational time, we do not re-estimate the covariance of the derived
$E$-mode 2PCFs, and we use the fiducial covariance matrix. The fiducial
constraint and the constraint with $E$-mode 2PCFs is shown in
Fig.~\ref{fig:inter_emode}\,. As shown, the difference between the two
constraints on $S_8$ is less than $0.5\sigma$, which indicates that the
$B$-mode residuals shown in Fig.~\ref{fig:data_bmodes} are not likely to cause
significant error on our cosmology constraint.

\section{COSMOLOGICAL CONSTRAINTS AND EXTERNAL ANALYSIS}
\label{sec:res_exter}

After having verified that the HSC-only constraints are robust, we check the
consistency of our constraints with other observations and quantify any
tension. The external observations include weak-lensing surveys such as DES
\citep{DESY3_CS_Secco2022, cosmicShear_DESY3_Amon2021} and KiDS
\citep{KiDS1000_CS_Asgari2020} (see Section~\ref{subsec:exter_wl}), the \textit
{Planck}-2018 CMB analysis \citep{cmb_Planck2018_Cosmology} (see
Section~\ref{subsec:exter_cmb}), and the eBOSS BAO analysis (see
Section~\ref{subsec:exter_bossbao}). We summarize the external observations as
follows
\begin{enumerate}
    \item \textbf{DES-Y3}: The DES-Y3 weak-lensing data contains about $100$
        million galaxies (with $n_\text{eff}$$\sim$$5.6~\mathrm{arcmin}^{-2}$) over
        more than $4,000$ square degress \citep{DESY3_catalog_Gatti2021}. We
        focus on the posterior from the cosmic shear 2PCFs presented in
        \citep{DESY3_CS_Secco2022, DESY3_CS_Amon2021}.
    \item \textbf{KiDS-1000:} The KiDS-1000 weak-lensing data contains 21 million
        galaxies (with $n_\text{eff}$$\sim$$6.2~\mathrm{arcmin}^{-2}$) over $1000$
        square degrees \citep{KiDS1000_catalog2021}. Their cosmic shear paper
        \citep{KiDS1000_CS_Asgari2020} present cosmic shear analyses using
        three different statistics (i.e., COSEBIs, 2PCFs, and pseudo-$C_\ell$).
        We use the analysis with COSEBIs, which is the fiducial result from
        KiDS, in this paper.
    \item \textbf{Planck-2018:} This is the final data release from the
        \textit{Planck} Cosmic Microwave Background (CMB) experiment
        \citep{cmb_Planck2018_Cosmology}. We incorporate the primary $TT$ data
        on scales $30<\ell < 2508$, and also the joint temperature and
        polarization measurements ($TT$‚ $TE$ ‚ $EE$ and $BB$) at scales: $2 <
        \ell < 30$\,.
    \item \textbf{eBOSS DR16:} We include spectroscopic baryon acoustic
        oscillation (BAO) measurements from the eBOSS galaxy sample. We
        recompute the posterior in our choice of cosmological parameter space
        (summarized in Table~\ref{tab:parameters}), and we combine it with our
        cosmic shear 2PCFs constraints assuming that these two measurements are
        independent.
\end{enumerate}

\subsection{Other Weak-lensing Analyses}
\label{subsec:exter_wl}

We first compare our fiducial constraints on $\OmegaM$ and $S_8$ with other
ongoing weak-lensing surveys (i.e., KiDS and DES). As shown in
Fig.~\ref{fig:ext_wl_cmb}, our result is consistent in general with the DES-Y3
and KiDS-1000 results. Specifically, the difference in the 2D plane of
$\OmegaM$ and $S_8$ is within the confidence region, even though the contour
size of HSC-Y3 is larger than DES-Y3 and KiDS-1000. We note that the larger
contour size is partly due to  the use of a flat prior on the photo-$z$ shifting
error parameters on the last two redshfit bins (i.e., $\Delta z_3$ and $\Delta
z_4$). If we employ the Gaussian informative prior with $\sigma(\Delta z_{3,4})
\sim \mathcal{O}(10^{-2})$ as derived in \citep{HSC3_photoz_Rau2022} on these
redshift error parameters, our constraint becomes $15\%$ tighter; however, as
shown in Section~\ref{subsec:inter_sys}, the posterior shows a non-negligible
shift toward the direction of larger $S_8$\,.

\subsection{\textit{Planck} CMB Analysis}
\label{subsec:exter_cmb}

\begin{figure*}[!ht]
\includegraphics[width=0.95\textwidth]{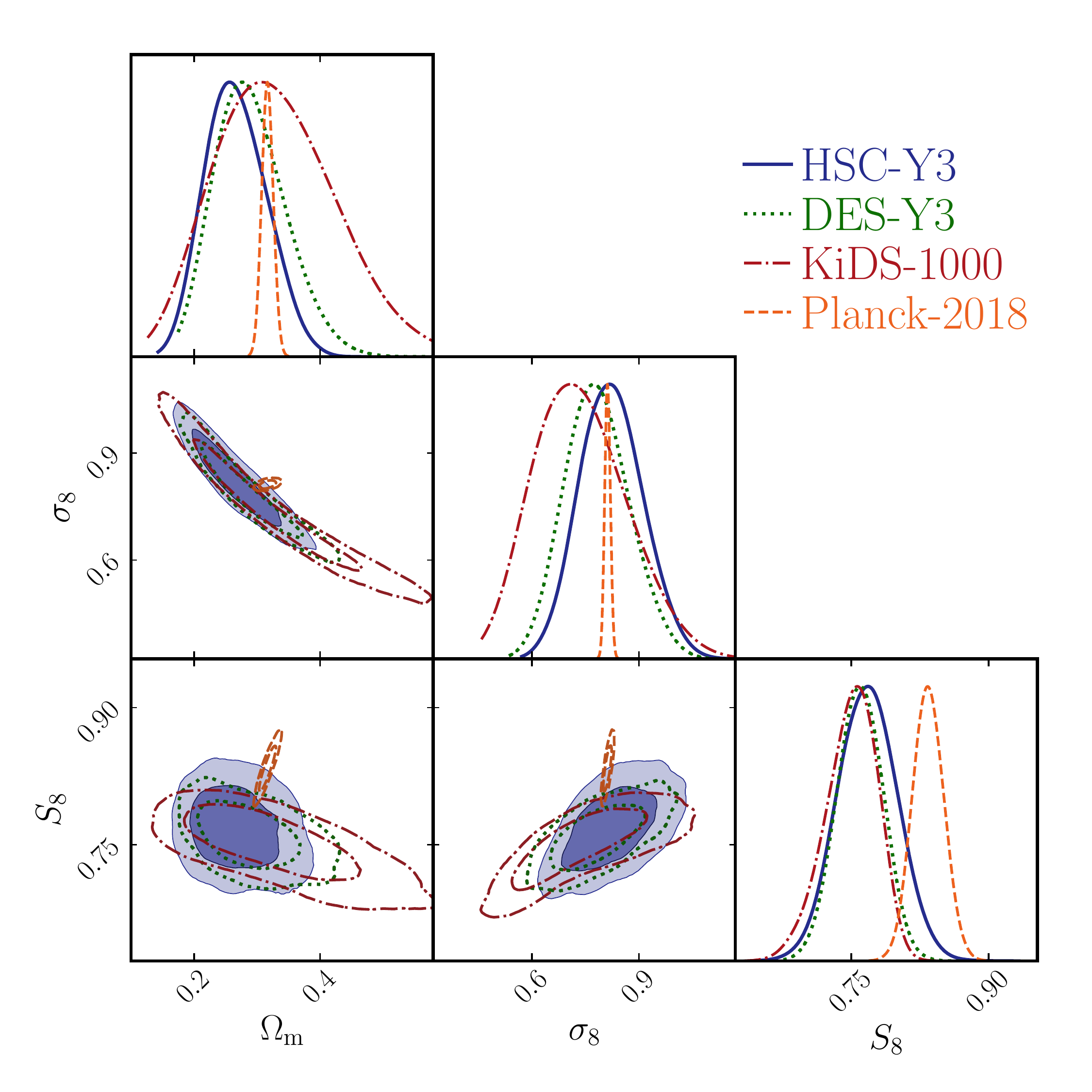}
\caption{
     Comparison between our fiducial cosmology constraint and contemporary
     weak-lensing observations (i.e., KiDS-1000 and DES-Y3) and
     \textit{Planck}-2018 CMB observation. The posterior data are plotted as
     published by each collaboration. These analyses have slightly different
     priors and astrophysical and systematic models.
    }
    \label{fig:ext_wl_cmb}
\end{figure*}

\begin{figure}[!ht]
\includegraphics[width=0.48\textwidth]{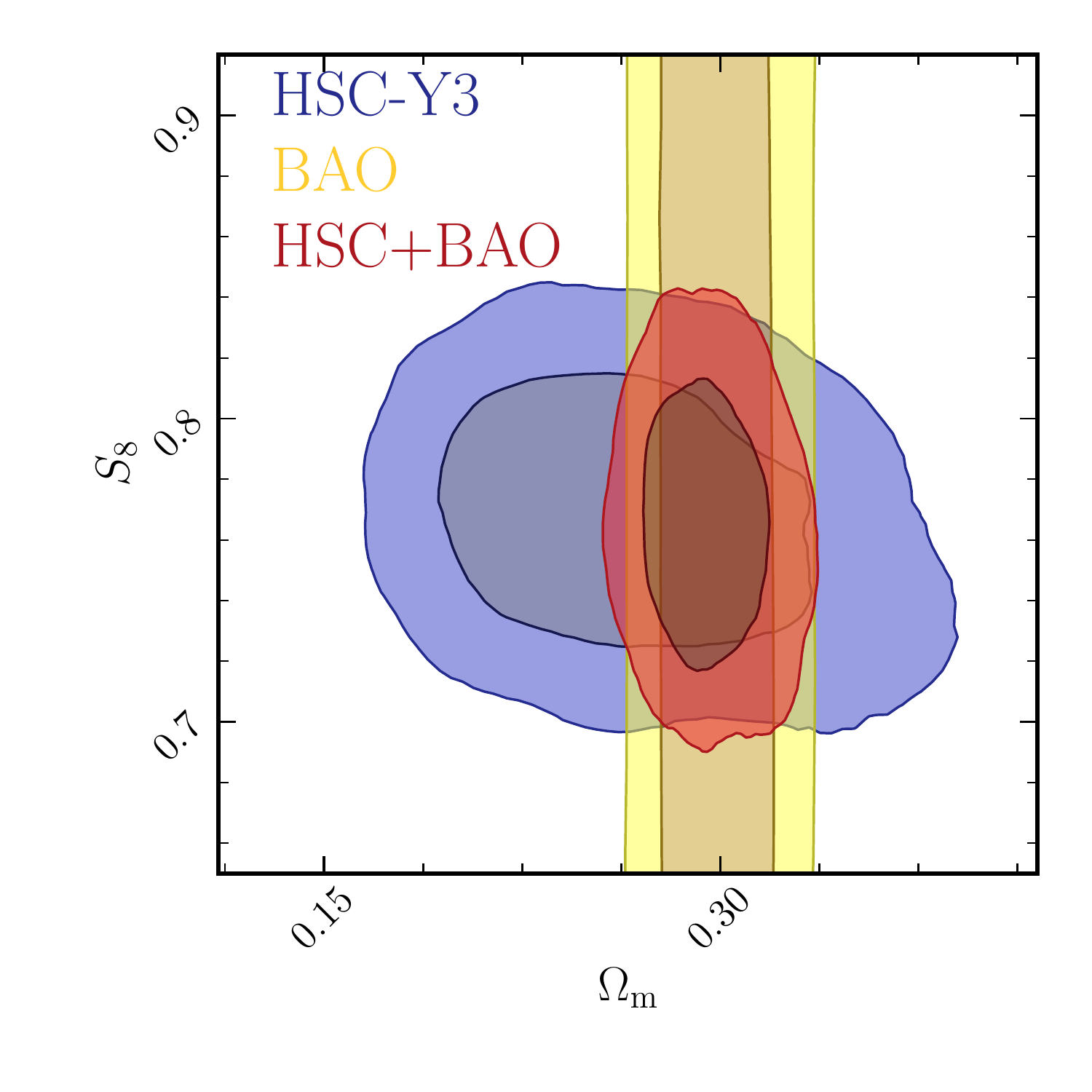}
\caption{
     Comparison between our fiducial constraint (blue contours) and the
     constraint on cosmology from the eBOSS BAO measurement (yellow contours),
     which is analyzed with the same priors in Table~\ref{tab:parameters}.
     The red contours are the joint estimation between the two observations.
     }
    \label{fig:ext_bao}
\end{figure}

We next compare our fiducial constraint with \textit{Planck}-2018. As shown in
Fig.~\ref{fig:ext_wl_cmb}, our constraint on $S_8$ appears to be in tension
with \textit{Planck}-2018. To quantify the tension, we perform  importance
sampling to generate chains of equal length from our fiducial and the
\textit{Planck}-2018 chains. We then assume that the cosmological constraints are
independent and create a new probability distribution with the difference in
$S_8$ between these two chains --- $\Delta S_8 = S_8(\text{HSC}) -
S_8(\textit{Planck})$ following \citet{cosmoTension2017}. The probability of
our fiducial analysis being in tension with \textit{Planck}-2018 is defined as
the posterior probability enclosed within the contour intersecting the point
$\Delta S_8 = 0$\,. We find a 95.3\% chance of being in tension with
\textit{Planck}-2018, corresponding to a $1.98\,\sigma$ tension.

In addition, we quantify the possible tension between our constraint and the
\textit{Planck}-2018 using the \texttt{eigentension} method developed in
\citet{Park2020} with the assumption that these two constraints are
independent. We first diagonalize the posterior covariance matrix in the space
spanned by the cosmological parameters to find the eigenvectors and the
corresponding eigenvalues. Since $n_s$, $\omega_b$ and $h_0$ in our analysis
are prior-dominated, we focus on the two parameters that are well-constrained
and not prior-dominated: $\OmegaM$, and $\sigma_8$\,. After diagonalizing the
covariance matrix of these parameters, we find the two eigenvectors, defined as
$e_1$ = $\sigma_8(\Omega_m)^{0.56}$ and $e_2$ = $\Omega_m(\sigma_8)^{-0.56}$.
We compute the posterior distribution of the difference in the eigenvector,
which is defined as
\begin{equation}
    (\Delta e_1, \Delta e_2) \equiv
    (e_1, e_2)_\text{HSC} - (e_1, e_2)_\textit{Planck}\,.
\end{equation}
By estimating the posterior probability above the contour intersecting the
point $(\Delta e_1, \Delta e_2) = (0, 0)$, we find our constraint has a 94.5\%
chance of being in tension with \textit{Planck}-2018, which corresponds to a
$1.92\,\sigma$ tension.

\subsection{eBOSS BAO Analysis}
\label{subsec:exter_bossbao}

Finally, we compare our fiducial constraint with the extended Baryon
Oscillation Spectroscopic Survey (eBOSS) DR16 analysis \citep{eBOSS_BAO2021}.
The eBOSS analysis uses galaxies as direct tracers of the density field to
measure baryon acoustic oscillation (BAO) up to $z\sim 3$. We re-analyze the
BAO measurements with the prior summarized in Table~\ref{tab:parameters} using
different types of galaxies, including SDSS main galaxy sample
\citep[MCGs;][]{mgs_Ross2015}, BOSS DR12 galaxies \citep{bossBao_Alam2017},
eBOSS galaxies (including luminous red galaxies
\citep[LRGs][]{ebossLRGs_Bautista2021}, emission line galaxies
\citep[ELGs][]{ebossELGs_deMattia2021}), quasars \citep{ebossQSOs_Neveux2020}
and Lyman-$\alpha$ Forest Samples \citep{ebossLya_duMas2020}. The analysis
adopts the likelihood implemented in \cosmosis{}. The projected 2D posteriors
for the BAO analysis is shown in Fig.~\ref{fig:ext_bao}.

Then we quantify the tension between our fiducial constraint and the BAO
constraint on both $S_8$ and $(e_1, e_2)$. We do not find significant tension
between these two analyses --- $0.23\,\sigma$ and $0.19\,\sigma$ tensions for
$S_8$ and $(e_1, e_2)$, respectively. Since these two constraints are
independent and do not show strong tension, we perform a joint analysis between
the cosmic shear 2PCFs and the BAO measurements assuming that the two
observations are independent. The joint HSC-eBOSS analysis is shown in
Fig.~\ref{fig:ext_bao}. We find our constraint on $S_8$ does not change since
BAO does not constrain $S_8$; however, the constraint on $\OmegaM$
significantly improves.

\section{SUMMARY AND OUTLOOK}

This paper presents the cosmological constraints from cosmic shear 2PCFs with
over $\sim$$1$ million galaxies from the three-year Hyper Suprime-Cam (HSC-Y3)
data, which covers $416~\mathrm{deg}^2$ up to redshift $z\sim$2\,.

By using our fiducial model to analyze different synthetic 2PCFs, we find the
modeling uncertainties on $S_8$ are less than $0.5 \sigma$ even for the
simulations with the most extreme baryonic feedback models. We model the cosmic
shear 2PCFs in the flat $\Lambda$CDM cosmology with the sum of neutrino mass
fixed to $0.06~\mathrm{eV}$, and constrain the lensing amplitude with 3.5\%
precision, finding $S_8 =\sigma_8 \sqrt{\Omega_\mathrm{m}/0.3} =
0.769_{-0.034}^{+0.031}$ (68\% CI). Additionally, the matter density is
constrained with 5\% precision: $\OmegaM = 0.256_{-0.044}^{+0.056}$ (68\% CI).
Systematic tests on synthetic data vectors show that the modeling errors on
$S_8$ do not exceed $0.5\sigma$, whereas the maximum modeling error on
$\OmegaM$ is about $1\sigma$ arising from projection of high-dimensional
posterior onto $1$D space and the modeling uncertainties in baryonic feedback.
To assess the robustness of our constraint, we conduct a number of blinded
internal consistency tests by analyzing different subsets of the data with
different systematic and astrophysical models under the context of the flat
$\Lambda$CDM cosmology. After unblinding, we compare our constraints on $S_8$
with other HSC-Y3 weak-lensing analyses
\citep[e.g.,][]{HSC3_cosmicShearFourier, HSC3_3x2pt_ls, HSC3_3x2pt_ss} and find
extremely good agreement between these analyses.

Furthermore, we compare our analysis with external dataset and find that our
results qualitatively agree well with weak-lensing analyses from the ongoing
surveys: KiDS-1000 \citep{KiDS1000_CS_Asgari2020}, DES-Y3
\citep{DESY3_CS_Amon2021}. However, these weak-lensing constraints on $S_8$ is
$\sim$$2\sigma$ lower than the constraint from \textit{Planck}-2018
\citep{cmb_Planck2018_Cosmology}.

For the final-year HSC dataset covering $\sim1,100~\mathrm{deg}^2$ on the
Northern sky, we expect the precision on $S_8$ measurement will be improved to
$\leq 0.025$ thanks to the increase in survey area. We will then be able to see
whether the $S_8$ tension remains. For the final-year analysis, controlling the
systematic errors will be more challenging. Below, we discuss a few places that
require improvements for the final year weak-lensing analyses.

\subsection{Modeling of Baryonic Feedback}

Our fiducial analysis models the baryonic feedback at small scales using
\hmcode{}~2016, and find a significant positive detection of baryonic feedback:
$A_\mathrm{b} = 2.34^{+0.40}_{-0.25}$, which is consistent with the HSC-Y3
Fourier space cosmic shear analysis ($2.43^{+0.46}_{-0.25}$)
\citep{HSC3_cosmicShearFourier}. We note that DES-Y3 adopts \halofit{}, a dark
matter only empirical model, to calculate the power spectrum since DES-Y3 goes to
larger angular scales than HSC-Y3, and they adopt a conservative small-scale
cut; therefore, they are less sensitive to baryonic feedback on small scales.
In addition, we conduct an analysis with \hmcode{}~2020, which models baryonic
feedback with $\Theta_\mathrm{AGN}$, and we find the difference in the $S_8$
constraint is less than $0.5\sigma$\,. Furthermore, we conduct a number of
tests using synthetic 2PCFs to confirm that, for our fiducial scale cut, our
constraint on $S_8$ is robust to modeling error in baryonic feedback. Given
this positive detection of baryonic feedback, future cosmology analyses,
especially ones aiming to use smaller scale data, will have to be careful in
understanding the modeling errors in baryonic feedback.

\subsection{$B$-modes at Large Scales}

We find significant $B$-modes in $\xi_+$ at scales $\theta>60~\mathrm{arcmin}$,
especially in the last two tomographic bins. To mitigate the $B$-mode leakage
into our cosmology analysis, we apply conservative scale cuts to remove angular
scales with $\theta>56$ arcmin in $\xi_+$\,. Similarly,
\citet{HSC3_cosmicShearFourier} find significant $B$-modes at scales
$\ell<300$. Note that the DES-Y3 2PCF analysis has a large-scale ct greater
than $200~\mathrm{arcmin}$. Since the DES-Y3 2PCF measurement is dominated by
data at very large scales, the modeling uncertainty from baryonic feedback,
which is significant at small scales, has a smaller influence on their analysis
than in our own. Therefore, controlling the $B$-modes in large angular scales
and including large-scale data in our analysis would not only improve the
accuracy but also reduce the modeling uncertainty from baryonic feedback. In
future analyses, we will further study the $B$-modes at large scales and
understand the cause of them.

\subsection{Redshift estimation errors}

Our fiducial analysis uses a conservative, wide, flat prior on the shifts in
the third and fourth redshift bins, and our results indicates significant
redshift error on these bins --- $\Delta z_3 = -0.115_{-0.058}^{0.052}$ and
$\Delta z_4 = -0.192_{-0.088}^{0.088}$, which do not agree with the Gaussian
priors on $\Delta z_{3,4}$ derived in \citet{HSC3_photoz_Rau2022}. In addition,
when applying the Gaussian priors of \citet{HSC3_photoz_Rau2022} on the last
two tomographic bins, we find that $S_8$ shifts to higher values by $\sim$$1.7
\sigma$\, compared to the fiducial analysis adopting flat priors. All of the
HSC-Y3 cosmology analyses \citep{HSC3_cosmicShearFourier, HSC3_3x2pt_ls,
HSC3_3x2pt_ss} find evidence for biased redshifts in these bins. Furthermore,
it should be noted that the study conducted by \citet{LWB2022} found a
correlation between the measured excess surface density and the mean source
redshift of the lensing survey, potentially resulting from redshift estimation
systematics. This was observed through a comparative analysis of the excess
surface density measurements of SDSS BOSS lens galaxies
(\citep{bossBao_Alam2017}) using background galaxies from several surveys
including CFHTLenS (\citep{CFHTLenS_Heymans2013}), CFHT Survey of Stripe 82
(\citep{LIL2017}), DES (\citep{DESY1_catalog_Zuntz2017}), KiDS
(\citep{KV450-Wright2019}), and HSC (\citep{HSC1_catalog}). Notably, our
findings on the redshift measurement errors are consistent with the trend
reported in \citet{LWB2022}.

We have taken the most conservative possible approach to these photo-$z$
errors. Due to the flat prior, the constraining power on $S_8$ of our HSC-Y3
analysis is similar to that of the HSC-Y1 analyses, although the sky coverage
is three times larger than the HSC-Y1 dataset. We will work on improving the
calibration of the source redshift distribution in the highest two tomographic
bins and try to improve the constraining power on $S_8$ .

\subsection{Future Improvements}

Our final-year HSC data release will cover $\sim1,100$ deg$^2$ of the Northern
sky with the same depth and image resolution. The data reduction will also be
performed by an updated version of LSST Science Pipelines
\citep{HSC1_pipeline}, with notable improvements being a multi-band deblender
\citep{scarlet_Melchior2018}, consistent selection of PSF stars and with a
state-of-the-art PSF modelling \citep{PIFF2021}. Due to the increase in the
data volume alone, the uncertainty on $S_8$ is expected to be reduced to about
2.5\%, and we will be able to see whether the significance of the tension
between HSC 2PCFs cosmic shear analysis and Planck increases. We will attempt
to improve the control of systematic errors for the final-year analyses with
the following approach:
\begin{enumerate}
    \item Include multi-band images in our galaxy image simulations with
        realistic galaxy color information \citep[for example,
        see][]{DC2_Abolfathi2021} to enable us to test photometric redshift
        estimation and calibrate the $n(z)$ estimation bias from
        redshift-dependent shear \citep{DESY3_BlendshearCalib_MacCrann2021,
        Skills_Li2023}.
    \item Update the shear estimation code to a state-of-art algorithm
        \citep{FPFS_Li2018, FPFS_Li2022, FPFS_Li2023} which uses correct for
        multiplicative bias from detection and selection below $0.5\%$ without
        relying on empirical calibration from external image simulation.
    \item Cross-check the high redshift $n(z)$ estimates with other galaxy
        samples with spectroscopic redshift estimations (e.g.,\ DESI
        \citep{DESI2016}) or with other measurments (e.g.,\ shear ratio test
        \citep{DESY3_shearRatio} and CMB lensing
        \citep{KiDS1000xCMBLens_Robertson2021}).
\end{enumerate}

The HSC survey is a pioneer survey for future Stage-IV imaging surveys which
have much larger sky coverage, higher resolution and/or deeper imaging. These
Stage-IV surveys include a ground-based survey: the Vera C. Rubin Observatory
Legacy Survey of Space and Time \citep[LSST; ][]{LSSTOverviwe2019}), and
space-based surveys: Euclid \citep{Euclid2011} and the Nancy Grace Roman Space
Telescope \citep[Roman; ][]{WFIRST15}. These datasets will allow us to better
constrain $S_8$ and understand the apparent tension between the
\textit{Planck}-2018 CMB observation.

\section*{Acknowledgements}

XL, TZ and RM are supported in part by the Department of Energy grant
DE-SC0010118 and in part by a grant from the Simons Foundation (Simons
Investigator in Astrophysics, Award ID 620789). RD acknowledges support from
the NSF Graduate Research Fellowship Program under Grant No.\ DGE-2039656. Any
opinions, findings, and conclusions or recommendations expressed in this
material are those of the authors and do not necessarily reflect the views of
the National Science Foundation. We thank the anonymous reviewers for their
careful reading of our manuscript and their many insightful comments and
suggestions.

This work was supported in part by World Premier International Research Center
Initiative (WPI Initiative), MEXT, Japan, and JSPS KAKENHI Grant Numbers
JP18H04350, JP18H04358, JP19H00677, JP19K14767, JP20H00181, JP20H01932,
JP20H04723, JP20H05850, JP20H05855, JP20H05856, JP20H05861, JP21J00011,
JP21H05456, JP21J10314, JP21H01081, JP21H05456, JP22H00130, JP22K03634,
JP22K03655 and JP22K21349 by Japan Science and Technology Agency (JST) CREST
JPMHCR1414, by JST AIP Acceleration Research Grant Number JP20317829, Japan,
and by Basic Research Grant (Super AI) of Institute for AI and Beyond of the
University of Tokyo. SS was supported in part by International Graduate Program
for Excellence in Earth-Space Science (IGPEES), WINGS Program, the University
of Tokyo. This work was supported by JSPS Core-to-Core Program (grant number:
JPJSCCA20210003)

The Hyper Suprime-Cam (HSC) collaboration includes the astronomical communities
of Japan and Taiwan, and Princeton University. The HSC instrumentation and
software were developed by the National Astronomical Observatory of Japan
(NAOJ), the Kavli Institute for the Physics and Mathematics of the Universe
(Kavli IPMU), the University of Tokyo, the High Energy Accelerator Research
Organization (KEK), the Academia Sinica Institute for Astronomy and
Astrophysics in Taiwan (ASIAA), and Princeton University. Funding was
contributed by the FIRST program from the Japanese Cabinet Office, the Ministry
of Education, Culture, Sports, Science and Technology (MEXT), the Japan Society
for the Promotion of Science (JSPS), Japan Science and Technology Agency (JST),
the Toray Science Foundation, NAOJ, Kavli IPMU, KEK, ASIAA, and Princeton
University.

This paper is based on data collected at the Subaru Telescope and retrieved
from the HSC data archive system, which is operated by the Subaru Telescope and
Astronomy Data Center (ADC) at NAOJ. Data analysis was in part carried out with
the cooperation of Center for Computational Astrophysics (CfCA), NAOJ. We are
honored and grateful for the opportunity of observing the Universe from
Maunakea, which has the cultural, historical and natural significance in
Hawaii.

The Pan-STARRS1 Surveys (PS1) and the PS1 public science archive have been made
possible through contributions by the Institute for Astronomy, the University
of Hawaii, the Pan-STARRS Project Office, the Max Planck Society and its
participating institutes, the Max Planck Institute for Astronomy, Heidelberg,
and the Max Planck Institute for Extraterrestrial Physics, Garching, The Johns
Hopkins University, Durham University, the University of Edinburgh, the Queen’s
University Belfast, the Harvard-Smithsonian Center for Astrophysics, the Las
Cumbres Observatory Global Telescope Network Incorporated, the National Central
University of Taiwan, the Space Telescope Science Institute, the National
Aeronautics and Space Administration under grant No. NNX08AR22G issued through
the Planetary Science Division of the NASA Science Mission Directorate, the
National Science Foundation grant No. AST-1238877, the University of Maryland,
Eotvos Lorand University (ELTE), the Los Alamos National Laboratory, and the
Gordon and Betty Moore Foundation.

\bibliography{citation-short}

\begin{thebibliography}{163}%
\makeatletter
\providecommand \@ifxundefined [1]{%
 \@ifx{#1\undefined}
}%
\providecommand \@ifnum [1]{%
 \ifnum #1\expandafter \@firstoftwo
 \else \expandafter \@secondoftwo
 \fi
}%
\providecommand \@ifx [1]{%
 \ifx #1\expandafter \@firstoftwo
 \else \expandafter \@secondoftwo
 \fi
}%
\providecommand \natexlab [1]{#1}%
\providecommand \enquote  [1]{``#1''}%
\providecommand \bibnamefont  [1]{#1}%
\providecommand \bibfnamefont [1]{#1}%
\providecommand \citenamefont [1]{#1}%
\providecommand \href@noop [0]{\@secondoftwo}%
\providecommand \href [0]{\begingroup \@sanitize@url \@href}%
\providecommand \@href[1]{\@@startlink{#1}\@@href}%
\providecommand \@@href[1]{\endgroup#1\@@endlink}%
\providecommand \@sanitize@url [0]{\catcode `\\12\catcode `\$12\catcode `\&12\catcode `\#12\catcode `\^12\catcode `\_12\catcode `\%12\relax}%
\providecommand \@@startlink[1]{}%
\providecommand \@@endlink[0]{}%
\providecommand \url  [0]{\begingroup\@sanitize@url \@url }%
\providecommand \@url [1]{\endgroup\@href {#1}{\urlprefix }}%
\providecommand \urlprefix  [0]{URL }%
\providecommand \Eprint [0]{\href }%
\providecommand \doibase [0]{http://dx.doi.org/}%
\providecommand \selectlanguage [0]{\@gobble}%
\providecommand \bibinfo  [0]{\@secondoftwo}%
\providecommand \bibfield  [0]{\@secondoftwo}%
\providecommand \translation [1]{[#1]}%
\providecommand \BibitemOpen [0]{}%
\providecommand \bibitemStop [0]{}%
\providecommand \bibitemNoStop [0]{.\EOS\space}%
\providecommand \EOS [0]{\spacefactor3000\relax}%
\providecommand \BibitemShut  [1]{\csname bibitem#1\endcsname}%
\let\auto@bib@innerbib\@empty
\bibitem [{\citenamefont {{Betoule}}\ \emph {et~al.}(2014)\citenamefont {{Betoule}}, \citenamefont {{Kessler}}, \citenamefont {{Guy}}, \citenamefont {{Mosher}}, \citenamefont {{Hardin}} \emph {et~al.}}]{IaSNA_SDSS_Betoule2014}%
  \BibitemOpen
  \bibfield  {author} {\bibinfo {author} {\bibfnamefont {M.}~\bibnamefont {{Betoule}}}, \bibinfo {author} {\bibfnamefont {R.}~\bibnamefont {{Kessler}}}, \bibinfo {author} {\bibfnamefont {J.}~\bibnamefont {{Guy}}}, \bibinfo {author} {\bibfnamefont {J.}~\bibnamefont {{Mosher}}}, \bibinfo {author} {\bibfnamefont {D.}~\bibnamefont {{Hardin}}},  \emph {et~al.},\ }\href {\doibase 10.1051/0004-6361/201423413} {\bibfield  {journal} {\bibinfo  {journal} {\aap}\ }\textbf {\bibinfo {volume} {568}},\ \bibinfo {eid} {A22} (\bibinfo {year} {2014})},\ \Eprint {http://arxiv.org/abs/1401.4064} {arXiv:1401.4064 [astro-ph.CO]} \BibitemShut {NoStop}%
\bibitem [{\citenamefont {{Fields}}\ \emph {et~al.}(2020)\citenamefont {{Fields}}, \citenamefont {{Olive}}, \citenamefont {{Yeh}},\ and\ \citenamefont {{Young}}}]{BBNPostplanck_Fields2020}%
  \BibitemOpen
  \bibfield  {author} {\bibinfo {author} {\bibfnamefont {B.~D.}\ \bibnamefont {{Fields}}}, \bibinfo {author} {\bibfnamefont {K.~A.}\ \bibnamefont {{Olive}}}, \bibinfo {author} {\bibfnamefont {T.-H.}\ \bibnamefont {{Yeh}}}, \ and\ \bibinfo {author} {\bibfnamefont {C.}~\bibnamefont {{Young}}},\ }\href {\doibase 10.1088/1475-7516/2020/03/010} {\bibfield  {journal} {\bibinfo  {journal} {\jcap}\ }\textbf {\bibinfo {volume} {2020}},\ \bibinfo {eid} {010} (\bibinfo {year} {2020})},\ \Eprint {http://arxiv.org/abs/1912.01132} {arXiv:1912.01132 [astro-ph.CO]} \BibitemShut {NoStop}%
\bibitem [{\citenamefont {Hinshaw}\ \emph {et~al.}(2013)\citenamefont {Hinshaw}, \citenamefont {Larson}, \citenamefont {Komatsu}, \citenamefont {Spergel}, \citenamefont {Bennett} \emph {et~al.}}]{cmb_WMAP9_Hinshaw2013}%
  \BibitemOpen
  \bibfield  {author} {\bibinfo {author} {\bibfnamefont {G.}~\bibnamefont {Hinshaw}}, \bibinfo {author} {\bibfnamefont {D.}~\bibnamefont {Larson}}, \bibinfo {author} {\bibfnamefont {E.}~\bibnamefont {Komatsu}}, \bibinfo {author} {\bibfnamefont {D.~N.}\ \bibnamefont {Spergel}}, \bibinfo {author} {\bibfnamefont {C.~L.}\ \bibnamefont {Bennett}},  \emph {et~al.},\ }\href {\doibase 10.1088/0067-0049/208/2/19} {\bibfield  {journal} {\bibinfo  {journal} {The Astrophysical Journal Supplement Series}\ }\textbf {\bibinfo {volume} {208}},\ \bibinfo {pages} {19} (\bibinfo {year} {2013})}\BibitemShut {NoStop}%
\bibitem [{\citenamefont {{Planck Collaboration}}\ \emph {et~al.}(2020)\citenamefont {{Planck Collaboration}}, \citenamefont {{Aghanim}}, \citenamefont {{Akrami}}, \citenamefont {{Ashdown}}, \citenamefont {{Aumont}} \emph {et~al.}}]{cmb_Planck2018_Cosmology}%
  \BibitemOpen
  \bibfield  {author} {\bibinfo {author} {\bibnamefont {{Planck Collaboration}}}, \bibinfo {author} {\bibfnamefont {N.}~\bibnamefont {{Aghanim}}}, \bibinfo {author} {\bibfnamefont {Y.}~\bibnamefont {{Akrami}}}, \bibinfo {author} {\bibfnamefont {M.}~\bibnamefont {{Ashdown}}}, \bibinfo {author} {\bibfnamefont {J.}~\bibnamefont {{Aumont}}},  \emph {et~al.},\ }\href {\doibase 10.1051/0004-6361/201833910} {\bibfield  {journal} {\bibinfo  {journal} {\aap}\ }\textbf {\bibinfo {volume} {641}},\ \bibinfo {eid} {A6} (\bibinfo {year} {2020})},\ \Eprint {http://arxiv.org/abs/1807.06209} {arXiv:1807.06209 [astro-ph.CO]} \BibitemShut {NoStop}%
\bibitem [{\citenamefont {{Hamana}}\ \emph {et~al.}(2020)\citenamefont {{Hamana}}, \citenamefont {{Shirasaki}}, \citenamefont {{Miyazaki}}, \citenamefont {{Hikage}}, \citenamefont {{Oguri}} \emph {et~al.}}]{cosmicShear_HSC1_Hamana2019}%
  \BibitemOpen
  \bibfield  {author} {\bibinfo {author} {\bibfnamefont {T.}~\bibnamefont {{Hamana}}}, \bibinfo {author} {\bibfnamefont {M.}~\bibnamefont {{Shirasaki}}}, \bibinfo {author} {\bibfnamefont {S.}~\bibnamefont {{Miyazaki}}}, \bibinfo {author} {\bibfnamefont {C.}~\bibnamefont {{Hikage}}}, \bibinfo {author} {\bibfnamefont {M.}~\bibnamefont {{Oguri}}},  \emph {et~al.},\ }\href {\doibase 10.1093/pasj/psz138} {\bibfield  {journal} {\bibinfo  {journal} {\pasj}\ }\textbf {\bibinfo {volume} {72}},\ \bibinfo {eid} {16} (\bibinfo {year} {2020})},\ \Eprint {http://arxiv.org/abs/1906.06041} {arXiv:1906.06041 [astro-ph.CO]} \BibitemShut {NoStop}%
\bibitem [{\citenamefont {{Asgari}}\ \emph {et~al.}(2021)\citenamefont {{Asgari}}, \citenamefont {{Lin}}, \citenamefont {{Joachimi}}, \citenamefont {{Giblin}}, \citenamefont {{Heymans}} \emph {et~al.}}]{KiDS1000_CS_Asgari2020}%
  \BibitemOpen
  \bibfield  {author} {\bibinfo {author} {\bibfnamefont {M.}~\bibnamefont {{Asgari}}}, \bibinfo {author} {\bibfnamefont {C.-A.}\ \bibnamefont {{Lin}}}, \bibinfo {author} {\bibfnamefont {B.}~\bibnamefont {{Joachimi}}}, \bibinfo {author} {\bibfnamefont {B.}~\bibnamefont {{Giblin}}}, \bibinfo {author} {\bibfnamefont {C.}~\bibnamefont {{Heymans}}},  \emph {et~al.},\ }\href {\doibase 10.1051/0004-6361/202039070} {\bibfield  {journal} {\bibinfo  {journal} {\aap}\ }\textbf {\bibinfo {volume} {645}},\ \bibinfo {eid} {A104} (\bibinfo {year} {2021})},\ \Eprint {http://arxiv.org/abs/2007.15633} {arXiv:2007.15633 [astro-ph.CO]} \BibitemShut {NoStop}%
\bibitem [{\citenamefont {{Secco}}\ \emph {et~al.}(2022)\citenamefont {{Secco}}, \citenamefont {{Samuroff}}, \citenamefont {{Krause}}, \citenamefont {{Jain}}, \citenamefont {{Blazek}}, \citenamefont {{Raveri}} \emph {et~al.}}]{DESY3_CS_Secco2022}%
  \BibitemOpen
  \bibfield  {author} {\bibinfo {author} {\bibfnamefont {L.~F.}\ \bibnamefont {{Secco}}}, \bibinfo {author} {\bibfnamefont {S.}~\bibnamefont {{Samuroff}}}, \bibinfo {author} {\bibfnamefont {E.}~\bibnamefont {{Krause}}}, \bibinfo {author} {\bibfnamefont {B.}~\bibnamefont {{Jain}}}, \bibinfo {author} {\bibfnamefont {J.}~\bibnamefont {{Blazek}}}, \bibinfo {author} {\bibfnamefont {M.}~\bibnamefont {{Raveri}}},  \emph {et~al.},\ }\href {\doibase 10.1103/PhysRevD.105.023515} {\bibfield  {journal} {\bibinfo  {journal} {\prd}\ }\textbf {\bibinfo {volume} {105}},\ \bibinfo {eid} {023515} (\bibinfo {year} {2022})},\ \Eprint {http://arxiv.org/abs/2105.13544} {arXiv:2105.13544 [astro-ph.CO]} \BibitemShut {NoStop}%
\bibitem [{\citenamefont {{Sugiyama}}\ \emph {et~al.}(2022)\citenamefont {{Sugiyama}}, \citenamefont {{Takada}}, \citenamefont {{Miyatake}}, \citenamefont {{Nishimichi}}, \citenamefont {{Shirasaki}}, \citenamefont {{Kobayashi}} \emph {et~al.}}]{HSC1_2x2pt_Sugiyama2022}%
  \BibitemOpen
  \bibfield  {author} {\bibinfo {author} {\bibfnamefont {S.}~\bibnamefont {{Sugiyama}}}, \bibinfo {author} {\bibfnamefont {M.}~\bibnamefont {{Takada}}}, \bibinfo {author} {\bibfnamefont {H.}~\bibnamefont {{Miyatake}}}, \bibinfo {author} {\bibfnamefont {T.}~\bibnamefont {{Nishimichi}}}, \bibinfo {author} {\bibfnamefont {M.}~\bibnamefont {{Shirasaki}}}, \bibinfo {author} {\bibfnamefont {Y.}~\bibnamefont {{Kobayashi}}},  \emph {et~al.},\ }\href {\doibase 10.1103/PhysRevD.105.123537} {\bibfield  {journal} {\bibinfo  {journal} {\prd}\ }\textbf {\bibinfo {volume} {105}},\ \bibinfo {eid} {123537} (\bibinfo {year} {2022})},\ \Eprint {http://arxiv.org/abs/2111.10966} {arXiv:2111.10966 [astro-ph.CO]} \BibitemShut {NoStop}%
\bibitem [{\citenamefont {{Miyatake}}\ \emph {et~al.}(2022)\citenamefont {{Miyatake}}, \citenamefont {{Sugiyama}}, \citenamefont {{Takada}}, \citenamefont {{Nishimichi}}, \citenamefont {{Shirasaki}} \emph {et~al.}}]{HSC1_2x2pt_Miyatake2022}%
  \BibitemOpen
  \bibfield  {author} {\bibinfo {author} {\bibfnamefont {H.}~\bibnamefont {{Miyatake}}}, \bibinfo {author} {\bibfnamefont {S.}~\bibnamefont {{Sugiyama}}}, \bibinfo {author} {\bibfnamefont {M.}~\bibnamefont {{Takada}}}, \bibinfo {author} {\bibfnamefont {T.}~\bibnamefont {{Nishimichi}}}, \bibinfo {author} {\bibfnamefont {M.}~\bibnamefont {{Shirasaki}}},  \emph {et~al.},\ }\href {\doibase 10.1103/PhysRevD.106.083520} {\bibfield  {journal} {\bibinfo  {journal} {\prd}\ }\textbf {\bibinfo {volume} {106}},\ \bibinfo {eid} {083520} (\bibinfo {year} {2022})},\ \Eprint {http://arxiv.org/abs/2111.02419} {arXiv:2111.02419 [astro-ph.CO]} \BibitemShut {NoStop}%
\bibitem [{\citenamefont {{Heymans}}\ \emph {et~al.}(2021)\citenamefont {{Heymans}}, \citenamefont {{Tr{\"o}ster}}, \citenamefont {{Asgari}}, \citenamefont {{Blake}}, \citenamefont {{Hildebrandt}}, \citenamefont {{Joachimi}} \emph {et~al.}}]{KiDS1000_3x2pt_Heymans2021}%
  \BibitemOpen
  \bibfield  {author} {\bibinfo {author} {\bibfnamefont {C.}~\bibnamefont {{Heymans}}}, \bibinfo {author} {\bibfnamefont {T.}~\bibnamefont {{Tr{\"o}ster}}}, \bibinfo {author} {\bibfnamefont {M.}~\bibnamefont {{Asgari}}}, \bibinfo {author} {\bibfnamefont {C.}~\bibnamefont {{Blake}}}, \bibinfo {author} {\bibfnamefont {H.}~\bibnamefont {{Hildebrandt}}}, \bibinfo {author} {\bibfnamefont {B.}~\bibnamefont {{Joachimi}}},  \emph {et~al.},\ }\href {\doibase 10.1051/0004-6361/202039063} {\bibfield  {journal} {\bibinfo  {journal} {\aap}\ }\textbf {\bibinfo {volume} {646}},\ \bibinfo {eid} {A140} (\bibinfo {year} {2021})},\ \Eprint {http://arxiv.org/abs/2007.15632} {arXiv:2007.15632 [astro-ph.CO]} \BibitemShut {NoStop}%
\bibitem [{\citenamefont {{Abbott}}\ \emph {et~al.}(2022)\citenamefont {{Abbott}}, \citenamefont {{Aguena}}, \citenamefont {{Alarcon}}, \citenamefont {{Allam}}, \citenamefont {{Alves}}, \citenamefont {{Amon}} \emph {et~al.}}]{DESY3_3x2pt2022}%
  \BibitemOpen
  \bibfield  {author} {\bibinfo {author} {\bibfnamefont {T.~M.~C.}\ \bibnamefont {{Abbott}}}, \bibinfo {author} {\bibfnamefont {M.}~\bibnamefont {{Aguena}}}, \bibinfo {author} {\bibfnamefont {A.}~\bibnamefont {{Alarcon}}}, \bibinfo {author} {\bibfnamefont {S.}~\bibnamefont {{Allam}}}, \bibinfo {author} {\bibfnamefont {O.}~\bibnamefont {{Alves}}}, \bibinfo {author} {\bibfnamefont {A.}~\bibnamefont {{Amon}}},  \emph {et~al.},\ }\href {\doibase 10.1103/PhysRevD.105.023520} {\bibfield  {journal} {\bibinfo  {journal} {\prd}\ }\textbf {\bibinfo {volume} {105}},\ \bibinfo {eid} {023520} (\bibinfo {year} {2022})},\ \Eprint {http://arxiv.org/abs/2105.13549} {arXiv:2105.13549 [astro-ph.CO]} \BibitemShut {NoStop}%
\bibitem [{\citenamefont {{Kobayashi}}\ \emph {et~al.}(2022)\citenamefont {{Kobayashi}}, \citenamefont {{Nishimichi}}, \citenamefont {{Takada}},\ and\ \citenamefont {{Miyatake}}}]{2022PhRvD.105h3517K}%
  \BibitemOpen
  \bibfield  {author} {\bibinfo {author} {\bibfnamefont {Y.}~\bibnamefont {{Kobayashi}}}, \bibinfo {author} {\bibfnamefont {T.}~\bibnamefont {{Nishimichi}}}, \bibinfo {author} {\bibfnamefont {M.}~\bibnamefont {{Takada}}}, \ and\ \bibinfo {author} {\bibfnamefont {H.}~\bibnamefont {{Miyatake}}},\ }\href {\doibase 10.1103/PhysRevD.105.083517} {\bibfield  {journal} {\bibinfo  {journal} {\prd}\ }\textbf {\bibinfo {volume} {105}},\ \bibinfo {eid} {083517} (\bibinfo {year} {2022})},\ \Eprint {http://arxiv.org/abs/2110.06969} {arXiv:2110.06969 [astro-ph.CO]} \BibitemShut {NoStop}%
\bibitem [{\citenamefont {{Abdalla}}\ \emph {et~al.}(2022)\citenamefont {{Abdalla}}, \citenamefont {{Abell{\'a}n}}, \citenamefont {{Aboubrahim}}, \citenamefont {{Agnello}}, \citenamefont {{Akarsu}} \emph {et~al.}}]{2022JHEAp..34...49A}%
  \BibitemOpen
  \bibfield  {author} {\bibinfo {author} {\bibfnamefont {E.}~\bibnamefont {{Abdalla}}}, \bibinfo {author} {\bibfnamefont {G.~F.}\ \bibnamefont {{Abell{\'a}n}}}, \bibinfo {author} {\bibfnamefont {A.}~\bibnamefont {{Aboubrahim}}}, \bibinfo {author} {\bibfnamefont {A.}~\bibnamefont {{Agnello}}}, \bibinfo {author} {\bibfnamefont {{\"O}.}~\bibnamefont {{Akarsu}}},  \emph {et~al.},\ }\href {\doibase 10.1016/j.jheap.2022.04.002} {\bibfield  {journal} {\bibinfo  {journal} {Journal of High Energy Astrophysics}\ }\textbf {\bibinfo {volume} {34}},\ \bibinfo {pages} {49} (\bibinfo {year} {2022})},\ \Eprint {http://arxiv.org/abs/2203.06142} {arXiv:2203.06142 [astro-ph.CO]} \BibitemShut {NoStop}%
\bibitem [{\citenamefont {Bartelmann}\ and\ \citenamefont {Schneider}(2001)}]{wlRevBartelmann}%
  \BibitemOpen
  \bibfield  {author} {\bibinfo {author} {\bibfnamefont {M.}~\bibnamefont {Bartelmann}}\ and\ \bibinfo {author} {\bibfnamefont {P.}~\bibnamefont {Schneider}},\ }\href {\doibase https://doi.org/10.1016/S0370-1573(00)00082-X} {\bibfield  {journal} {\bibinfo  {journal} {Physics Reports}\ }\textbf {\bibinfo {volume} {340}},\ \bibinfo {pages} {291} (\bibinfo {year} {2001})}\BibitemShut {NoStop}%
\bibitem [{\citenamefont {{Kilbinger}}(2015)}]{rev_cosmicShear_Kilbinger15}%
  \BibitemOpen
  \bibfield  {author} {\bibinfo {author} {\bibfnamefont {M.}~\bibnamefont {{Kilbinger}}},\ }\href {\doibase 10.1088/0034-4885/78/8/086901} {\bibfield  {journal} {\bibinfo  {journal} {Reports on Progress in Physics}\ }\textbf {\bibinfo {volume} {78}},\ \bibinfo {eid} {086901} (\bibinfo {year} {2015})},\ \Eprint {http://arxiv.org/abs/1411.0115} {arXiv:1411.0115} \BibitemShut {NoStop}%
\bibitem [{\citenamefont {{de Jong}}\ \emph {et~al.}(2013)\citenamefont {{de Jong}}, \citenamefont {{Verdoes Kleijn}}, \citenamefont {{Kuijken}},\ and\ \citenamefont {{Valentijn}}}]{KIDS13}%
  \BibitemOpen
  \bibfield  {author} {\bibinfo {author} {\bibfnamefont {J.~T.~A.}\ \bibnamefont {{de Jong}}}, \bibinfo {author} {\bibfnamefont {G.~A.}\ \bibnamefont {{Verdoes Kleijn}}}, \bibinfo {author} {\bibfnamefont {K.~H.}\ \bibnamefont {{Kuijken}}}, \ and\ \bibinfo {author} {\bibfnamefont {E.~A.}\ \bibnamefont {{Valentijn}}},\ }\href {\doibase 10.1007/s10686-012-9306-1} {\bibfield  {journal} {\bibinfo  {journal} {Experimental Astronomy}\ }\textbf {\bibinfo {volume} {35}},\ \bibinfo {pages} {25} (\bibinfo {year} {2013})},\ \Eprint {http://arxiv.org/abs/1206.1254} {arXiv:1206.1254 [astro-ph.CO]} \BibitemShut {NoStop}%
\bibitem [{\citenamefont {{Dark Energy Survey Collaboration}}\ \emph {et~al.}(2016)\citenamefont {{Dark Energy Survey Collaboration}}, \citenamefont {{Abbott}}, \citenamefont {{Abdalla}}, \citenamefont {{Aleksi{\'c}}}, \citenamefont {{Allam}} \emph {et~al.}}]{DES_overview_2016}%
  \BibitemOpen
  \bibfield  {author} {\bibinfo {author} {\bibnamefont {{Dark Energy Survey Collaboration}}}, \bibinfo {author} {\bibfnamefont {T.}~\bibnamefont {{Abbott}}}, \bibinfo {author} {\bibfnamefont {F.~B.}\ \bibnamefont {{Abdalla}}}, \bibinfo {author} {\bibfnamefont {J.}~\bibnamefont {{Aleksi{\'c}}}}, \bibinfo {author} {\bibfnamefont {S.}~\bibnamefont {{Allam}}},  \emph {et~al.},\ }\href {\doibase 10.1093/mnras/stw641} {\bibfield  {journal} {\bibinfo  {journal} {\mnras}\ }\textbf {\bibinfo {volume} {460}},\ \bibinfo {pages} {1270} (\bibinfo {year} {2016})},\ \Eprint {http://arxiv.org/abs/1601.00329} {arXiv:1601.00329} \BibitemShut {NoStop}%
\bibitem [{\citenamefont {{Aihara}}\ \emph {et~al.}(2018)\citenamefont {{Aihara}}, \citenamefont {{Arimoto}}, \citenamefont {{Armstrong}}, \citenamefont {{Arnouts}}, \citenamefont {{Bahcall}}, \citenamefont {{Bickerton}}, \citenamefont {{Bosch}} \emph {et~al.}}]{HSC_SSP2018}%
  \BibitemOpen
  \bibfield  {author} {\bibinfo {author} {\bibfnamefont {H.}~\bibnamefont {{Aihara}}}, \bibinfo {author} {\bibfnamefont {N.}~\bibnamefont {{Arimoto}}}, \bibinfo {author} {\bibfnamefont {R.}~\bibnamefont {{Armstrong}}}, \bibinfo {author} {\bibfnamefont {S.}~\bibnamefont {{Arnouts}}}, \bibinfo {author} {\bibfnamefont {N.~A.}\ \bibnamefont {{Bahcall}}}, \bibinfo {author} {\bibfnamefont {S.}~\bibnamefont {{Bickerton}}}, \bibinfo {author} {\bibfnamefont {J.}~\bibnamefont {{Bosch}}},  \emph {et~al.},\ }\href {\doibase 10.1093/pasj/psx066} {\bibfield  {journal} {\bibinfo  {journal} {\pasj}\ }\textbf {\bibinfo {volume} {70}},\ \bibinfo {eid} {S4} (\bibinfo {year} {2018})},\ \Eprint {http://arxiv.org/abs/1704.05858} {arXiv:1704.05858 [astro-ph.IM]} \BibitemShut {NoStop}%
\bibitem [{\citenamefont {{Miyazaki}}\ \emph {et~al.}(2018)\citenamefont {{Miyazaki}}, \citenamefont {{Komiyama}}, \citenamefont {{Kawanomoto}}, \citenamefont {{Doi}}, \citenamefont {{Furusawa}}, \citenamefont {{Hamana}} \emph {et~al.}}]{HSC_hardware_Miyazaki2018}%
  \BibitemOpen
  \bibfield  {author} {\bibinfo {author} {\bibfnamefont {S.}~\bibnamefont {{Miyazaki}}}, \bibinfo {author} {\bibfnamefont {Y.}~\bibnamefont {{Komiyama}}}, \bibinfo {author} {\bibfnamefont {S.}~\bibnamefont {{Kawanomoto}}}, \bibinfo {author} {\bibfnamefont {Y.}~\bibnamefont {{Doi}}}, \bibinfo {author} {\bibfnamefont {H.}~\bibnamefont {{Furusawa}}}, \bibinfo {author} {\bibfnamefont {T.}~\bibnamefont {{Hamana}}},  \emph {et~al.},\ }\href {\doibase 10.1093/pasj/psx063} {\bibfield  {journal} {\bibinfo  {journal} {\pasj}\ }\textbf {\bibinfo {volume} {70}},\ \bibinfo {eid} {S1} (\bibinfo {year} {2018})}\BibitemShut {NoStop}%
\bibitem [{\citenamefont {{Komiyama}}\ \emph {et~al.}(2018)\citenamefont {{Komiyama}}, \citenamefont {{Obuchi}}, \citenamefont {{Nakaya}}, \citenamefont {{Kamata}}, \citenamefont {{Kawanomoto}}, \citenamefont {{Utsumi}} \emph {et~al.}}]{HSC_hardware_Komiyama2018}%
  \BibitemOpen
  \bibfield  {author} {\bibinfo {author} {\bibfnamefont {Y.}~\bibnamefont {{Komiyama}}}, \bibinfo {author} {\bibfnamefont {Y.}~\bibnamefont {{Obuchi}}}, \bibinfo {author} {\bibfnamefont {H.}~\bibnamefont {{Nakaya}}}, \bibinfo {author} {\bibfnamefont {Y.}~\bibnamefont {{Kamata}}}, \bibinfo {author} {\bibfnamefont {S.}~\bibnamefont {{Kawanomoto}}}, \bibinfo {author} {\bibfnamefont {Y.}~\bibnamefont {{Utsumi}}},  \emph {et~al.},\ }\href {\doibase 10.1093/pasj/psx069} {\bibfield  {journal} {\bibinfo  {journal} {\pasj}\ }\textbf {\bibinfo {volume} {70}},\ \bibinfo {eid} {S2} (\bibinfo {year} {2018})}\BibitemShut {NoStop}%
\bibitem [{\citenamefont {{Kawanomoto}}\ \emph {et~al.}(2018)\citenamefont {{Kawanomoto}}, \citenamefont {{Uraguchi}}, \citenamefont {{Komiyama}}, \citenamefont {{Miyazaki}}, \citenamefont {{Furusawa}} \emph {et~al.}}]{HSC_hardware_Kawanomoto2018}%
  \BibitemOpen
  \bibfield  {author} {\bibinfo {author} {\bibfnamefont {S.}~\bibnamefont {{Kawanomoto}}}, \bibinfo {author} {\bibfnamefont {F.}~\bibnamefont {{Uraguchi}}}, \bibinfo {author} {\bibfnamefont {Y.}~\bibnamefont {{Komiyama}}}, \bibinfo {author} {\bibfnamefont {S.}~\bibnamefont {{Miyazaki}}}, \bibinfo {author} {\bibfnamefont {H.}~\bibnamefont {{Furusawa}}},  \emph {et~al.},\ }\href {\doibase 10.1093/pasj/psy056} {\bibfield  {journal} {\bibinfo  {journal} {\pasj}\ }\textbf {\bibinfo {volume} {70}},\ \bibinfo {eid} {66} (\bibinfo {year} {2018})}\BibitemShut {NoStop}%
\bibitem [{\citenamefont {{Furusawa}}\ \emph {et~al.}(2018)\citenamefont {{Furusawa}}, \citenamefont {{Koike}}, \citenamefont {{Takata}}, \citenamefont {{Okura}}, \citenamefont {{Miyatake}}, \citenamefont {{Lupton}} \emph {et~al.}}]{HSC_hardware_Furusawa2018}%
  \BibitemOpen
  \bibfield  {author} {\bibinfo {author} {\bibfnamefont {H.}~\bibnamefont {{Furusawa}}}, \bibinfo {author} {\bibfnamefont {M.}~\bibnamefont {{Koike}}}, \bibinfo {author} {\bibfnamefont {T.}~\bibnamefont {{Takata}}}, \bibinfo {author} {\bibfnamefont {Y.}~\bibnamefont {{Okura}}}, \bibinfo {author} {\bibfnamefont {H.}~\bibnamefont {{Miyatake}}}, \bibinfo {author} {\bibfnamefont {R.~H.}\ \bibnamefont {{Lupton}}},  \emph {et~al.},\ }\href {\doibase 10.1093/pasj/psx079} {\bibfield  {journal} {\bibinfo  {journal} {\pasj}\ }\textbf {\bibinfo {volume} {70}},\ \bibinfo {eid} {S3} (\bibinfo {year} {2018})}\BibitemShut {NoStop}%
\bibitem [{\citenamefont {{Li}}\ \emph {et~al.}(2021)\citenamefont {{Li}}, \citenamefont {{Miyatake}}, \citenamefont {{Luo}}, \citenamefont {{More}}, \citenamefont {{Oguri}}, \citenamefont {{Hamana}}, \citenamefont {{Mandelbaum}} \emph {et~al.}}]{HSC3_catalog_Li2021}%
  \BibitemOpen
  \bibfield  {author} {\bibinfo {author} {\bibfnamefont {X.}~\bibnamefont {{Li}}}, \bibinfo {author} {\bibfnamefont {H.}~\bibnamefont {{Miyatake}}}, \bibinfo {author} {\bibfnamefont {W.}~\bibnamefont {{Luo}}}, \bibinfo {author} {\bibfnamefont {S.}~\bibnamefont {{More}}}, \bibinfo {author} {\bibfnamefont {M.}~\bibnamefont {{Oguri}}}, \bibinfo {author} {\bibfnamefont {T.}~\bibnamefont {{Hamana}}}, \bibinfo {author} {\bibfnamefont {R.}~\bibnamefont {{Mandelbaum}}},  \emph {et~al.},\ }\href@noop {} {\bibfield  {journal} {\bibinfo  {journal} {arXiv e-prints}\ ,\ \bibinfo {eid} {arXiv:2107.00136}} (\bibinfo {year} {2021})},\ \Eprint {http://arxiv.org/abs/2107.00136} {arXiv:2107.00136 [astro-ph.CO]} \BibitemShut {NoStop}%
\bibitem [{\citenamefont {{Rau}}\ \emph {et~al.}(2022)\citenamefont {{Rau}}, \citenamefont {{Dalal}}, \citenamefont {{Zhang}}, \citenamefont {{Li}}, \citenamefont {{Nishizawa}} \emph {et~al.}}]{HSC3_photoz_Rau2022}%
  \BibitemOpen
  \bibfield  {author} {\bibinfo {author} {\bibfnamefont {M.~M.}\ \bibnamefont {{Rau}}}, \bibinfo {author} {\bibfnamefont {R.}~\bibnamefont {{Dalal}}}, \bibinfo {author} {\bibfnamefont {T.}~\bibnamefont {{Zhang}}}, \bibinfo {author} {\bibfnamefont {X.}~\bibnamefont {{Li}}}, \bibinfo {author} {\bibfnamefont {A.~J.}\ \bibnamefont {{Nishizawa}}},  \emph {et~al.},\ }\href {\doibase 10.48550/arXiv.2211.16516} {\bibfield  {journal} {\bibinfo  {journal} {arXiv e-prints}\ ,\ \bibinfo {eid} {arXiv:2211.16516}} (\bibinfo {year} {2022})},\ \Eprint {http://arxiv.org/abs/2211.16516} {arXiv:2211.16516 [astro-ph.CO]} \BibitemShut {NoStop}%
\bibitem [{\citenamefont {Oguri}(2014)}]{CAMIRA_Oguri2014}%
  \BibitemOpen
  \bibfield  {author} {\bibinfo {author} {\bibfnamefont {M.}~\bibnamefont {Oguri}},\ }\href {\doibase 10.1093/mnras/stu1446} {\bibfield  {journal} {\bibinfo  {journal} {Monthly Notices of the Royal Astronomical Society}\ }\textbf {\bibinfo {volume} {444}},\ \bibinfo {pages} {147} (\bibinfo {year} {2014})},\ \Eprint {http://arxiv.org/abs/1407.4693} {arXiv:1407.4693 [astro-ph.CO]} \BibitemShut {NoStop}%
\bibitem [{\citenamefont {Oguri}\ \emph {et~al.}(2018)\citenamefont {Oguri}, \citenamefont {Lin}, \citenamefont {Lin}, \citenamefont {Nishizawa}, \citenamefont {More} \emph {et~al.}}]{CAMIRA_HSC_Oguri2018}%
  \BibitemOpen
  \bibfield  {author} {\bibinfo {author} {\bibfnamefont {M.}~\bibnamefont {Oguri}}, \bibinfo {author} {\bibfnamefont {Y.-T.}\ \bibnamefont {Lin}}, \bibinfo {author} {\bibfnamefont {S.-C.}\ \bibnamefont {Lin}}, \bibinfo {author} {\bibfnamefont {A.~J.}\ \bibnamefont {Nishizawa}}, \bibinfo {author} {\bibfnamefont {A.}~\bibnamefont {More}},  \emph {et~al.},\ }\href {\doibase 10.1093/pasj/psx042} {\bibfield  {journal} {\bibinfo  {journal} {Publications of the Astronomical Society of Japan}\ }\textbf {\bibinfo {volume} {70}},\ \bibinfo {pages} {S20} (\bibinfo {year} {2018})},\ \Eprint {http://arxiv.org/abs/1701.00818} {arXiv:1701.00818} \BibitemShut {NoStop}%
\bibitem [{\citenamefont {{Oguri}}\ \emph {et~al.}(2018)\citenamefont {{Oguri}}, \citenamefont {{Miyazaki}}, \citenamefont {{Hikage}}, \citenamefont {{Mandelbaum}}, \citenamefont {{Utsumi}} \emph {et~al.}}]{HSC1_2D3Dmassmap}%
  \BibitemOpen
  \bibfield  {author} {\bibinfo {author} {\bibfnamefont {M.}~\bibnamefont {{Oguri}}}, \bibinfo {author} {\bibfnamefont {S.}~\bibnamefont {{Miyazaki}}}, \bibinfo {author} {\bibfnamefont {C.}~\bibnamefont {{Hikage}}}, \bibinfo {author} {\bibfnamefont {R.}~\bibnamefont {{Mandelbaum}}}, \bibinfo {author} {\bibfnamefont {Y.}~\bibnamefont {{Utsumi}}},  \emph {et~al.},\ }\href {\doibase 10.1093/pasj/psx070} {\bibfield  {journal} {\bibinfo  {journal} {\pasj}\ }\textbf {\bibinfo {volume} {70}},\ \bibinfo {eid} {S26} (\bibinfo {year} {2018})},\ \Eprint {http://arxiv.org/abs/1705.06792} {arXiv:1705.06792 [astro-ph.CO]} \BibitemShut {NoStop}%
\bibitem [{\citenamefont {{Zhang}}\ \emph {et~al.}(2022)\citenamefont {{Zhang}}, \citenamefont {{Li}}, \citenamefont {{Dalal}}, \citenamefont {{Mandelbaum}}, \citenamefont {{Strauss}}, \citenamefont {{Kannawadi}}, \citenamefont {{Miyatake}} \emph {et~al.}}]{HSC3_PSF}%
  \BibitemOpen
  \bibfield  {author} {\bibinfo {author} {\bibfnamefont {T.}~\bibnamefont {{Zhang}}}, \bibinfo {author} {\bibfnamefont {X.}~\bibnamefont {{Li}}}, \bibinfo {author} {\bibfnamefont {R.}~\bibnamefont {{Dalal}}}, \bibinfo {author} {\bibfnamefont {R.}~\bibnamefont {{Mandelbaum}}}, \bibinfo {author} {\bibfnamefont {M.~A.}\ \bibnamefont {{Strauss}}}, \bibinfo {author} {\bibfnamefont {A.}~\bibnamefont {{Kannawadi}}}, \bibinfo {author} {\bibfnamefont {H.}~\bibnamefont {{Miyatake}}},  \emph {et~al.},\ }\href {\doibase 10.48550/arXiv.2212.03257} {\bibfield  {journal} {\bibinfo  {journal} {arXiv e-prints}\ ,\ \bibinfo {eid} {arXiv:2212.03257}} (\bibinfo {year} {2022})},\ \Eprint {http://arxiv.org/abs/2212.03257} {arXiv:2212.03257 [astro-ph.CO]} \BibitemShut {NoStop}%
\bibitem [{\citenamefont {{Mandelbaum}}(2018)}]{rev_wlsys_Mandelbaum2017}%
  \BibitemOpen
  \bibfield  {author} {\bibinfo {author} {\bibfnamefont {R.}~\bibnamefont {{Mandelbaum}}},\ }\href {\doibase 10.1146/annurev-astro-081817-051928} {\bibfield  {journal} {\bibinfo  {journal} {\araa}\ }\textbf {\bibinfo {volume} {56}},\ \bibinfo {pages} {393} (\bibinfo {year} {2018})},\ \Eprint {http://arxiv.org/abs/1710.03235} {arXiv:1710.03235 [astro-ph.CO]} \BibitemShut {NoStop}%
\bibitem [{\citenamefont {{Mead}}\ \emph {et~al.}(2021)\citenamefont {{Mead}}, \citenamefont {{Brieden}}, \citenamefont {{Tr{\"o}ster}},\ and\ \citenamefont {{Heymans}}}]{halofit_mead21}%
  \BibitemOpen
  \bibfield  {author} {\bibinfo {author} {\bibfnamefont {A.~J.}\ \bibnamefont {{Mead}}}, \bibinfo {author} {\bibfnamefont {S.}~\bibnamefont {{Brieden}}}, \bibinfo {author} {\bibfnamefont {T.}~\bibnamefont {{Tr{\"o}ster}}}, \ and\ \bibinfo {author} {\bibfnamefont {C.}~\bibnamefont {{Heymans}}},\ }\href {\doibase 10.1093/mnras/stab082} {\bibfield  {journal} {\bibinfo  {journal} {\mnras}\ }\textbf {\bibinfo {volume} {502}},\ \bibinfo {pages} {1401} (\bibinfo {year} {2021})},\ \Eprint {http://arxiv.org/abs/2009.01858} {arXiv:2009.01858 [astro-ph.CO]} \BibitemShut {NoStop}%
\bibitem [{\citenamefont {{Moran}}\ \emph {et~al.}(2022)\citenamefont {{Moran}}, \citenamefont {{Heitmann}}, \citenamefont {{Lawrence}}, \citenamefont {{Habib}}, \citenamefont {{Bingham}} \emph {et~al.}}]{cosmicEmu2022}%
  \BibitemOpen
  \bibfield  {author} {\bibinfo {author} {\bibfnamefont {K.~R.}\ \bibnamefont {{Moran}}}, \bibinfo {author} {\bibfnamefont {K.}~\bibnamefont {{Heitmann}}}, \bibinfo {author} {\bibfnamefont {E.}~\bibnamefont {{Lawrence}}}, \bibinfo {author} {\bibfnamefont {S.}~\bibnamefont {{Habib}}}, \bibinfo {author} {\bibfnamefont {D.}~\bibnamefont {{Bingham}}},  \emph {et~al.},\ }\href {\doibase 10.1093/mnras/stac3452} {\bibfield  {journal} {\bibinfo  {journal} {\mnras}\ } (\bibinfo {year} {2022}),\ 10.1093/mnras/stac3452},\ \Eprint {http://arxiv.org/abs/2207.12345} {arXiv:2207.12345 [astro-ph.CO]} \BibitemShut {NoStop}%
\bibitem [{\citenamefont {{Osato}}\ \emph {et~al.}(2015)\citenamefont {{Osato}}, \citenamefont {{Shirasaki}},\ and\ \citenamefont {{Yoshida}}}]{baryon_osato15}%
  \BibitemOpen
  \bibfield  {author} {\bibinfo {author} {\bibfnamefont {K.}~\bibnamefont {{Osato}}}, \bibinfo {author} {\bibfnamefont {M.}~\bibnamefont {{Shirasaki}}}, \ and\ \bibinfo {author} {\bibfnamefont {N.}~\bibnamefont {{Yoshida}}},\ }\href {\doibase 10.1088/0004-637X/806/2/186} {\bibfield  {journal} {\bibinfo  {journal} {\apj}\ }\textbf {\bibinfo {volume} {806}},\ \bibinfo {eid} {186} (\bibinfo {year} {2015})},\ \Eprint {http://arxiv.org/abs/1501.02055} {arXiv:1501.02055 [astro-ph.CO]} \BibitemShut {NoStop}%
\bibitem [{\citenamefont {{Chen}}\ \emph {et~al.}(2022)\citenamefont {{Chen}}, \citenamefont {{Aric{\`o}}}, \citenamefont {{Huterer}}, \citenamefont {{Angulo}}, \citenamefont {{Weaverdyck}}, \citenamefont {{Friedrich}} \emph {et~al.}}]{baryon_chen22}%
  \BibitemOpen
  \bibfield  {author} {\bibinfo {author} {\bibfnamefont {A.}~\bibnamefont {{Chen}}}, \bibinfo {author} {\bibfnamefont {G.}~\bibnamefont {{Aric{\`o}}}}, \bibinfo {author} {\bibfnamefont {D.}~\bibnamefont {{Huterer}}}, \bibinfo {author} {\bibfnamefont {R.}~\bibnamefont {{Angulo}}}, \bibinfo {author} {\bibfnamefont {N.}~\bibnamefont {{Weaverdyck}}}, \bibinfo {author} {\bibfnamefont {O.}~\bibnamefont {{Friedrich}}},  \emph {et~al.},\ }\href@noop {} {\bibfield  {journal} {\bibinfo  {journal} {arXiv e-prints}\ ,\ \bibinfo {eid} {arXiv:2206.08591}} (\bibinfo {year} {2022})},\ \Eprint {http://arxiv.org/abs/2206.08591} {arXiv:2206.08591 [astro-ph.CO]} \BibitemShut {NoStop}%
\bibitem [{\citenamefont {{Tr{\"o}ster}}\ \emph {et~al.}(2022{\natexlab{a}})\citenamefont {{Tr{\"o}ster}}, \citenamefont {{Mead}}, \citenamefont {{Heymans}}, \citenamefont {{Yan}}, \citenamefont {{Alonso}} \emph {et~al.}}]{baryon_troster22}%
  \BibitemOpen
  \bibfield  {author} {\bibinfo {author} {\bibfnamefont {T.}~\bibnamefont {{Tr{\"o}ster}}}, \bibinfo {author} {\bibfnamefont {A.~J.}\ \bibnamefont {{Mead}}}, \bibinfo {author} {\bibfnamefont {C.}~\bibnamefont {{Heymans}}}, \bibinfo {author} {\bibfnamefont {Z.}~\bibnamefont {{Yan}}}, \bibinfo {author} {\bibfnamefont {D.}~\bibnamefont {{Alonso}}},  \emph {et~al.},\ }\href {\doibase 10.1051/0004-6361/202142197} {\bibfield  {journal} {\bibinfo  {journal} {\aap}\ }\textbf {\bibinfo {volume} {660}},\ \bibinfo {eid} {A27} (\bibinfo {year} {2022}{\natexlab{a}})},\ \Eprint {http://arxiv.org/abs/2109.04458} {arXiv:2109.04458 [astro-ph.CO]} \BibitemShut {NoStop}%
\bibitem [{\citenamefont {{Hirata}}\ \emph {et~al.}(2007)\citenamefont {{Hirata}}, \citenamefont {{Mandelbaum}}, \citenamefont {{Ishak}}, \citenamefont {{Seljak}}, \citenamefont {{Nichol}}, \citenamefont {{Pimbblet}} \emph {et~al.}}]{nla_hirata07}%
  \BibitemOpen
  \bibfield  {author} {\bibinfo {author} {\bibfnamefont {C.~M.}\ \bibnamefont {{Hirata}}}, \bibinfo {author} {\bibfnamefont {R.}~\bibnamefont {{Mandelbaum}}}, \bibinfo {author} {\bibfnamefont {M.}~\bibnamefont {{Ishak}}}, \bibinfo {author} {\bibfnamefont {U.}~\bibnamefont {{Seljak}}}, \bibinfo {author} {\bibfnamefont {R.}~\bibnamefont {{Nichol}}}, \bibinfo {author} {\bibfnamefont {K.~A.}\ \bibnamefont {{Pimbblet}}},  \emph {et~al.},\ }\href {\doibase 10.1111/j.1365-2966.2007.12312.x} {\bibfield  {journal} {\bibinfo  {journal} {\mnras}\ }\textbf {\bibinfo {volume} {381}},\ \bibinfo {pages} {1197} (\bibinfo {year} {2007})},\ \Eprint {http://arxiv.org/abs/astro-ph/0701671} {arXiv:astro-ph/0701671 [astro-ph]} \BibitemShut {NoStop}%
\bibitem [{\citenamefont {{Bridle}}\ and\ \citenamefont {{King}}(2007)}]{nla_bridle07}%
  \BibitemOpen
  \bibfield  {author} {\bibinfo {author} {\bibfnamefont {S.}~\bibnamefont {{Bridle}}}\ and\ \bibinfo {author} {\bibfnamefont {L.}~\bibnamefont {{King}}},\ }\href {\doibase 10.1088/1367-2630/9/12/444} {\bibfield  {journal} {\bibinfo  {journal} {New Journal of Physics}\ }\textbf {\bibinfo {volume} {9}},\ \bibinfo {pages} {444} (\bibinfo {year} {2007})},\ \Eprint {http://arxiv.org/abs/arXiv:0705.0166} {arXiv:0705.0166} \BibitemShut {NoStop}%
\bibitem [{\citenamefont {{Blazek}}\ \emph {et~al.}(2019)\citenamefont {{Blazek}}, \citenamefont {{MacCrann}}, \citenamefont {{Troxel}},\ and\ \citenamefont {{Fang}}}]{tatt_blazek17}%
  \BibitemOpen
  \bibfield  {author} {\bibinfo {author} {\bibfnamefont {J.~A.}\ \bibnamefont {{Blazek}}}, \bibinfo {author} {\bibfnamefont {N.}~\bibnamefont {{MacCrann}}}, \bibinfo {author} {\bibfnamefont {M.~A.}\ \bibnamefont {{Troxel}}}, \ and\ \bibinfo {author} {\bibfnamefont {X.}~\bibnamefont {{Fang}}},\ }\href {\doibase 10.1103/PhysRevD.100.103506} {\bibfield  {journal} {\bibinfo  {journal} {\prd}\ }\textbf {\bibinfo {volume} {100}},\ \bibinfo {eid} {103506} (\bibinfo {year} {2019})},\ \Eprint {http://arxiv.org/abs/1708.09247} {arXiv:1708.09247 [astro-ph.CO]} \BibitemShut {NoStop}%
\bibitem [{\citenamefont {{Amon}}\ \emph {et~al.}(2021{\natexlab{a}})\citenamefont {{Amon}}, \citenamefont {{Gruen}}, \citenamefont {{Troxel}}, \citenamefont {{MacCrann}}, \citenamefont {{Dodelson}}, \citenamefont {{Choi}} \emph {et~al.}}]{cosmicShear_DESY3_Amon2021}%
  \BibitemOpen
  \bibfield  {author} {\bibinfo {author} {\bibfnamefont {A.}~\bibnamefont {{Amon}}}, \bibinfo {author} {\bibfnamefont {D.}~\bibnamefont {{Gruen}}}, \bibinfo {author} {\bibfnamefont {M.~A.}\ \bibnamefont {{Troxel}}}, \bibinfo {author} {\bibfnamefont {N.}~\bibnamefont {{MacCrann}}}, \bibinfo {author} {\bibfnamefont {S.}~\bibnamefont {{Dodelson}}}, \bibinfo {author} {\bibfnamefont {A.}~\bibnamefont {{Choi}}},  \emph {et~al.},\ }\href@noop {} {\bibfield  {journal} {\bibinfo  {journal} {arXiv e-prints}\ ,\ \bibinfo {eid} {arXiv:2105.13543}} (\bibinfo {year} {2021}{\natexlab{a}})},\ \Eprint {http://arxiv.org/abs/2105.13543} {arXiv:2105.13543 [astro-ph.CO]} \BibitemShut {NoStop}%
\bibitem [{\citenamefont {Dalal}\ \emph {et~al.}(2023)\citenamefont {Dalal}, \citenamefont {Li}, \citenamefont {Nicola}, \citenamefont {Zuntz}, \citenamefont {Strauss} \emph {et~al.}}]{HSC3_cosmicShearFourier}%
  \BibitemOpen
  \bibfield  {author} {\bibinfo {author} {\bibfnamefont {R.}~\bibnamefont {Dalal}}, \bibinfo {author} {\bibfnamefont {X.}~\bibnamefont {Li}}, \bibinfo {author} {\bibfnamefont {A.}~\bibnamefont {Nicola}}, \bibinfo {author} {\bibfnamefont {J.}~\bibnamefont {Zuntz}}, \bibinfo {author} {\bibfnamefont {M.~A.}\ \bibnamefont {Strauss}},  \emph {et~al.},\ }\href@noop {} {\  (\bibinfo {year} {2023})},\ \Eprint {http://arxiv.org/abs/2304.00701} {arXiv:2304.00701 [astro-ph.CO]} \BibitemShut {NoStop}%
\bibitem [{\citenamefont {More}\ \emph {et~al.}(2023)\citenamefont {More}, \citenamefont {Sugiyama}, \citenamefont {Miyatake}, \citenamefont {Rau}, \citenamefont {Shirasaki} \emph {et~al.}}]{HSC3_3x2pt_meas}%
  \BibitemOpen
  \bibfield  {author} {\bibinfo {author} {\bibfnamefont {S.}~\bibnamefont {More}}, \bibinfo {author} {\bibfnamefont {S.}~\bibnamefont {Sugiyama}}, \bibinfo {author} {\bibfnamefont {H.}~\bibnamefont {Miyatake}}, \bibinfo {author} {\bibfnamefont {M.~M.}\ \bibnamefont {Rau}}, \bibinfo {author} {\bibfnamefont {M.}~\bibnamefont {Shirasaki}},  \emph {et~al.},\ }\href@noop {} {\  (\bibinfo {year} {2023})},\ \Eprint {http://arxiv.org/abs/2304.00703} {arXiv:2304.00703 [astro-ph.CO]} \BibitemShut {NoStop}%
\bibitem [{\citenamefont {Sugiyama}\ \emph {et~al.}(2023)\citenamefont {Sugiyama}, \citenamefont {Miyatake}, \citenamefont {More}, \citenamefont {Li}, \citenamefont {Shirasaki} \emph {et~al.}}]{HSC3_3x2pt_ls}%
  \BibitemOpen
  \bibfield  {author} {\bibinfo {author} {\bibfnamefont {S.}~\bibnamefont {Sugiyama}}, \bibinfo {author} {\bibfnamefont {H.}~\bibnamefont {Miyatake}}, \bibinfo {author} {\bibfnamefont {S.}~\bibnamefont {More}}, \bibinfo {author} {\bibfnamefont {X.}~\bibnamefont {Li}}, \bibinfo {author} {\bibfnamefont {M.}~\bibnamefont {Shirasaki}},  \emph {et~al.},\ }\href@noop {} {\  (\bibinfo {year} {2023})},\ \Eprint {http://arxiv.org/abs/2304.00705} {arXiv:2304.00705 [astro-ph.CO]} \BibitemShut {NoStop}%
\bibitem [{\citenamefont {Miyatake}\ \emph {et~al.}(2023)\citenamefont {Miyatake}, \citenamefont {Sugiyama}, \citenamefont {Takada}, \citenamefont {Nishimichi}, \citenamefont {Li}, \citenamefont {Shirasaki}, \citenamefont {More} \emph {et~al.}}]{HSC3_3x2pt_ss}%
  \BibitemOpen
  \bibfield  {author} {\bibinfo {author} {\bibfnamefont {H.}~\bibnamefont {Miyatake}}, \bibinfo {author} {\bibfnamefont {S.}~\bibnamefont {Sugiyama}}, \bibinfo {author} {\bibfnamefont {M.}~\bibnamefont {Takada}}, \bibinfo {author} {\bibfnamefont {T.}~\bibnamefont {Nishimichi}}, \bibinfo {author} {\bibfnamefont {X.}~\bibnamefont {Li}}, \bibinfo {author} {\bibfnamefont {M.}~\bibnamefont {Shirasaki}}, \bibinfo {author} {\bibfnamefont {S.}~\bibnamefont {More}},  \emph {et~al.},\ }\href@noop {} {\  (\bibinfo {year} {2023})},\ \Eprint {http://arxiv.org/abs/2304.00704} {arXiv:2304.00704 [astro-ph.CO]} \BibitemShut {NoStop}%
\bibitem [{\citenamefont {{Nishizawa}}\ \emph {et~al.}(2020)\citenamefont {{Nishizawa}}, \citenamefont {{Hsieh}}, \citenamefont {{Tanaka}},\ and\ \citenamefont {{Takata}}}]{HSC3_photoz_Nishizawa2020}%
  \BibitemOpen
  \bibfield  {author} {\bibinfo {author} {\bibfnamefont {A.~J.}\ \bibnamefont {{Nishizawa}}}, \bibinfo {author} {\bibfnamefont {B.-C.}\ \bibnamefont {{Hsieh}}}, \bibinfo {author} {\bibfnamefont {M.}~\bibnamefont {{Tanaka}}}, \ and\ \bibinfo {author} {\bibfnamefont {T.}~\bibnamefont {{Takata}}},\ }\href@noop {} {\bibfield  {journal} {\bibinfo  {journal} {arXiv e-prints}\ ,\ \bibinfo {eid} {arXiv:2003.01511}} (\bibinfo {year} {2020})},\ \Eprint {http://arxiv.org/abs/2003.01511} {arXiv:2003.01511 [astro-ph.GA]} \BibitemShut {NoStop}%
\bibitem [{\citenamefont {{Chang}}\ \emph {et~al.}(2013)\citenamefont {{Chang}}, \citenamefont {{Jarvis}}, \citenamefont {{Jain}}, \citenamefont {{Kahn}}, \citenamefont {{Kirkby}} \emph {et~al.}}]{WLsurvey_neffective_Chang2013}%
  \BibitemOpen
  \bibfield  {author} {\bibinfo {author} {\bibfnamefont {C.}~\bibnamefont {{Chang}}}, \bibinfo {author} {\bibfnamefont {M.}~\bibnamefont {{Jarvis}}}, \bibinfo {author} {\bibfnamefont {B.}~\bibnamefont {{Jain}}}, \bibinfo {author} {\bibfnamefont {S.~M.}\ \bibnamefont {{Kahn}}}, \bibinfo {author} {\bibfnamefont {D.}~\bibnamefont {{Kirkby}}},  \emph {et~al.},\ }\href {\doibase 10.1093/mnras/stt1156} {\bibfield  {journal} {\bibinfo  {journal} {\mnras}\ }\textbf {\bibinfo {volume} {434}},\ \bibinfo {pages} {2121} (\bibinfo {year} {2013})},\ \Eprint {http://arxiv.org/abs/1305.0793} {arXiv:1305.0793 [astro-ph.CO]} \BibitemShut {NoStop}%
\bibitem [{\citenamefont {{Hildebrandt}}\ \emph {et~al.}(2017)\citenamefont {{Hildebrandt}}, \citenamefont {{Viola}}, \citenamefont {{Heymans}}, \citenamefont {{Joudaki}}, \citenamefont {{Kuijken}}, \citenamefont {{Blake}} \emph {et~al.}}]{KiDS450_cs_Hildebrandt2017}%
  \BibitemOpen
  \bibfield  {author} {\bibinfo {author} {\bibfnamefont {H.}~\bibnamefont {{Hildebrandt}}}, \bibinfo {author} {\bibfnamefont {M.}~\bibnamefont {{Viola}}}, \bibinfo {author} {\bibfnamefont {C.}~\bibnamefont {{Heymans}}}, \bibinfo {author} {\bibfnamefont {S.}~\bibnamefont {{Joudaki}}}, \bibinfo {author} {\bibfnamefont {K.}~\bibnamefont {{Kuijken}}}, \bibinfo {author} {\bibfnamefont {C.}~\bibnamefont {{Blake}}},  \emph {et~al.},\ }\href {\doibase 10.1093/mnras/stw2805} {\bibfield  {journal} {\bibinfo  {journal} {\mnras}\ }\textbf {\bibinfo {volume} {465}},\ \bibinfo {pages} {1454} (\bibinfo {year} {2017})},\ \Eprint {http://arxiv.org/abs/1606.05338} {arXiv:1606.05338 [astro-ph.CO]} \BibitemShut {NoStop}%
\bibitem [{\citenamefont {{Hirata}}\ and\ \citenamefont {{Seljak}}(2003)}]{Regaussianization}%
  \BibitemOpen
  \bibfield  {author} {\bibinfo {author} {\bibfnamefont {C.}~\bibnamefont {{Hirata}}}\ and\ \bibinfo {author} {\bibfnamefont {U.}~\bibnamefont {{Seljak}}},\ }\href {\doibase 10.1046/j.1365-8711.2003.06683.x} {\bibfield  {journal} {\bibinfo  {journal} {\mnras}\ }\textbf {\bibinfo {volume} {343}},\ \bibinfo {pages} {459} (\bibinfo {year} {2003})},\ \Eprint {http://arxiv.org/abs/astro-ph/0301054} {astro-ph/0301054} \BibitemShut {NoStop}%
\bibitem [{\citenamefont {{Bosch}}\ \emph {et~al.}(2018)\citenamefont {{Bosch}}, \citenamefont {{Armstrong}}, \citenamefont {{Bickerton}}, \citenamefont {{Furusawa}}, \citenamefont {{Ikeda}}, \citenamefont {{Koike}} \emph {et~al.}}]{HSC1_pipeline}%
  \BibitemOpen
  \bibfield  {author} {\bibinfo {author} {\bibfnamefont {J.}~\bibnamefont {{Bosch}}}, \bibinfo {author} {\bibfnamefont {R.}~\bibnamefont {{Armstrong}}}, \bibinfo {author} {\bibfnamefont {S.}~\bibnamefont {{Bickerton}}}, \bibinfo {author} {\bibfnamefont {H.}~\bibnamefont {{Furusawa}}}, \bibinfo {author} {\bibfnamefont {H.}~\bibnamefont {{Ikeda}}}, \bibinfo {author} {\bibfnamefont {M.}~\bibnamefont {{Koike}}},  \emph {et~al.},\ }\href {\doibase 10.1093/pasj/psx080} {\bibfield  {journal} {\bibinfo  {journal} {\pasj}\ }\textbf {\bibinfo {volume} {70}},\ \bibinfo {eid} {S5} (\bibinfo {year} {2018})},\ \Eprint {http://arxiv.org/abs/1705.06766} {arXiv:1705.06766 [astro-ph.IM]} \BibitemShut {NoStop}%
\bibitem [{\citenamefont {{Leauthaud}}\ \emph {et~al.}(2007)\citenamefont {{Leauthaud}}, \citenamefont {{Massey}}, \citenamefont {{Kneib}}, \citenamefont {{Rhodes}}, \citenamefont {{Johnston}}, \citenamefont {{Capak}} \emph {et~al.}}]{HST_shapeCatalog_Alexie2007}%
  \BibitemOpen
  \bibfield  {author} {\bibinfo {author} {\bibfnamefont {A.}~\bibnamefont {{Leauthaud}}}, \bibinfo {author} {\bibfnamefont {R.}~\bibnamefont {{Massey}}}, \bibinfo {author} {\bibfnamefont {J.-P.}\ \bibnamefont {{Kneib}}}, \bibinfo {author} {\bibfnamefont {J.}~\bibnamefont {{Rhodes}}}, \bibinfo {author} {\bibfnamefont {D.~E.}\ \bibnamefont {{Johnston}}}, \bibinfo {author} {\bibfnamefont {P.}~\bibnamefont {{Capak}}},  \emph {et~al.},\ }\href {\doibase 10.1086/516598} {\bibfield  {journal} {\bibinfo  {journal} {\apjs}\ }\textbf {\bibinfo {volume} {172}},\ \bibinfo {pages} {219} (\bibinfo {year} {2007})},\ \Eprint {http://arxiv.org/abs/astro-ph/0702359} {astro-ph/0702359} \BibitemShut {NoStop}%
\bibitem [{\citenamefont {{Mandelbaum}}\ \emph {et~al.}(2018{\natexlab{a}})\citenamefont {{Mandelbaum}}, \citenamefont {{Lanusse}}, \citenamefont {{Leauthaud}}, \citenamefont {{Armstrong}}, \citenamefont {{Simet}}, \citenamefont {{Miyatake}}, \citenamefont {{Meyers}} \emph {et~al.}}]{HSC1-GREAT3Sim}%
  \BibitemOpen
  \bibfield  {author} {\bibinfo {author} {\bibfnamefont {R.}~\bibnamefont {{Mandelbaum}}}, \bibinfo {author} {\bibfnamefont {F.}~\bibnamefont {{Lanusse}}}, \bibinfo {author} {\bibfnamefont {A.}~\bibnamefont {{Leauthaud}}}, \bibinfo {author} {\bibfnamefont {R.}~\bibnamefont {{Armstrong}}}, \bibinfo {author} {\bibfnamefont {M.}~\bibnamefont {{Simet}}}, \bibinfo {author} {\bibfnamefont {H.}~\bibnamefont {{Miyatake}}}, \bibinfo {author} {\bibfnamefont {J.~E.}\ \bibnamefont {{Meyers}}},  \emph {et~al.},\ }\href {\doibase 10.1093/mnras/sty2420} {\bibfield  {journal} {\bibinfo  {journal} {\mnras}\ }\textbf {\bibinfo {volume} {481}},\ \bibinfo {pages} {3170} (\bibinfo {year} {2018}{\natexlab{a}})},\ \Eprint {http://arxiv.org/abs/1710.00885} {arXiv:1710.00885 [astro-ph.CO]} \BibitemShut {NoStop}%
\bibitem [{\citenamefont {{Krause}}\ \emph {et~al.}(2021)\citenamefont {{Krause}}, \citenamefont {{Fang}}, \citenamefont {{Pandey}}, \citenamefont {{Secco}}, \citenamefont {{Alves}} \emph {et~al.}}]{DESY3_highOrder_Krause2021}%
  \BibitemOpen
  \bibfield  {author} {\bibinfo {author} {\bibfnamefont {E.}~\bibnamefont {{Krause}}}, \bibinfo {author} {\bibfnamefont {X.}~\bibnamefont {{Fang}}}, \bibinfo {author} {\bibfnamefont {S.}~\bibnamefont {{Pandey}}}, \bibinfo {author} {\bibfnamefont {L.~F.}\ \bibnamefont {{Secco}}}, \bibinfo {author} {\bibfnamefont {O.}~\bibnamefont {{Alves}}},  \emph {et~al.},\ }\href {\doibase 10.48550/arXiv.2105.13548} {\bibfield  {journal} {\bibinfo  {journal} {arXiv e-prints}\ ,\ \bibinfo {eid} {arXiv:2105.13548}} (\bibinfo {year} {2021})},\ \Eprint {http://arxiv.org/abs/2105.13548} {arXiv:2105.13548 [astro-ph.CO]} \BibitemShut {NoStop}%
\bibitem [{\citenamefont {{Hsieh}}\ and\ \citenamefont {{Yee}}(2014)}]{dempz_Hsieh2014}%
  \BibitemOpen
  \bibfield  {author} {\bibinfo {author} {\bibfnamefont {B.~C.}\ \bibnamefont {{Hsieh}}}\ and\ \bibinfo {author} {\bibfnamefont {H.~K.~C.}\ \bibnamefont {{Yee}}},\ }\href {\doibase 10.1088/0004-637X/792/2/102} {\bibfield  {journal} {\bibinfo  {journal} {\apj}\ }\textbf {\bibinfo {volume} {792}},\ \bibinfo {eid} {102} (\bibinfo {year} {2014})},\ \Eprint {http://arxiv.org/abs/1407.5151} {arXiv:1407.5151 [astro-ph.GA]} \BibitemShut {NoStop}%
\bibitem [{\citenamefont {{Tanaka}}(2015)}]{mizuki_Tanaka2015}%
  \BibitemOpen
  \bibfield  {author} {\bibinfo {author} {\bibfnamefont {M.}~\bibnamefont {{Tanaka}}},\ }\href {\doibase 10.1088/0004-637X/801/1/20} {\bibfield  {journal} {\bibinfo  {journal} {\apj}\ }\textbf {\bibinfo {volume} {801}},\ \bibinfo {eid} {20} (\bibinfo {year} {2015})},\ \Eprint {http://arxiv.org/abs/1501.02047} {arXiv:1501.02047 [astro-ph.GA]} \BibitemShut {NoStop}%
\bibitem [{\citenamefont {{Bruzual}}\ and\ \citenamefont {{Charlot}}(2003)}]{SPS_Bruzual2003}%
  \BibitemOpen
  \bibfield  {author} {\bibinfo {author} {\bibfnamefont {G.}~\bibnamefont {{Bruzual}}}\ and\ \bibinfo {author} {\bibfnamefont {S.}~\bibnamefont {{Charlot}}},\ }\href {\doibase 10.1046/j.1365-8711.2003.06897.x} {\bibfield  {journal} {\bibinfo  {journal} {\mnras}\ }\textbf {\bibinfo {volume} {344}},\ \bibinfo {pages} {1000} (\bibinfo {year} {2003})},\ \Eprint {http://arxiv.org/abs/astro-ph/0309134} {arXiv:astro-ph/0309134 [astro-ph]} \BibitemShut {NoStop}%
\bibitem [{\citenamefont {{Chabrier}}(2003)}]{IMF_Chabrier2003}%
  \BibitemOpen
  \bibfield  {author} {\bibinfo {author} {\bibfnamefont {G.}~\bibnamefont {{Chabrier}}},\ }\href {\doibase 10.1086/376392} {\bibfield  {journal} {\bibinfo  {journal} {\pasp}\ }\textbf {\bibinfo {volume} {115}},\ \bibinfo {pages} {763} (\bibinfo {year} {2003})},\ \Eprint {http://arxiv.org/abs/astro-ph/0304382} {arXiv:astro-ph/0304382 [astro-ph]} \BibitemShut {NoStop}%
\bibitem [{\citenamefont {{Inoue}}(2011)}]{metaGal_Inoue2011}%
  \BibitemOpen
  \bibfield  {author} {\bibinfo {author} {\bibfnamefont {A.~K.}\ \bibnamefont {{Inoue}}},\ }\href {\doibase 10.1111/j.1365-2966.2011.18906.x} {\bibfield  {journal} {\bibinfo  {journal} {\mnras}\ }\textbf {\bibinfo {volume} {415}},\ \bibinfo {pages} {2920} (\bibinfo {year} {2011})},\ \Eprint {http://arxiv.org/abs/1102.5150} {arXiv:1102.5150 [astro-ph.CO]} \BibitemShut {NoStop}%
\bibitem [{\citenamefont {{Calzetti}}\ \emph {et~al.}(2000)\citenamefont {{Calzetti}}, \citenamefont {{Armus}}, \citenamefont {{Bohlin}}, \citenamefont {{Kinney}}, \citenamefont {{Koornneef}},\ and\ \citenamefont {{Storchi-Bergmann}}}]{StarFormGal_Calzetti2000}%
  \BibitemOpen
  \bibfield  {author} {\bibinfo {author} {\bibfnamefont {D.}~\bibnamefont {{Calzetti}}}, \bibinfo {author} {\bibfnamefont {L.}~\bibnamefont {{Armus}}}, \bibinfo {author} {\bibfnamefont {R.~C.}\ \bibnamefont {{Bohlin}}}, \bibinfo {author} {\bibfnamefont {A.~L.}\ \bibnamefont {{Kinney}}}, \bibinfo {author} {\bibfnamefont {J.}~\bibnamefont {{Koornneef}}}, \ and\ \bibinfo {author} {\bibfnamefont {T.~o.}\ \bibnamefont {{Storchi-Bergmann}}},\ }\href {\doibase 10.1086/308692} {\bibfield  {journal} {\bibinfo  {journal} {\apj}\ }\textbf {\bibinfo {volume} {533}},\ \bibinfo {pages} {682} (\bibinfo {year} {2000})},\ \Eprint {http://arxiv.org/abs/astro-ph/9911459} {arXiv:astro-ph/9911459 [astro-ph]} \BibitemShut {NoStop}%
\bibitem [{\citenamefont {{Bordoloi}}\ \emph {et~al.}(2010)\citenamefont {{Bordoloi}}, \citenamefont {{Lilly}},\ and\ \citenamefont {{Amara}}}]{specXphot_Bordoloi2010}%
  \BibitemOpen
  \bibfield  {author} {\bibinfo {author} {\bibfnamefont {R.}~\bibnamefont {{Bordoloi}}}, \bibinfo {author} {\bibfnamefont {S.~J.}\ \bibnamefont {{Lilly}}}, \ and\ \bibinfo {author} {\bibfnamefont {A.}~\bibnamefont {{Amara}}},\ }\href {\doibase 10.1111/j.1365-2966.2010.16765.x} {\bibfield  {journal} {\bibinfo  {journal} {\mnras}\ }\textbf {\bibinfo {volume} {406}},\ \bibinfo {pages} {881} (\bibinfo {year} {2010})},\ \Eprint {http://arxiv.org/abs/0910.5735} {arXiv:0910.5735 [astro-ph.CO]} \BibitemShut {NoStop}%
\bibitem [{\citenamefont {{Shirasaki}}\ \emph {et~al.}(2019)\citenamefont {{Shirasaki}}, \citenamefont {{Hamana}}, \citenamefont {{Takada}}, \citenamefont {{Takahashi}},\ and\ \citenamefont {{Miyatake}}}]{HSC1_mock_Shirasaki2019}%
  \BibitemOpen
  \bibfield  {author} {\bibinfo {author} {\bibfnamefont {M.}~\bibnamefont {{Shirasaki}}}, \bibinfo {author} {\bibfnamefont {T.}~\bibnamefont {{Hamana}}}, \bibinfo {author} {\bibfnamefont {M.}~\bibnamefont {{Takada}}}, \bibinfo {author} {\bibfnamefont {R.}~\bibnamefont {{Takahashi}}}, \ and\ \bibinfo {author} {\bibfnamefont {H.}~\bibnamefont {{Miyatake}}},\ }\href {\doibase 10.1093/mnras/stz791} {\bibfield  {journal} {\bibinfo  {journal} {\mnras}\ }\textbf {\bibinfo {volume} {486}},\ \bibinfo {pages} {52} (\bibinfo {year} {2019})},\ \Eprint {http://arxiv.org/abs/1901.09488} {arXiv:1901.09488 [astro-ph.CO]} \BibitemShut {NoStop}%
\bibitem [{\citenamefont {Takahashi}\ \emph {et~al.}(2017)\citenamefont {Takahashi}, \citenamefont {Hamana}, \citenamefont {Shirasaki}, \citenamefont {Namikawa}, \citenamefont {Nishimichi} \emph {et~al.}}]{raytracingTakahashi2017}%
  \BibitemOpen
  \bibfield  {author} {\bibinfo {author} {\bibfnamefont {R.}~\bibnamefont {Takahashi}}, \bibinfo {author} {\bibfnamefont {T.}~\bibnamefont {Hamana}}, \bibinfo {author} {\bibfnamefont {M.}~\bibnamefont {Shirasaki}}, \bibinfo {author} {\bibfnamefont {T.}~\bibnamefont {Namikawa}}, \bibinfo {author} {\bibfnamefont {T.}~\bibnamefont {Nishimichi}},  \emph {et~al.},\ }\href {\doibase 10.3847/1538-4357/aa943d} {\bibfield  {journal} {\bibinfo  {journal} {The Astrophysical Journal}\ }\textbf {\bibinfo {volume} {850}},\ \bibinfo {pages} {24} (\bibinfo {year} {2017})}\BibitemShut {NoStop}%
\bibitem [{\citenamefont {{Hamana}}\ \emph {et~al.}(2015)\citenamefont {{Hamana}}, \citenamefont {{Sakurai}}, \citenamefont {{Koike}},\ and\ \citenamefont {{Miller}}}]{clusterCount_Hamana2015}%
  \BibitemOpen
  \bibfield  {author} {\bibinfo {author} {\bibfnamefont {T.}~\bibnamefont {{Hamana}}}, \bibinfo {author} {\bibfnamefont {J.}~\bibnamefont {{Sakurai}}}, \bibinfo {author} {\bibfnamefont {M.}~\bibnamefont {{Koike}}}, \ and\ \bibinfo {author} {\bibfnamefont {L.}~\bibnamefont {{Miller}}},\ }\href {\doibase 10.1093/pasj/psv034} {\bibfield  {journal} {\bibinfo  {journal} {\pasj}\ }\textbf {\bibinfo {volume} {67}},\ \bibinfo {eid} {34} (\bibinfo {year} {2015})},\ \Eprint {http://arxiv.org/abs/1503.01851} {arXiv:1503.01851 [astro-ph.CO]} \BibitemShut {NoStop}%
\bibitem [{\citenamefont {{Shirasaki}}\ \emph {et~al.}(2015)\citenamefont {{Shirasaki}}, \citenamefont {{Hamana}},\ and\ \citenamefont {{Yoshida}}}]{clusterCount_Shirasaki2015}%
  \BibitemOpen
  \bibfield  {author} {\bibinfo {author} {\bibfnamefont {M.}~\bibnamefont {{Shirasaki}}}, \bibinfo {author} {\bibfnamefont {T.}~\bibnamefont {{Hamana}}}, \ and\ \bibinfo {author} {\bibfnamefont {N.}~\bibnamefont {{Yoshida}}},\ }\href {\doibase 10.1093/mnras/stv1854} {\bibfield  {journal} {\bibinfo  {journal} {\mnras}\ }\textbf {\bibinfo {volume} {453}},\ \bibinfo {pages} {3043} (\bibinfo {year} {2015})},\ \Eprint {http://arxiv.org/abs/1504.05672} {arXiv:1504.05672 [astro-ph.CO]} \BibitemShut {NoStop}%
\bibitem [{\citenamefont {{Schneider}}\ \emph {et~al.}(2002)\citenamefont {{Schneider}}, \citenamefont {{van Waerbeke}},\ and\ \citenamefont {{Mellier}}}]{bmode_zcluster_schneider2002}%
  \BibitemOpen
  \bibfield  {author} {\bibinfo {author} {\bibfnamefont {P.}~\bibnamefont {{Schneider}}}, \bibinfo {author} {\bibfnamefont {L.}~\bibnamefont {{van Waerbeke}}}, \ and\ \bibinfo {author} {\bibfnamefont {Y.}~\bibnamefont {{Mellier}}},\ }\href {\doibase 10.1051/0004-6361:20020626} {\bibfield  {journal} {\bibinfo  {journal} {\aap}\ }\textbf {\bibinfo {volume} {389}},\ \bibinfo {pages} {729} (\bibinfo {year} {2002})},\ \Eprint {http://arxiv.org/abs/astro-ph/0112441} {arXiv:astro-ph/0112441 [astro-ph]} \BibitemShut {NoStop}%
\bibitem [{\citenamefont {{Amon}}\ \emph {et~al.}(2021{\natexlab{b}})\citenamefont {{Amon}}, \citenamefont {{Gruen}}, \citenamefont {{Troxel}}, \citenamefont {{MacCrann}}, \citenamefont {{Dodelson}}, \citenamefont {{Choi}} \emph {et~al.}}]{DESY3_CS_Amon2021}%
  \BibitemOpen
  \bibfield  {author} {\bibinfo {author} {\bibfnamefont {A.}~\bibnamefont {{Amon}}}, \bibinfo {author} {\bibfnamefont {D.}~\bibnamefont {{Gruen}}}, \bibinfo {author} {\bibfnamefont {M.~A.}\ \bibnamefont {{Troxel}}}, \bibinfo {author} {\bibfnamefont {N.}~\bibnamefont {{MacCrann}}}, \bibinfo {author} {\bibfnamefont {S.}~\bibnamefont {{Dodelson}}}, \bibinfo {author} {\bibfnamefont {A.}~\bibnamefont {{Choi}}},  \emph {et~al.},\ }\href@noop {} {\bibfield  {journal} {\bibinfo  {journal} {arXiv e-prints}\ ,\ \bibinfo {eid} {arXiv:2105.13543}} (\bibinfo {year} {2021}{\natexlab{b}})},\ \Eprint {http://arxiv.org/abs/2105.13543} {arXiv:2105.13543 [astro-ph.CO]} \BibitemShut {NoStop}%
\bibitem [{\citenamefont {{Hartlap}}\ \emph {et~al.}(2007)\citenamefont {{Hartlap}}, \citenamefont {{Simon}},\ and\ \citenamefont {{Schneider}}}]{covariance_Hartlap2007}%
  \BibitemOpen
  \bibfield  {author} {\bibinfo {author} {\bibfnamefont {J.}~\bibnamefont {{Hartlap}}}, \bibinfo {author} {\bibfnamefont {P.}~\bibnamefont {{Simon}}}, \ and\ \bibinfo {author} {\bibfnamefont {P.}~\bibnamefont {{Schneider}}},\ }\href {\doibase 10.1051/0004-6361:20066170} {\bibfield  {journal} {\bibinfo  {journal} {\aap}\ }\textbf {\bibinfo {volume} {464}},\ \bibinfo {pages} {399} (\bibinfo {year} {2007})},\ \Eprint {http://arxiv.org/abs/astro-ph/0608064} {arXiv:astro-ph/0608064 [astro-ph]} \BibitemShut {NoStop}%
\bibitem [{\citenamefont {{Camacho}}\ \emph {et~al.}(2021)\citenamefont {{Camacho}}, \citenamefont {{Andrade-Oliveira}}, \citenamefont {{Troja}}, \citenamefont {{Rosenfeld}}, \citenamefont {{Faga}} \emph {et~al.}}]{pseudoCl_Camacho2021}%
  \BibitemOpen
  \bibfield  {author} {\bibinfo {author} {\bibfnamefont {H.}~\bibnamefont {{Camacho}}}, \bibinfo {author} {\bibfnamefont {F.}~\bibnamefont {{Andrade-Oliveira}}}, \bibinfo {author} {\bibfnamefont {A.}~\bibnamefont {{Troja}}}, \bibinfo {author} {\bibfnamefont {R.}~\bibnamefont {{Rosenfeld}}}, \bibinfo {author} {\bibfnamefont {L.}~\bibnamefont {{Faga}}},  \emph {et~al.},\ }\href@noop {} {\bibfield  {journal} {\bibinfo  {journal} {arXiv e-prints}\ ,\ \bibinfo {eid} {arXiv:2111.07203}} (\bibinfo {year} {2021})},\ \Eprint {http://arxiv.org/abs/2111.07203} {arXiv:2111.07203 [astro-ph.CO]} \BibitemShut {NoStop}%
\bibitem [{\citenamefont {{Nicola}}\ \emph {et~al.}(2021)\citenamefont {{Nicola}}, \citenamefont {{Garc{\'\i}a-Garc{\'\i}a}}, \citenamefont {{Alonso}}, \citenamefont {{Dunkley}}, \citenamefont {{Ferreira}}, \citenamefont {{Slosar}},\ and\ \citenamefont {{Spergel}}}]{pseudoCl_Nicola2021}%
  \BibitemOpen
  \bibfield  {author} {\bibinfo {author} {\bibfnamefont {A.}~\bibnamefont {{Nicola}}}, \bibinfo {author} {\bibfnamefont {C.}~\bibnamefont {{Garc{\'\i}a-Garc{\'\i}a}}}, \bibinfo {author} {\bibfnamefont {D.}~\bibnamefont {{Alonso}}}, \bibinfo {author} {\bibfnamefont {J.}~\bibnamefont {{Dunkley}}}, \bibinfo {author} {\bibfnamefont {P.~G.}\ \bibnamefont {{Ferreira}}}, \bibinfo {author} {\bibfnamefont {A.}~\bibnamefont {{Slosar}}}, \ and\ \bibinfo {author} {\bibfnamefont {D.~N.~o.}\ \bibnamefont {{Spergel}}},\ }\href {\doibase 10.1088/1475-7516/2021/03/067} {\bibfield  {journal} {\bibinfo  {journal} {\jcap}\ }\textbf {\bibinfo {volume} {2021}},\ \bibinfo {eid} {067} (\bibinfo {year} {2021})},\ \Eprint {http://arxiv.org/abs/2010.09717} {arXiv:2010.09717 [astro-ph.CO]} \BibitemShut {NoStop}%
\bibitem [{\citenamefont {{Singh}}(2021)}]{pseudoCl_Singh2021}%
  \BibitemOpen
  \bibfield  {author} {\bibinfo {author} {\bibfnamefont {S.}~\bibnamefont {{Singh}}},\ }\href {\doibase 10.48550/arXiv.2105.04548} {\bibfield  {journal} {\bibinfo  {journal} {\mnras}\ }\textbf {\bibinfo {volume} {508}},\ \bibinfo {pages} {1632} (\bibinfo {year} {2021})},\ \Eprint {http://arxiv.org/abs/2105.04548} {arXiv:2105.04548 [astro-ph.CO]} \BibitemShut {NoStop}%
\bibitem [{\citenamefont {{Schneider}}\ \emph {et~al.}(2010)\citenamefont {{Schneider}}, \citenamefont {{Eifler}},\ and\ \citenamefont {{Krause}}}]{Schneider2010}%
  \BibitemOpen
  \bibfield  {author} {\bibinfo {author} {\bibfnamefont {P.}~\bibnamefont {{Schneider}}}, \bibinfo {author} {\bibfnamefont {T.}~\bibnamefont {{Eifler}}}, \ and\ \bibinfo {author} {\bibfnamefont {E.}~\bibnamefont {{Krause}}},\ }\href {\doibase 10.1051/0004-6361/201014235} {\bibfield  {journal} {\bibinfo  {journal} {\aap}\ }\textbf {\bibinfo {volume} {520}},\ \bibinfo {eid} {A116} (\bibinfo {year} {2010})},\ \Eprint {http://arxiv.org/abs/1002.2136} {arXiv:1002.2136 [astro-ph.CO]} \BibitemShut {NoStop}%
\bibitem [{\citenamefont {{Hamana}}\ \emph {et~al.}(2022)\citenamefont {{Hamana}}, \citenamefont {{Hikage}}, \citenamefont {{Oguri}}, \citenamefont {{Shirasaki}},\ and\ \citenamefont {{More}}}]{HSC1_CS_cosebis2022}%
  \BibitemOpen
  \bibfield  {author} {\bibinfo {author} {\bibfnamefont {T.}~\bibnamefont {{Hamana}}}, \bibinfo {author} {\bibfnamefont {C.}~\bibnamefont {{Hikage}}}, \bibinfo {author} {\bibfnamefont {M.}~\bibnamefont {{Oguri}}}, \bibinfo {author} {\bibfnamefont {M.}~\bibnamefont {{Shirasaki}}}, \ and\ \bibinfo {author} {\bibfnamefont {S.~o.}\ \bibnamefont {{More}}},\ }\href {\doibase 10.1093/pasj/psac046} {\bibfield  {journal} {\bibinfo  {journal} {\pasj}\ }\textbf {\bibinfo {volume} {74}},\ \bibinfo {pages} {923} (\bibinfo {year} {2022})},\ \Eprint {http://arxiv.org/abs/2201.12698} {arXiv:2201.12698 [astro-ph.CO]} \BibitemShut {NoStop}%
\bibitem [{\citenamefont {{Takada}}\ and\ \citenamefont {{Hu}}(2013)}]{2013PhRvD..87l3504T}%
  \BibitemOpen
  \bibfield  {author} {\bibinfo {author} {\bibfnamefont {M.}~\bibnamefont {{Takada}}}\ and\ \bibinfo {author} {\bibfnamefont {W.}~\bibnamefont {{Hu}}},\ }\href {\doibase 10.1103/PhysRevD.87.123504} {\bibfield  {journal} {\bibinfo  {journal} {\prd}\ }\textbf {\bibinfo {volume} {87}},\ \bibinfo {eid} {123504} (\bibinfo {year} {2013})},\ \Eprint {http://arxiv.org/abs/1302.6994} {arXiv:1302.6994 [astro-ph.CO]} \BibitemShut {NoStop}%
\bibitem [{\citenamefont {{Kodwani}}\ \emph {et~al.}(2019)\citenamefont {{Kodwani}}, \citenamefont {{Alonso}},\ and\ \citenamefont {{Ferreira}}}]{cosmoCov_Kodwani2019}%
  \BibitemOpen
  \bibfield  {author} {\bibinfo {author} {\bibfnamefont {D.}~\bibnamefont {{Kodwani}}}, \bibinfo {author} {\bibfnamefont {D.}~\bibnamefont {{Alonso}}}, \ and\ \bibinfo {author} {\bibfnamefont {P.}~\bibnamefont {{Ferreira}}},\ }\href {\doibase 10.21105/astro.1811.11584} {\bibfield  {journal} {\bibinfo  {journal} {The Open Journal of Astrophysics}\ }\textbf {\bibinfo {volume} {2}},\ \bibinfo {eid} {3} (\bibinfo {year} {2019})},\ \Eprint {http://arxiv.org/abs/1811.11584} {arXiv:1811.11584 [astro-ph.CO]} \BibitemShut {NoStop}%
\bibitem [{\citenamefont {{Krause}}\ and\ \citenamefont {{Hirata}}(2010)}]{highorderDeflect_Krause2010}%
  \BibitemOpen
  \bibfield  {author} {\bibinfo {author} {\bibfnamefont {E.}~\bibnamefont {{Krause}}}\ and\ \bibinfo {author} {\bibfnamefont {C.~M.}\ \bibnamefont {{Hirata}}},\ }\href {\doibase 10.1051/0004-6361/200913524} {\bibfield  {journal} {\bibinfo  {journal} {\aap}\ }\textbf {\bibinfo {volume} {523}},\ \bibinfo {eid} {A28} (\bibinfo {year} {2010})},\ \Eprint {http://arxiv.org/abs/0910.3786} {arXiv:0910.3786 [astro-ph.CO]} \BibitemShut {NoStop}%
\bibitem [{\citenamefont {{Zuntz}}\ \emph {et~al.}(2015)\citenamefont {{Zuntz}}, \citenamefont {{Paterno}}, \citenamefont {{Jennings}}, \citenamefont {{Rudd}}, \citenamefont {{Manzotti}} \emph {et~al.}}]{cosmosis_Zuntz2015}%
  \BibitemOpen
  \bibfield  {author} {\bibinfo {author} {\bibfnamefont {J.}~\bibnamefont {{Zuntz}}}, \bibinfo {author} {\bibfnamefont {M.}~\bibnamefont {{Paterno}}}, \bibinfo {author} {\bibfnamefont {E.}~\bibnamefont {{Jennings}}}, \bibinfo {author} {\bibfnamefont {D.}~\bibnamefont {{Rudd}}}, \bibinfo {author} {\bibfnamefont {A.}~\bibnamefont {{Manzotti}}},  \emph {et~al.},\ }\href {\doibase 10.1016/j.ascom.2015.05.005} {\bibfield  {journal} {\bibinfo  {journal} {Astronomy and Computing}\ }\textbf {\bibinfo {volume} {12}},\ \bibinfo {pages} {45} (\bibinfo {year} {2015})},\ \Eprint {http://arxiv.org/abs/1409.3409} {arXiv:1409.3409 [astro-ph.CO]} \BibitemShut {NoStop}%
\bibitem [{\citenamefont {{Fang}}\ \emph {et~al.}(2020)\citenamefont {{Fang}}, \citenamefont {{Eifler}},\ and\ \citenamefont {{Krause}}}]{fftlog2020}%
  \BibitemOpen
  \bibfield  {author} {\bibinfo {author} {\bibfnamefont {X.}~\bibnamefont {{Fang}}}, \bibinfo {author} {\bibfnamefont {T.}~\bibnamefont {{Eifler}}}, \ and\ \bibinfo {author} {\bibfnamefont {E.}~\bibnamefont {{Krause}}},\ }\href {\doibase 10.1093/mnras/staa1726} {\bibfield  {journal} {\bibinfo  {journal} {\mnras}\ }\textbf {\bibinfo {volume} {497}},\ \bibinfo {pages} {2699} (\bibinfo {year} {2020})},\ \Eprint {http://arxiv.org/abs/2004.04833} {arXiv:2004.04833 [astro-ph.CO]} \BibitemShut {NoStop}%
\bibitem [{\citenamefont {{Troxel}}\ and\ \citenamefont {{Ishak}}(2015)}]{IA_rev_Troxel2015}%
  \BibitemOpen
  \bibfield  {author} {\bibinfo {author} {\bibfnamefont {M.~A.}\ \bibnamefont {{Troxel}}}\ and\ \bibinfo {author} {\bibfnamefont {M.}~\bibnamefont {{Ishak}}},\ }\href {\doibase 10.1016/j.physrep.2014.11.001} {\bibfield  {journal} {\bibinfo  {journal} {\physrep}\ }\textbf {\bibinfo {volume} {558}},\ \bibinfo {pages} {1} (\bibinfo {year} {2015})},\ \Eprint {http://arxiv.org/abs/1407.6990} {arXiv:1407.6990 [astro-ph.CO]} \BibitemShut {NoStop}%
\bibitem [{\citenamefont {{Limber}}(1953)}]{Limber1953}%
  \BibitemOpen
  \bibfield  {author} {\bibinfo {author} {\bibfnamefont {D.~N.}\ \bibnamefont {{Limber}}},\ }\href {\doibase 10.1086/145672} {\bibfield  {journal} {\bibinfo  {journal} {\apj}\ }\textbf {\bibinfo {volume} {117}},\ \bibinfo {pages} {134} (\bibinfo {year} {1953})}\BibitemShut {NoStop}%
\bibitem [{\citenamefont {{LoVerde}}\ and\ \citenamefont {{Afshordi}}(2008)}]{Loverde2008}%
  \BibitemOpen
  \bibfield  {author} {\bibinfo {author} {\bibfnamefont {M.}~\bibnamefont {{LoVerde}}}\ and\ \bibinfo {author} {\bibfnamefont {N.}~\bibnamefont {{Afshordi}}},\ }\href {\doibase 10.1103/PhysRevD.78.123506} {\bibfield  {journal} {\bibinfo  {journal} {\prd}\ }\textbf {\bibinfo {volume} {78}},\ \bibinfo {eid} {123506} (\bibinfo {year} {2008})},\ \Eprint {http://arxiv.org/abs/0809.5112} {arXiv:0809.5112} \BibitemShut {NoStop}%
\bibitem [{\citenamefont {{Lewis}}\ \emph {et~al.}(2000)\citenamefont {{Lewis}}, \citenamefont {{Challinor}},\ and\ \citenamefont {{Lasenby}}}]{CAMB2000}%
  \BibitemOpen
  \bibfield  {author} {\bibinfo {author} {\bibfnamefont {A.}~\bibnamefont {{Lewis}}}, \bibinfo {author} {\bibfnamefont {A.}~\bibnamefont {{Challinor}}}, \ and\ \bibinfo {author} {\bibfnamefont {A.}~\bibnamefont {{Lasenby}}},\ }\href {\doibase 10.1086/309179} {\bibfield  {journal} {\bibinfo  {journal} {\apj}\ }\textbf {\bibinfo {volume} {538}},\ \bibinfo {pages} {473} (\bibinfo {year} {2000})},\ \Eprint {http://arxiv.org/abs/astro-ph/9911177} {arXiv:astro-ph/9911177 [astro-ph]} \BibitemShut {NoStop}%
\bibitem [{\citenamefont {{Lesgourgues}}(2011)}]{CLASSI2011}%
  \BibitemOpen
  \bibfield  {author} {\bibinfo {author} {\bibfnamefont {J.}~\bibnamefont {{Lesgourgues}}},\ }\href@noop {} {\bibfield  {journal} {\bibinfo  {journal} {arXiv e-prints}\ ,\ \bibinfo {eid} {arXiv:1104.2932}} (\bibinfo {year} {2011})},\ \Eprint {http://arxiv.org/abs/1104.2932} {arXiv:1104.2932 [astro-ph.IM]} \BibitemShut {NoStop}%
\bibitem [{\citenamefont {{Blas}}\ \emph {et~al.}(2011)\citenamefont {{Blas}}, \citenamefont {{Lesgourgues}},\ and\ \citenamefont {{Tram}}}]{CLASSII2011}%
  \BibitemOpen
  \bibfield  {author} {\bibinfo {author} {\bibfnamefont {D.}~\bibnamefont {{Blas}}}, \bibinfo {author} {\bibfnamefont {J.}~\bibnamefont {{Lesgourgues}}}, \ and\ \bibinfo {author} {\bibfnamefont {T.}~\bibnamefont {{Tram}}},\ }\href {\doibase 10.1088/1475-7516/2011/07/034} {\bibfield  {journal} {\bibinfo  {journal} {\jcap}\ }\textbf {\bibinfo {volume} {2011}},\ \bibinfo {eid} {034} (\bibinfo {year} {2011})},\ \Eprint {http://arxiv.org/abs/1104.2933} {arXiv:1104.2933 [astro-ph.CO]} \BibitemShut {NoStop}%
\bibitem [{\citenamefont {{Takahashi}}\ \emph {et~al.}(2012)\citenamefont {{Takahashi}}, \citenamefont {{Sato}}, \citenamefont {{Nishimichi}}, \citenamefont {{Taruya}},\ and\ \citenamefont {{Oguri}}}]{halofitT2012}%
  \BibitemOpen
  \bibfield  {author} {\bibinfo {author} {\bibfnamefont {R.}~\bibnamefont {{Takahashi}}}, \bibinfo {author} {\bibfnamefont {M.}~\bibnamefont {{Sato}}}, \bibinfo {author} {\bibfnamefont {T.}~\bibnamefont {{Nishimichi}}}, \bibinfo {author} {\bibfnamefont {A.}~\bibnamefont {{Taruya}}}, \ and\ \bibinfo {author} {\bibfnamefont {M.~o.}\ \bibnamefont {{Oguri}}},\ }\href {\doibase 10.1088/0004-637X/761/2/152} {\bibfield  {journal} {\bibinfo  {journal} {\apj}\ }\textbf {\bibinfo {volume} {761}},\ \bibinfo {eid} {152} (\bibinfo {year} {2012})},\ \Eprint {http://arxiv.org/abs/1208.2701} {arXiv:1208.2701 [astro-ph.CO]} \BibitemShut {NoStop}%
\bibitem [{\citenamefont {{Mead}}\ \emph {et~al.}(2015)\citenamefont {{Mead}}, \citenamefont {{Peacock}}, \citenamefont {{Heymans}}, \citenamefont {{Joudaki}},\ and\ \citenamefont {{Heavens}}}]{halofit_mead15}%
  \BibitemOpen
  \bibfield  {author} {\bibinfo {author} {\bibfnamefont {A.~J.}\ \bibnamefont {{Mead}}}, \bibinfo {author} {\bibfnamefont {J.~A.}\ \bibnamefont {{Peacock}}}, \bibinfo {author} {\bibfnamefont {C.}~\bibnamefont {{Heymans}}}, \bibinfo {author} {\bibfnamefont {S.}~\bibnamefont {{Joudaki}}}, \ and\ \bibinfo {author} {\bibfnamefont {A.~F.}\ \bibnamefont {{Heavens}}},\ }\href {\doibase 10.1093/mnras/stv2036} {\bibfield  {journal} {\bibinfo  {journal} {\mnras}\ }\textbf {\bibinfo {volume} {454}},\ \bibinfo {pages} {1958} (\bibinfo {year} {2015})},\ \Eprint {http://arxiv.org/abs/1505.07833} {arXiv:1505.07833} \BibitemShut {NoStop}%
\bibitem [{\citenamefont {{Mead}}\ \emph {et~al.}(2016)\citenamefont {{Mead}}, \citenamefont {{Heymans}}, \citenamefont {{Lombriser}}, \citenamefont {{Peacock}}, \citenamefont {{Steele}},\ and\ \citenamefont {{Winther}}}]{halofit_mead16}%
  \BibitemOpen
  \bibfield  {author} {\bibinfo {author} {\bibfnamefont {A.~J.}\ \bibnamefont {{Mead}}}, \bibinfo {author} {\bibfnamefont {C.}~\bibnamefont {{Heymans}}}, \bibinfo {author} {\bibfnamefont {L.}~\bibnamefont {{Lombriser}}}, \bibinfo {author} {\bibfnamefont {J.~A.}\ \bibnamefont {{Peacock}}}, \bibinfo {author} {\bibfnamefont {O.~I.}\ \bibnamefont {{Steele}}}, \ and\ \bibinfo {author} {\bibfnamefont {H.~A.~o.}\ \bibnamefont {{Winther}}},\ }\href {\doibase 10.1093/mnras/stw681} {\bibfield  {journal} {\bibinfo  {journal} {\mnras}\ }\textbf {\bibinfo {volume} {459}},\ \bibinfo {pages} {1468} (\bibinfo {year} {2016})},\ \Eprint {http://arxiv.org/abs/1602.02154} {arXiv:1602.02154} \BibitemShut {NoStop}%
\bibitem [{\citenamefont {{Aric{\`o}}}\ \emph {et~al.}(2021)\citenamefont {{Aric{\`o}}}, \citenamefont {{Angulo}},\ and\ \citenamefont {{Zennaro}}}]{baccoEmuLin_Arico2021}%
  \BibitemOpen
  \bibfield  {author} {\bibinfo {author} {\bibfnamefont {G.}~\bibnamefont {{Aric{\`o}}}}, \bibinfo {author} {\bibfnamefont {R.~E.}\ \bibnamefont {{Angulo}}}, \ and\ \bibinfo {author} {\bibfnamefont {M.}~\bibnamefont {{Zennaro}}},\ }\href {\doibase 10.48550/arXiv.2104.14568} {\bibfield  {journal} {\bibinfo  {journal} {arXiv e-prints}\ ,\ \bibinfo {eid} {arXiv:2104.14568}} (\bibinfo {year} {2021})},\ \Eprint {http://arxiv.org/abs/2104.14568} {arXiv:2104.14568 [astro-ph.CO]} \BibitemShut {NoStop}%
\bibitem [{\citenamefont {{G{\"u}nther}}\ \emph {et~al.}(2022)\citenamefont {{G{\"u}nther}}, \citenamefont {{Lesgourgues}}, \citenamefont {{Samaras}}, \citenamefont {{Sch{\"o}neberg}}, \citenamefont {{Stadtmann}} \emph {et~al.}}]{cosmicNet2022}%
  \BibitemOpen
  \bibfield  {author} {\bibinfo {author} {\bibfnamefont {S.}~\bibnamefont {{G{\"u}nther}}}, \bibinfo {author} {\bibfnamefont {J.}~\bibnamefont {{Lesgourgues}}}, \bibinfo {author} {\bibfnamefont {G.}~\bibnamefont {{Samaras}}}, \bibinfo {author} {\bibfnamefont {N.}~\bibnamefont {{Sch{\"o}neberg}}}, \bibinfo {author} {\bibfnamefont {F.}~\bibnamefont {{Stadtmann}}},  \emph {et~al.},\ }\href {\doibase 10.1088/1475-7516/2022/11/035} {\bibfield  {journal} {\bibinfo  {journal} {\jcap}\ }\textbf {\bibinfo {volume} {2022}},\ \bibinfo {eid} {035} (\bibinfo {year} {2022})},\ \Eprint {http://arxiv.org/abs/2207.05707} {arXiv:2207.05707 [astro-ph.CO]} \BibitemShut {NoStop}%
\bibitem [{\citenamefont {{Nishimichi}}\ \emph {et~al.}(2019)\citenamefont {{Nishimichi}}, \citenamefont {{Takada}}, \citenamefont {{Takahashi}}, \citenamefont {{Osato}}, \citenamefont {{Shirasaki}} \emph {et~al.}}]{darkemu_Nishimichi2019}%
  \BibitemOpen
  \bibfield  {author} {\bibinfo {author} {\bibfnamefont {T.}~\bibnamefont {{Nishimichi}}}, \bibinfo {author} {\bibfnamefont {M.}~\bibnamefont {{Takada}}}, \bibinfo {author} {\bibfnamefont {R.}~\bibnamefont {{Takahashi}}}, \bibinfo {author} {\bibfnamefont {K.}~\bibnamefont {{Osato}}}, \bibinfo {author} {\bibfnamefont {M.}~\bibnamefont {{Shirasaki}}},  \emph {et~al.},\ }\href {\doibase 10.3847/1538-4357/ab3719} {\bibfield  {journal} {\bibinfo  {journal} {\apj}\ }\textbf {\bibinfo {volume} {884}},\ \bibinfo {eid} {29} (\bibinfo {year} {2019})},\ \Eprint {http://arxiv.org/abs/1811.09504} {arXiv:1811.09504 [astro-ph.CO]} \BibitemShut {NoStop}%
\bibitem [{\citenamefont {{Angulo}}\ \emph {et~al.}(2021)\citenamefont {{Angulo}}, \citenamefont {{Zennaro}}, \citenamefont {{Contreras}}, \citenamefont {{Aric{\`o}}}, \citenamefont {{Pellejero-Iba{\~n}ez}} \emph {et~al.}}]{baccoEmuNL_Angulo2021}%
  \BibitemOpen
  \bibfield  {author} {\bibinfo {author} {\bibfnamefont {R.~E.}\ \bibnamefont {{Angulo}}}, \bibinfo {author} {\bibfnamefont {M.}~\bibnamefont {{Zennaro}}}, \bibinfo {author} {\bibfnamefont {S.}~\bibnamefont {{Contreras}}}, \bibinfo {author} {\bibfnamefont {G.}~\bibnamefont {{Aric{\`o}}}}, \bibinfo {author} {\bibfnamefont {M.}~\bibnamefont {{Pellejero-Iba{\~n}ez}}},  \emph {et~al.},\ }\href {\doibase 10.1093/mnras/stab2018} {\bibfield  {journal} {\bibinfo  {journal} {\mnras}\ }\textbf {\bibinfo {volume} {507}},\ \bibinfo {pages} {5869} (\bibinfo {year} {2021})},\ \Eprint {http://arxiv.org/abs/2004.06245} {arXiv:2004.06245 [astro-ph.CO]} \BibitemShut {NoStop}%
\bibitem [{\citenamefont {{Euclid Collaboration}}\ \emph {et~al.}(2021)\citenamefont {{Euclid Collaboration}}, \citenamefont {{Knabenhans}}, \citenamefont {{Stadel}}, \citenamefont {{Potter}}, \citenamefont {{Dakin}} \emph {et~al.}}]{euclidEmu2}%
  \BibitemOpen
  \bibfield  {author} {\bibinfo {author} {\bibnamefont {{Euclid Collaboration}}}, \bibinfo {author} {\bibfnamefont {M.}~\bibnamefont {{Knabenhans}}}, \bibinfo {author} {\bibfnamefont {J.}~\bibnamefont {{Stadel}}}, \bibinfo {author} {\bibfnamefont {D.}~\bibnamefont {{Potter}}}, \bibinfo {author} {\bibfnamefont {J.}~\bibnamefont {{Dakin}}},  \emph {et~al.},\ }\href {\doibase 10.1093/mnras/stab1366} {\bibfield  {journal} {\bibinfo  {journal} {\mnras}\ }\textbf {\bibinfo {volume} {505}},\ \bibinfo {pages} {2840} (\bibinfo {year} {2021})},\ \Eprint {http://arxiv.org/abs/2010.11288} {arXiv:2010.11288 [astro-ph.CO]} \BibitemShut {NoStop}%
\bibitem [{\citenamefont {{Joachimi}}\ \emph {et~al.}(2021{\natexlab{a}})\citenamefont {{Joachimi}}, \citenamefont {{Lin}}, \citenamefont {{Asgari}}, \citenamefont {{Tr{\"o}ster}}, \citenamefont {{Heymans}}, \citenamefont {{Hildebrandt}}, \citenamefont {{K{\"o}hlinger}} \emph {et~al.}}]{Joachimi2020}%
  \BibitemOpen
  \bibfield  {author} {\bibinfo {author} {\bibfnamefont {B.}~\bibnamefont {{Joachimi}}}, \bibinfo {author} {\bibfnamefont {C.~A.}\ \bibnamefont {{Lin}}}, \bibinfo {author} {\bibfnamefont {M.}~\bibnamefont {{Asgari}}}, \bibinfo {author} {\bibfnamefont {T.}~\bibnamefont {{Tr{\"o}ster}}}, \bibinfo {author} {\bibfnamefont {C.}~\bibnamefont {{Heymans}}}, \bibinfo {author} {\bibfnamefont {H.}~\bibnamefont {{Hildebrandt}}}, \bibinfo {author} {\bibfnamefont {F.}~\bibnamefont {{K{\"o}hlinger}}},  \emph {et~al.},\ }\href {\doibase 10.1051/0004-6361/202038831} {\bibfield  {journal} {\bibinfo  {journal} {\aap}\ }\textbf {\bibinfo {volume} {646}},\ \bibinfo {eid} {A129} (\bibinfo {year} {2021}{\natexlab{a}})},\ \Eprint {http://arxiv.org/abs/2007.01844} {arXiv:2007.01844 [astro-ph.CO]} \BibitemShut {NoStop}%
\bibitem [{\citenamefont {{Schneider}}\ \emph {et~al.}(2019)\citenamefont {{Schneider}}, \citenamefont {{Teyssier}}, \citenamefont {{Stadel}}, \citenamefont {{Chisari}}, \citenamefont {{Le Brun}} \emph {et~al.}}]{baronCorrect_Schneider2019}%
  \BibitemOpen
  \bibfield  {author} {\bibinfo {author} {\bibfnamefont {A.}~\bibnamefont {{Schneider}}}, \bibinfo {author} {\bibfnamefont {R.}~\bibnamefont {{Teyssier}}}, \bibinfo {author} {\bibfnamefont {J.}~\bibnamefont {{Stadel}}}, \bibinfo {author} {\bibfnamefont {N.~E.}\ \bibnamefont {{Chisari}}}, \bibinfo {author} {\bibfnamefont {A.~M.~C.}\ \bibnamefont {{Le Brun}}},  \emph {et~al.},\ }\href {\doibase 10.1088/1475-7516/2019/03/020} {\bibfield  {journal} {\bibinfo  {journal} {\jcap}\ }\textbf {\bibinfo {volume} {2019}},\ \bibinfo {eid} {020} (\bibinfo {year} {2019})},\ \Eprint {http://arxiv.org/abs/1810.08629} {arXiv:1810.08629 [astro-ph.CO]} \BibitemShut {NoStop}%
\bibitem [{\citenamefont {{Huang}}\ \emph {et~al.}(2021)\citenamefont {{Huang}}, \citenamefont {{Eifler}}, \citenamefont {{Mandelbaum}}, \citenamefont {{Bernstein}}, \citenamefont {{Chen}} \emph {et~al.}}]{baryonPCA_Huang2021}%
  \BibitemOpen
  \bibfield  {author} {\bibinfo {author} {\bibfnamefont {H.-J.}\ \bibnamefont {{Huang}}}, \bibinfo {author} {\bibfnamefont {T.}~\bibnamefont {{Eifler}}}, \bibinfo {author} {\bibfnamefont {R.}~\bibnamefont {{Mandelbaum}}}, \bibinfo {author} {\bibfnamefont {G.~M.}\ \bibnamefont {{Bernstein}}}, \bibinfo {author} {\bibfnamefont {A.}~\bibnamefont {{Chen}}},  \emph {et~al.},\ }\href {\doibase 10.1093/mnras/stab357} {\bibfield  {journal} {\bibinfo  {journal} {\mnras}\ }\textbf {\bibinfo {volume} {502}},\ \bibinfo {pages} {6010} (\bibinfo {year} {2021})},\ \Eprint {http://arxiv.org/abs/2007.15026} {arXiv:2007.15026 [astro-ph.CO]} \BibitemShut {NoStop}%
\bibitem [{\citenamefont {{McEwen}}\ \emph {et~al.}(2016)\citenamefont {{McEwen}}, \citenamefont {{Fang}}, \citenamefont {{Hirata}},\ and\ \citenamefont {{Blazek}}}]{fastpt2016}%
  \BibitemOpen
  \bibfield  {author} {\bibinfo {author} {\bibfnamefont {J.~E.}\ \bibnamefont {{McEwen}}}, \bibinfo {author} {\bibfnamefont {X.}~\bibnamefont {{Fang}}}, \bibinfo {author} {\bibfnamefont {C.~M.}\ \bibnamefont {{Hirata}}}, \ and\ \bibinfo {author} {\bibfnamefont {J.~A.}\ \bibnamefont {{Blazek}}},\ }\href {\doibase 10.1088/1475-7516/2016/09/015} {\bibfield  {journal} {\bibinfo  {journal} {\jcap}\ }\textbf {\bibinfo {volume} {2016}},\ \bibinfo {eid} {015} (\bibinfo {year} {2016})},\ \Eprint {http://arxiv.org/abs/1603.04826} {arXiv:1603.04826 [astro-ph.CO]} \BibitemShut {NoStop}%
\bibitem [{\citenamefont {{Fang}}\ \emph {et~al.}(2017)\citenamefont {{Fang}}, \citenamefont {{Blazek}}, \citenamefont {{McEwen}},\ and\ \citenamefont {{Hirata}}}]{fastpt2017}%
  \BibitemOpen
  \bibfield  {author} {\bibinfo {author} {\bibfnamefont {X.}~\bibnamefont {{Fang}}}, \bibinfo {author} {\bibfnamefont {J.~A.}\ \bibnamefont {{Blazek}}}, \bibinfo {author} {\bibfnamefont {J.~E.}\ \bibnamefont {{McEwen}}}, \ and\ \bibinfo {author} {\bibfnamefont {C.~M.}\ \bibnamefont {{Hirata}}},\ }\href {\doibase 10.1088/1475-7516/2017/02/030} {\bibfield  {journal} {\bibinfo  {journal} {\jcap}\ }\textbf {\bibinfo {volume} {2017}},\ \bibinfo {eid} {030} (\bibinfo {year} {2017})},\ \Eprint {http://arxiv.org/abs/1609.05978} {arXiv:1609.05978 [astro-ph.CO]} \BibitemShut {NoStop}%
\bibitem [{\citenamefont {{Brown}}\ \emph {et~al.}(2002)\citenamefont {{Brown}}, \citenamefont {{Taylor}}, \citenamefont {{Hambly}},\ and\ \citenamefont {{Dye}}}]{supercosmos2002}%
  \BibitemOpen
  \bibfield  {author} {\bibinfo {author} {\bibfnamefont {M.~L.}\ \bibnamefont {{Brown}}}, \bibinfo {author} {\bibfnamefont {A.~N.}\ \bibnamefont {{Taylor}}}, \bibinfo {author} {\bibfnamefont {N.~C.}\ \bibnamefont {{Hambly}}}, \ and\ \bibinfo {author} {\bibfnamefont {S.}~\bibnamefont {{Dye}}},\ }\href {\doibase 10.1046/j.1365-8711.2002.05354.x} {\bibfield  {journal} {\bibinfo  {journal} {\mnras}\ }\textbf {\bibinfo {volume} {333}},\ \bibinfo {pages} {501} (\bibinfo {year} {2002})},\ \Eprint {http://arxiv.org/abs/astro-ph/0009499} {arXiv:astro-ph/0009499 [astro-ph]} \BibitemShut {NoStop}%
\bibitem [{\citenamefont {{Singh}}\ \emph {et~al.}(2015)\citenamefont {{Singh}}, \citenamefont {{Mandelbaum}},\ and\ \citenamefont {{More}}}]{iaSDSS_Singh2015}%
  \BibitemOpen
  \bibfield  {author} {\bibinfo {author} {\bibfnamefont {S.}~\bibnamefont {{Singh}}}, \bibinfo {author} {\bibfnamefont {R.}~\bibnamefont {{Mandelbaum}}}, \ and\ \bibinfo {author} {\bibfnamefont {S.}~\bibnamefont {{More}}},\ }\href {\doibase 10.1093/mnras/stv778} {\bibfield  {journal} {\bibinfo  {journal} {\mnras}\ }\textbf {\bibinfo {volume} {450}},\ \bibinfo {pages} {2195} (\bibinfo {year} {2015})},\ \Eprint {http://arxiv.org/abs/1411.1755} {arXiv:1411.1755 [astro-ph.CO]} \BibitemShut {NoStop}%
\bibitem [{\citenamefont {{Jagvaral}}\ \emph {et~al.}(2022)\citenamefont {{Jagvaral}}, \citenamefont {{Singh}},\ and\ \citenamefont {{Mandelbaum}}}]{iaBulgeDisk_Jagvaral2022}%
  \BibitemOpen
  \bibfield  {author} {\bibinfo {author} {\bibfnamefont {Y.}~\bibnamefont {{Jagvaral}}}, \bibinfo {author} {\bibfnamefont {S.}~\bibnamefont {{Singh}}}, \ and\ \bibinfo {author} {\bibfnamefont {R.}~\bibnamefont {{Mandelbaum}}},\ }\href {\doibase 10.1093/mnras/stac1424} {\bibfield  {journal} {\bibinfo  {journal} {\mnras}\ }\textbf {\bibinfo {volume} {514}},\ \bibinfo {pages} {1021} (\bibinfo {year} {2022})},\ \Eprint {http://arxiv.org/abs/2202.08849} {arXiv:2202.08849 [astro-ph.GA]} \BibitemShut {NoStop}%
\bibitem [{\citenamefont {{Campos}}\ \emph {et~al.}(2022)\citenamefont {{Campos}}, \citenamefont {{Samuroff}},\ and\ \citenamefont {{Mandelbaum}}}]{IAmodel_Campos2022}%
  \BibitemOpen
  \bibfield  {author} {\bibinfo {author} {\bibfnamefont {A.}~\bibnamefont {{Campos}}}, \bibinfo {author} {\bibfnamefont {S.}~\bibnamefont {{Samuroff}}}, \ and\ \bibinfo {author} {\bibfnamefont {R.}~\bibnamefont {{Mandelbaum}}},\ }\href {\doibase 10.48550/arXiv.2211.02800} {\bibfield  {journal} {\bibinfo  {journal} {arXiv e-prints}\ ,\ \bibinfo {eid} {arXiv:2211.02800}} (\bibinfo {year} {2022})},\ \Eprint {http://arxiv.org/abs/2211.02800} {arXiv:2211.02800 [astro-ph.CO]} \BibitemShut {NoStop}%
\bibitem [{\citenamefont {{Ishikawa}}\ \emph {et~al.}(2021)\citenamefont {{Ishikawa}}, \citenamefont {{Okumura}}, \citenamefont {{Oguri}},\ and\ \citenamefont {{Lin}}}]{CAMIRA3}%
  \BibitemOpen
  \bibfield  {author} {\bibinfo {author} {\bibfnamefont {S.}~\bibnamefont {{Ishikawa}}}, \bibinfo {author} {\bibfnamefont {T.}~\bibnamefont {{Okumura}}}, \bibinfo {author} {\bibfnamefont {M.}~\bibnamefont {{Oguri}}}, \ and\ \bibinfo {author} {\bibfnamefont {S.-C.}\ \bibnamefont {{Lin}}},\ }\href {\doibase 10.3847/1538-4357/ac1f90} {\bibfield  {journal} {\bibinfo  {journal} {\apj}\ }\textbf {\bibinfo {volume} {922}},\ \bibinfo {eid} {23} (\bibinfo {year} {2021})},\ \Eprint {http://arxiv.org/abs/2103.08628} {arXiv:2103.08628 [astro-ph.GA]} \BibitemShut {NoStop}%
\bibitem [{\citenamefont {{Zhang}}\ \emph {et~al.}(2023)\citenamefont {{Zhang}}, \citenamefont {{Rau}}, \citenamefont {{Mandelbaum}}, \citenamefont {{Li}},\ and\ \citenamefont {{Moews}}}]{HSCY3_NZ_ERR}%
  \BibitemOpen
  \bibfield  {author} {\bibinfo {author} {\bibfnamefont {T.}~\bibnamefont {{Zhang}}}, \bibinfo {author} {\bibfnamefont {M.~M.}\ \bibnamefont {{Rau}}}, \bibinfo {author} {\bibfnamefont {R.}~\bibnamefont {{Mandelbaum}}}, \bibinfo {author} {\bibfnamefont {X.}~\bibnamefont {{Li}}}, \ and\ \bibinfo {author} {\bibfnamefont {B.~o.}\ \bibnamefont {{Moews}}},\ }\href {\doibase 10.1093/mnras/stac3090} {\bibfield  {journal} {\bibinfo  {journal} {\mnras}\ }\textbf {\bibinfo {volume} {518}},\ \bibinfo {pages} {709} (\bibinfo {year} {2023})},\ \Eprint {http://arxiv.org/abs/2206.10169} {arXiv:2206.10169 [astro-ph.CO]} \BibitemShut {NoStop}%
\bibitem [{\citenamefont {{Bernstein}}(2010)}]{modelBias_Bernstein10}%
  \BibitemOpen
  \bibfield  {author} {\bibinfo {author} {\bibfnamefont {G.~M.}\ \bibnamefont {{Bernstein}}},\ }\href {\doibase 10.1111/j.1365-2966.2010.16883.x} {\bibfield  {journal} {\bibinfo  {journal} {\mnras}\ }\textbf {\bibinfo {volume} {406}},\ \bibinfo {pages} {2793} (\bibinfo {year} {2010})},\ \Eprint {http://arxiv.org/abs/1001.2333} {arXiv:1001.2333 [astro-ph.IM]} \BibitemShut {NoStop}%
\bibitem [{\citenamefont {{Refregier}}\ \emph {et~al.}(2012)\citenamefont {{Refregier}}, \citenamefont {{Kacprzak}}, \citenamefont {{Amara}}, \citenamefont {{Bridle}},\ and\ \citenamefont {{Rowe}}}]{noiseBiasRefregier2012}%
  \BibitemOpen
  \bibfield  {author} {\bibinfo {author} {\bibfnamefont {A.}~\bibnamefont {{Refregier}}}, \bibinfo {author} {\bibfnamefont {T.}~\bibnamefont {{Kacprzak}}}, \bibinfo {author} {\bibfnamefont {A.}~\bibnamefont {{Amara}}}, \bibinfo {author} {\bibfnamefont {S.}~\bibnamefont {{Bridle}}}, \ and\ \bibinfo {author} {\bibfnamefont {B.}~\bibnamefont {{Rowe}}},\ }\href {\doibase 10.1111/j.1365-2966.2012.21483.x} {\bibfield  {journal} {\bibinfo  {journal} {\mnras}\ }\textbf {\bibinfo {volume} {425}},\ \bibinfo {pages} {1951} (\bibinfo {year} {2012})},\ \Eprint {http://arxiv.org/abs/1203.5050} {arXiv:1203.5050} \BibitemShut {NoStop}%
\bibitem [{\citenamefont {Kaiser}(2000)}]{kaiserFlow}%
  \BibitemOpen
  \bibfield  {author} {\bibinfo {author} {\bibfnamefont {N.}~\bibnamefont {Kaiser}},\ }\href {\doibase 10.1086/309041} {\bibfield  {journal} {\bibinfo  {journal} {The Astrophysical Journal}\ }\textbf {\bibinfo {volume} {537}},\ \bibinfo {pages} {555} (\bibinfo {year} {2000})}\BibitemShut {NoStop}%
\bibitem [{\citenamefont {{Sheldon}}\ \emph {et~al.}(2020)\citenamefont {{Sheldon}}, \citenamefont {{Becker}}, \citenamefont {{MacCrann}},\ and\ \citenamefont {{Jarvis}}}]{metaDet_Sheldon2020}%
  \BibitemOpen
  \bibfield  {author} {\bibinfo {author} {\bibfnamefont {E.~S.}\ \bibnamefont {{Sheldon}}}, \bibinfo {author} {\bibfnamefont {M.~R.}\ \bibnamefont {{Becker}}}, \bibinfo {author} {\bibfnamefont {N.}~\bibnamefont {{MacCrann}}}, \ and\ \bibinfo {author} {\bibfnamefont {M.}~\bibnamefont {{Jarvis}}},\ }\href {\doibase 10.3847/1538-4357/abb595} {\bibfield  {journal} {\bibinfo  {journal} {\apj}\ }\textbf {\bibinfo {volume} {902}},\ \bibinfo {eid} {138} (\bibinfo {year} {2020})},\ \Eprint {http://arxiv.org/abs/1911.02505} {arXiv:1911.02505 [astro-ph.CO]} \BibitemShut {NoStop}%
\bibitem [{\citenamefont {{Foreman-Mackey}}\ \emph {et~al.}(2013)\citenamefont {{Foreman-Mackey}}, \citenamefont {{Hogg}}, \citenamefont {{Lang}},\ and\ \citenamefont {{Goodman}}}]{emcee_Foreman2013}%
  \BibitemOpen
  \bibfield  {author} {\bibinfo {author} {\bibfnamefont {D.}~\bibnamefont {{Foreman-Mackey}}}, \bibinfo {author} {\bibfnamefont {D.~W.}\ \bibnamefont {{Hogg}}}, \bibinfo {author} {\bibfnamefont {D.}~\bibnamefont {{Lang}}}, \ and\ \bibinfo {author} {\bibfnamefont {J.}~\bibnamefont {{Goodman}}},\ }\href {\doibase 10.1086/670067} {\bibfield  {journal} {\bibinfo  {journal} {\pasp}\ }\textbf {\bibinfo {volume} {125}},\ \bibinfo {pages} {306} (\bibinfo {year} {2013})},\ \Eprint {http://arxiv.org/abs/1202.3665} {arXiv:1202.3665 [astro-ph.IM]} \BibitemShut {NoStop}%
\bibitem [{\citenamefont {{Feroz}}\ \emph {et~al.}(2009)\citenamefont {{Feroz}}, \citenamefont {{Hobson}},\ and\ \citenamefont {{Bridges}}}]{multinest_Feroz2009}%
  \BibitemOpen
  \bibfield  {author} {\bibinfo {author} {\bibfnamefont {F.}~\bibnamefont {{Feroz}}}, \bibinfo {author} {\bibfnamefont {M.~P.}\ \bibnamefont {{Hobson}}}, \ and\ \bibinfo {author} {\bibfnamefont {M.}~\bibnamefont {{Bridges}}},\ }\href {\doibase 10.1111/j.1365-2966.2009.14548.x} {\bibfield  {journal} {\bibinfo  {journal} {\mnras}\ }\textbf {\bibinfo {volume} {398}},\ \bibinfo {pages} {1601} (\bibinfo {year} {2009})},\ \Eprint {http://arxiv.org/abs/0809.3437} {arXiv:0809.3437 [astro-ph]} \BibitemShut {NoStop}%
\bibitem [{\citenamefont {{Handley}}\ \emph {et~al.}(2015)\citenamefont {{Handley}}, \citenamefont {{Hobson}},\ and\ \citenamefont {{Lasenby}}}]{polychord_Handley2015}%
  \BibitemOpen
  \bibfield  {author} {\bibinfo {author} {\bibfnamefont {W.~J.}\ \bibnamefont {{Handley}}}, \bibinfo {author} {\bibfnamefont {M.~P.}\ \bibnamefont {{Hobson}}}, \ and\ \bibinfo {author} {\bibfnamefont {A.~N.}\ \bibnamefont {{Lasenby}}},\ }\href {\doibase 10.1093/mnrasl/slv047} {\bibfield  {journal} {\bibinfo  {journal} {\mnras}\ }\textbf {\bibinfo {volume} {450}},\ \bibinfo {pages} {L61} (\bibinfo {year} {2015})},\ \Eprint {http://arxiv.org/abs/1502.01856} {arXiv:1502.01856 [astro-ph.CO]} \BibitemShut {NoStop}%
\bibitem [{\citenamefont {{Lemos}}\ \emph {et~al.}(2022)\citenamefont {{Lemos}}, \citenamefont {{Weaverdyck}}, \citenamefont {{Rollins}}, \citenamefont {{Muir}}, \citenamefont {{Fert{\'e}}}, \citenamefont {{Liddle}}, \citenamefont {{Campos}} \emph {et~al.}}]{DESY3_sampler_Lemos}%
  \BibitemOpen
  \bibfield  {author} {\bibinfo {author} {\bibfnamefont {P.}~\bibnamefont {{Lemos}}}, \bibinfo {author} {\bibfnamefont {N.}~\bibnamefont {{Weaverdyck}}}, \bibinfo {author} {\bibfnamefont {R.~P.}\ \bibnamefont {{Rollins}}}, \bibinfo {author} {\bibfnamefont {J.}~\bibnamefont {{Muir}}}, \bibinfo {author} {\bibfnamefont {A.}~\bibnamefont {{Fert{\'e}}}}, \bibinfo {author} {\bibfnamefont {A.~R.}\ \bibnamefont {{Liddle}}}, \bibinfo {author} {\bibfnamefont {A.}~\bibnamefont {{Campos}}},  \emph {et~al.},\ }\href {\doibase 10.1093/mnras/stac2786} {\bibfield  {journal} {\bibinfo  {journal} {\mnras}\ } (\bibinfo {year} {2022}),\ 10.1093/mnras/stac2786},\ \Eprint {http://arxiv.org/abs/2202.08233} {arXiv:2202.08233 [astro-ph.CO]} \BibitemShut {NoStop}%
\bibitem [{\citenamefont {{Higson}}\ \emph {et~al.}(2019)\citenamefont {{Higson}}, \citenamefont {{Handley}}, \citenamefont {{Hobson}},\ and\ \citenamefont {{Lasenby}}}]{nestcheck2019}%
  \BibitemOpen
  \bibfield  {author} {\bibinfo {author} {\bibfnamefont {E.}~\bibnamefont {{Higson}}}, \bibinfo {author} {\bibfnamefont {W.}~\bibnamefont {{Handley}}}, \bibinfo {author} {\bibfnamefont {M.}~\bibnamefont {{Hobson}}}, \ and\ \bibinfo {author} {\bibfnamefont {A.}~\bibnamefont {{Lasenby}}},\ }\href {\doibase 10.1093/mnras/sty3090} {\bibfield  {journal} {\bibinfo  {journal} {\mnras}\ }\textbf {\bibinfo {volume} {483}},\ \bibinfo {pages} {2044} (\bibinfo {year} {2019})},\ \Eprint {http://arxiv.org/abs/1804.06406} {arXiv:1804.06406 [stat.CO]} \BibitemShut {NoStop}%
\bibitem [{\citenamefont {{Joachimi}}\ \emph {et~al.}(2021{\natexlab{b}})\citenamefont {{Joachimi}}, \citenamefont {{Lin}}, \citenamefont {{Asgari}}, \citenamefont {{Tr{\"o}ster}}, \citenamefont {{Heymans}}, \citenamefont {{Hildebrandt}}, \citenamefont {{K{\"o}hlinger}} \emph {et~al.}}]{KiDS1000_3x2pt_model}%
  \BibitemOpen
  \bibfield  {author} {\bibinfo {author} {\bibfnamefont {B.}~\bibnamefont {{Joachimi}}}, \bibinfo {author} {\bibfnamefont {C.~A.}\ \bibnamefont {{Lin}}}, \bibinfo {author} {\bibfnamefont {M.}~\bibnamefont {{Asgari}}}, \bibinfo {author} {\bibfnamefont {T.}~\bibnamefont {{Tr{\"o}ster}}}, \bibinfo {author} {\bibfnamefont {C.}~\bibnamefont {{Heymans}}}, \bibinfo {author} {\bibfnamefont {H.}~\bibnamefont {{Hildebrandt}}}, \bibinfo {author} {\bibfnamefont {F.}~\bibnamefont {{K{\"o}hlinger}}},  \emph {et~al.},\ }\href {\doibase 10.1051/0004-6361/202038831} {\bibfield  {journal} {\bibinfo  {journal} {\aap}\ }\textbf {\bibinfo {volume} {646}},\ \bibinfo {eid} {A129} (\bibinfo {year} {2021}{\natexlab{b}})},\ \Eprint {http://arxiv.org/abs/2007.01844} {arXiv:2007.01844 [astro-ph.CO]} \BibitemShut {NoStop}%
\bibitem [{\citenamefont {{Schaye}}\ \emph {et~al.}(2010)\citenamefont {{Schaye}}, \citenamefont {{Dalla Vecchia}}, \citenamefont {{Booth}}, \citenamefont {{Wiersma}}, \citenamefont {{Theuns}} \emph {et~al.}}]{OWLS_Schaye2010}%
  \BibitemOpen
  \bibfield  {author} {\bibinfo {author} {\bibfnamefont {J.}~\bibnamefont {{Schaye}}}, \bibinfo {author} {\bibfnamefont {C.}~\bibnamefont {{Dalla Vecchia}}}, \bibinfo {author} {\bibfnamefont {C.~M.}\ \bibnamefont {{Booth}}}, \bibinfo {author} {\bibfnamefont {R.~P.~C.}\ \bibnamefont {{Wiersma}}}, \bibinfo {author} {\bibfnamefont {T.}~\bibnamefont {{Theuns}}},  \emph {et~al.},\ }\href {\doibase 10.1111/j.1365-2966.2009.16029.x} {\bibfield  {journal} {\bibinfo  {journal} {\mnras}\ }\textbf {\bibinfo {volume} {402}},\ \bibinfo {pages} {1536} (\bibinfo {year} {2010})},\ \Eprint {http://arxiv.org/abs/0909.5196} {arXiv:0909.5196 [astro-ph.CO]} \BibitemShut {NoStop}%
\bibitem [{\citenamefont {{van Daalen}}\ \emph {et~al.}(2011)\citenamefont {{van Daalen}}, \citenamefont {{Schaye}}, \citenamefont {{Booth}},\ and\ \citenamefont {{Dalla Vecchia}}}]{OWLS_vanDaalen2011}%
  \BibitemOpen
  \bibfield  {author} {\bibinfo {author} {\bibfnamefont {M.~P.}\ \bibnamefont {{van Daalen}}}, \bibinfo {author} {\bibfnamefont {J.}~\bibnamefont {{Schaye}}}, \bibinfo {author} {\bibfnamefont {C.~M.}\ \bibnamefont {{Booth}}}, \ and\ \bibinfo {author} {\bibfnamefont {C.}~\bibnamefont {{Dalla Vecchia}}},\ }\href {\doibase 10.1111/j.1365-2966.2011.18981.x} {\bibfield  {journal} {\bibinfo  {journal} {\mnras}\ }\textbf {\bibinfo {volume} {415}},\ \bibinfo {pages} {3649} (\bibinfo {year} {2011})},\ \Eprint {http://arxiv.org/abs/1104.1174} {arXiv:1104.1174 [astro-ph.CO]} \BibitemShut {NoStop}%
\bibitem [{\citenamefont {{Le Brun}}\ \emph {et~al.}(2014)\citenamefont {{Le Brun}}, \citenamefont {{McCarthy}}, \citenamefont {{Schaye}},\ and\ \citenamefont {{Ponman}}}]{cowls_LeBrun2014}%
  \BibitemOpen
  \bibfield  {author} {\bibinfo {author} {\bibfnamefont {A.~M.~C.}\ \bibnamefont {{Le Brun}}}, \bibinfo {author} {\bibfnamefont {I.~G.}\ \bibnamefont {{McCarthy}}}, \bibinfo {author} {\bibfnamefont {J.}~\bibnamefont {{Schaye}}}, \ and\ \bibinfo {author} {\bibfnamefont {T.~J.}\ \bibnamefont {{Ponman}}},\ }\href {\doibase 10.1093/mnras/stu608} {\bibfield  {journal} {\bibinfo  {journal} {\mnras}\ }\textbf {\bibinfo {volume} {441}},\ \bibinfo {pages} {1270} (\bibinfo {year} {2014})},\ \Eprint {http://arxiv.org/abs/1312.5462} {arXiv:1312.5462 [astro-ph.CO]} \BibitemShut {NoStop}%
\bibitem [{\citenamefont {{Vogelsberger}}\ \emph {et~al.}(2014)\citenamefont {{Vogelsberger}}, \citenamefont {{Genel}}, \citenamefont {{Springel}}, \citenamefont {{Torrey}}, \citenamefont {{Sijacki}} \emph {et~al.}}]{Illustris_Vogelsberger2014}%
  \BibitemOpen
  \bibfield  {author} {\bibinfo {author} {\bibfnamefont {M.}~\bibnamefont {{Vogelsberger}}}, \bibinfo {author} {\bibfnamefont {S.}~\bibnamefont {{Genel}}}, \bibinfo {author} {\bibfnamefont {V.}~\bibnamefont {{Springel}}}, \bibinfo {author} {\bibfnamefont {P.}~\bibnamefont {{Torrey}}}, \bibinfo {author} {\bibfnamefont {D.}~\bibnamefont {{Sijacki}}},  \emph {et~al.},\ }\href {\doibase 10.1093/mnras/stu1536} {\bibfield  {journal} {\bibinfo  {journal} {\mnras}\ }\textbf {\bibinfo {volume} {444}},\ \bibinfo {pages} {1518} (\bibinfo {year} {2014})},\ \Eprint {http://arxiv.org/abs/1405.2921} {arXiv:1405.2921 [astro-ph.CO]} \BibitemShut {NoStop}%
\bibitem [{\citenamefont {{Khandai}}\ \emph {et~al.}(2015)\citenamefont {{Khandai}}, \citenamefont {{Di Matteo}}, \citenamefont {{Croft}}, \citenamefont {{Wilkins}}, \citenamefont {{Feng}} \emph {et~al.}}]{massiveBlackII_Khandai2015}%
  \BibitemOpen
  \bibfield  {author} {\bibinfo {author} {\bibfnamefont {N.}~\bibnamefont {{Khandai}}}, \bibinfo {author} {\bibfnamefont {T.}~\bibnamefont {{Di Matteo}}}, \bibinfo {author} {\bibfnamefont {R.}~\bibnamefont {{Croft}}}, \bibinfo {author} {\bibfnamefont {S.}~\bibnamefont {{Wilkins}}}, \bibinfo {author} {\bibfnamefont {Y.}~\bibnamefont {{Feng}}},  \emph {et~al.},\ }\href {\doibase 10.1093/mnras/stv627} {\bibfield  {journal} {\bibinfo  {journal} {\mnras}\ }\textbf {\bibinfo {volume} {450}},\ \bibinfo {pages} {1349} (\bibinfo {year} {2015})},\ \Eprint {http://arxiv.org/abs/1402.0888} {arXiv:1402.0888 [astro-ph.CO]} \BibitemShut {NoStop}%
\bibitem [{\citenamefont {{Crain}}\ \emph {et~al.}(2015)\citenamefont {{Crain}}, \citenamefont {{Schaye}}, \citenamefont {{Bower}}, \citenamefont {{Furlong}}, \citenamefont {{Schaller}} \emph {et~al.}}]{eagle_Crain2015}%
  \BibitemOpen
  \bibfield  {author} {\bibinfo {author} {\bibfnamefont {R.~A.}\ \bibnamefont {{Crain}}}, \bibinfo {author} {\bibfnamefont {J.}~\bibnamefont {{Schaye}}}, \bibinfo {author} {\bibfnamefont {R.~G.}\ \bibnamefont {{Bower}}}, \bibinfo {author} {\bibfnamefont {M.}~\bibnamefont {{Furlong}}}, \bibinfo {author} {\bibfnamefont {M.}~\bibnamefont {{Schaller}}},  \emph {et~al.},\ }\href {\doibase 10.1093/mnras/stv725} {\bibfield  {journal} {\bibinfo  {journal} {\mnras}\ }\textbf {\bibinfo {volume} {450}},\ \bibinfo {pages} {1937} (\bibinfo {year} {2015})},\ \Eprint {http://arxiv.org/abs/1501.01311} {arXiv:1501.01311 [astro-ph.GA]} \BibitemShut {NoStop}%
\bibitem [{\citenamefont {{Kaviraj}}\ \emph {et~al.}(2017)\citenamefont {{Kaviraj}}, \citenamefont {{Laigle}}, \citenamefont {{Kimm}}, \citenamefont {{Devriendt}}, \citenamefont {{Dubois}} \emph {et~al.}}]{horizonAgn_Kaviraj2017}%
  \BibitemOpen
  \bibfield  {author} {\bibinfo {author} {\bibfnamefont {S.}~\bibnamefont {{Kaviraj}}}, \bibinfo {author} {\bibfnamefont {C.}~\bibnamefont {{Laigle}}}, \bibinfo {author} {\bibfnamefont {T.}~\bibnamefont {{Kimm}}}, \bibinfo {author} {\bibfnamefont {J.~E.~G.}\ \bibnamefont {{Devriendt}}}, \bibinfo {author} {\bibfnamefont {Y.}~\bibnamefont {{Dubois}}},  \emph {et~al.},\ }\href {\doibase 10.1093/mnras/stx126} {\bibfield  {journal} {\bibinfo  {journal} {\mnras}\ }\textbf {\bibinfo {volume} {467}},\ \bibinfo {pages} {4739} (\bibinfo {year} {2017})},\ \Eprint {http://arxiv.org/abs/1605.09379} {arXiv:1605.09379 [astro-ph.GA]} \BibitemShut {NoStop}%
\bibitem [{\citenamefont {{Nelson}}\ \emph {et~al.}(2019)\citenamefont {{Nelson}}, \citenamefont {{Springel}}, \citenamefont {{Pillepich}}, \citenamefont {{Rodriguez-Gomez}}, \citenamefont {{Torrey}}, \citenamefont {{Genel}} \emph {et~al.}}]{IllustrisTNG_Nelson2019}%
  \BibitemOpen
  \bibfield  {author} {\bibinfo {author} {\bibfnamefont {D.}~\bibnamefont {{Nelson}}}, \bibinfo {author} {\bibfnamefont {V.}~\bibnamefont {{Springel}}}, \bibinfo {author} {\bibfnamefont {A.}~\bibnamefont {{Pillepich}}}, \bibinfo {author} {\bibfnamefont {V.}~\bibnamefont {{Rodriguez-Gomez}}}, \bibinfo {author} {\bibfnamefont {P.}~\bibnamefont {{Torrey}}}, \bibinfo {author} {\bibfnamefont {S.}~\bibnamefont {{Genel}}},  \emph {et~al.},\ }\href {\doibase 10.1186/s40668-019-0028-x} {\bibfield  {journal} {\bibinfo  {journal} {Computational Astrophysics and Cosmology}\ }\textbf {\bibinfo {volume} {6}},\ \bibinfo {eid} {2} (\bibinfo {year} {2019})},\ \Eprint {http://arxiv.org/abs/1812.05609} {arXiv:1812.05609 [astro-ph.GA]} \BibitemShut {NoStop}%
\bibitem [{\citenamefont {Nelder}\ and\ \citenamefont {Mead}(1965)}]{NeldMead65}%
  \BibitemOpen
  \bibfield  {author} {\bibinfo {author} {\bibfnamefont {J.~A.}\ \bibnamefont {Nelder}}\ and\ \bibinfo {author} {\bibfnamefont {R.}~\bibnamefont {Mead}},\ }\href@noop {} {\bibfield  {journal} {\bibinfo  {journal} {computer journal}\ }\textbf {\bibinfo {volume} {7}},\ \bibinfo {pages} {308} (\bibinfo {year} {1965})}\BibitemShut {NoStop}%
\bibitem [{\citenamefont {Hinton}(2016)}]{ChainConsumer}%
  \BibitemOpen
  \bibfield  {author} {\bibinfo {author} {\bibfnamefont {S.}~\bibnamefont {Hinton}},\ }\href {\doibase 10.21105/joss.00045} {\bibfield  {journal} {\bibinfo  {journal} {Journal of Open Source Software}\ }\textbf {\bibinfo {volume} {1}},\ \bibinfo {pages} {45} (\bibinfo {year} {2016})}\BibitemShut {NoStop}%
\bibitem [{\citenamefont {{Lewis}}(2019)}]{GetDist2019}%
  \BibitemOpen
  \bibfield  {author} {\bibinfo {author} {\bibfnamefont {A.}~\bibnamefont {{Lewis}}},\ }\href {\doibase 10.48550/arXiv.1910.13970} {\bibfield  {journal} {\bibinfo  {journal} {arXiv e-prints}\ ,\ \bibinfo {eid} {arXiv:1910.13970}} (\bibinfo {year} {2019})},\ \Eprint {http://arxiv.org/abs/1910.13970} {arXiv:1910.13970 [astro-ph.IM]} \BibitemShut {NoStop}%
\bibitem [{\citenamefont {{Hikage}}\ \emph {et~al.}(2019)\citenamefont {{Hikage}}, \citenamefont {{Oguri}}, \citenamefont {{Hamana}}, \citenamefont {{More}}, \citenamefont {{Mandelbaum}}, \citenamefont {{Takada}}, \citenamefont {{K{\"o}hlinger}} \emph {et~al.}}]{cosmicShear_HSC1_Chiaki2019}%
  \BibitemOpen
  \bibfield  {author} {\bibinfo {author} {\bibfnamefont {C.}~\bibnamefont {{Hikage}}}, \bibinfo {author} {\bibfnamefont {M.}~\bibnamefont {{Oguri}}}, \bibinfo {author} {\bibfnamefont {T.}~\bibnamefont {{Hamana}}}, \bibinfo {author} {\bibfnamefont {S.}~\bibnamefont {{More}}}, \bibinfo {author} {\bibfnamefont {R.}~\bibnamefont {{Mandelbaum}}}, \bibinfo {author} {\bibfnamefont {M.}~\bibnamefont {{Takada}}}, \bibinfo {author} {\bibfnamefont {F.}~\bibnamefont {{K{\"o}hlinger}}},  \emph {et~al.},\ }\href {\doibase 10.1093/pasj/psz010} {\bibfield  {journal} {\bibinfo  {journal} {\pasj}\ }\textbf {\bibinfo {volume} {71}},\ \bibinfo {eid} {43} (\bibinfo {year} {2019})},\ \Eprint {http://arxiv.org/abs/1809.09148} {arXiv:1809.09148 [astro-ph.CO]} \BibitemShut {NoStop}%
\bibitem [{\citenamefont {{Longley}}\ \emph {et~al.}(2023)\citenamefont {{Longley}}, \citenamefont {{Chang}}, \citenamefont {{Walter}}, \citenamefont {{Zuntz}}, \citenamefont {{Ishak}}, \citenamefont {{Mandelbaum}} \emph {et~al.}}]{wlreana_catalogs_Longley2023}%
  \BibitemOpen
  \bibfield  {author} {\bibinfo {author} {\bibfnamefont {E.~P.}\ \bibnamefont {{Longley}}}, \bibinfo {author} {\bibfnamefont {C.}~\bibnamefont {{Chang}}}, \bibinfo {author} {\bibfnamefont {C.~W.}\ \bibnamefont {{Walter}}}, \bibinfo {author} {\bibfnamefont {J.}~\bibnamefont {{Zuntz}}}, \bibinfo {author} {\bibfnamefont {M.}~\bibnamefont {{Ishak}}}, \bibinfo {author} {\bibfnamefont {R.}~\bibnamefont {{Mandelbaum}}},  \emph {et~al.},\ }\href {\doibase 10.1093/mnras/stad246} {\bibfield  {journal} {\bibinfo  {journal} {\mnras}\ } (\bibinfo {year} {2023}),\ 10.1093/mnras/stad246},\ \Eprint {http://arxiv.org/abs/2208.07179} {arXiv:2208.07179 [astro-ph.CO]} \BibitemShut {NoStop}%
\bibitem [{\citenamefont {{Tr{\"o}ster}}\ \emph {et~al.}(2022{\natexlab{b}})\citenamefont {{Tr{\"o}ster}}, \citenamefont {{Mead}}, \citenamefont {{Heymans}}, \citenamefont {{Yan}}, \citenamefont {{Alonso}} \emph {et~al.}}]{pyhmcode_Troster2022}%
  \BibitemOpen
  \bibfield  {author} {\bibinfo {author} {\bibfnamefont {T.}~\bibnamefont {{Tr{\"o}ster}}}, \bibinfo {author} {\bibfnamefont {A.~J.}\ \bibnamefont {{Mead}}}, \bibinfo {author} {\bibfnamefont {C.}~\bibnamefont {{Heymans}}}, \bibinfo {author} {\bibfnamefont {Z.}~\bibnamefont {{Yan}}}, \bibinfo {author} {\bibfnamefont {D.}~\bibnamefont {{Alonso}}},  \emph {et~al.},\ }\href {\doibase 10.1051/0004-6361/202142197} {\bibfield  {journal} {\bibinfo  {journal} {\aap}\ }\textbf {\bibinfo {volume} {660}},\ \bibinfo {eid} {A27} (\bibinfo {year} {2022}{\natexlab{b}})},\ \Eprint {http://arxiv.org/abs/2109.04458} {arXiv:2109.04458 [astro-ph.CO]} \BibitemShut {NoStop}%
\bibitem [{\citenamefont {{Bakx}}\ \emph {et~al.}(2023)\citenamefont {{Bakx}}, \citenamefont {{Kurita}}, \citenamefont {{Chisari}}, \citenamefont {{Vlah}},\ and\ \citenamefont {{Schmidt}}}]{ia_EFT_Bakx2023}%
  \BibitemOpen
  \bibfield  {author} {\bibinfo {author} {\bibfnamefont {T.}~\bibnamefont {{Bakx}}}, \bibinfo {author} {\bibfnamefont {T.}~\bibnamefont {{Kurita}}}, \bibinfo {author} {\bibfnamefont {N.~E.}\ \bibnamefont {{Chisari}}}, \bibinfo {author} {\bibfnamefont {Z.}~\bibnamefont {{Vlah}}}, \ and\ \bibinfo {author} {\bibfnamefont {F.~o.}\ \bibnamefont {{Schmidt}}},\ }\href {\doibase 10.48550/arXiv.2303.15565} {\bibfield  {journal} {\bibinfo  {journal} {arXiv e-prints}\ ,\ \bibinfo {eid} {arXiv:2303.15565}} (\bibinfo {year} {2023})},\ \Eprint {http://arxiv.org/abs/2303.15565} {arXiv:2303.15565 [astro-ph.CO]} \BibitemShut {NoStop}%
\bibitem [{\citenamefont {{Leonard}}\ \emph {et~al.}(2018)\citenamefont {{Leonard}}, \citenamefont {{Mandelbaum}},\ and\ \citenamefont {{LSST Dark Energy Science Collaboration}}}]{ia_multiEst_Leonard2018}%
  \BibitemOpen
  \bibfield  {author} {\bibinfo {author} {\bibfnamefont {C.~D.}\ \bibnamefont {{Leonard}}}, \bibinfo {author} {\bibfnamefont {R.}~\bibnamefont {{Mandelbaum}}}, \ and\ \bibinfo {author} {\bibnamefont {{LSST Dark Energy Science Collaboration}}},\ }\href {\doibase 10.1093/mnras/sty1444} {\bibfield  {journal} {\bibinfo  {journal} {\mnras}\ }\textbf {\bibinfo {volume} {479}},\ \bibinfo {pages} {1412} (\bibinfo {year} {2018})},\ \Eprint {http://arxiv.org/abs/1802.08263} {arXiv:1802.08263 [astro-ph.CO]} \BibitemShut {NoStop}%
\bibitem [{\citenamefont {{MacMahon}}\ and\ \citenamefont {{Leonard}}(2023)}]{ia_multiEst_MacMahon2023}%
  \BibitemOpen
  \bibfield  {author} {\bibinfo {author} {\bibfnamefont {C.~M.~B.}\ \bibnamefont {{MacMahon}}}\ and\ \bibinfo {author} {\bibfnamefont {C.~D.}\ \bibnamefont {{Leonard}}},\ }\href {\doibase 10.48550/arXiv.2306.11428} {\bibfield  {journal} {\bibinfo  {journal} {arXiv e-prints}\ ,\ \bibinfo {eid} {arXiv:2306.11428}} (\bibinfo {year} {2023})},\ \Eprint {http://arxiv.org/abs/2306.11428} {arXiv:2306.11428 [astro-ph.CO]} \BibitemShut {NoStop}%
\bibitem [{\citenamefont {Zhang}(2010)}]{ia_selfcal_Zhang2010}%
  \BibitemOpen
  \bibfield  {author} {\bibinfo {author} {\bibfnamefont {P.}~\bibnamefont {Zhang}},\ }\href {\doibase 10.1088/0004-637X/720/2/1090} {\bibfield  {journal} {\bibinfo  {journal} {The Astrophysical Journal}\ }\textbf {\bibinfo {volume} {720}},\ \bibinfo {pages} {1090} (\bibinfo {year} {2010})}\BibitemShut {NoStop}%
\bibitem [{\citenamefont {{Yao}}\ \emph {et~al.}(2019)\citenamefont {{Yao}}, \citenamefont {{Ishak}}, \citenamefont {{Troxel}},\ and\ \citenamefont {{LSST Dark Energy Science Collaboration}}}]{ia_selfcal_Yao2019}%
  \BibitemOpen
  \bibfield  {author} {\bibinfo {author} {\bibfnamefont {J.}~\bibnamefont {{Yao}}}, \bibinfo {author} {\bibfnamefont {M.}~\bibnamefont {{Ishak}}}, \bibinfo {author} {\bibfnamefont {M.~A.}\ \bibnamefont {{Troxel}}}, \ and\ \bibinfo {author} {\bibnamefont {{LSST Dark Energy Science Collaboration}}},\ }\href {\doibase 10.1093/mnras/sty3188} {\bibfield  {journal} {\bibinfo  {journal} {\mnras}\ }\textbf {\bibinfo {volume} {483}},\ \bibinfo {pages} {276} (\bibinfo {year} {2019})},\ \Eprint {http://arxiv.org/abs/1809.07273} {arXiv:1809.07273 [astro-ph.CO]} \BibitemShut {NoStop}%
\bibitem [{\citenamefont {{Capozzi}}\ \emph {et~al.}(2014)\citenamefont {{Capozzi}}, \citenamefont {{Fogli}}, \citenamefont {{Lisi}}, \citenamefont {{Marrone}}, \citenamefont {{Montanino}} \emph {et~al.}}]{neutrino_Capozzi2014}%
  \BibitemOpen
  \bibfield  {author} {\bibinfo {author} {\bibfnamefont {F.}~\bibnamefont {{Capozzi}}}, \bibinfo {author} {\bibfnamefont {G.~L.}\ \bibnamefont {{Fogli}}}, \bibinfo {author} {\bibfnamefont {E.}~\bibnamefont {{Lisi}}}, \bibinfo {author} {\bibfnamefont {A.}~\bibnamefont {{Marrone}}}, \bibinfo {author} {\bibfnamefont {D.}~\bibnamefont {{Montanino}}},  \emph {et~al.},\ }\href {\doibase 10.1103/PhysRevD.89.093018} {\bibfield  {journal} {\bibinfo  {journal} {\prd}\ }\textbf {\bibinfo {volume} {89}},\ \bibinfo {eid} {093018} (\bibinfo {year} {2014})},\ \Eprint {http://arxiv.org/abs/1312.2878} {arXiv:1312.2878 [hep-ph]} \BibitemShut {NoStop}%
\bibitem [{\citenamefont {{Esteban}}\ \emph {et~al.}(2019)\citenamefont {{Esteban}}, \citenamefont {{Gonzalez-Garcia}}, \citenamefont {{Hernandez-Cabezudo}}, \citenamefont {{Maltoni}},\ and\ \citenamefont {{Schwetz}}}]{neutrino_Esteban2019}%
  \BibitemOpen
  \bibfield  {author} {\bibinfo {author} {\bibfnamefont {I.}~\bibnamefont {{Esteban}}}, \bibinfo {author} {\bibfnamefont {M.~C.}\ \bibnamefont {{Gonzalez-Garcia}}}, \bibinfo {author} {\bibfnamefont {A.}~\bibnamefont {{Hernandez-Cabezudo}}}, \bibinfo {author} {\bibfnamefont {M.}~\bibnamefont {{Maltoni}}}, \ and\ \bibinfo {author} {\bibfnamefont {T.}~\bibnamefont {{Schwetz}}},\ }\href {\doibase 10.1007/JHEP01(2019)106} {\bibfield  {journal} {\bibinfo  {journal} {Journal of High Energy Physics}\ }\textbf {\bibinfo {volume} {2019}},\ \bibinfo {eid} {106} (\bibinfo {year} {2019})},\ \Eprint {http://arxiv.org/abs/1811.05487} {arXiv:1811.05487 [hep-ph]} \BibitemShut {NoStop}%
\bibitem [{\citenamefont {{Ichiki}}\ \emph {et~al.}(2009)\citenamefont {{Ichiki}}, \citenamefont {{Takada}},\ and\ \citenamefont {{Takahashi}}}]{neutrinoWLCMB_Ichiki2009}%
  \BibitemOpen
  \bibfield  {author} {\bibinfo {author} {\bibfnamefont {K.}~\bibnamefont {{Ichiki}}}, \bibinfo {author} {\bibfnamefont {M.}~\bibnamefont {{Takada}}}, \ and\ \bibinfo {author} {\bibfnamefont {T.}~\bibnamefont {{Takahashi}}},\ }\href {\doibase 10.1103/PhysRevD.79.023520} {\bibfield  {journal} {\bibinfo  {journal} {\prd}\ }\textbf {\bibinfo {volume} {79}},\ \bibinfo {eid} {023520} (\bibinfo {year} {2009})},\ \Eprint {http://arxiv.org/abs/0810.4921} {arXiv:0810.4921 [astro-ph]} \BibitemShut {NoStop}%
\bibitem [{\citenamefont {Malz}\ and\ \citenamefont {Hogg}(2022)}]{stackpz_Malz2020}%
  \BibitemOpen
  \bibfield  {author} {\bibinfo {author} {\bibfnamefont {A.~I.}\ \bibnamefont {Malz}}\ and\ \bibinfo {author} {\bibfnamefont {D.~W.}\ \bibnamefont {Hogg}},\ }\href {\doibase 10.3847/1538-4357/ac062f} {\bibfield  {journal} {\bibinfo  {journal} {Astrophys. J.}\ }\textbf {\bibinfo {volume} {928}},\ \bibinfo {pages} {127} (\bibinfo {year} {2022})},\ \Eprint {http://arxiv.org/abs/2007.12178} {arXiv:2007.12178 [astro-ph.CO]} \BibitemShut {NoStop}%
\bibitem [{\citenamefont {{Gatti}}\ \emph {et~al.}(2021)\citenamefont {{Gatti}}, \citenamefont {{Sheldon}}, \citenamefont {{Amon}}, \citenamefont {{Becker}}, \citenamefont {{Troxel}} \emph {et~al.}}]{DESY3_catalog_Gatti2021}%
  \BibitemOpen
  \bibfield  {author} {\bibinfo {author} {\bibfnamefont {M.}~\bibnamefont {{Gatti}}}, \bibinfo {author} {\bibfnamefont {E.}~\bibnamefont {{Sheldon}}}, \bibinfo {author} {\bibfnamefont {A.}~\bibnamefont {{Amon}}}, \bibinfo {author} {\bibfnamefont {M.}~\bibnamefont {{Becker}}}, \bibinfo {author} {\bibfnamefont {M.}~\bibnamefont {{Troxel}}},  \emph {et~al.},\ }\href {\doibase 10.1093/mnras/stab918} {\bibfield  {journal} {\bibinfo  {journal} {\mnras}\ }\textbf {\bibinfo {volume} {504}},\ \bibinfo {pages} {4312} (\bibinfo {year} {2021})},\ \Eprint {http://arxiv.org/abs/2011.03408} {arXiv:2011.03408 [astro-ph.CO]} \BibitemShut {NoStop}%
\bibitem [{\citenamefont {{Giblin}}\ \emph {et~al.}(2021)\citenamefont {{Giblin}}, \citenamefont {{Heymans}}, \citenamefont {{Asgari}}, \citenamefont {{Hildebrandt}}, \citenamefont {{Hoekstra}}, \citenamefont {{Joachimi}} \emph {et~al.}}]{KiDS1000_catalog2021}%
  \BibitemOpen
  \bibfield  {author} {\bibinfo {author} {\bibfnamefont {B.}~\bibnamefont {{Giblin}}}, \bibinfo {author} {\bibfnamefont {C.}~\bibnamefont {{Heymans}}}, \bibinfo {author} {\bibfnamefont {M.}~\bibnamefont {{Asgari}}}, \bibinfo {author} {\bibfnamefont {H.}~\bibnamefont {{Hildebrandt}}}, \bibinfo {author} {\bibfnamefont {H.}~\bibnamefont {{Hoekstra}}}, \bibinfo {author} {\bibfnamefont {B.}~\bibnamefont {{Joachimi}}},  \emph {et~al.},\ }\href {\doibase 10.1051/0004-6361/202038850} {\bibfield  {journal} {\bibinfo  {journal} {\aap}\ }\textbf {\bibinfo {volume} {645}},\ \bibinfo {eid} {A105} (\bibinfo {year} {2021})},\ \Eprint {http://arxiv.org/abs/2007.01845} {arXiv:2007.01845 [astro-ph.CO]} \BibitemShut {NoStop}%
\bibitem [{\citenamefont {{Charnock}}\ \emph {et~al.}(2017)\citenamefont {{Charnock}}, \citenamefont {{Battye}},\ and\ \citenamefont {{Moss}}}]{cosmoTension2017}%
  \BibitemOpen
  \bibfield  {author} {\bibinfo {author} {\bibfnamefont {T.}~\bibnamefont {{Charnock}}}, \bibinfo {author} {\bibfnamefont {R.~A.}\ \bibnamefont {{Battye}}}, \ and\ \bibinfo {author} {\bibfnamefont {A.}~\bibnamefont {{Moss}}},\ }\href {\doibase 10.1103/PhysRevD.95.123535} {\bibfield  {journal} {\bibinfo  {journal} {\prd}\ }\textbf {\bibinfo {volume} {95}},\ \bibinfo {eid} {123535} (\bibinfo {year} {2017})},\ \Eprint {http://arxiv.org/abs/1703.05959} {arXiv:1703.05959 [astro-ph.CO]} \BibitemShut {NoStop}%
\bibitem [{\citenamefont {{Park}}\ and\ \citenamefont {{Rozo}}(2020)}]{Park2020}%
  \BibitemOpen
  \bibfield  {author} {\bibinfo {author} {\bibfnamefont {Y.}~\bibnamefont {{Park}}}\ and\ \bibinfo {author} {\bibfnamefont {E.}~\bibnamefont {{Rozo}}},\ }\href {\doibase 10.1093/mnras/staa2647} {\bibfield  {journal} {\bibinfo  {journal} {\mnras}\ }\textbf {\bibinfo {volume} {499}},\ \bibinfo {pages} {4638} (\bibinfo {year} {2020})},\ \Eprint {http://arxiv.org/abs/1907.05798} {arXiv:1907.05798 [astro-ph.CO]} \BibitemShut {NoStop}%
\bibitem [{\citenamefont {{Alam}}\ \emph {et~al.}(2021)\citenamefont {{Alam}}, \citenamefont {{Aubert}}, \citenamefont {{Avila}}, \citenamefont {{Balland}}, \citenamefont {{Bautista}} \emph {et~al.}}]{eBOSS_BAO2021}%
  \BibitemOpen
  \bibfield  {author} {\bibinfo {author} {\bibfnamefont {S.}~\bibnamefont {{Alam}}}, \bibinfo {author} {\bibfnamefont {M.}~\bibnamefont {{Aubert}}}, \bibinfo {author} {\bibfnamefont {S.}~\bibnamefont {{Avila}}}, \bibinfo {author} {\bibfnamefont {C.}~\bibnamefont {{Balland}}}, \bibinfo {author} {\bibfnamefont {J.~E.}\ \bibnamefont {{Bautista}}},  \emph {et~al.},\ }\href {\doibase 10.1103/PhysRevD.103.083533} {\bibfield  {journal} {\bibinfo  {journal} {\prd}\ }\textbf {\bibinfo {volume} {103}},\ \bibinfo {eid} {083533} (\bibinfo {year} {2021})},\ \Eprint {http://arxiv.org/abs/2007.08991} {arXiv:2007.08991 [astro-ph.CO]} \BibitemShut {NoStop}%
\bibitem [{\citenamefont {{Ross}}\ \emph {et~al.}(2015)\citenamefont {{Ross}}, \citenamefont {{Samushia}}, \citenamefont {{Howlett}}, \citenamefont {{Percival}}, \citenamefont {{Burden}} \emph {et~al.}}]{mgs_Ross2015}%
  \BibitemOpen
  \bibfield  {author} {\bibinfo {author} {\bibfnamefont {A.~J.}\ \bibnamefont {{Ross}}}, \bibinfo {author} {\bibfnamefont {L.}~\bibnamefont {{Samushia}}}, \bibinfo {author} {\bibfnamefont {C.}~\bibnamefont {{Howlett}}}, \bibinfo {author} {\bibfnamefont {W.~J.}\ \bibnamefont {{Percival}}}, \bibinfo {author} {\bibfnamefont {A.}~\bibnamefont {{Burden}}},  \emph {et~al.},\ }\href {\doibase 10.1093/mnras/stv154} {\bibfield  {journal} {\bibinfo  {journal} {\mnras}\ }\textbf {\bibinfo {volume} {449}},\ \bibinfo {pages} {835} (\bibinfo {year} {2015})},\ \Eprint {http://arxiv.org/abs/1409.3242} {arXiv:1409.3242 [astro-ph.CO]} \BibitemShut {NoStop}%
\bibitem [{\citenamefont {{Alam}}\ \emph {et~al.}(2017)\citenamefont {{Alam}}, \citenamefont {{Ata}}, \citenamefont {{Bailey}}, \citenamefont {{Beutler}}, \citenamefont {{Bizyaev}}, \citenamefont {{Blazek}} \emph {et~al.}}]{bossBao_Alam2017}%
  \BibitemOpen
  \bibfield  {author} {\bibinfo {author} {\bibfnamefont {S.}~\bibnamefont {{Alam}}}, \bibinfo {author} {\bibfnamefont {M.}~\bibnamefont {{Ata}}}, \bibinfo {author} {\bibfnamefont {S.}~\bibnamefont {{Bailey}}}, \bibinfo {author} {\bibfnamefont {F.}~\bibnamefont {{Beutler}}}, \bibinfo {author} {\bibfnamefont {D.}~\bibnamefont {{Bizyaev}}}, \bibinfo {author} {\bibfnamefont {J.~A.}\ \bibnamefont {{Blazek}}},  \emph {et~al.},\ }\href {\doibase 10.1093/mnras/stx721} {\bibfield  {journal} {\bibinfo  {journal} {\mnras}\ }\textbf {\bibinfo {volume} {470}},\ \bibinfo {pages} {2617} (\bibinfo {year} {2017})},\ \Eprint {http://arxiv.org/abs/1607.03155} {arXiv:1607.03155 [astro-ph.CO]} \BibitemShut {NoStop}%
\bibitem [{\citenamefont {{Bautista}}\ \emph {et~al.}(2021)\citenamefont {{Bautista}}, \citenamefont {{Paviot}}, \citenamefont {{Vargas Maga{\~n}a}}, \citenamefont {{de la Torre}}, \citenamefont {{Fromenteau}}, \citenamefont {{Gil-Mar{\'\i}n}} \emph {et~al.}}]{ebossLRGs_Bautista2021}%
  \BibitemOpen
  \bibfield  {author} {\bibinfo {author} {\bibfnamefont {J.~E.}\ \bibnamefont {{Bautista}}}, \bibinfo {author} {\bibfnamefont {R.}~\bibnamefont {{Paviot}}}, \bibinfo {author} {\bibfnamefont {M.}~\bibnamefont {{Vargas Maga{\~n}a}}}, \bibinfo {author} {\bibfnamefont {S.}~\bibnamefont {{de la Torre}}}, \bibinfo {author} {\bibfnamefont {S.}~\bibnamefont {{Fromenteau}}}, \bibinfo {author} {\bibfnamefont {H.}~\bibnamefont {{Gil-Mar{\'\i}n}}},  \emph {et~al.},\ }\href {\doibase 10.1093/mnras/staa2800} {\bibfield  {journal} {\bibinfo  {journal} {\mnras}\ }\textbf {\bibinfo {volume} {500}},\ \bibinfo {pages} {736} (\bibinfo {year} {2021})},\ \Eprint {http://arxiv.org/abs/2007.08993} {arXiv:2007.08993 [astro-ph.CO]} \BibitemShut {NoStop}%
\bibitem [{\citenamefont {{de Mattia}}\ \emph {et~al.}(2021)\citenamefont {{de Mattia}}, \citenamefont {{Ruhlmann-Kleider}}, \citenamefont {{Raichoor}}, \citenamefont {{Ross}}, \citenamefont {{Tamone}} \emph {et~al.}}]{ebossELGs_deMattia2021}%
  \BibitemOpen
  \bibfield  {author} {\bibinfo {author} {\bibfnamefont {A.}~\bibnamefont {{de Mattia}}}, \bibinfo {author} {\bibfnamefont {V.}~\bibnamefont {{Ruhlmann-Kleider}}}, \bibinfo {author} {\bibfnamefont {A.}~\bibnamefont {{Raichoor}}}, \bibinfo {author} {\bibfnamefont {A.~J.}\ \bibnamefont {{Ross}}}, \bibinfo {author} {\bibfnamefont {A.}~\bibnamefont {{Tamone}}},  \emph {et~al.},\ }\href {\doibase 10.1093/mnras/staa3891} {\bibfield  {journal} {\bibinfo  {journal} {\mnras}\ }\textbf {\bibinfo {volume} {501}},\ \bibinfo {pages} {5616} (\bibinfo {year} {2021})},\ \Eprint {http://arxiv.org/abs/2007.09008} {arXiv:2007.09008 [astro-ph.CO]} \BibitemShut {NoStop}%
\bibitem [{\citenamefont {{Neveux}}\ \emph {et~al.}(2020)\citenamefont {{Neveux}}, \citenamefont {{Burtin}}, \citenamefont {{de Mattia}}, \citenamefont {{Smith}}, \citenamefont {{Ross}} \emph {et~al.}}]{ebossQSOs_Neveux2020}%
  \BibitemOpen
  \bibfield  {author} {\bibinfo {author} {\bibfnamefont {R.}~\bibnamefont {{Neveux}}}, \bibinfo {author} {\bibfnamefont {E.}~\bibnamefont {{Burtin}}}, \bibinfo {author} {\bibfnamefont {A.}~\bibnamefont {{de Mattia}}}, \bibinfo {author} {\bibfnamefont {A.}~\bibnamefont {{Smith}}}, \bibinfo {author} {\bibfnamefont {A.~J.}\ \bibnamefont {{Ross}}},  \emph {et~al.},\ }\href {\doibase 10.1093/mnras/staa2780} {\bibfield  {journal} {\bibinfo  {journal} {\mnras}\ }\textbf {\bibinfo {volume} {499}},\ \bibinfo {pages} {210} (\bibinfo {year} {2020})},\ \Eprint {http://arxiv.org/abs/2007.08999} {arXiv:2007.08999 [astro-ph.CO]} \BibitemShut {NoStop}%
\bibitem [{\citenamefont {{du Mas des Bourboux}}\ \emph {et~al.}(2020)\citenamefont {{du Mas des Bourboux}}, \citenamefont {{Rich}}, \citenamefont {{Font-Ribera}}, \citenamefont {{de Sainte Agathe}}, \citenamefont {{Farr}} \emph {et~al.}}]{ebossLya_duMas2020}%
  \BibitemOpen
  \bibfield  {author} {\bibinfo {author} {\bibfnamefont {H.}~\bibnamefont {{du Mas des Bourboux}}}, \bibinfo {author} {\bibfnamefont {J.}~\bibnamefont {{Rich}}}, \bibinfo {author} {\bibfnamefont {A.}~\bibnamefont {{Font-Ribera}}}, \bibinfo {author} {\bibfnamefont {V.}~\bibnamefont {{de Sainte Agathe}}}, \bibinfo {author} {\bibfnamefont {J.}~\bibnamefont {{Farr}}},  \emph {et~al.},\ }\href {\doibase 10.3847/1538-4357/abb085} {\bibfield  {journal} {\bibinfo  {journal} {\apj}\ }\textbf {\bibinfo {volume} {901}},\ \bibinfo {eid} {153} (\bibinfo {year} {2020})},\ \Eprint {http://arxiv.org/abs/2007.08995} {arXiv:2007.08995 [astro-ph.CO]} \BibitemShut {NoStop}%
\bibitem [{\citenamefont {{Leauthaud}}\ \emph {et~al.}(2022)\citenamefont {{Leauthaud}}, \citenamefont {{Amon}}, \citenamefont {{Singh}}, \citenamefont {{Gruen}}, \citenamefont {{Lange}} \emph {et~al.}}]{LWB2022}%
  \BibitemOpen
  \bibfield  {author} {\bibinfo {author} {\bibfnamefont {A.}~\bibnamefont {{Leauthaud}}}, \bibinfo {author} {\bibfnamefont {A.}~\bibnamefont {{Amon}}}, \bibinfo {author} {\bibfnamefont {S.}~\bibnamefont {{Singh}}}, \bibinfo {author} {\bibfnamefont {D.}~\bibnamefont {{Gruen}}}, \bibinfo {author} {\bibfnamefont {J.~U.}\ \bibnamefont {{Lange}}},  \emph {et~al.},\ }\href {\doibase 10.1093/mnras/stab3586} {\bibfield  {journal} {\bibinfo  {journal} {\mnras}\ }\textbf {\bibinfo {volume} {510}},\ \bibinfo {pages} {6150} (\bibinfo {year} {2022})},\ \Eprint {http://arxiv.org/abs/2111.13805} {arXiv:2111.13805 [astro-ph.CO]} \BibitemShut {NoStop}%
\bibitem [{\citenamefont {{Heymans}}\ \emph {et~al.}(2013)\citenamefont {{Heymans}}, \citenamefont {{Grocutt}}, \citenamefont {{Heavens}}, \citenamefont {{Kilbinger}}, \citenamefont {{Kitching}}, \citenamefont {{Simpson}} \emph {et~al.}}]{CFHTLenS_Heymans2013}%
  \BibitemOpen
  \bibfield  {author} {\bibinfo {author} {\bibfnamefont {C.}~\bibnamefont {{Heymans}}}, \bibinfo {author} {\bibfnamefont {E.}~\bibnamefont {{Grocutt}}}, \bibinfo {author} {\bibfnamefont {A.}~\bibnamefont {{Heavens}}}, \bibinfo {author} {\bibfnamefont {M.}~\bibnamefont {{Kilbinger}}}, \bibinfo {author} {\bibfnamefont {T.~D.}\ \bibnamefont {{Kitching}}}, \bibinfo {author} {\bibfnamefont {F.}~\bibnamefont {{Simpson}}},  \emph {et~al.},\ }\href {\doibase 10.1093/mnras/stt601} {\bibfield  {journal} {\bibinfo  {journal} {\mnras}\ }\textbf {\bibinfo {volume} {432}},\ \bibinfo {pages} {2433} (\bibinfo {year} {2013})},\ \Eprint {http://arxiv.org/abs/1303.1808} {arXiv:1303.1808 [astro-ph.CO]} \BibitemShut {NoStop}%
\bibitem [{\citenamefont {{Leauthaud}}\ \emph {et~al.}(2017)\citenamefont {{Leauthaud}}, \citenamefont {{Saito}}, \citenamefont {{Hilbert}}, \citenamefont {{Barreira}}, \citenamefont {{More}} \emph {et~al.}}]{LIL2017}%
  \BibitemOpen
  \bibfield  {author} {\bibinfo {author} {\bibfnamefont {A.}~\bibnamefont {{Leauthaud}}}, \bibinfo {author} {\bibfnamefont {S.}~\bibnamefont {{Saito}}}, \bibinfo {author} {\bibfnamefont {S.}~\bibnamefont {{Hilbert}}}, \bibinfo {author} {\bibfnamefont {A.}~\bibnamefont {{Barreira}}}, \bibinfo {author} {\bibfnamefont {S.}~\bibnamefont {{More}}},  \emph {et~al.},\ }\href {\doibase 10.1093/mnras/stx258} {\bibfield  {journal} {\bibinfo  {journal} {\mnras}\ }\textbf {\bibinfo {volume} {467}},\ \bibinfo {pages} {3024} (\bibinfo {year} {2017})},\ \Eprint {http://arxiv.org/abs/1611.08606} {arXiv:1611.08606 [astro-ph.CO]} \BibitemShut {NoStop}%
\bibitem [{\citenamefont {{Zuntz}}\ \emph {et~al.}(2017)\citenamefont {{Zuntz}}, \citenamefont {{Sheldon}}, \citenamefont {{Samuroff}}, \citenamefont {{Troxel}}, \citenamefont {{Jarvis}}, \citenamefont {{MacCrann}} \emph {et~al.}}]{DESY1_catalog_Zuntz2017}%
  \BibitemOpen
  \bibfield  {author} {\bibinfo {author} {\bibfnamefont {J.}~\bibnamefont {{Zuntz}}}, \bibinfo {author} {\bibfnamefont {E.}~\bibnamefont {{Sheldon}}}, \bibinfo {author} {\bibfnamefont {S.}~\bibnamefont {{Samuroff}}}, \bibinfo {author} {\bibfnamefont {M.~A.}\ \bibnamefont {{Troxel}}}, \bibinfo {author} {\bibfnamefont {M.}~\bibnamefont {{Jarvis}}}, \bibinfo {author} {\bibfnamefont {N.}~\bibnamefont {{MacCrann}}},  \emph {et~al.},\ }\href@noop {} {\bibfield  {journal} {\bibinfo  {journal} {ArXiv e-prints}\ } (\bibinfo {year} {2017})},\ \Eprint {http://arxiv.org/abs/1708.01533} {arXiv:1708.01533} \BibitemShut {NoStop}%
\bibitem [{\citenamefont {{Wright}}\ \emph {et~al.}(2019)\citenamefont {{Wright}}, \citenamefont {{Hildebrandt}}, \citenamefont {{Kuijken}}, \citenamefont {{Erben}}, \citenamefont {{Blake}} \emph {et~al.}}]{KV450-Wright2019}%
  \BibitemOpen
  \bibfield  {author} {\bibinfo {author} {\bibfnamefont {A.~H.}\ \bibnamefont {{Wright}}}, \bibinfo {author} {\bibfnamefont {H.}~\bibnamefont {{Hildebrandt}}}, \bibinfo {author} {\bibfnamefont {K.}~\bibnamefont {{Kuijken}}}, \bibinfo {author} {\bibfnamefont {T.}~\bibnamefont {{Erben}}}, \bibinfo {author} {\bibfnamefont {R.}~\bibnamefont {{Blake}}},  \emph {et~al.},\ }\href {\doibase 10.1051/0004-6361/201834879} {\bibfield  {journal} {\bibinfo  {journal} {\aap}\ }\textbf {\bibinfo {volume} {632}},\ \bibinfo {eid} {A34} (\bibinfo {year} {2019})},\ \Eprint {http://arxiv.org/abs/1812.06077} {arXiv:1812.06077 [astro-ph.CO]} \BibitemShut {NoStop}%
\bibitem [{\citenamefont {{Mandelbaum}}\ \emph {et~al.}(2018{\natexlab{b}})\citenamefont {{Mandelbaum}}, \citenamefont {{Miyatake}}, \citenamefont {{Hamana}}, \citenamefont {{Oguri}}, \citenamefont {{Simet}}, \citenamefont {{Armstrong}} \emph {et~al.}}]{HSC1_catalog}%
  \BibitemOpen
  \bibfield  {author} {\bibinfo {author} {\bibfnamefont {R.}~\bibnamefont {{Mandelbaum}}}, \bibinfo {author} {\bibfnamefont {H.}~\bibnamefont {{Miyatake}}}, \bibinfo {author} {\bibfnamefont {T.}~\bibnamefont {{Hamana}}}, \bibinfo {author} {\bibfnamefont {M.}~\bibnamefont {{Oguri}}}, \bibinfo {author} {\bibfnamefont {M.}~\bibnamefont {{Simet}}}, \bibinfo {author} {\bibfnamefont {R.}~\bibnamefont {{Armstrong}}},  \emph {et~al.},\ }\href {\doibase 10.1093/pasj/psx130} {\bibfield  {journal} {\bibinfo  {journal} {\pasj}\ }\textbf {\bibinfo {volume} {70}},\ \bibinfo {eid} {S25} (\bibinfo {year} {2018}{\natexlab{b}})},\ \Eprint {http://arxiv.org/abs/1705.06745} {arXiv:1705.06745} \BibitemShut {NoStop}%
\bibitem [{\citenamefont {{Melchior}}\ \emph {et~al.}(2018)\citenamefont {{Melchior}}, \citenamefont {{Moolekamp}}, \citenamefont {{Jerdee}}, \citenamefont {{Armstrong}}, \citenamefont {{Sun}}, \citenamefont {{Bosch}},\ and\ \citenamefont {{Lupton}}}]{scarlet_Melchior2018}%
  \BibitemOpen
  \bibfield  {author} {\bibinfo {author} {\bibfnamefont {P.}~\bibnamefont {{Melchior}}}, \bibinfo {author} {\bibfnamefont {F.}~\bibnamefont {{Moolekamp}}}, \bibinfo {author} {\bibfnamefont {M.}~\bibnamefont {{Jerdee}}}, \bibinfo {author} {\bibfnamefont {R.}~\bibnamefont {{Armstrong}}}, \bibinfo {author} {\bibfnamefont {A.~L.}\ \bibnamefont {{Sun}}}, \bibinfo {author} {\bibfnamefont {J.}~\bibnamefont {{Bosch}}}, \ and\ \bibinfo {author} {\bibfnamefont {R.~o.}\ \bibnamefont {{Lupton}}},\ }\href {\doibase 10.1016/j.ascom.2018.07.001} {\bibfield  {journal} {\bibinfo  {journal} {Astronomy and Computing}\ }\textbf {\bibinfo {volume} {24}},\ \bibinfo {eid} {129} (\bibinfo {year} {2018})},\ \Eprint {http://arxiv.org/abs/1802.10157} {arXiv:1802.10157 [astro-ph.IM]} \BibitemShut {NoStop}%
\bibitem [{\citenamefont {{Jarvis}}\ \emph {et~al.}(2021)\citenamefont {{Jarvis}}, \citenamefont {{Meyers}}, \citenamefont {{Leget}},\ and\ \citenamefont {{Davis}}}]{PIFF2021}%
  \BibitemOpen
  \bibfield  {author} {\bibinfo {author} {\bibfnamefont {M.}~\bibnamefont {{Jarvis}}}, \bibinfo {author} {\bibfnamefont {J.}~\bibnamefont {{Meyers}}}, \bibinfo {author} {\bibfnamefont {P.-F.}\ \bibnamefont {{Leget}}}, \ and\ \bibinfo {author} {\bibfnamefont {C.}~\bibnamefont {{Davis}}},\ }\href@noop {} {\enquote {\bibinfo {title} {{Piff: PSFs In the Full FOV}},}\ }\bibinfo {howpublished} {Astrophysics Source Code Library, record ascl:2102.024} (\bibinfo {year} {2021}),\ \Eprint {http://arxiv.org/abs/2102.024} {ascl:2102.024} \BibitemShut {NoStop}%
\bibitem [{\citenamefont {{LSST Dark Energy Science Collaboration (LSST DESC)}}\ \emph {et~al.}(2021)\citenamefont {{LSST Dark Energy Science Collaboration (LSST DESC)}}, \citenamefont {{Abolfathi}}, \citenamefont {{Alonso}}, \citenamefont {{Armstrong}}, \citenamefont {{Aubourg}}, \citenamefont {{Awan}} \emph {et~al.}}]{DC2_Abolfathi2021}%
  \BibitemOpen
  \bibfield  {author} {\bibinfo {author} {\bibnamefont {{LSST Dark Energy Science Collaboration (LSST DESC)}}}, \bibinfo {author} {\bibfnamefont {B.}~\bibnamefont {{Abolfathi}}}, \bibinfo {author} {\bibfnamefont {D.}~\bibnamefont {{Alonso}}}, \bibinfo {author} {\bibfnamefont {R.}~\bibnamefont {{Armstrong}}}, \bibinfo {author} {\bibfnamefont {{\'E}.}~\bibnamefont {{Aubourg}}}, \bibinfo {author} {\bibfnamefont {H.}~\bibnamefont {{Awan}}},  \emph {et~al.},\ }\href {\doibase 10.3847/1538-4365/abd62c} {\bibfield  {journal} {\bibinfo  {journal} {\apjs}\ }\textbf {\bibinfo {volume} {253}},\ \bibinfo {eid} {31} (\bibinfo {year} {2021})},\ \Eprint {http://arxiv.org/abs/2010.05926} {arXiv:2010.05926 [astro-ph.IM]} \BibitemShut {NoStop}%
\bibitem [{\citenamefont {{MacCrann}}\ \emph {et~al.}(2022)\citenamefont {{MacCrann}}, \citenamefont {{Becker}}, \citenamefont {{McCullough}}, \citenamefont {{Amon}}, \citenamefont {{Gruen}}, \citenamefont {{Jarvis}} \emph {et~al.}}]{DESY3_BlendshearCalib_MacCrann2021}%
  \BibitemOpen
  \bibfield  {author} {\bibinfo {author} {\bibfnamefont {N.}~\bibnamefont {{MacCrann}}}, \bibinfo {author} {\bibfnamefont {M.~R.}\ \bibnamefont {{Becker}}}, \bibinfo {author} {\bibfnamefont {J.}~\bibnamefont {{McCullough}}}, \bibinfo {author} {\bibfnamefont {A.}~\bibnamefont {{Amon}}}, \bibinfo {author} {\bibfnamefont {D.}~\bibnamefont {{Gruen}}}, \bibinfo {author} {\bibfnamefont {M.}~\bibnamefont {{Jarvis}}},  \emph {et~al.},\ }\href {\doibase 10.1093/mnras/stab2870} {\bibfield  {journal} {\bibinfo  {journal} {\mnras}\ }\textbf {\bibinfo {volume} {509}},\ \bibinfo {pages} {3371} (\bibinfo {year} {2022})},\ \Eprint {http://arxiv.org/abs/2012.08567} {arXiv:2012.08567 [astro-ph.CO]} \BibitemShut {NoStop}%
\bibitem [{\citenamefont {{Li}}\ \emph {et~al.}(2023)\citenamefont {{Li}}, \citenamefont {{Kuijken}}, \citenamefont {{Hoekstra}}, \citenamefont {{Miller}}, \citenamefont {{Heymans}}, \citenamefont {{Hildebrandt}} \emph {et~al.}}]{Skills_Li2023}%
  \BibitemOpen
  \bibfield  {author} {\bibinfo {author} {\bibfnamefont {S.-S.}\ \bibnamefont {{Li}}}, \bibinfo {author} {\bibfnamefont {K.}~\bibnamefont {{Kuijken}}}, \bibinfo {author} {\bibfnamefont {H.}~\bibnamefont {{Hoekstra}}}, \bibinfo {author} {\bibfnamefont {L.}~\bibnamefont {{Miller}}}, \bibinfo {author} {\bibfnamefont {C.}~\bibnamefont {{Heymans}}}, \bibinfo {author} {\bibfnamefont {H.}~\bibnamefont {{Hildebrandt}}},  \emph {et~al.},\ }\href {\doibase 10.1051/0004-6361/202245210} {\bibfield  {journal} {\bibinfo  {journal} {\aap}\ }\textbf {\bibinfo {volume} {670}},\ \bibinfo {eid} {A100} (\bibinfo {year} {2023})},\ \Eprint {http://arxiv.org/abs/2210.07163} {arXiv:2210.07163 [astro-ph.CO]} \BibitemShut {NoStop}%
\bibitem [{\citenamefont {{Li}}\ \emph {et~al.}(2018)\citenamefont {{Li}}, \citenamefont {{Katayama}}, \citenamefont {{Oguri}},\ and\ \citenamefont {{More}}}]{FPFS_Li2018}%
  \BibitemOpen
  \bibfield  {author} {\bibinfo {author} {\bibfnamefont {X.}~\bibnamefont {{Li}}}, \bibinfo {author} {\bibfnamefont {N.}~\bibnamefont {{Katayama}}}, \bibinfo {author} {\bibfnamefont {M.}~\bibnamefont {{Oguri}}}, \ and\ \bibinfo {author} {\bibfnamefont {S.}~\bibnamefont {{More}}},\ }\href {\doibase 10.1093/mnras/sty2548} {\bibfield  {journal} {\bibinfo  {journal} {\mnras}\ }\textbf {\bibinfo {volume} {481}},\ \bibinfo {pages} {4445} (\bibinfo {year} {2018})},\ \Eprint {http://arxiv.org/abs/1805.08514} {arXiv:1805.08514 [astro-ph.CO]} \BibitemShut {NoStop}%
\bibitem [{\citenamefont {{Li}}\ \emph {et~al.}(2022)\citenamefont {{Li}}, \citenamefont {{Li}},\ and\ \citenamefont {{Massey}}}]{FPFS_Li2022}%
  \BibitemOpen
  \bibfield  {author} {\bibinfo {author} {\bibfnamefont {X.}~\bibnamefont {{Li}}}, \bibinfo {author} {\bibfnamefont {Y.}~\bibnamefont {{Li}}}, \ and\ \bibinfo {author} {\bibfnamefont {R.}~\bibnamefont {{Massey}}},\ }\href {\doibase 10.1093/mnras/stac342} {\bibfield  {journal} {\bibinfo  {journal} {\mnras}\ }\textbf {\bibinfo {volume} {511}},\ \bibinfo {pages} {4850} (\bibinfo {year} {2022})},\ \Eprint {http://arxiv.org/abs/2110.01214} {arXiv:2110.01214 [astro-ph.CO]} \BibitemShut {NoStop}%
\bibitem [{\citenamefont {{Li}}\ and\ \citenamefont {{Mandelbaum}}(2022)}]{FPFS_Li2023}%
  \BibitemOpen
  \bibfield  {author} {\bibinfo {author} {\bibfnamefont {X.}~\bibnamefont {{Li}}}\ and\ \bibinfo {author} {\bibfnamefont {R.}~\bibnamefont {{Mandelbaum}}},\ }\href {\doibase 10.48550/arXiv.2208.10522} {\bibfield  {journal} {\bibinfo  {journal} {arXiv e-prints}\ ,\ \bibinfo {eid} {arXiv:2208.10522}} (\bibinfo {year} {2022})},\ \Eprint {http://arxiv.org/abs/2208.10522} {arXiv:2208.10522 [astro-ph.CO]} \BibitemShut {NoStop}%
\bibitem [{\citenamefont {{DESI Collaboration}}\ \emph {et~al.}(2016)\citenamefont {{DESI Collaboration}}, \citenamefont {{Aghamousa}}, \citenamefont {{Aguilar}}, \citenamefont {{Ahlen}}, \citenamefont {{Alam}} \emph {et~al.}}]{DESI2016}%
  \BibitemOpen
  \bibfield  {author} {\bibinfo {author} {\bibnamefont {{DESI Collaboration}}}, \bibinfo {author} {\bibfnamefont {A.}~\bibnamefont {{Aghamousa}}}, \bibinfo {author} {\bibfnamefont {J.}~\bibnamefont {{Aguilar}}}, \bibinfo {author} {\bibfnamefont {S.}~\bibnamefont {{Ahlen}}}, \bibinfo {author} {\bibfnamefont {S.}~\bibnamefont {{Alam}}},  \emph {et~al.},\ }\href {\doibase 10.48550/arXiv.1611.00036} {\bibfield  {journal} {\bibinfo  {journal} {arXiv e-prints}\ ,\ \bibinfo {eid} {arXiv:1611.00036}} (\bibinfo {year} {2016})},\ \Eprint {http://arxiv.org/abs/1611.00036} {arXiv:1611.00036 [astro-ph.IM]} \BibitemShut {NoStop}%
\bibitem [{\citenamefont {{S{\'a}nchez}}\ \emph {et~al.}(2022)\citenamefont {{S{\'a}nchez}}, \citenamefont {{Prat}}, \citenamefont {{Zacharegkas}}, \citenamefont {{Pandey}}, \citenamefont {{Baxter}}, \citenamefont {{Bernstein}} \emph {et~al.}}]{DESY3_shearRatio}%
  \BibitemOpen
  \bibfield  {author} {\bibinfo {author} {\bibfnamefont {C.}~\bibnamefont {{S{\'a}nchez}}}, \bibinfo {author} {\bibfnamefont {J.}~\bibnamefont {{Prat}}}, \bibinfo {author} {\bibfnamefont {G.}~\bibnamefont {{Zacharegkas}}}, \bibinfo {author} {\bibfnamefont {S.}~\bibnamefont {{Pandey}}}, \bibinfo {author} {\bibfnamefont {E.}~\bibnamefont {{Baxter}}}, \bibinfo {author} {\bibfnamefont {G.~M.}\ \bibnamefont {{Bernstein}}},  \emph {et~al.},\ }\href {\doibase 10.1103/PhysRevD.105.083529} {\bibfield  {journal} {\bibinfo  {journal} {\prd}\ }\textbf {\bibinfo {volume} {105}},\ \bibinfo {eid} {083529} (\bibinfo {year} {2022})},\ \Eprint {http://arxiv.org/abs/2105.13542} {arXiv:2105.13542 [astro-ph.CO]} \BibitemShut {NoStop}%
\bibitem [{\citenamefont {{Robertson}}\ \emph {et~al.}(2021)\citenamefont {{Robertson}}, \citenamefont {{Alonso}}, \citenamefont {{Harnois-D{\'e}raps}}, \citenamefont {{Darwish}}, \citenamefont {{Kannawadi}}, \citenamefont {{Amon}}, \citenamefont {{Asgari}} \emph {et~al.}}]{KiDS1000xCMBLens_Robertson2021}%
  \BibitemOpen
  \bibfield  {author} {\bibinfo {author} {\bibfnamefont {N.~C.}\ \bibnamefont {{Robertson}}}, \bibinfo {author} {\bibfnamefont {D.}~\bibnamefont {{Alonso}}}, \bibinfo {author} {\bibfnamefont {J.}~\bibnamefont {{Harnois-D{\'e}raps}}}, \bibinfo {author} {\bibfnamefont {O.}~\bibnamefont {{Darwish}}}, \bibinfo {author} {\bibfnamefont {A.}~\bibnamefont {{Kannawadi}}}, \bibinfo {author} {\bibfnamefont {A.}~\bibnamefont {{Amon}}}, \bibinfo {author} {\bibfnamefont {M.}~\bibnamefont {{Asgari}}},  \emph {et~al.},\ }\href {\doibase 10.1051/0004-6361/202039975} {\bibfield  {journal} {\bibinfo  {journal} {\aap}\ }\textbf {\bibinfo {volume} {649}},\ \bibinfo {eid} {A146} (\bibinfo {year} {2021})},\ \Eprint {http://arxiv.org/abs/2011.11613} {arXiv:2011.11613 [astro-ph.CO]} \BibitemShut {NoStop}%
\bibitem [{\citenamefont {{Ivezi{\'c}}}\ \emph {et~al.}(2019)\citenamefont {{Ivezi{\'c}}}, \citenamefont {{Kahn}}, \citenamefont {{Tyson}}, \citenamefont {{Abel}}, \citenamefont {{Acosta}} \emph {et~al.}}]{LSSTOverviwe2019}%
  \BibitemOpen
  \bibfield  {author} {\bibinfo {author} {\bibfnamefont {{\v{Z}}.}~\bibnamefont {{Ivezi{\'c}}}}, \bibinfo {author} {\bibfnamefont {S.~M.}\ \bibnamefont {{Kahn}}}, \bibinfo {author} {\bibfnamefont {J.~A.}\ \bibnamefont {{Tyson}}}, \bibinfo {author} {\bibfnamefont {B.}~\bibnamefont {{Abel}}}, \bibinfo {author} {\bibfnamefont {E.}~\bibnamefont {{Acosta}}},  \emph {et~al.},\ }\href {\doibase 10.3847/1538-4357/ab042c} {\bibfield  {journal} {\bibinfo  {journal} {\apj}\ }\textbf {\bibinfo {volume} {873}},\ \bibinfo {eid} {111} (\bibinfo {year} {2019})},\ \Eprint {http://arxiv.org/abs/0805.2366} {arXiv:0805.2366 [astro-ph]} \BibitemShut {NoStop}%
\bibitem [{\citenamefont {{Laureijs}}\ \emph {et~al.}(2011)\citenamefont {{Laureijs}}, \citenamefont {{Amiaux}}, \citenamefont {{Arduini}}, \citenamefont {{Augu{\`e}res}}, \citenamefont {{Brinchmann}} \emph {et~al.}}]{Euclid2011}%
  \BibitemOpen
  \bibfield  {author} {\bibinfo {author} {\bibfnamefont {R.}~\bibnamefont {{Laureijs}}}, \bibinfo {author} {\bibfnamefont {J.}~\bibnamefont {{Amiaux}}}, \bibinfo {author} {\bibfnamefont {S.}~\bibnamefont {{Arduini}}}, \bibinfo {author} {\bibfnamefont {J.~.}\ \bibnamefont {{Augu{\`e}res}}}, \bibinfo {author} {\bibfnamefont {J.}~\bibnamefont {{Brinchmann}}},  \emph {et~al.},\ }\href@noop {} {\bibfield  {journal} {\bibinfo  {journal} {ArXiv e-prints}\ } (\bibinfo {year} {2011})},\ \Eprint {http://arxiv.org/abs/1110.3193} {arXiv:1110.3193 [astro-ph.CO]} \BibitemShut {NoStop}%
\bibitem [{\citenamefont {{Spergel}}\ \emph {et~al.}(2015)\citenamefont {{Spergel}}, \citenamefont {{Gehrels}}, \citenamefont {{Baltay}}, \citenamefont {{Bennett}}, \citenamefont {{Breckinridge}}, \citenamefont {{Donahue}} \emph {et~al.}}]{WFIRST15}%
  \BibitemOpen
  \bibfield  {author} {\bibinfo {author} {\bibfnamefont {D.}~\bibnamefont {{Spergel}}}, \bibinfo {author} {\bibfnamefont {N.}~\bibnamefont {{Gehrels}}}, \bibinfo {author} {\bibfnamefont {C.}~\bibnamefont {{Baltay}}}, \bibinfo {author} {\bibfnamefont {D.}~\bibnamefont {{Bennett}}}, \bibinfo {author} {\bibfnamefont {J.}~\bibnamefont {{Breckinridge}}}, \bibinfo {author} {\bibfnamefont {M.}~\bibnamefont {{Donahue}}},  \emph {et~al.},\ }\href@noop {} {\bibfield  {journal} {\bibinfo  {journal} {ArXiv e-prints}\ } (\bibinfo {year} {2015})},\ \Eprint {http://arxiv.org/abs/1503.03757} {arXiv:1503.03757 [astro-ph.IM]} \BibitemShut {NoStop}%
\end{thebibliography}%

\appendix
\section{Biases in \texttt{ChainConsumer}}
\label{app:chainconsumer}

\begin{figure*}
\includegraphics[width=1.0\textwidth]{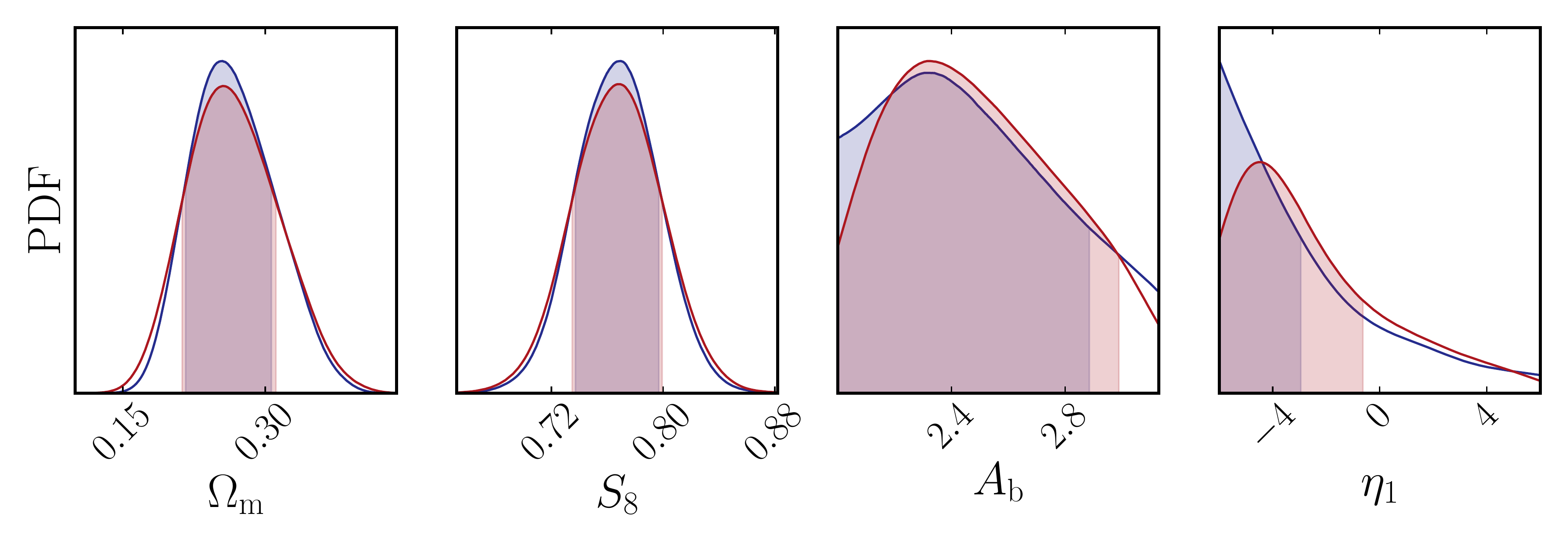}
\caption{
    This figure shows the normalized 1D marginalized posterior of our fiducial
    constraints on four parameters. The red lines are before the corrections
    for the boundary effect and multiplicative bias due to the smoothing  in
    \texttt{ChainConsumer}, and the blue lines are after the bias corrections.
    This illustrates how the posteriors for prior-dominated nuisance parameters
    are noticeably modified at the boundaries, whereas those for the
    cosmological parameters simply change by $\sim10$\% in width while
    remaining centered at the same parameter values.
    }
    \label{fig:chainconsumer_problem}
\end{figure*}

In this appendix, we show the biases caused by the boundary effect and the
smoothing of MC samples when analyzing the 1D marginalized posteriors using
\texttt{ChainConsumer}. Boundary bias arises near the boundaries of the
projected 1D sample. A traditional KDE assumes that the sample extends
infinitely, which is not true for real-world data. As a result, density
estimates close to the edges may be biased downwards since the kernel function
extends beyond the data range, effectively underrepresenting the true density.
\citep{GetDist2019} uses first-order boundary correction and a multiplicative
bias correction for higher-order bias caused by the KDE smoothing.

Fig.~\ref{fig:chainconsumer_problem} presents the marginalized 1D posteriors
from the fiducial analysis, both before and after bias corrections. It reveals
that for parameters inadequately constrained by cosmic shear data, the boundary
posteriors are underestimated due to the boundary effect. Furthermore, for
parameters less affected by the top-hat priors, the errors are overestimated
owing to the multiplicative bias introduced by KDE smoothing.

\section{2D posteriors for internal tests}
\label{app:inter}
\begin{figure*}
\includegraphics[width=1.0\textwidth]{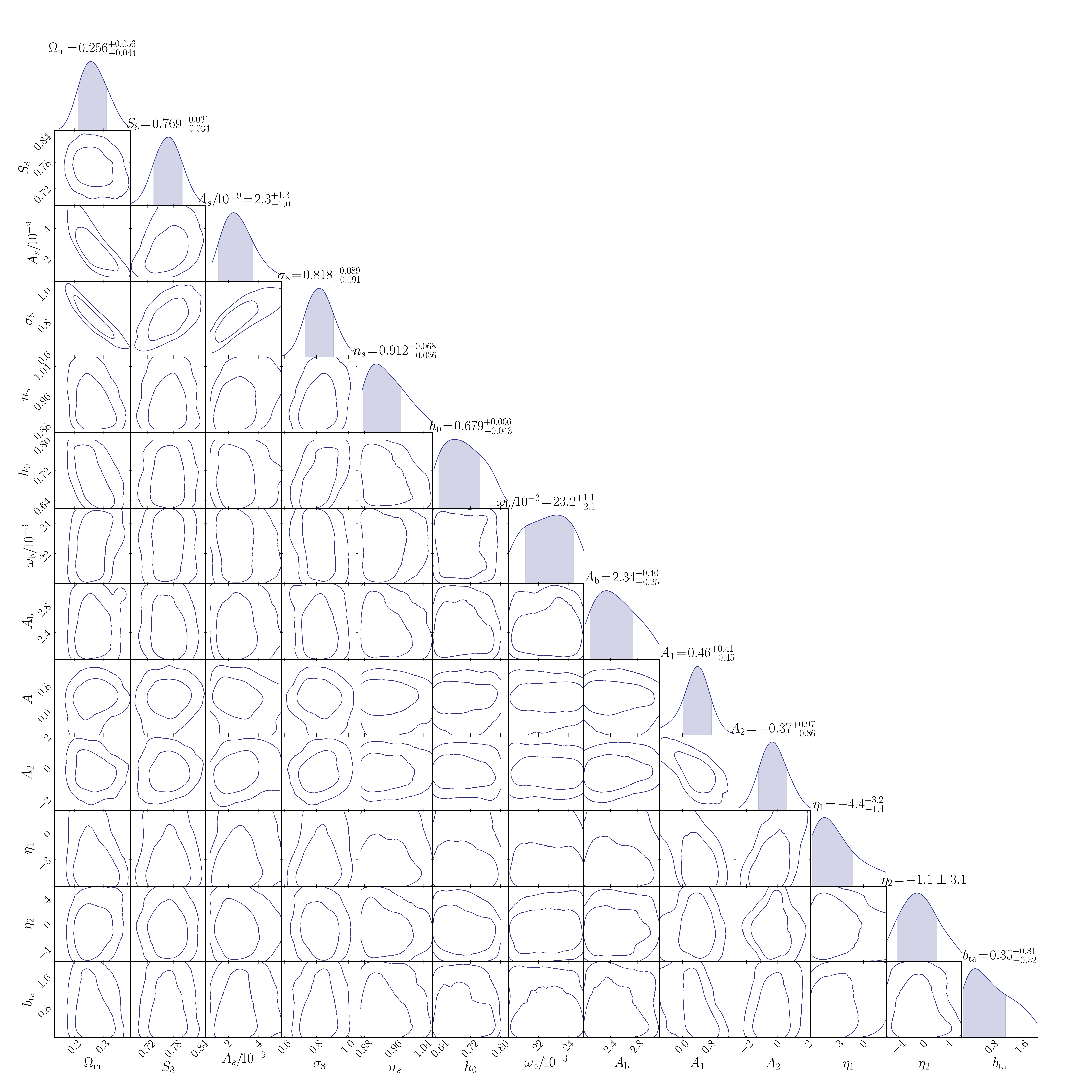}
\caption{
    Marginalized posteriors of the fiducial analysis for the various parameters
    in our analysis, including the cosmological ($\OmegaM$, $S_8$, $A_s$,
    $\sigma_8$, $n_s$, $h_0$, $\omega_\mathrm{b}$), the parameter from
    \hmcode{}~2016 encoding baryonic feedback ($A_\mathrm{b}$), and the TATT
    intrinsic alignment ($A_1$, $A_2$, $\eta_1$, $\eta_2$, $b_\mathrm{ta}$)
    parameters.
    }
    \label{fig:app_inter_all}
\end{figure*}

In this appendix, we show the 1D and 2D marginalized posteriors of our fiducial
constant and the constraints of our internal consistent tests. First, in
Fig.~\ref{fig:app_inter_all}, we show the corner plot for marginalized 2D
posteriors of cosmological parameters and astronomical parameters in our
fiducial analysis. As shown, only the matter density, the matter amplitude and
the amplitudes of intrinsic alignment parameters are well constrained by our
cosmic shear 2PCFs analysis. Then the posteriors for our internal consistent
tests are shown in the following subsections.

\subsection{Priors}
\label{app:inter_prior}

\begin{figure*}
\begin{minipage}{0.48\textwidth}
    \includegraphics[width=1.0\textwidth]{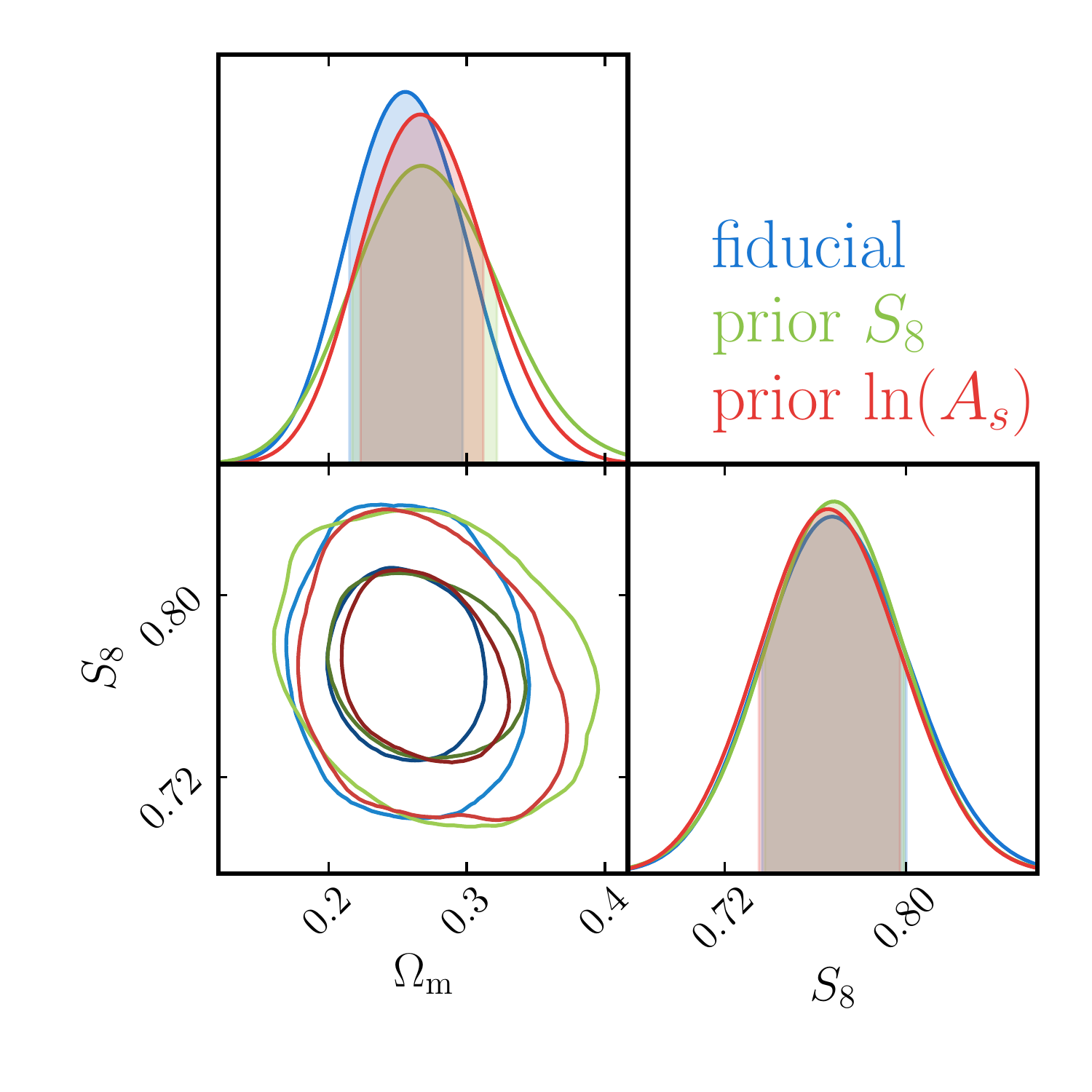}
    \caption{
        The marginalized 2D posteriors analyzed with different flat priors on
        $A_s$ (fiducial), $S_8$ and $\mathrm{ln}(A_s)$ sampled with
        \multinest{}.
    }
    \label{fig:app_prior_mul}
\end{minipage}
\qquad
\begin{minipage}{0.48\textwidth}
    \includegraphics[width=1.0\textwidth]{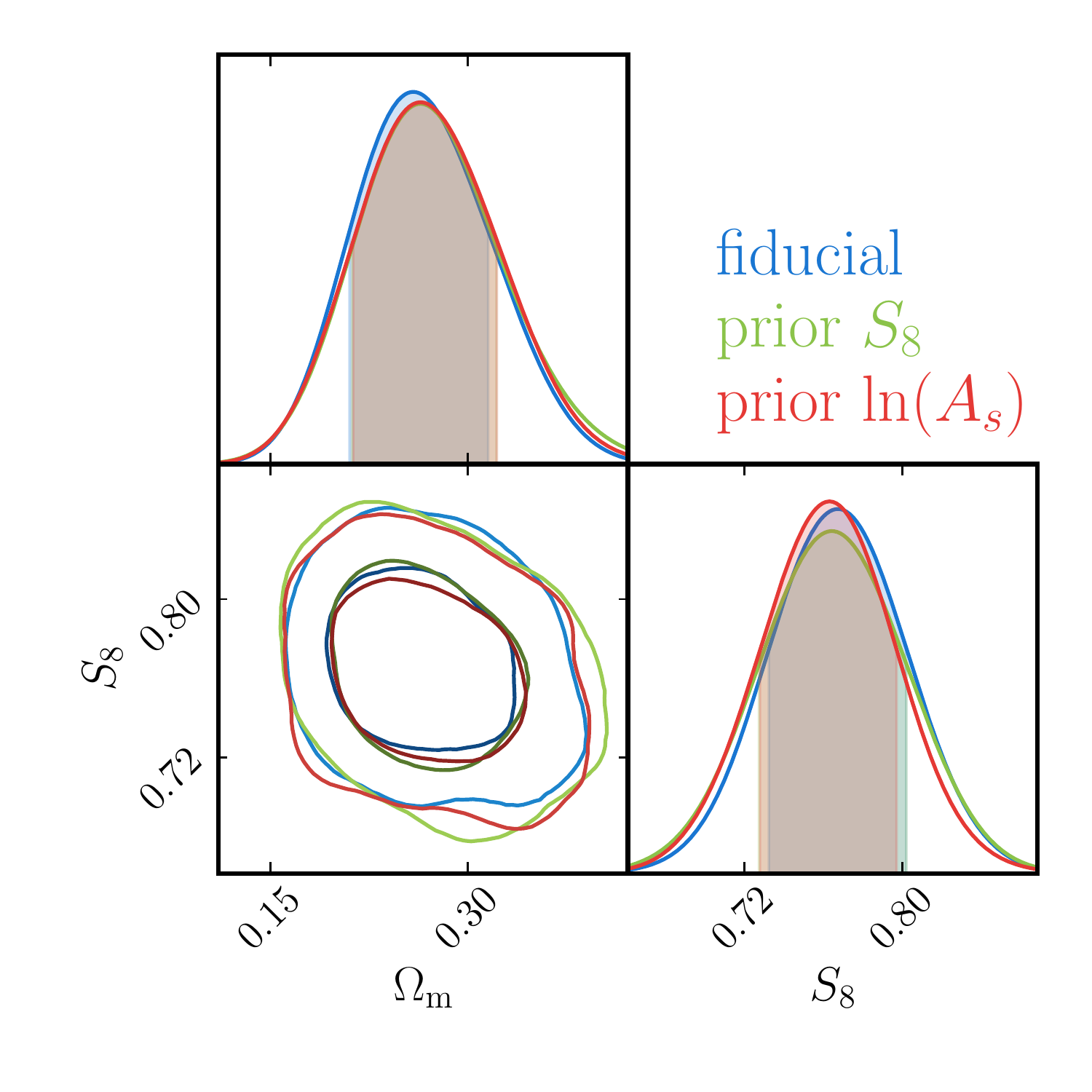}
    \caption{
        The marginalized 2D posteriors analyzed with different flat priors on
        $A_s$ (fiducial), $S_8$ and $\mathrm{ln}(A_s)$ sampled with
        \polychord{}.
    }
    \label{fig:app_prior_poly}
\end{minipage}
\end{figure*}

We show the marginalized 2D posteriors in the $(\OmegaM, S_8)$ plane for
analyses with different flat priors on $A_s$ $\mathrm{ln}(A_s)$ and $S_8$ in
Fig.~\ref{fig:app_prior_mul} and Fig.~\ref{fig:app_prior_poly}, which are
sampled with the \multinest{} and \polychord{} samplers, respectively. We refer
the readers to section~\ref{subsec:inter_prior} for a detailed discussion.

\subsection{Physical Models}
\label{app:inter_phys}

\begin{figure*}
\begin{minipage}{0.48\textwidth}
\includegraphics[width=1.0\textwidth]{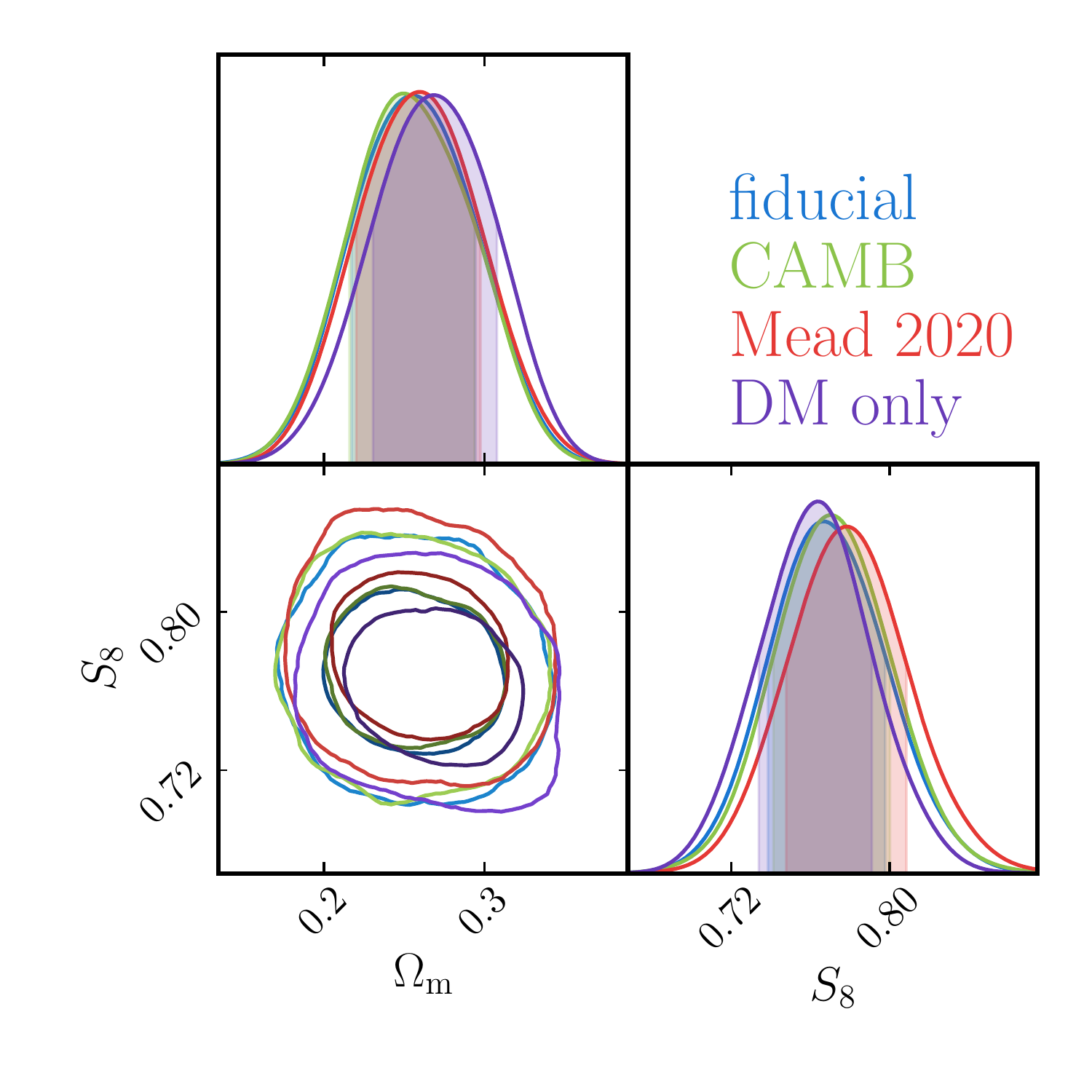}
\caption{
    The marginalized 2D posteriors analyzed with different models for cold
    matter power spectrum.
    }
    \label{fig:app_phys}
\end{minipage}
\qquad
\begin{minipage}{0.48\textwidth}
\includegraphics[width=1.0\textwidth]{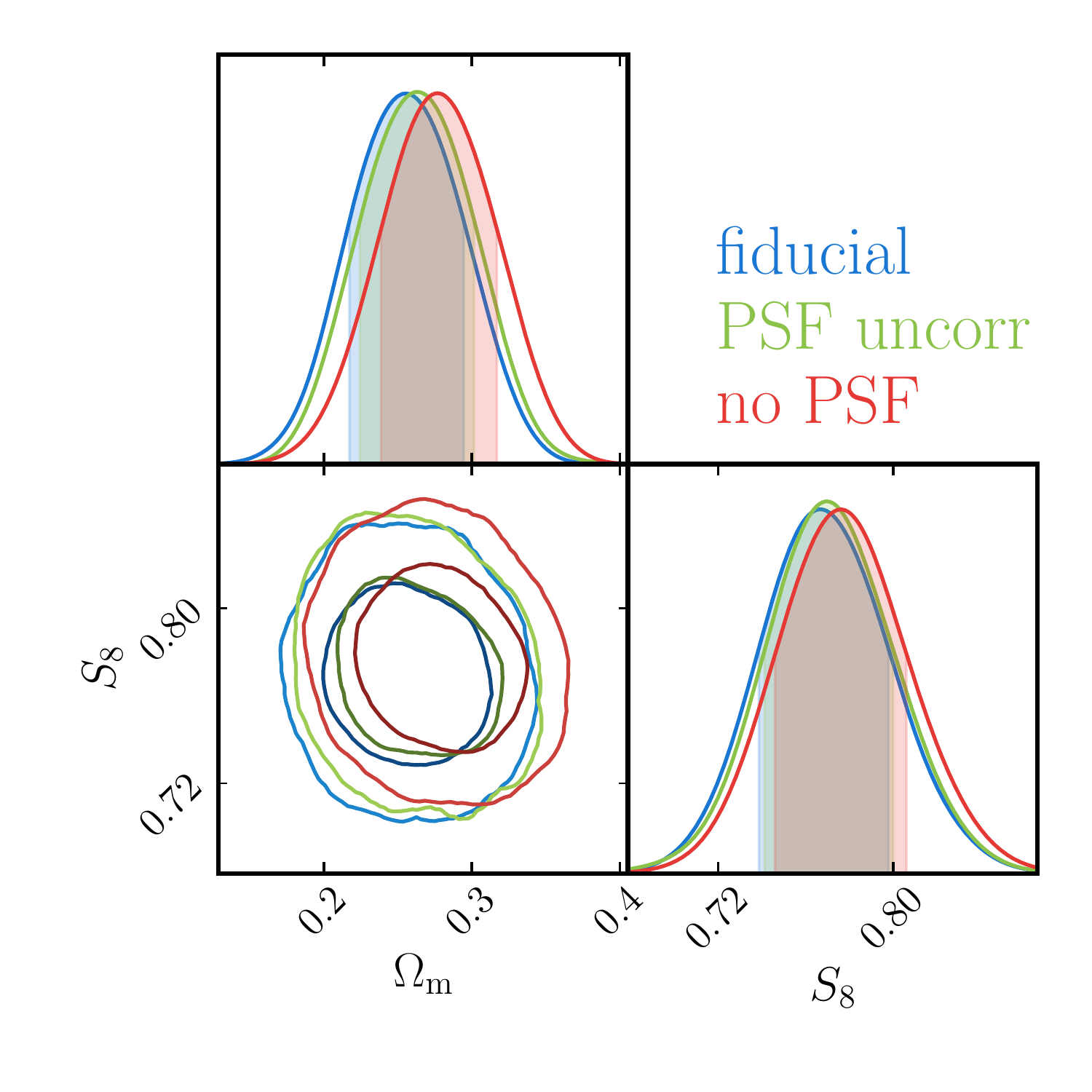}
\caption{
    The marginalized 2D posteriors analyzed with different systematic models.
    }
    \label{fig:app_sys}
\end{minipage}
\end{figure*}

We show the marginalized 2D posteriors in the $(\OmegaM, S_8)$ plane for
analyses with different physical models in Fig.~\ref{fig:app_phys}. The
posteriors are sampled with \multinest{}. We refer the readers to
section~\ref{subsec:inter_phys} for a detailed discussion.

\subsection{Systematic Models}
\label{app:inter_sys}

We show the marginalized 2D posteriors in the $(\OmegaM, S_8)$ plane for
analyses with different systematic models in Fig.~\ref{fig:app_sys}. The
posteriors are sampled with \multinest{}. We refer the readers to
section~\ref{subsec:inter_sys} for a detailed discussion.

\subsection{Subfields}
\label{app:inter_fields}

\begin{figure*}
\begin{minipage}{0.48\textwidth}
\includegraphics[width=1.0\textwidth]{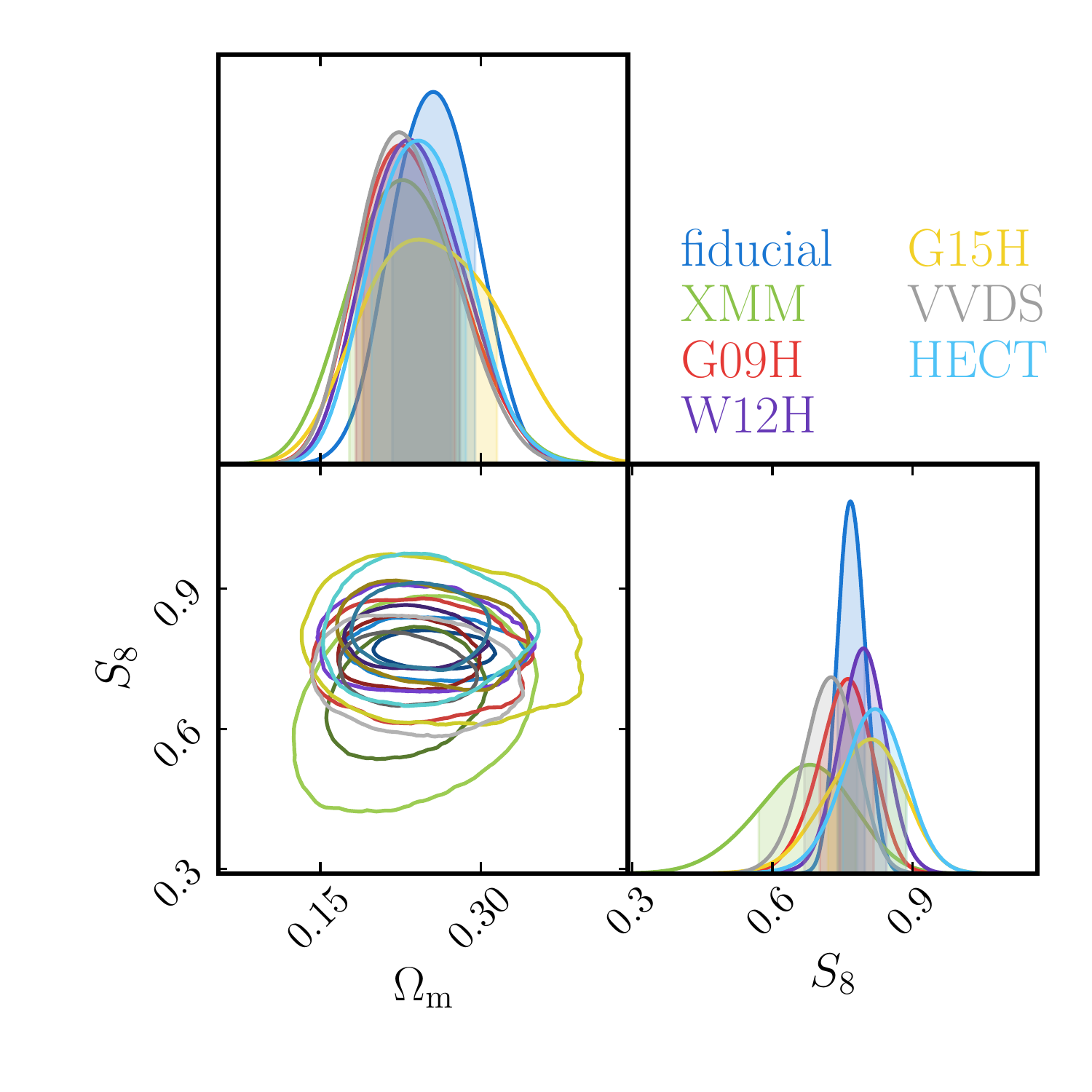}
\caption{
    The marginalized 2D  posteriors in the ($\OmegaM$, $\sigma_8$) plane for six
    different subfields.
    }
    \label{fig:app_fields}
\end{minipage}
\qquad
\begin{minipage}{0.48\textwidth}
\includegraphics[width=1.0\textwidth]{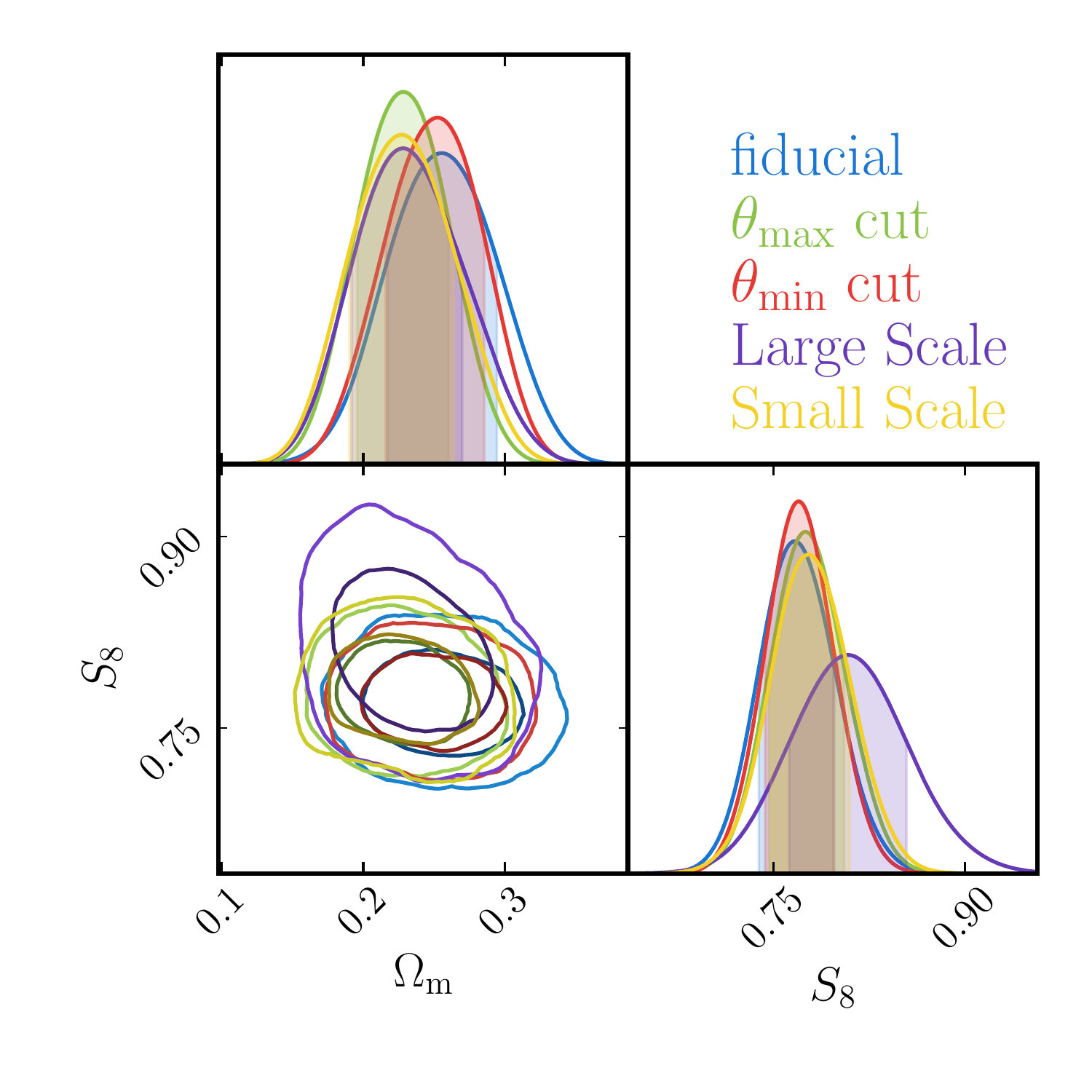}
\caption{
    The marginalized 2D posteriors in the ($\OmegaM$, $\sigma_8$) plane
    with six different cuts on angular scales.
    }
    \label{fig:app_scales}
\end{minipage}
\end{figure*}

We show the marginalized 2D posteriors in the $(\OmegaM, S_8)$ plane for analyses
on different HSC-Y3 subfields in Fig.~\ref{fig:app_fields}. The posteriors are
sampled with \multinest{}. We refer the readers to
section~\ref{subsec:inter_data} for a detailed discussion.

\subsection{Scales}
\label{app:inter_scales}

We show the marginalized 2D posteriors in the $(\OmegaM, S_8)$ plane for analyses
with different angular scale cuts in Fig.~\ref{fig:app_scales}. The posteriors
are sampled with \multinest{}. We refer the readers to
section~\ref{subsec:inter_data} for a detailed discussion.

\subsection{Tomographic Bins}
\label{app:inter_zbins}

\begin{figure*}
\begin{minipage}{0.48\textwidth}
\includegraphics[width=1.0\textwidth]{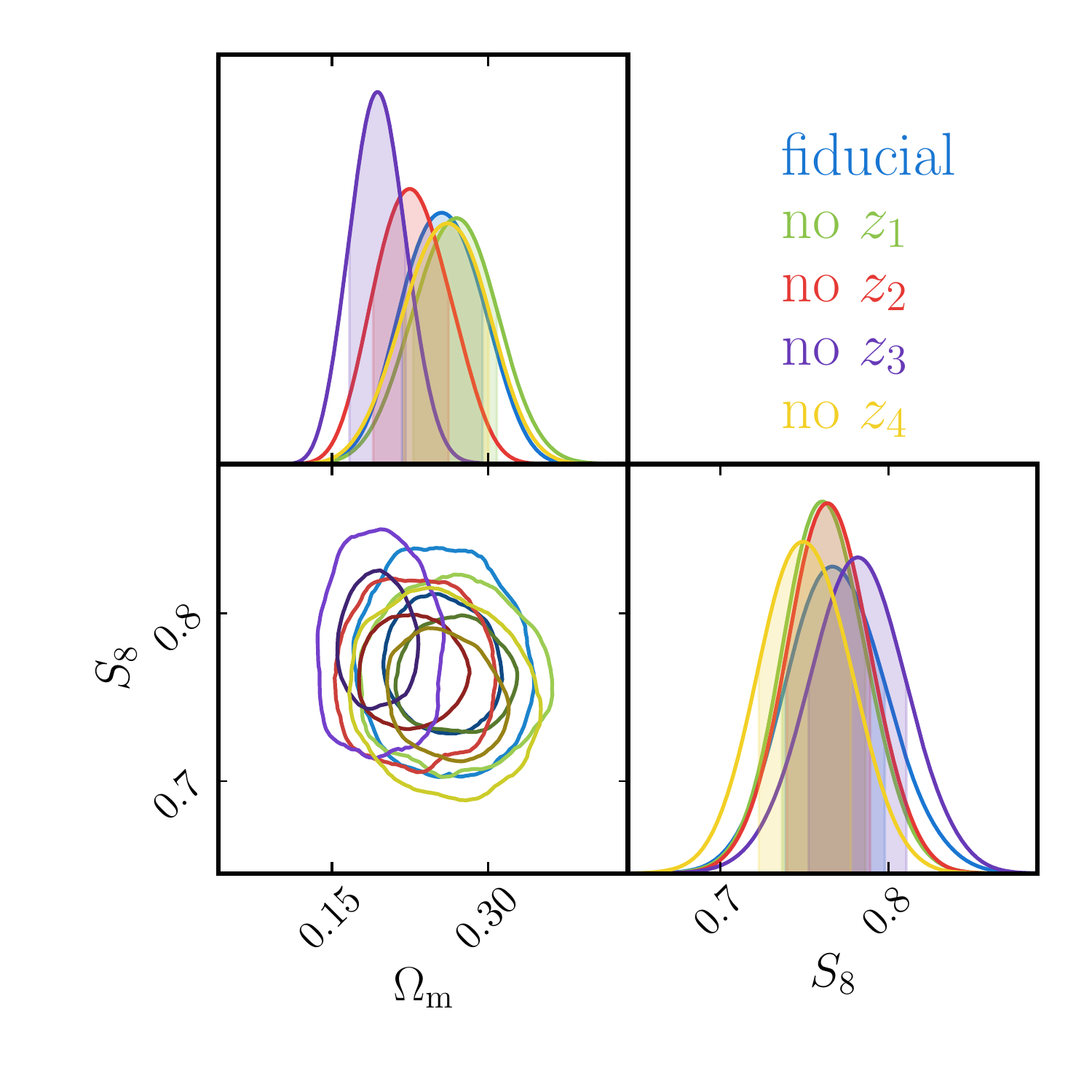}
\caption{
    The marginalized 2D posteriors in the ($\OmegaM$, $\sigma_8$) plane when
    removing one of the redshift bins.
    }
    \label{fig:app_zbins}
\end{minipage}
\qquad
\begin{minipage}{0.48\textwidth}
\includegraphics[width=1.0\textwidth]{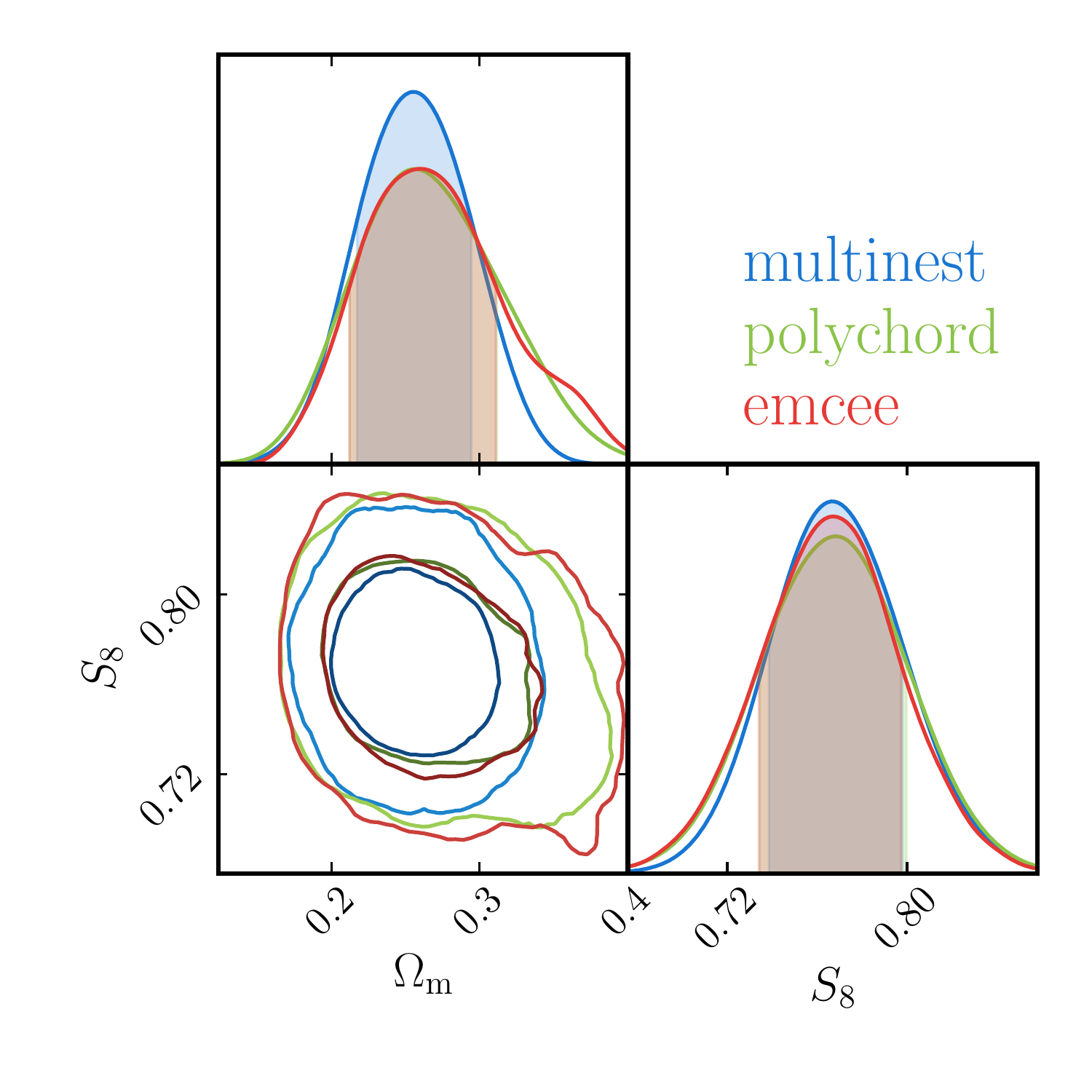}
\caption{
    The marginalized 2D posteriors in the ($\OmegaM$, $\sigma_8$) plane
    for different samplers: \multinest{}, \polychord{} and
    \emcee{}.
    }
    \label{fig:app_samplers}
\end{minipage}
\end{figure*}

We show the marginalized 2D posteriors in the $(\OmegaM, S_8)$ plane for analyses
with removals of one of the four tomographic bins in Fig.~\ref{fig:app_zbins}.
The posteriors are sampled with \multinest{}. We refer the readers to
section~\ref{subsec:inter_data} for a detailed discussion.

\subsection{Samplers}
\label{app:inter_samplers}

We show the marginalized 2D posteriors in the $(\OmegaM, S_8)$ plane for the
analyses with fiducial setup but sampled with different samplers in
Fig.~\ref{fig:app_samplers}. We refer the readers to
section~\ref{subsec:inter_sampler} for a detailed discussion.

\end{document}